\newif\ifabstract
\abstracttrue
\newif\iffull
\ifabstract \fullfalse \else \fulltrue \fi
\documentclass[11pt]{article}
\usepackage{amsfonts}
\usepackage{amssymb}
\usepackage{amstext}
\usepackage{amsmath}
\usepackage{xspace}
\usepackage{ntheorem}
\usepackage{graphicx}
\usepackage{url}
\usepackage{graphics}
\usepackage{colordvi}
\usepackage{hyperref}
\usepackage{subfigure}
\usepackage{xcolor}
\usepackage{cleveref}

\advance \oddsidemargin by -0.8in
\newcommand{\myparskip}{3pt}
\parskip \myparskip

\newcommand{\heap}{\mathsf{Heap}}
\newcommand{\pred}{\mathsf{pred}}

\newcommand{\procpathpeel}{\ensuremath{\mathsf{ProcPathPeel}}\xspace}
\newcommand{\supernodes}{\mbox{\sf{supernodes}}}
\newcommand{\Otilde}{\widetilde O}
\newcommand{\otilde}{\widetilde O}

\newcommand{\clusterlist}{\mbox{\sf{ClusterList}}}

\newcommand{\coveringcluster}{\mbox{\sf{CoveringCluster}}}
\newcommand{\specialcluster}{\mbox{\sf{SpecialCluster}}}
\newcommand{\clustercover}{\mbox{\sf{CoveringCluster}}}
\newcommand{\SSSP}{\mbox{\sf{SSSP}}\xspace}
\newcommand{\APSP}{\mbox{\sf{APSP}}\xspace}
\newcommand{\maintainemulator}{\mbox{\sf{MantainEmulator}}\xspace}

\newcommand{\delvertex}{\ensuremath{\operatorname{DeleteVertex}}\xspace}
\newcommand{\addsupernode}{\ensuremath{\operatorname{AddSuperNode}}\xspace}
\newcommand{\csplit}{\ensuremath{\operatorname{ClusterSplit}}\xspace}
\newcommand{\algtransformpath}{\ensuremath{\operatorname{AlgTransformPath}}\xspace}
\newcommand{\algemulator}{\ensuremath{\operatorname{AlgEmulator}}\xspace}
\newcommand{\transformpath}{\ensuremath{\mbox{\sf{TransformPath}}}\xspace}

\newcommand{\alg}{\ensuremath{\mbox{\sf{Alg}}}\xspace}

\newcommand{\anc}{\operatorname{Ancestor}}

\newcommand{\DS}{\mathsf{DS}}

\newcommand{\rep}{\operatorname{rep}}
\newcommand{\ceil}[1]{\ensuremath{\left\lceil#1\right\rceil}}
\newcommand{\floor}[1]{\ensuremath{\left\lfloor#1\right\rfloor}}

\newcommand{\CMG}{{\sf{Cut-Matching Game}}\xspace}
\newcommand{\DMG}{{\sf{Distanced Matching Game}}\xspace}
\newcommand{\set}[1]{\left\{ #1 \right\}}

\newcommand{\tset}{{\mathcal T}}
\newcommand{\iset}{{\mathcal{J}}}
\newcommand{\pset}{{\mathcal{P}}}
\newcommand{\qset}{{\mathcal{Q}}}
\newcommand{\lset}{{\mathcal{L}}}

\newcommand{\aset}{{\mathcal{A}}}
\newcommand{\cset}{{\mathcal{C}}}

\newcommand{\mset}{{\mathcal M}}
\newcommand{\jset}{{\mathcal{K}}}

\newcommand{\wset}{{\mathcal{W}}}

\newcommand{\rset}{{\mathcal{R}}}

\newcommand{\hset}{{\mathcal{H}}}

\newcommand{\sset}{{\mathcal{S}}}

\newcommand{\uset}{{\mathcal{U}}}
\newcommand{\dset}{{\mathcal{D}}}
\newcommand{\oset}{{\mathcal{O}}}

\newcommand{\be}{\begin{enumerate}}
\newcommand{\ee}{\end{enumerate}}
\newcommand{\bd}{\begin{description}}
\newcommand{\ed}{\end{description}}
\newcommand{\bi}{\begin{itemize}}
\newcommand{\ei}{\end{itemize}}

\newcommand{\HSS}{Hierarchical Support Structure\xspace}
\newcommand{\td}{\tilde d}

\newtheorem{theorem}{Theorem}[section]
\newtheorem{lemma}[theorem]{Lemma}
\newtheorem{observation}[theorem]{Observation}
\newtheorem{corollary}[theorem]{Corollary}
\newtheorem{claim}[theorem]{Claim}

\newtheorem{assumption}[theorem]{Assumption}
\newtheorem{definition}[theorem]{Definition}
\newenvironment{proof}{\par \smallskip{\bf Proof:}}{\hfill\stopproof}
\def\stopproof{\square}
\def\square{\vbox{\hrule height.2pt\hbox{\vrule width.2pt height5pt \kern5pt
\vrule width.2pt} \hrule height.2pt}}

\newenvironment{proofof}[1]{\noindent{\bf Proof of #1.}}%
        {\hfill\stopproof}

\newcommand{\scale}{\operatorname{scale}}

\newcommand{\DSFS}{\ensuremath{\mathsf{DS_{FlatSets}}}\xspace}
\newcommand{\TFS}{\ensuremath{\mathsf{T_{FS}}}}
\newcommand{\Tbasic}{\ensuremath{\mathsf{T_{basic}}}}
\newcommand{\DSbasic}{\ensuremath{\mathsf{DS_{Basic}}}\xspace}
\newcommand{\DSimp}{\ensuremath{\mathsf{DS_{Imp}}}\xspace}

\newenvironment{prog}[1]{
\begin{minipage}{5.8 in}
\begin{center}
{\sc #1}
\end{center}
}
{
\end{minipage}
}

\renewcommand{\phi}{\varphi}
\newcommand{\eps}{\epsilon}

\newcommand{\poly}{\operatorname{poly}}
\newcommand{\dist}{\mbox{\sf dist}}
\newcommand{\diam}{\mbox{\sf diam}}

\newenvironment{properties}[2][0]
{
\begin{enumerate} \setcounter{enumi}{#1}}{\end{enumerate}}

\newcommand{\mynote}[2][red]{\textcolor{red}{\sc\bf{[#2]}}}
\newcommand{\level}{\operatorname{Level}}
\newcommand{\attime}[1][\tau]{^{(#1)}}

\newcommand{\shortpath}{\mbox{\sf{short-path}}}
\newcommand{\shortestpathquery}{\mbox{\sf{shortest-path-query}}}

\newcommand{\spquery}{\mbox{\sf{short-path-query}}\xspace}
\newcommand{\shortestpath}{\mbox{\sf{SSSP-query}}\xspace}

\newcommand{\maintainNC}{\ensuremath{\mathsf{MaintainNC}}\xspace}
\newcommand{\maintaincluster}{\ensuremath{\mathsf{MaintainCluster}}\xspace}

\newcommand{\recdynnc}{\ensuremath{\mathsf{RecDynNC}}\xspace}
\newcommand{\recdynNC}{\ensuremath{\mathsf{RecDynNC}}\xspace}

\newcommand{\RecDynNC}{\ensuremath{\mathsf{RecDynNC}}\xspace}
\newcommand{\dquery}{\mbox{\sf{dist-query}}\xspace}
\newcommand{\pathquery}{\mbox{\sf{shortest-path-query}}\xspace}
\newcommand{\distquery}{\mbox{\sf{dist-query}}\xspace}

\newcommand{\CONNSF}{\mbox{\sf{CONN-SF}}}

\newcommand{\EST}{\mbox{\sf{ES-Tree}}\xspace}

\newcommand{\conn}{\mbox{\sf{conn}}}

\newcommand{\tH}{\tilde H}
\begin{document}

\begin{titlepage}
	
	\title{A New Deterministic Algorithm for Fully Dynamic All-Pairs Shortest Paths\footnote{to appear in STOC 2023}}
	
	\author{Julia Chuzhoy\thanks{Toyota Technological Institute at Chicago. Email: {\tt cjulia@ttic.edu}. Supported in part by NSF grants CCF-1616584 and CCF-2006464.}\and Ruimin Zhang\thanks{University of Chicago. Email: {\tt ruimin@uchicago.edu}.}}
	\maketitle
\pagenumbering{gobble}
	
	\thispagestyle{empty}

\begin{abstract}
	We study the fully dynamic All-Pairs Shortest Paths (APSP) problem in undirected edge-weighted graphs. Given an $n$-vertex graph $G$ with non-negative edge lengths, that undergoes an online sequence of edge insertions and deletions, the goal is to support approximate distance queries and shortest-path queries. We provide a deterministic algorithm for this problem, that, for a given precision parameter $\eps$, achieves approximation factor $(\log\log n)^{2^{O(1/\eps^3)}}$, and has amortized update time $O(n^{\eps}\log L)$ per operation, where $L$ is the ratio of longest to shortest edge length. Query time for distance-query is $O(2^{O(1/\eps)}\cdot \log n\cdot \log\log L)$, and query time for shortest-path query is $O(|E(P)|+2^{O(1/\eps)}\cdot \log n\cdot \log\log L)$, where $P$ is the path that the algorithm returns.
	To the best of our knowledge, even allowing any $o(n)$-approximation factor, no adaptive-update algorithms with better than $\Theta(m)$ amortized update time and better than $\Theta(n)$ query time were known prior to this work. We also note that our guarantees are stronger than the best current guarantees for APSP  in decremental graphs in the adaptive-adversary setting.
	
	In order to obtain these results, we consider an intermediate problem, called Recursive Dynamic Neighborhood Cover (RecDynNC), that was formally introduced in [Chuzhoy, STOC '21]. At a high level, given an undirected edge-weighted graph $G$ undergoing an online sequence of edge deletions, together with a distance parameter $D$, the goal is to maintain a sparse $D$-neighborhood cover of $G$, with some additional technical requirements. Our main technical contribution is twofolds. First, we provide a black-box reduction from  APSP in fully dynamic graphs to the RecDynNC problem. Second, we provide a new deterministic algorithm for the RecDynNC problem, that, for a given precision parameter $\eps$, achieves approximation factor $(\log\log m)^{2^{O(1/\eps^2)}}$, with total update time $O(m^{1+\eps})$, where $m$ is the total number of edges ever present in $G$. This improves the previous algorithm of [Chuzhoy, STOC '21], that achieved approximation factor $(\log m)^{2^{O(1/\eps)}}$ with similar total update time.
	Combining these two results immediately leads to the deterministic algorithm for fully-dynamic APSP with the guarantees stated above. 
\end{abstract}

\newpage

\tableofcontents{}
\end{titlepage}

\pagenumbering{arabic}

\section{Introduction}
\label{sec: intro}
We study the fully dynamic All-Pairs Shortest-Paths (\APSP) problem  in weighted undirected graphs. 
In this problem, the input is an undirected $n$-vertex graph $G$ with lengths $\ell(e)\geq 1$ on its edges, that undergoes an online sequence of edge insertions and deletions. The goal is to support (approximate) shortest-path queries $\shortestpathquery(x,y)$: given a pair $x,y$ of vertices of $G$, return a path connecting $x$ to $y$, whose length is within factor $\alpha$ of the length of the shortest $x$-$y$ path in $G$, where $\alpha$ is the \emph{approximation factor} of the algorithm. We also consider approximate distance queries, $\distquery(x,y)$: given a pair $x,y$ of vertices of $G$, return an estimate $\dist'(x,y)$ on the distance $\dist_G(x,y)$ between $x$ and $y$ in $G$, such that $\dist_G(x,y)\leq \dist'(x,y)\leq \alpha\cdot \dist_G(x,y)$. 
Throughout, we denote $|V(G)|=n$, and we denote by $m$ the total number of edges that are ever present in $G$; if an edge is deleted from $G$ and then inserted into $G$ multiple times, we count these as different edges. We also denote by $\Lambda$ the ratio of longest to shortest edge length.

\APSP is one of the most fundamental problems in graph algorithms, both in the dynamic and the static settings. Algorithms for this problem often serve as building blocks for designing algorithms for a range of other graph problems and beyond. Interestingly, algorithms for dynamic \APSP turned out to be extremely useful in the design of fast algorithms for classical cut, flow, and other graph problems in the static setting. 
Not surprisingly, this problem has been the subject of extensive study, from many different angles and in various regimes. 

A central goal in this area is to obtain algorithms with the strongest possible guarantees for the problem. Specifically, we would like the approximation factor $\alpha$ that the algorithm achieves to be low, and its total update time\footnote{In the context of fully dynamic algorithms, it is customary to focus on amortized update time per operation, which, in our case, is simply the total update time divided by $m$. We will use total update time and amortized update time per operation interchangeably, but we will try to clearly distinguish between them to avoid confusion.} -- the time required to maintain its data structures -- as close as possible to linear in $m$. In addition to the approximation factor and the total update time, another important parameter is  \emph{query time} -- the time it takes to process a single query. Ideally, we would like  the query time for $\distquery$ to be
$O(\poly\log(n\cdot \Lambda))$, and the query time for 
$\shortestpathquery$ to be close to $ O(|E(P)|)$, where $P$ is the path that the algorithm returns, which is  close to the best query time we can hope for.   Lastly, we distinguish between the \emph{oblivious-adversary} setting, where the sequence of updates to graph $G$ is constructed in advance and may not depend on the algorithm's behavior, and the \emph{adaptive-adversary} setting, where each update to graph $G$ may  depend arbitrarily on the algorithm's inner state and past behavior, such as responses to queries. While the oblivious-adversary setting appears significantly easier to handle algorithmically, many applications that rely on algorithms for dynamic \APSP require that the algorithm works in the adaptive-adversary setting. It is well known that deterministic algorithms always work against an adaptive adversary. Seeing that the \APSP problem itself is used as a building block in many different other setting,  designing a deterministic algorithm for the problem is especially desirable.

A straightforward algorithm for the fully-dynamic \APSP problem is the following: every time a query $\shortestpathquery(x,y)$ arrives, compute the shortest $x$-$y$ path in $G$ from scratch. This algorithm solves the problem exactly, but it has query time $\Theta(m)$. Another approach is to rely on \emph{spanners}. A spanner of a dynamic graph $G$ is another dynamic graph $H\subseteq G$, with $V(H)=V(G)$, such that the distances between the vertices of $G$ are approximately preserved in $H$; ideally a spanner $H$ should be very sparse. For example, a work of
\cite{BaswanaKS12} provides a randomized algorithm that maintains a spanner of a fully dynamic $n$-vertex graph $G$, that, for any parameter $k\leq O(\log n)$, achieves approximation factor $(2k-1)$, has expected amortized update time $O(k^2\log^2n)$ per update operation, and expected spanner size $O(kn^{1+1/k}\log n)$. Unfortunately, this algorithm only works against an oblivious adversary.
A recent work of
\cite{new-spanner} provides a randomized algorithm for maintaining a spanner of a fully dynamic $n$-vertex graph $G$ that can withstand an adaptive adversary. The algorithm achieves approximation factor $O(\poly\log n)$ and total update time $\Otilde(m)$, and it ensures that the number of edges in the spanner $H$ is always bounded by $O(n\poly\log n)$.  An algorithm for the \APSP problem can naturally build on such  constructions of spanners: given a query $\shortestpathquery(x,y)$ or $\distquery(x,y)$, we simply compute the shortest $x$-$y$ path in the spanner $H$. For example, the algorithm for graph spanners of \cite{new-spanner} implies a randomized $\poly\log n$-approximation algorithm for \APSP that has $O(m\poly\log n)$ total update time.
A recent work of \cite{newest-spanner} provides additional spanner-based algorithms for \APSP. 
Unfortunately, it seems inevitable that this straightforward spanner-based approach to \APSP  must have query time $\Omega(n)$ for both $\shortestpathquery$ and \distquery, and, with current state of the art algorithms, cannot lead to a better than logarithmic approximation. 

In this paper, our focus is on developing algorithms for the \APSP problem, whose query time is $ O(|E(P)|\cdot \poly\log(n\cdot \Lambda))$ for $\shortestpathquery$, where $P$ is the path that the query returns, and $O(\poly\log(n\cdot \Lambda))$ for $\distquery$. %While, in the worst case, it is possible that $P$ contains $\Theta(n)$ edges, this may not be the case typically. 
There are several reasons to strive for these faster query times. First, we typically want responses to the queries to be computed as fast as possible, and the above query times are close to the fastest possible. Second, ensuring that query time for $\shortestpathquery$ is bounded by $ O(|E(P)|\cdot \poly\log(n\cdot \Lambda))$ is often crucial to obtaining fast algorithms for other static graph problems, that use algorithms for \APSP as a subroutine.

%We distinguish between dynamic algorithms that work against an \emph{oblivious adversary}, where the input sequence of edge deletions is fixed in advance and may not depend on the algorithm's past behavior, and algorithms that work against an \emph{adaptive adversary}, where the input update sequence may depend on the algorithm's past responses and inner states arbitrarily. We refer to the former as \emph{oblivious-update} and to the latter as \emph{adaptive-update} algorithms. We note that any deterministic algorithm for the \APSP problem is an adaptive-update algorithm by definition.

As mentioned already, there are several parameters of interest that we would like to optimize in algorithms for \APSP: namely, query time, total update time, and the approximation factor. Additionally, we would like the algorithm to withstand an adaptive adversary, and ideally to be deterministic. There is a huge body of work that studies the \APSP problem, in both the dynamic and the static settings, that tries to optimize or achieve various tradeoffs among these different parameters. Some of this work also only focuses on supporting $\distquery$ queries, and not $\shortestpathquery$. We do not attempt to survey all of this work here, partially because this seems impossible, and partially because it may lead to confusion due to the large number of different settings considered. Instead, we will restrict our attention to the adaptive-adversary setting, where the query time for $\shortestpathquery$ is  $ O(|E(P)|\cdot \poly\log(n\cdot \Lambda))$, where $P$ is the returned path, and query time for  $\distquery(x,y)$ is $O(\poly\log(n\cdot \Lambda))$. We will try to survey the most relevant results for this setting, in order to put our results in context with previous work. We will also include some results for \APSP in \emph{decremental} graphs, where only edge-deletion updates are allowed.

{\bf Low-approximation regime.}
One major direction of study is to obtain algorithms for \APSP whose approximation factor  is very close to $1$. 
The classical data structure of Even and Shiloach \cite{EvenS,Dinitz,HenzingerKing}, that we refer to as \EST throughout the paper, implies an exact deterministic
algorithm for decremental unweighted $\APSP$ with $O(mn^2)$ total update time, and the desired $O(|E(P)|)$ query time for $\shortestpathquery$, where $P$ is the returned path. 
Short of obtaining an exact algorithm for \APSP, the best possible approximation factor one may hope for is $(1+\eps)$, for any $\eps$. A long line of work is dedicated to this direction in the decremental setting \cite{BaswanaHS07,rodittyZ2,henzinger16,bernstein16} 
and in the fully dynamic setting \cite{demetrescu2004new,thorup2004fully,brand2021fast}. In the decremental setting, the fastest algorithms in this line of work, due to \cite{henzinger16} and \cite{bernstein16}, achieve total update time $\tilde{O}(mn/\epsilon)$; the former algorithm is deterministic but only works in unweighted undirected graphs, while the latter algorithm works in directed weighted graphs, with an overhead of $\log \Lambda$ in the total update time, but can only handle an oblivious adversary. In the fully-dynamic setting, all algorithms cited above have amortized update time per operation at least $\Omega(n^2)$. 
A very recent result of \cite{bernstein2022deterministic} obtained a $(2+\eps)$-approximation for fully-dynamic \APSP, with amortized update time $O(m^{1+o(1)})$ per operation.
The high running times of the above mentioned algorithms are perhaps not surprising in view of strong lower bounds that are known for the low-approximation setting.

{\bf Lower Bounds.} 
A number of lower bounds are known for dynamic \APSP with low approximation factor. For example, Dor, Halperin and Zwick \cite{DorHZ00}, and Roddity and Zwick \cite{RodittyZ11} showed that, assuming the Boolean Matrix Multiplication (BMM) conjecture\footnote{The conjecture states that
	there is no ``combinatorial'' algorithm for multiplying two Boolean matrices of size $n\times n$ in time $n^{3-\delta}$ for any constant $\delta>0$.}, for any $\alpha,\beta\geq 1$ with $2\alpha+\beta<4$, no combinatorial algorithm for \APSP achieves a multiplicative $\alpha$ and additive $\beta$ approximation, with total update time
$O(n^{3-\delta})$ and query time $O(n^{1-\delta})$ for \distquery, for any constant $0<\delta<1$. This result was generalized by \cite{HenzingerKNS15}, who showed the same lower bounds for all algorithms and not just combinatorial ones, assuming the Online Boolean Matrix-Vector Multiplication (OMV) conjecture\footnote{The conjecture assumes that there is no $n^{3-\delta}$-time algorithm, for any constant $0<\delta<1$, for the OMV problem, in which the input is a Bollean $(n\times n)$ matrix, with $n$ Boolean dimension-$n$ vectors $v_1,\ldots,v_n$ arriving online. The algorithm needs to output $Mv_i$ immediately after $v_i$ arrives.}. 
The work of Vassilevska Williams and Williams \cite{williams2018subcubic}, combined with the work of  Roddity and Zwick \cite{RodittyZ11}, implies that obtaining such an algorithm would lead to subcubic-time algorithms for a number of important static problems on graphs and matrices.
A very recent result of \cite{abboud2022hardness} provides new lower bounds for the dynamic \APSP problem, in the regime where only \distquery queries need to be supported, under either the 3-SUM conjecture or the APSP conjecture. Let $k\geq 4$ be an integer, let $\eps,\delta>0$ be parameters, and let $c=\frac{4}{3-\omega}$ and $d=\frac{2\omega-2}{3-\omega}$, where $\omega$ is the exponent of matrix multiplication. Then \cite{abboud2022hardness}  show that, assuming either the 3-SUM Conjecture or the APSP Conjecture, there is no $(k-\delta)$-approximation algorithm for decremental \APSP with total update time  $O(m^{1+\frac{1}{ck-d}-\eps})$ and query time for \distquery bounded by $O(m^{\frac{1}{ck-d}-\eps})$. They also show that there is no $(k-\delta)$-approximation algorithm for fully dynamic \APSP that has   $O(n^3)$ preprocessing time, and then supports (fully dynamic) updates and \distquery queries in $O(m^{\frac{1}{ck-d}-\eps})$ time.
Due to these lower bounds, it is natural to focus on somewhat higher approximation factors.

{\bf Higher approximation factor.}
In the regime of higher approximation factors, a long line of work \cite{BernsteinR11,henzinger16,abraham2014fully,HenzingerKN14_focs} focused on the decremental setting with an oblivious adversary. This direction recently culminated with an algorithm of Chechik~\cite{chechik}, that, for any integer $k\ge1$ and parameter $0<\eps<1$, obtains a
$((2+\epsilon)k-1)$-approximation, with total update time $O(mn^{1/k+o(1)}\cdot \log  \Lambda)$, when the input graph is  weighted and undirected. This result is near-optimal, as all its parameters almost match
the best static algorithm of \cite{TZ}. This result was recently slightly improved by \cite{lkacki2020near}, who obtain total update time $O(mn^{1/k}\cdot \log \Lambda)$, and improve query time for $\dquery$.

The best currently known results for the fully dynamic setting with an oblivious adversary are significantly weaker. 
%
%Several algorithms are also known for fully dynamic \APSP in the oblivious setting, that only support approximate \distquery.
For unweighted graphs, the algorithm of \cite{ForsterG19} achieves approximation factor $n^{o(1)}$, with amortized update time $n^{1/2+o(1)}$ per operation  on unweighted graphs, while
the algorithm of \cite{abraham2014fully} achieves a constant approximation factor with expected $o(m)$ amortized update time per operation. In fact the latter paper provides a more general tradeoff between the approximation factor and update time, but in all regimes the expected amortized update time is at least $\Theta(\sqrt m)$ per operation. Lastly, the 
 the algorithm of \cite{ForsterGH20}, based on low-stretch trees, achieves $O(m^{\eps})$ update time per operation, with a factor  $(\log n)^{O(1/\eps)}$-approximation in weighted graphs. 
 We note that a very recent independent work \cite{forster2023bootstrapping} provides an algorithm for distance oracles in the fully dynamic setting with an oblivious adversary, whose approximation factor is $(1/\eps)^{O(1/\eps)}$, amortized update time is $\tilde O(n^{\eps})$, and query time is $\tilde O(n^{\eps/8})$ for \distquery.
 All of the above mentioned algorithms for fully-dynamic \APSP with oblivious adversary only support \distquery. We are not aware of algorithms that can additionally support \shortestpathquery.

In contrast, progress in the adaptive-update setting has been much slower. Until very recently, the fastest algorithm for decremental unweighted graphs 
\cite{henzinger16,GutenbergW20} only achieved an $\tilde{O}(mn/\epsilon)$ total update time (for approximation factor $(1+\eps)$), and the work of 
\cite{APSP-old}, for any parameter $1\le k\le o(\log^{1/8}n)$, achieved a multiplicative $3\cdot 2^{k}$ and additive $2^{(O(k\log^{3/4}n)}$ approximation, with query time $O(|E(P)|\cdot n^{o(1)})$ for $\shortestpathquery$, and total update time $n^{2.5+2/k+o(1)}$. Until very recently, the fastest  adaptive-update algorithms for {\bf weighted} graphs had total update time $O\left(\frac{n^{3}\log \Lambda}{\eps}\right )$ and approximation factor $(1+\eps)$ (see \cite{reliable-hubs}), even in the {\bf decremental} setting. 

To summarize, to the best of our knowledge, until very recently, even if we allowed an $o(n)$-approximation factor,
no adaptive-update algorithms with better than $\Theta(n^{3})$ total
update time and better than $\Theta(n)$ query time for $\shortestpathquery$ and $\distquery$ were known for weighted undirected graphs, 
and no adaptive-update algorithms with better than $\Theta(n^{2.5})$ total update time  and better than $\Theta(n)$ query time were known for unweighted undirected graphs, even in the {\bf decremental} setting.

Two very recent results\footnote{To the best of our knowledge, the two results are independent.} provided significantly stronger algorithms for decremental \APSP in weighted graphs: \cite{APSP-previous} designed a deterministic algorithm, that, for any $\Omega(1/\log\log m)<\eps<1$, achieves approximation factor 
$(\log m)^{2^{O(1/\eps)}}$, and has total update time $O\left (m^{1+O(\eps)}\cdot (\log m)^{O(1/\eps^2)}\cdot \log \Lambda\right )$. The query time is  $O(\log m \log\log \Lambda)$ for $\distquery$, and   $O(|E(P)|+\log m \log\log \Lambda)$ for $\shortestpathquery$, where $P$ is the returned path.
The main focus of \cite{bernstein2022deterministic} was mostly on a special case of \APSP called Single Source Shortest Paths (\SSSP), but they also obtained a deterministic algorithm for decremental \APSP with approximation factor $m^{o(1)}$ and total update time $O(m^{1+o(1)})$; unfortunately, the tradeoff between the approximation factor and the total update time is not stated explicitly, though they mention that the approximation factor is super-logarithmic. As mentioned already, they also obtain new results in the low-approximation regime for the fully dynamic setting of \APSP: a $(2+\eps)$-approximation with amortized update time $O(m^{1+o(1)})$ per operation.

In this paper we improve the results of \cite{APSP-previous} in two ways. First, we extend the algorithm to the fully-dynamic setting, and second, we improve the approximation factor to $(\log\log m)^{2^{O(1/\eps^3)}}$.
Altogether, we obtain a deterministic algorithm for fully-dynamic \APSP, that, given  a precision parameter $\frac{2}{(\log n)^{1/200}}<\eps<1/400$, achieves approximation factor $\alpha=(\log\log n)^{2^{O(1/\eps^2)}}$, and has amortized update time $O\left (n^{O(\eps)}\cdot \log \Lambda\right )$ per operation (if starting from an empty graph). Query time for \distquery is $O\left (2^{O(1/\eps)}\cdot \log n\cdot \log \log \Lambda\right )$, and query time for $\shortestpathquery$ is:  

$$O\left (|E(P)|+2^{O(1/\eps)}\cdot \log n\cdot \log\log \Lambda\right ),$$ where $P$ is the path that the algorithm returns (note that, if we choose $\eps\geq 1/\log\log n$, then query time for $\distquery$ becomes $O(\poly\log(n\cdot \Lambda)$, and query time for $\shortestpathquery$ becomes $ O(|E(P)|+\poly\log(n\cdot \Lambda)$).
An important intermediate problem that we study is Sparse Neighborhood Cover, and its generalization called Recursive Dynamic Neighborhood Cover (\recdynnc) that we discuss next.

{\bf Sparse Neighborhood Cover and \recdynnc problem.} 
Given a graph $G$ with lengths on edges, a vertex $v\in V(G)$, and a distance parameter $D$, we denote by $B_G(v,D)$ the \emph{ball of radius $D$ around $v$}, that is, the set of all vertices $u$ with $\dist_G(v,u)\leq D$.
Suppose we are given a static graph $G$ with non-negative edge lengths, a distance parameter $D$, and a desired approximation factor $\alpha$. A $(D,\alpha\cdot D)$-\emph{neighborhood cover} for $G$ is a collection $\cset$ of vertex-induced subgraphs of $G$ (that we call \emph{clusters}), such that, for every vertex $v\in V(G)$, there is some cluster $C\in \cset$ with $B_G(v,D)\subseteq V(C)$. Additionally, we require that, for every cluster $C\in \cset$, for every pair $x,y\in V(C)$ of its vertices, $\dist_G(x,y)\leq \alpha\cdot D$; if this property holds, then we say that $\cset$ is a \emph{weak} $(D,\alpha\cdot D)$-neighborhood cover of $G$. If, additionally, the diameter of every cluster $C\in \cset$ is bounded by $\alpha\cdot D$, then we say that  $\cset$ is a \emph{strong} $(D,\alpha\cdot D)$-neighborhood cover of $G$. Ideally, it is also desirable that the neighborhood cover is \emph{sparse}, that is, every edge (or every vertex) of $G$ only lies in a small number of clusters of $\cset$. For this static setting of the problem, the work of \cite{neighborhood-cover2,neighborhood-cover1} provides a deterministic algorithm that produces a strong $(D,O(D\log n))$-neighborhood cover of graph $G$, where every edge lies in at most $O(\log n)$ clusters, with running time $\Otilde(|E(G)|+|V(G)|)$.

In \cite{APSP-previous} a new problem, called Recursive Dynamic Neighborhood Cover (\recdynnc) was introduced. The problem can be viewed as an adaptation of Sparse Neighborhood Covers to the dynamic (decremental) setting, but with additional constraints that make it easy to use as a building block in other dynamic algorithms.
The input to this problem is a bipartite graph $H=(V,U,E)$, with non-negative lengths $\ell(e)$ on edges $e\in E$, and a distance parameter $D$. Vertices in set $V$ are called \emph{regular vertices}, while vertices in set $U$ are called \emph{supernodes}. Graph $H$ undergoes an online sequence $\Sigma$ of updates, each of which must be of one of the following three kinds: (i) edge deletion; or (ii) isolated vertex deletion; or (iii) supernode splitting. In the latter kind of update, we are given a supernode $u\in U$, and a collection $E'\subseteq \delta_H(u)$ of its incident edges. We need to insert a new supernode $u'$ into $H$, and, for every edge $e=(u,v)\in E'$, insert an edge $(u',v)$ into $H$. We note that, while, in general, graph $H$ is decremental, the supernode-splitting update allows us to insert edges into it, in a limited fashion. For conciseness, we will refer to an input $\iset=\left(H=(V,U,E),\set{\ell(e)}_{e\in E},D \right )$  as described above, as \emph{valid input structure}, and to edge-deletion, isolated vertex-deletion, and supernode-splitting updates as \emph{valid update operations}. Since edges may be inserted into graph $H$ via supernode-splitting updates, in order to control the size of the resulting graph, another parameter called \emph{dynamic degree bound} is used. We say that the dynamic degree bound of valid input structure $\iset$ that undergoes a sequence $\Sigma$ of valid update operations is $\mu$ if, for every regular vertex $v$, the total number of edges that are ever present in $H$ and are incident to $v$, is bounded by $\mu$.

The goal in the \recdynnc problem is to maintain a weak $(D,\alpha\cdot D)$-neighborhood cover $\cset$ of the graph $H$. However, we require that the clusters in $\cset$ are only updated in a specific fashion: once an initial neighborhood cover $\cset$ of $H$  is computed, we can only update clusters via \emph{allowed changes}: for each cluster $C$, we can delete edges or vertices from $C$, and, additionally, if some supernode $u\in V(C)$ just underwent a supernode-splitting update, we can insert the resulting new supernode $u'$ and all edges connecting it to other vertices of $C$, into cluster $C$. A new cluster $C'$ may only be added to $\cset$, if there is a cluster $C\in \cset$ with $C'\subseteq C$. In this case, we say that cluster $C$ underwent a \emph{cluster-splitting} update.  The algorithm must also maintain, for every regular vertex $v$ of $H$, a cluster $C=\coveringcluster(v)\in\cset$, with $B_H(v,D)\subseteq V(C)$. Additionally, we require that the neighborhood cover is \emph{sparse}, namely, for every regular vertex $v$ of $H$, the total number of clusters of $\cset$ to which $v$ may ever belong over the course of the algorithm is small. 
Lastly, we require that the algorithm supports  queries $\spquery(C,v,v')$: given two  vertices $v,v'\in V$, and a cluster $C\in \cset$ with $v,v'\in C$, return a path $P$ in the current graph $H$, of length at most $\alpha\cdot D$ connecting $v$ to $v'$ in $G$, in time $O(|E(P)|)$, where $\alpha$ is the approximation factor of the algorithm.

Given any edge-weighted decremental graph $G$ and a distance bound $D$, it is easy to transform $G$ into a valid input structure: we simply view the vertices of $G$ as supernodes, and we subdivide its edges with new vertices, that become regular vertices in the resulting bipartite graph $H$. An algorithm for solving the \recdynnc problem on the resulting valid input structure $\iset$ (that only undergoes edge-deletion updates) then naturally allows us to maintain a sparse neighborhood cover in the original graph $G$. However, the specific definition of the \recdynnc problem makes it more versatile, and more specifically, we can naturally  compose instances of the problem recursively with one another.

A typical way to exploit this composability property is the following. Suppose we solve the \recdynnc problem on a bipartite graph $H$, with some distance bound $D$. Let $\cset$ be the collection of clusters that the resulting algorithm maintains. Assume now that we would like to solve the same problem on graph $H$, with a larger distance bound $D'>D$. We can then construct another graph $H'$, whose set of regular vertices is the same as that in $H$, and the set of supernodes is $\set{u(C)\mid C\in \cset}$. We add an edge $(v,u(C))$ to the graph if and only if regular vertex $v$ lies in cluster $C\in \cset$, and we set the lengths of the resulting edges to be $D$. As the clusters in $\cset$ evolve, we can maintain graph $H'$ via valid update operations: when some cluster $C\in \cset$ undergoes cluster-splitting, and a new cluster $C'\subseteq C$ is created, we can apply supernode-splitting to supernode $u(C)$ in order to update graph $H'$ accordingly. It is not hard to verify that the resulting graph $H'$ is an emulator for $H$, with respect to distances that are greater than $D$. We can then scale all edge lengths down by factor $D$, and solve the problem on graph $H'$, with a new, significantly smaller, distance parameter $D'/D$. If neighborhood cover $\cset$ is sparse, and every regular vertex of $H$ ever belongs to at most $\Delta$ clusters of $\cset$, then the dynamic degree bound for graph $H'$ is bounded by $\Delta$, so graph $H'$ itself is sparse.

We note that, while the \recdynnc problem was first formally defined in \cite{APSP-previous}, the idea of using clustering of a dynamic graph $G$ in order to construct an emulator was exploited before numerous times (see e.g. the constructions of \cite{ForsterG19,chechikLowStretch,ForsterGH20} of dynamic low-stretch spanning trees). In several of these works, a family of clusters of a dynamic graph $G$ is constructed and maintained, and the restrictions on the allowed updates to the cluster family are similar to the ones that we impose; it is also observed in several of these works that with such restrictions one can naturally compose the resulting emulators recursively -- an approach that we follow here as well.
While neither of these algorithms provide neighborhood covers (as can be observed from the fact that one can view the sets of clusters that are maintained for each distance scale as disjoint, something that cannot be achieved in neighborhood covers), a connection between low-diameter decompositions (that often serve as the basis of low-stretch spanning trees) and neighborhood covers has been noticed in prior work. For example, \cite{miller2015improved}, provide a construction of neighborhood covers from low-diameter decompositions.
Additionally, all the above-mentioned algorithms are randomized and assume an oblivious adversary. On the other hand,  \cite{henzinger16,GutenbergW20} implicitly provide a deterministic algorithm for maintaining a neighborhood cover of a dynamic graph. However, these algorithms have a number of drawbacks: first, the running time for maintaining the neighborhood cover is too prohibitive (the total update time is $O(mn)$). Second, the neighborhood cover maintained is not necessarily sparse; in fact a vertex may lie in a very large number of resulting clusters. Lastly, clusters that join the neighborhood cover as the algorithm progresses may be arbitrary. The restriction that, for every cluster $C$ added to the neighborhood cover $\cset$, there must be a cluster $C'$ containing $C$ that already belongs to $\cset$, seems crucial in order to allow an easy recursive composition of emulators obtained from the neighborhood covers, and the requirement that the neighborhood cover is sparse is essential for bounding the sizes of the graphs that arise as the result of such recursive compositions. We also note that
a similar approach of recursive composition of emulators was used in numerous algorithms for \APSP  (see, e.g. \cite{chechik}), and a similar approach to handling cluster-splitting in an emulator that is based on clustering was used before in numerous works, including, e.g., \cite{BernsteinChechik,Bernstein,fast-vertex-sparsest,chechikLowStretch}.

It is not hard to verify that an algorithm for the \recdynnc problem immediately implies an algorithm for {\bf decremental} \APSP with the same approximation factor, and the same total update time (to within $O(\log \Lambda)$-factor). In \cite{APSP-previous}, a deterministic algorithm for the \recdynnc problem  was provided, with approximation factor $\alpha=O\left ((\log m)^{2^{O(1/\eps)}}\right )$, and total update time: $$O\left (m^{1+O(\eps)}\cdot (\log m)^{O(1/\eps^2)}\right ).$$ The algorithm ensured that, for every regular vertex $v\in V(H)$, the total number of clusters of $ \cset$ that $v$ ever belongs to is bounded by $m^{O(1/\log\log m)}$. 

In this work, we improve the results of \cite{APSP-previous} in two ways. First, we provide a black-box reduction from fully dynamic \APSP to  the $\recdynnc$ problem. Second, we provide an improved algorithm for the \recdynnc problem. The algorithm, given a valid input structure  $\iset=\left(H,\set{\ell(e)}_{e\in E(H)},D \right )$ undergoing a sequence of valid update operations, with dynamic degree bound $\mu$, together with parameters $\hat W$ and  $1/(\log \hat W)^{1/100}\leq \eps<1/400$,  such that, if we denote by $N$ the number of regular vertices in $H$ at the beginning of the algorithm, then $N\cdot \mu= \hat W$ holds, achieves approximation factor $\alpha=(\log\log \hat W)^{2^{O(1/\eps^2)}}$,  with total update time $O(N^{1+O(\eps)}\cdot \mu^{O(1/\eps)})$. The algorithm also ensures that, for every regular vertex $v$, the total number of clusters in the weak neighborhood cover $\cset$ that the algorithm maintains, to which $v$ ever belongs over the course of the algorithm, is bounded by $\hat W^{4\eps^3}$. By combining these two results, we obtain a deterministic algorithm for the fully dynamic \APSP problem, that, given a precision parameter $\frac{2}{(\log n)^{1/200}}<\eps<1/400$, achieves approximation factor $\alpha=(\log\log n)^{2^{O(1/\eps^2)}}$, and has amortized update time $O\left (n^{O(\eps)}\cdot \log \Lambda\right )$ per operation (if starting from an empty graph), with query time $O\left (2^{O(1/\eps)}\cdot \log n\cdot \log \log \Lambda\right )$ for \distquery  and query time   $O\left (|E(P)|+2^{O(1/\eps)}\cdot \log n\cdot \log\log \Lambda\right )$ for \pathquery, where $P$ is the path that the algorithm returns.
We now state our results more formally, and discuss the techniques that we employ, while pointing out specific remaining bottlenecks for obtaining a better tradeoff between the approximation factor and the update time of the algorithm.

\subsection{Our Results}

As mentioned already, a problem that plays a central role in this work is \recdynnc. We do not repeat the definition of the problem from above; a formal (and equivalent) definition can be found in \Cref{sec: valid inputs}. However, the definition that we provided above omitted one technical detail: the Consistent Covering requirement.

Let $\iset=\left(H=(V,U,E),\set{\ell(e)}_{e\in E},D \right )$ be a given valid input structure that undergoes an online sequence $\Sigma$ of valid update operations. Let $\tset$ be the time horizon associated with $\Sigma$. In order to define the Consistent Covering property, we first need to define the notion of \emph{ancestor-clusters}. This notion is defined in a natural way. If $C$ is a cluster that is present in $\cset$ at the beginning of the algorithm, then for all $\tau\in \tset$, $\anc\attime(C)=C$, so $C$ is its own ancestor. Assume now that $C'$ is a cluster that was added to set $\cset$ at some time $\tau'>0$, by applying a cluster-splitting update to a cluster $C\in \cset$. Then for all $\tau\in \tset$, if $\tau<\tau'$, $\anc\attime(C')=\anc\attime(C)$, and otherwise $\anc\attime(C')=C'$.

We are now ready to define the Consistent Covering property. Consider an algorithm for the \recdynnc problem on input $(\iset,\Sigma)$, and let $\cset$ be the collection of cluster that it maintains. We say that the algorithm obeys the Consistent Covering property, if, for every regular vertex $v\in V(H)$, for every pair $\tau'<\tau\in \tset$ of time points, if $C=\coveringcluster(v)$ at time $\tau$, and $\anc\attime[\tau'](C)=C'$, then, at time $\tau'$, $B_{H}(v,D)\subseteq V(C')$ held. We require that algorithms for the \recdynnc problem obey the Consistent Covering property.
Our first result is a reduction from a variant of fully-dynamic \APSP to  \recdynnc.

\subsubsection{Reduction from Fully-Dynamic \APSP to  \recdynnc}

We provide a black-box reduction from fully-dynamic \APSP to \recdynnc.
Our reduction shows that, if there exists an algorithm for the \recdynnc problem with some general set of parameters, then we can convert it into an algorithm for the fully-dynamic \APSP problem. The assumption on the existence of an algorithm for \recdynnc, that serves as the starting point of the reduction, is the following.

\begin{assumption}\label{assumption: alg for recdynnc2}
	There is a deterministic algorithm for \recdynNC, that,  given a valid input structure $\iset=\left(H=(V,U,E),\set{\ell(e)}_{e\in E},D \right )$ undergoing a sequence of valid update operations, with dynamic degree bound $\mu$, together with parameters $\hat W$ and  $1/(\log \hat W)^{1/100}\leq \eps<1/400$,  such that, if we denote by $N$ the number of regular vertices in $H$ at the beginning of the algorithm, then $N\cdot \mu\leq \hat W$ holds, achieves approximation factor $\alpha(\hat W)$,  with total update time $O(N^{1+O(\eps)}\cdot \mu^{O(1/\eps)})$. Moreover, the algorithm ensures that, for every regular vertex $v\in V$, the total number of clusters in the weak neighborhood cover $\cset$ that the algorithm maintains, to which vertex $v$ ever belongs over the course of the algorithm, is bounded by $\hat W^{4\eps^3}$. Here, $\alpha(\cdot)$ is a non-decreasing function.
\end{assumption}

If \Cref{assumption: alg for recdynnc2} holds, then it is quite easy to obtain an algorithm for decremental \APSP (see Section 3.4.2 in the full version of \cite{APSP-previous}), that, on an input graph $G$ that initially has $m$ edges, has total update time $O(m^{1+O(\eps)+o(1)} (\log m)^{O(1/\eps)} \log \Lambda)$, and achieves an approximation factor roughly $\alpha(m)$. One of the main contributions of this work is showing that an algorithm for the \RecDynNC problem implies an algorithm for {\bf fully-dynamic} \APSP. 
Specifically, we show that, if \Cref{assumption: alg for recdynnc2} holds, then there is an algorithm for a problem that is very similar to, but is slightly different from fully-dynamic \APSP. We call this problem $D^*$-restricted \APSP, and define it next. For a dynamic graph $G$ and time $\tau$, we denote by $G\attime$ the graph $G$ at time $\tau$.

\begin{definition}[$D^*$-restricted \APSP problem]\label{def: bounded APSP}
	The input to the $D^*$-restricted \APSP problem is 
	an $n$-vertex graph $G$ with integral lengths $\ell(e)\geq 1$ on its edges $e\in E(G)$,
	that undergoes an online sequence $\Sigma$ of edge deletions and insertions, together with a precision parameter $\frac{1}{(\log n)^{1/200}}<\eps<1/400$, and a distance parameter $D^*>0$. The goal is to support approximate \shortpath\ queries: given a pair $x,y\in V(G)$ of vertices, the algorithm needs to respond ``YES'' or ''NO'', in time $O\left(2^{O(1/\eps)}\cdot \log n\right )$. If the response is ``NO'', then $\dist_G(x,y)>D^*$ must hold. If the response is ``YES'', then the algorithm should be able, additionally, to compute a path $P$ in the current graph $G$, connecting $x$ to $y$, of length at most $\alpha'\cdot D^*$, in time $O(|E(P)|)$, where $\alpha'$ is the \emph{approximation factor} of the algorithm.
\end{definition}

The following theorem summarizes our reduction from $D^*$-restricted \APSP to \recdynnc.

\begin{theorem}\label{thm: main fully-dynamic APSP single scale}
	Suppose \Cref{assumption: alg for recdynnc2} holds. Then there is a deterministic algorithm for the $D^*$-restricted $\APSP$ problem,  that achieves approximation factor $\alpha'=(\alpha(n^3))^{O(1/\eps)}$, and has amortized update time at most $n^{O(\eps)}$ per operation, if starting from an empty graph.
\end{theorem}

\subsubsection{New Algorithm for \recdynnc}
Our next result is an improved algorithm for the \recdynnc problem, that is summarized in the following theorem.

\begin{theorem}\label{thm: main final dynamic NC algorithm}
	There is a deterministic algorithm for the \recdynNC problem, that,  given a valid input structure  $\iset=\left(H,\set{\ell(e)}_{e\in E(H)},D \right )$ undergoing a sequence of valid update operations, with dynamic degree bound $\mu$, together with parameters $\hat W$ and  $1/(\log \hat W)^{1/100}\leq \eps<1/400$,  such that, if we denote by $N$ the number of regular vertices in $H\attime[0]$, then $N\cdot \mu\leq  \hat W$ holds, achieves approximation factor $\alpha=(\log\log \hat W)^{2^{O(1/\eps^2)}}$,  with total update time $O(N^{1+O(\eps)}\cdot \mu^{O(1/\eps)})$. The algorithm ensures that, for every regular vertex $v\in V(H)$, the total number of clusters in the weak neighborhood cover $\cset$ that the algorithm maintains, to which vertex $v$ ever belongs over the course of the algorithm, is bounded by $\hat W^{4\eps^3}$.
\end{theorem}

By combining \Cref{thm: main fully-dynamic APSP single scale} and  \Cref{thm: main final dynamic NC algorithm}, we immediately obtain the following corollary, whose proof appears in Section \ref{appx: proof main APSP} of Appendix.

\begin{corollary}\label{cor: fully APSP}
 There is a deterministic algorithm for fully-dynamic \APSP, that, given an $n$-vertex graph $G$ undergoing an online sequence of edge insertions and deletions, and a precision parameter $\frac{1}{(\log n)^{1/200}}<\eps<1/400$, achieves approximation factor $\alpha'=(\log\log n)^{2^{O(1/\eps^2)}}$, and has amortized update time $O\left (n^{O(\eps)}\cdot \log \Lambda\right )$ per operation if starting from an empty graph, where $\Lambda$ is the ratio of longest to shortest edge length. Query time  for \distquery  is $O\left (2^{O(1/\eps)}\cdot \log n\cdot \log\log \Lambda\right )$ and  for  \pathquery it is   $O\left (|E(P)|+2^{O(1/\eps)}\cdot \log n\cdot \log\log \Lambda\right )$, where $P$ is the path that the algorithm returns.
\end{corollary}

\subsection{Our Techniques}
We provide a brief overview of our techniques, starting with the proof of 
\Cref{thm: main fully-dynamic APSP single scale}.

{\bf Reduction from $D^*$-restricted \APSP to  \recdynnc.} 
The description that we provide here is somewhat over-simplified, and is intended for intuition only.
We assume that we are given a fully dynamic graph $G$, that undergoes an online sequence $\Sigma$ of edge-insertions and deletions, such that $|E(G\attime[0])|+|V(G)|+|\Sigma|=m$, together with a distance parameter $D^*$, and a precision parameter $\eps$. At a high level, we use a rather natural approach. This high-level approach was used before in multiple reductions from fully-dynamic to decremental algorithms (see e.g. \cite{henzinger2001maintaining,HenzingerKing,abraham2014fully,ForsterG19,ForsterGH20}), but due to the specific setting of the problem that we consider, the use of this approach in our setting gives rise to a number of new technical challenges that we highlight below. We also provide a brief comparison with previous results where a similar approach was used.  Assume for simplicity that $q=1/\eps$ is an integer, and that so is $M=m^{\eps}$. Assume further that the distance parameter $D^*$ is an integral power of $2$. The data structures that we maintain are partitioned into $q+1$ \emph{levels}. We also define a hierarchical partition of the time horizon $\tset$ into phases.

For level $L=0$, there is a single level-$0$ phase, that spans the entire time horizon $\tset$. We maintain a level-$0$ graph $H^0$, that is constructed as follows. Let $G'$ be the dynamic graph that is obtained from the input graph $G$, by ignoring all edge insertions, and only executing edge-deletion updates. Graph $H^0$ is a bipartite graph, that has a regular vertex $v(x)$ for every vertex $x\in V(G)$, and a regular vertex $v(e)$ for every edge $e\in E(G')$. Additionally, it has a supernode $u(x)$ for every vertex $x\in V(G)$, that connects, with an edge of length $1$, to the corresponding regular vertex $v(x)$. For every edge $e=(x,y)\in E(G')$, we also connect $v(e)$ to $u(x)$ and $u(y)$, with edges of length $\ell(e)$. As graph $G'$ undergoes edge-deletions, the corresponding bipartite graph $H^0$ undergoes edge-deletions as well. For every integer $0\leq i\leq \log D^*$, we can view graph $H^0$ as an instance of the \recdynnc problem, with distance bound $D_i=2^i$. We apply the algorithm for \recdynnc from \Cref{assumption: alg for recdynnc2} to this instance, and we denote by $\cset^0_i$ the resulting collection of clusters that it maintains. For every cluster $C\in \cset^0_i$, we say that the \emph{scale} of $C$ is $i$, and we denote $\scale(C)=i$. Let $\cset^0=\bigcup_{i=0}^{\log D^*}$ be the collection of all level-$0$ clusters.

Consider now some level $0<L\leq q$. We partition the time horizon $\tset$ into at most $M^{L}$ \emph{level-$L$ phases}. Each level-$L$ phase $\Phi^L_k$ spans exactly $M^{q-L}$ consecutive edge-insertion updates from the update sequence $\Sigma$ for graph $G$, except for possibly the last phase that may contain fewer edge insertions. We define this hierarchical partition of the time horizon so that, for all $0<L\leq q$, every level-$L$ phase is contained in some level-$(L-1)$ phase.

Consider now some level $0<L\leq q$ and a level-$L$ phase $\Phi^L_k$. Let $\Phi^{(L-1)}_{k'}$ be the unique level-$(L-1)$ phase that contains $\Phi^L_k$. We associate, with phase $\Phi^L_k$, a collection $A^L_k$ of edges of $G$, that the level-$L$ data structure will be ``responsible'' for during phase $\Phi^L_k$. These are all the edges that were inserted into $G$ since the beginning of level-$(L-1)$ phase $\Phi^{(L-1)}_{k'}$, but before the beginning of level-$L$ phase $\Phi^L_k$. It is easy to see  that the number of such edges must be bounded by $M^{q-L+1}$. We also denote by $S^L_k$ the collection of vertices of $G$ that serve as endpoints of the edges of $A^L_k$.

We are now guaranteed that, at all times $\tau\in \tset$, for every edge $e\in E(G)$, either $e\in E(G\attime[0])$ (in which we say that it lies at level $0$); or there is some level $0<L\leq q$, such that $e\in A^L_k$ currently holds, where $k$ is the index of the current level-$L$ phase (in which case we say that the level of $e$ is $L$). For every path $P$ in graph $G$, we also define the \emph{level} of path $P$ to be the largest level of any of its edges.

Consider now some level $0<L\leq q$, and recall that there are at most $M^L$ level-$L$ phases. At the beginning of every level-$L$ phase, we construct level-$L$ data structures from scratch. These data structures consist of a dynamic level-$L$ bipartite graph $H^L$, that is viewed as an input to the \recdynnc problem. The set of regular vertices of $H^L$ is $S^L_k$, where $k$ is the index of the current level-$L$ phase. Intuitively, graph $H^L$ will be ``responsible'' for all level-$L$ paths in graph $G$. We describe the sets of supernodes and of edges of $H^L$ below. For all $0\leq i\leq \log D^*$, we view graph $H^L$, together with distance parameter $D_i=2^i$, as an instance of the \recdynnc problem, and we apply the algorithm from \Cref{assumption: alg for recdynnc2} to this instance, denoting the resulting collection of clusters by $\cset^L_i$. We say that all clusters in $\cset^L_i$ have scale $i$, and we denote  $\cset^L=\bigcup_{i=0}^{D^*}\cset^L_i$. The supernodes of graph $H^L$ are vertices $u(C)$ corresponding to some of the clusters  $C\in\bigcup_{L'<L}\cset^{L'}$. As the clusters in set $\bigcup_{L'<L}\cset^{L'}$ evolve, we maintain the corresponding dynamic graph $H^L$ via valid update operations, where, for example, a cluster-splitting update of a cluster $C\in\bigcup_{L'<L}\cset^{L'}$  can be implemented via a supernode-splitting update applied to supernode $u(C)$.

 Notice that, while the number of level-$L$ phase may be as large as $M^L$, the number of regular vertices in the level-$L$ graph $H^L$ is bounded by $2M^{q-L+1}$. Therefore, even though we need to recompute a level-$L$ data structure from scratch at the beginning of each level-$L$ phase, the size of the corresponding graph is sufficiently small that we can afford it. The level-$q$ data structure is computed from scratch after every edge-insertion update, though the number of regular vertices in the corresponding graph $H^q$ is bounded by $M\leq m^{\eps}$.
 
 While the high-level idea described above is quite natural, and was used multiple times in the past (see e.g. \cite{henzinger2001maintaining,HenzingerKing,abraham2014fully,ForsterG19,ForsterGH20}), it poses a number of challenges. The main challenge is the coordination between the different levels that is needed in order to support \shortpath\ queries. Consider, for example, a \shortpath\ query between a pair $x,y$ of vertices of $G$, and assume that there is a path $P$ in $G$ connecting $x$ to $y$, whose length is $D<D^*$. Notice, however, that the edges of $P$ may belong to different levels, and there may not be a single level $L$, such that all vertices of $P$ lie in the graph $H^L$. Assume that the level of path $P$ is $L$. Then we would like the level-$L$ data structure to be ``responsible'' for this query. In other words, we would like some path $P'$, whose length is comparable to $D$, to represent path $P$ in graph $H^L$. But it is possible that the endpoints $x$ and $y$ of $P$ do not even lie in graph $H^L$, so it is not clear which path in $H^L$ we should use as a representative of path $P$.

This issue seems especially challenging in the setting of  \APSP with adaptive adversary, where it is required that approximate \spquery queries are supported. For comparison, \cite{abraham2014fully}
and \cite{ForsterGH20} use a very similar high-level idea of a hierarchical partition of the time horizon and the set of edges. In \cite{ForsterGH20}, the algorithm is only required to maintain a low-stretch probabilistic tree embedding of the graph. This allows them to combine the trees maintained at different levels into a single tree that has a relatively low height, thereby circumventing the problem of coordinating between graphs from different levels. In order to respond to \distquery between a pair of vertices, they simply compute the length of the path between the two vertices in the tree that they maintain. Their data structure however cannot support approximate \shortestpathquery. If we tried to similarly combine graphs from different levels in order to overcome the challenge of coordinating between them, we would obtain another fully dynamic (non-tree) graph, and it is not clear how to support approximate \shortestpathquery\ in this graph. A different approach was taken by \cite{abraham2014fully}, whose algorithm exploits specific properties of the distance oracles of \cite{TZ,rodittyZ2}. The latter constructions however are randomized and can only withstand an oblivious adversary.

 In order to resolve this issue of coordination between levels, we associate, to every cluster $C\in \bigcup_{L=0}^q\cset^L$, a set $V^F(C)\subseteq V(G)$ of vertices, and we think of cluster $C$ as representing this collection of vertices of $G$. For a level $0\le L\leq q$, we include in graph $H^L$ supernodes $u(C)$ for all clusters $C\in \bigcup_{L'<L}\cset^{L'}$ with $S^L_k\cap V^F(C)\neq \emptyset$, where $k$ is the index of the current level-$L$ phase. For every vertex $x\in S^L_k$, and supernode $u(C)$ with $x\in V^F(C)$, we add an edge $(v(x),u(C))$ to graph $H^L$, whose length is $2^{\scale(C)}$. The main challenge in this construction is to define the sets $V^F(C)$ of vertices for clusters $C\in \bigcup_{L=0}^q\cset^L$. On the one hand, we would like to make these sets broad enough, so that the resulting graphs $H^L$ are rich enough in order to allow us to support approximate \shortpath\ queries. On the other hand, in order to ensure that the algorithm is efficient, these sets cannot be too large.
 
 In order to support \shortpath\ queries between pairs of vertices $x,y\in V(G)$, we employ a notion of ``covering chains'' -- structures that span multiple levels. Suppose the shortest path $P$ connecting $x$ to $y$ in $G$ has length $D\leq D^*$, and belongs to level $L$. Using the covering chains, we compute small collections $R(x),R(y)\subseteq S^L_k$ of vertices associated with $x$ and $y$ respectively, such that there exists a vertex $x'\in R(x)$ and a vertex $y'\in R(y)$, together with a path $P'$ in graph $H^L$ connecting $v(x')$ to $v(y')$, whose length is comparable to $D$. Conversely, we show that any such path in $H^L$ can be transformed into a path in graph $G$ that connects $x$ to $y$, and has length that is not much larger than $D$.

 Next, we provide a high-level overview of the proof of \Cref{thm: main final dynamic NC algorithm}. We also  point out the main remaining bottlenecks to obtaining a better approximation.
 
 {\bf Improved algorithm for \recdynnc.}
  The \recdynnc problem can be effectively partitioned into two subproblems. The first subproblem, called \maintaincluster problem, is responsible for maintaining a single cluster. Suppose we are given any such cluster $C\subseteq H$, where $H$ is the current graph, and a distance parameter $D^*>D$. Cluster $C$ will undergo a sequence $\Sigma_C$ of valid update operations that correspond to the updates applied to $H$, possibly with some additional edge-deletions and isolated vertex-deletions. The goal of the \maintaincluster problem is to support \spquery queries: given a pair $x,y$ of regular vertices of $C$, compute a path $P$ of length at most $\alpha\cdot D^*$ connecting them in graph $C$, in time $ O(|E(P)|)$, where $\alpha$ is the approximation factor that the algorithm achieves. Whenever the diameter of cluster $C$ becomes too large, the algorithm may raise a flag $F_C$, and to provide a pair $x,y$ of regular vertices of $C$ (that we call a \emph{witness pair}), such that $\dist_C(x,y)> D^*$. After that, the algorithm will receive, as part of the update sequence $\Sigma_C$, a sequence of edge-deletions and isolated vertex-deletions (that we call a \emph{flag-lowering sequence}), following which at least one of the vertices $x,y$ is deleted from $C$, and flag $F_C$ is lowered. If the diameter of $C$ remains too large, the algorithm can raise the flag again immediately. Queries $\spquery$ may not be asked when flag $F_C$ is up. \maintaincluster problem was defined in \cite{APSP-previous}, and we employ the same definition here.
 
 The second problem is \maintainNC problem. This problem is responsible for managing the neighborhood cover $\cset$ itself. Initially, we start with $\cset$ containing a single cluster - cluster $H$. The clusters in $\cset$ may only undergo allowed operations that are defined exactly like in the \recdynnc problem. The algorithm also needs to maintain, for every regular vertex $v\in V(H)$, a cluster $\coveringcluster(v)\in \cset$, that contains all vertices of $B_H(v,D)$, so that the Consistent Covering property holds. The algorithm does not need to support any queries. But, at any time, it may receive a cluster $C$ and a pair $x,y$ of vertices of $C$, such that $\dist_C(x,y)>D^*$ holds (for a parameter $D^*$ that we specify below). It must then produce a flag-lowering sequence $\Sigma'$ for $C$ (that is, a sequence of edge- and isolated vertex-deletions, after which at least one of $x,y$ is deleted from $C$). All updates from $\Sigma'$ must be then applied to cluster $C$, but they may be interspersed with cluster-splitting operations, when new clusters $C'\subseteq C$ are added to $\cset$. The algorithm must also ensure that every regular vertex of $H$ only belongs to a small number of clusters over the course of the time horizon.
 
 By combining the algorithms for the  \maintaincluster and the \maintainNC problems, it is easy to obtain an algorithm for the \recdynnc problem, whose approximation factor is $\alpha \cdot D^*/D$, where $\alpha$ is the approximation factor of the algorithm for \maintaincluster, and $D^*$ is the threshold parameter for raising the flags $\set{F_C}_{C\in \cset}$.

  While \cite{APSP-previous} did not explicitly define the \maintainNC problem, they effectively provided a simple algorithm for it, that relies on a variation of the standard ball-growing technique, and uses parameter $D^*=\Omega(D\cdot \log N)$, where $N$ is the number of regular vertices in graph $H$. This overhead of $O(\log N)$ factor is one of the reasons for the $(\log N)^{2^{O(1/\eps)}}$-approximation factor that their algorithm achieves, and it is one of the barriers to obtaining a better approximation. We provide a different algorithm for the  \maintainNC problem, that allows us to set $D^*=O(D\cdot \log \log N)$. The overhead of factor  $O(\log\log N)$ in this algorithm is the only remaining barrier to obtaining an improved algorithm for the \recdynnc problem, and for \APSP. For example, if we could ensure that $D^*=O\left (2^{2^{O(1/\poly(\eps))}}\cdot D\right )$ is sufficient, we would obtain an algorithm for \recdynnc and for fully dynamic \APSP with approximation factor $2^{2^{O(1/\poly(\eps))}}$ and the same update time immediately.

In the remainder of this overview, we focus on the \maintaincluster problem. We first provide a brief overview of the algorithm from \cite{APSP-previous}, and then describe our improvements.

The central concept that \cite{APSP-previous} use in designing an algorithm for the \maintaincluster problem is that of a \emph{balanced pseudocut}, which they also introduced. Let $N$ be the number of regular vertices in $H\attime[0]$, and let $\mu$ be the dynamic degree bound. Recall that, as input to the \maintaincluster problem, we are given a cluster $C$ of $H$, that undergoes a sequence $\Sigma_C$ of valid update operations with dynamic degree bound $\mu$, and a distance parameter $D^*$. We use an additional parameter $\rho$; it may be convenient to think of $\rho=N^{\eps}$. Let $\hat D>D^*$ be another distance parameter; its specific value is not important for this technical overview, but it is close to $D^*$. A $(\hat D,\rho)$-pseudocut in graph $C$ is a collection $T$ of regular vertices of $C$, such that, for every regular vertex $v\in V(C)\setminus T$, $B_{C\setminus T}(v,\hat D)$ contains at most $N/\rho$ regular vertices. This notion can be viewed as a generalization of the balanced vertex multicut, that can be defined as a collection $T$ of vertices, such that every connected component of $C\setminus T$ contains at most $N/\rho$ vertices. Intuitively, once the vertices of the pseudocut (or of a balanced multicut) are deleted from $C$, we can break it into significantly smaller clusters, while still maintaining the covering properties of the neighborhood cover $\cset$. However, balanced pseudocuts have one additional crucial property: \cite{APSP-previous} provided an algorithm, that, given a $(\hat D,\rho)$-pseudocut $T$ in cluster $C$, either (i) computes an expander graph $X$, with $V(X)\subseteq T$, such that $|V(X)|$ is comparable to $|T|$, together with an embedding of $X$ into $C$ via short paths that cause a relatively low congestion; or (ii) computes another $(\hat D,\rho)$-pseudocut $T'$ in $C$, with $|T'|\ll |T|$. We denote this algorithm $\alg$. This algorithm is the core technical part in the algorithm of \cite{APSP-previous} for the \maintaincluster problem, and our main technical contribution to the \maintaincluster problem essentially replaces algorithm $\alg$ with a different algorithm. We now provide a very brief description of algorithm $\alg$. 

{\bf Algorithm $\alg$.}
A central observation that is needed for the algorithm is the following: let $T$ be a $(\hat D,\rho)$-pseudocut in graph $C$, and suppose we have computed a relatively small subset $E'$ of edges of $C$, and a collection $T_1,T_2,\ldots,T_{\rho+1}$ of subsets of vertices of $T$, such that each such subset $T_i$ is sufficiently large, and, for all $1\leq i<j\leq \rho+1$, $\dist_{C\setminus E'}(T_i,T_j)>4\hat D$. Then we can compute a pseudocut $T'$ for graph $C$ with $|T'|\ll|T|$. The idea is that there must be some index $1\leq i\leq \rho+1$, such that $B_{C\setminus E'}(T_i,2\hat D)$ contains at most $N/\rho$ regular vertices. By replacing set $T_i$ in the pseudocut $T$ with the endpoints of the edges in $E'$, we obtain a significantly smaller pseudocut $T'$. Algorithm $\alg$ starts with the given pseudocut $T$, and then attempts to compute an expander $X$ over a large subset of vertices of $T$, and to embed it into $C$ via the Cut-Matching Game of \cite{KRV} (in fact, they need to use a weaker variant of the game from \cite{detbalanced}, who provide a deterministic algorithm for the cut player, but unfortunately only ensure a rather weak expansion in the resulting graph $X$, which also contributes to the relatively high approximation factor of \cite{APSP-previous}).
If the Cut-Matching Game fails to construct the expander $X$ and embed it into $C$ as required, then it produces two large subsets $T',T''\subseteq T$ of vertices, and a relatively small subset $E'$ of edges, such that $\dist_{C\setminus E'}(T',T'')>4\hat D$. Then they recursively apply the same algorithm to $T'$ and to $T''$. After $\rho$ such iterations, if the algorithm failed to construct the desired expander $X$ and its embedding, we  obtain large vertex subsets $T_1,\ldots,T_{\rho+1}\subseteq T$, and a subset $E'$ of edges of $C$, that allow us to compute a much smaller pseudocut, as described above. Even though they perform $\rho$ iterations of the algorithm for the Cut-Matching Game, since, in case of a failure, the subsets $T',T''\subseteq T$ of vertices that it produces are very large compared to $|T|$, the resulting subsets $T_1,\ldots,T_{\rho+1}$ of vertices are still sufficiently large to make progress.
We now complete the description of the algorithm of \cite{APSP-previous} for the \maintaincluster problem.

The algorithm is partitioned into phases. Initially, we construct a pseudocut $T$ that contains all regular vertices of $C$. At the beginning of each phase, we use Algorithm $\alg$ (possibly iteratively), in order to compute a pseudocut $T'$, and an expander $X$ defined over a large subset of vertices of $T'$, together with an embedding of $X$ into $C$ via short path that cause a low congestion. Assume first that $|T'|>N^{1-\Theta(\eps)}$. The algorithm of \cite{APSP-previous} employs an algorithm for \APSP in expanders on graph $X$. This algorithm can maintain a large ``core'' $S\subseteq V(X)$, over the course of a large number of edge-deletions from $C$ (the  number that is roughly comparable to $|T'|$). It can also support queries in which, given a pair $x,y\in V(S)$ of vertices, a path of length at most roughly $(\log N)^{O(1/\poly(\eps))}$ connecting $x$ to $y$ in $X$ is returned. This path can then be transformed into a path of comparable length connecting $x$ to $y$ in $C$, using the embedding of $X$ into $C$. Additionally, they maintain an \EST in graph $C$, that is rooted at the vertices of $S$. This tree can be used in order to ensure that all regular vertices of $C$ are sufficiently close to the core $S$, and, whenever this is not the case, flag $F_C$ is raised. Once the algorithm for \APSP in expanders can no longer maintain the core $S$ (after roughly $|T'|$ deletions of edges from $C$), the phase terminates. It is easy to verify that, as long as the cardinality of the pseudocut $T'$ is sufficiently large (say at least $N^{1-\Theta(\eps)}$), the number of phases remains relatively small, and the algorithm can be executed efficiently. Once the cardinality of the pseudocut $T'$ becomes too small, the last phase begins, during which the pseudocut $T'$ remains unchanged. We omit here the description of this phase, since our implementation of this part is essentially identical to that of \cite{APSP-previous}. We only note that this phase solves the \recdynnc problem recursively on two instances, whose sizes are significantly smaller than that of $H$. The two instances are then composed in a natural way, which eventually leads to the doubly-exponential dependence of the approximation factor on $1/\poly(\eps)$.

The algorithm of \cite{APSP-previous} for the \maintaincluster problem loses a super-logarithmic in $N$ approximation factor via this approach, that contributes to the  final $(\log N)^{2^{1/\poly(\eps)}}$-approximation factor for the \recdynnc problem. This loss is largely  due to the use of expander graphs. In addition to the issues that we have mentioned with the implementation of the Cut-Matching game via a deterministic algorithm, all currently known algorithms for \APSP in expanders only achieve a superlogarithmic approximation factor, and even if they are improved, the loss of at least a polylogarithmic approximation factor seems inevitable. It is typical for this issue to arise when relying on expander graphs for distance-based problems, such as \APSP. A recent work of \cite{DMG} suggested a method to overcome this difficulty, by replacing expander graphs with \emph{well-connected} graphs. Intuitively, if $G$ is a graph, and $S$ is large subset of its vertices, we say that $G$ is well-connected with respect to $S$ (or just well-connected) if, for every pair $A,B\subseteq S$ of disjoint equal-cardinality subsets of vertices of $S$, there is a collection $\pset$ of paths in $G$, that connects every vertex of $A$ to a distinct vertex of $B$, such that the paths in $\pset$ are short, and they cause a low congestion. In a typical setting, if $G$ is an $n$-vertex graph, then the lengths of the paths in $\pset$ are bounded by $2^{\poly(1/\eps)}$, and the congestion that they cause is bounded by $n^{O(\eps)}$. \cite{DMG} also developed a toolkit of algorithmic techniques around well-connected graphs, that mirror those known for expanders. For example, they provide an analogue of the Cut-Matching Game, that, given a graph $C$ and a set $T$ of its vertices, either computes a large set $S\subseteq T$ of vertices, and a graph $X$ with $V(X)\subseteq T$, that is well-connected with respect to $S$, together with an embedding of $X$ into $C$ via short paths that cause a low congestion; or it computes two relatively large sets $T',T''\subseteq T$ of vertices, and a small set $E'$ of edges, such  that $\dist_{C\setminus E'}(T',T'')$ is large. Additionally, they provide an algorithm for \APSP in well-connected graphs, that has similar properties to the above mentioned algorithm for \APSP in expanders, but achieves a much better approximation factor of $2^{1/\poly(\eps)}$. By replacing expander graphs with well-connected graphs in the algorithm for \maintaincluster problem of \cite{APSP-previous}, we avoid the superlogarithmic loss in the approximation factor that their algorithm incurred. We note however that replacing expanders with well-connected graphs in algorithm $\alg$ is quite challenging technically, for the following reason. Recall that, in the approach that used the Cut-Matching Game, if the algorithm fails to compute an expander $X$ containing a large number of vertices from the given set $T$ and embed it into $C$, it provides two very large subsets $T',T''\subseteq T$ of vertices, together with a small set $E'$ of edges, such that $\dist_{C\setminus E'}(T',T'')>4\hat D$. Unfortunately, the analogous algorithm of \cite{DMG}, in case of a failure to embed a well-connected graph, provides vertex sets $T',T''$, whose cardinalities are significantly smaller than that of $T$. Specifically, it only ensures that $|T'|,|T''|\geq |T|^{1-4\eps^3}/4$. Since we need to continue applying this algorithm recursively, until $\rho+1$ subsets $T_1,\ldots,T_{\rho+1}$ of vertices of $T$ are constructed, we can no longer guarantee that, for all $i$, $|T_i|$ is sufficiently large. As a result, if our algorithm fails to compute a well-connected graph $X$ and its embedding into $C$, we can no longer compute a new pseudocut whose cardinality is significantly lower than that of $T$. Since the time required to execute this algorithm is super-linear in $|E(C)|$, we cannot afford to execute it many times, so it is critical for us that the cardinality of the pseudocut $T$ decreases significantly with every execution. Our main technical contribution to the algorithm for the \maintaincluster problem is overcoming this hurdle, and designing an analogue of Algorithm $\alg$ that works with well-connected graphs instead of expanders.

\paragraph{Organization.}
We start with preliminaries in \Cref{sec: prelims}. In \Cref{sec: valid inputs}, we formally define valid input structure, valid update operations, and the \recdynnc problem. We also provide the statement our main technical result for the \recdynnc problem -- an algorithm whose guarantees are somewhat weaker than those in \Cref{thm: main final dynamic NC algorithm}, which however allows us to prove \Cref{thm: main final dynamic NC algorithm}. \Cref{sec: fully APSP inner main}  is dedicated to the reduction from fully dynamic \APSP to  \recdynnc and the proof of \Cref{thm: main fully-dynamic APSP single scale}.
In \Cref{sec: proof of recdynnc inner} we formally define the \maintainNC and \maintaincluster problems, and state our main results for them. We also complete our algorithm for the \recdynnc problem using these results. We provide our algorithms for the \maintainNC and the \maintaincluster problems in \Cref{sec: alg for maintainNC} and \Cref{sec: balanced pseudocut}, respectively.

%In \Cref{sec: proof of recdynnc inner} we formally define the \maintainNC and \maintaincluster problems, and state our main results for them. We also complete our algorithm for the \recdynnc problem using these results. We provide our algorithms for the \maintainNC and the \maintaincluster problems in \Cref{sec: alg for maintainNC} and \Cref{sec: balanced pseudocut}, respectively.

\section{Preliminaries}\label{sec: prelims}

All logarithms in this paper are to the base of $2$. 
Throughout the paper, we use a $\tilde O(\cdot)$ notation to hide multiplicative factors that are polynomial in $\log m$ and $\log n$, where $m$ and $n$ are the number of edges and vertices, respectively, in the initial input graph. All graphs in this paper are simple, so they may not contain loops or parallel edges. We explicitly refer to graphs with parallel edges as \emph{multigraphs}.

\subsection{Graph-Theoretic Notation, Clusters and Routings}
We follow standard graph-theoretic notation.   Given a graph $G=(V,E)$ and two disjoint subsets $A,B$ of its vertices, we denote by $E_G(A,B)$ the set of all edges with one endpoint in $A$ and another in $B$, and by $E_G(A)$ the set of all edges with both endpoints in $A$. We also denote by $\delta_G(A)$ the set of all edges with exactly one endpoint in $A$. For a vertex $v\in V(G)$, we denote by $\delta_G(v)$ the set of all edges incident to $v$ in $G$, and by $\deg_G(v)$ the degree of $v$ in $G$.
We may omit the subscript $G$ when clear from context. Given a subset $S\subseteq V$ of vertices of $G$, we denote by $G[S]$ the subgraph of $G$ induced by $S$.

Given a graph $G$, we say that a graph $C$ is a \emph{cluster} of $G$, if $C$ is a connected vertex-induced subgraph of $G$.

%Given a graph $G$ and a weight function $w: V(G)\rightarrow \reals$ on its vertices, for a subset $V'\subseteq V(G)$ of its vertices, we denote by $W(V')=\sum_{v\in V'}w(v')$ the total weight of all vertices in $V'$. Abusing the notation, for a subgraph $C\subseteq G$, we denote the weight of the subgraph $W(C)=\sum_{v\in V(C)}w(v)$.

\paragraph{Matchings and routings.}
If $G$ is a graph, and $\pset$ is a collection of paths in $G$, we say that the paths in $\pset$ cause congestion $\eta$, if every edge $e\in E(G)$ participates in at most $\eta$ paths in $\pset$, and some edge $e\in E(G)$ participates in exactly $\eta$ such paths.

Let $G$ be a graph, and let $\mset=\set{(s_1,t_1),\ldots,(s_k,t_k)}$ be a collection of pairs of vertices of $G$. We say that $\mset$ is a \emph{matching} if every vertex $v\in V(G)$ participates in at most one pair in $\mset$, and for every pair $(s_i,t_i)\in \mset$, $s_i\neq t_i$. Note that we do not require that the pairs $(s_i,t_i)\in \mset$ correspond to edges of $G$. We say that a collection $\pset$ of paths  in graph $G$ is a \emph{routing} of the pairs in $\mset$, if $|\pset|=k$, the paths in $\pset$ are simple paths, and, for every pair $(s_i,t_i)\in \mset$ of vertices, there is a path $P_i\in \pset$ whose endpoints are $s_i$ and $t_i$.

Assume now that we are given a graph $G$, two disjoint sets $S,T$ of its vertices, and a collection $\pset$ of paths. We say that the paths in $\pset$ \emph{route} vertices of $S$ to vertices of $T$, or that $\pset$ is a \emph{routing} of $S$ to $T$, if $\pset=\set{P(s)\mid s\in S}$, and, for all $s\in S$, path $P(s)$ originates at vertex $s$ and terminates at some vertex of $T$. We further require that every vertex of $S\cup T$ serves as an endpoint of at most one path in $\pset$.

\paragraph{Embeddings of Graphs.}
Let $G$, $X$ be two graphs with $V(X)\subseteq V(G)$. An \emph{embedding} of $X$ into $G$ is a collection $\pset=\set{P(e)\mid e\in E(X)}$ of paths in graph $G$, such that, for every edge $e=(x,y)\in E(X)$, path $P(e)$ connects $x$ to $y$. The \emph{congestion} of the embedding is the maximum, over all edges $e'\in E(G)$, of the number of paths in $\pset$ containing $e'$.

\subsection{Distances, Balls, and Neighborhood Cover}
Suppose we are given a graph $G$ with lengths $\ell(e)> 0$ on its edges $e\in E(G)$. For a path $P$ in $G$, we denote its length by $\ell_G(P)=\sum_{e\in E(P)}\ell(e)$. For a pair of vertices $u,v\in V(G)$, we denote by $\dist_G(u,v)$ the \emph{distance} between $u$ and $v$ in $G$: the smallest length $\ell_G(P)$ of any path $P$ connecting $u$ to $v$ in $G$.  For a pair $S,T$ of subsets of vertices of $G$, we define the distance between $S$ and $T$ to be $\dist_G(S,T)=\min_{s\in S,t\in T}\set{\dist_G(s,t)}$.
For a vertex $v\in V(G)$, and a subset $S\subseteq V(G)$ of vertices, we also define the distance between $v$ and $S$ as $\dist_G(v,S)=\min_{u\in S}\set{\dist_G(v,u)}$.
 The \emph{diameter} of the graph $G$, denoted by $\diam(G)$, is the maximum distance between any pair of vertices in $G$.

 Consider now some vertex $v\in V(G)$, and a distance parameter $D\geq 0$. The \emph{ball of radius $D$ around $v$} is defines as: $B_G(v,D)=\set{u\in V(G)\mid \dist_G(u,v)\leq D}$.
Similarly, for a subset $S\subseteq V(G)$ of vertices, we let the ball of radius $D$ around $S$ be $B_G(S,D)=\set{u\in V(G)\mid \dist_G(u,S)\leq D}$.
 We will sometimes omit the subscript $G$ when clear from context.

Given a graph $G$ with length $\ell(e)$ on its edges $e\in E(G)$, we may sometimes consider subgraphs $H\subseteq G$, or graphs that are derived from $G$ in some fashion, that also have lengths $\ell(e)$ on their edges. In such cases, for clarity of exposition, we will sometimes denote by $\ell_G(e)$ the length of edge $e$ in graph $G$, and by $\ell_H(G)$ the length of $e$ in $H$. When the graph is unambiguously clear, we may omit the corresponding subscript.

\paragraph{Dijkstra's Algorithm / Weighted BFS.}
We will sometimes employ  Dijkstra's algorithm, that we also refer to as \emph{weighted BFS}. Given an $n$-vertex graph $G$ with  lengths $\ell(e)>0$ on its edges $e\in E(G)$, the algorithm performs a weighted BFS in $G$, starting from some give subset $T$ of vertices, up to some pre-specified depth $D$.
Recall that Dijkstra's algorithm maintains a set $S$ of ``discovered'' vertices of $G$, where at the beginning $S=T$. Throughout the algorithm, for every vertex $y\in S$, it maintains the distance $\dist_G(T,y)$, and a neighbor vertex $a_y$ of $y$ that does not lie in $S$, and minimizes the length of the edge $(y,a_y)$. In every step, we select a vertex $y\in S$, for which $\dist_G(T,y)+\ell(y,a_y)$ is minimized, and add vertex $a_y$ to $S$. We are then guaranteed that $\dist_G(T,a_y)=\dist_G(T,y)+\ell_G(y,a_y)$. 
Assume that we are given, for every vertex $y\in V(G)$, a list $\lambda(y)$ of its neighbors $a$, sorted according to the length $\ell_G(a,y)$ of the corresponding edge, from smallest to largest. Then Dijkstra's algorithm can be implemented so that, if $S_i$ is the set $S$ after the $i$th step, then the total running time of the algorithm up to, and including iteration $i$ is $O(|E_G(S_i)|\cdot \log n)$. In order to do so, we maintain, for every vertex $y\in S$, a pointer $p_y$ to the vertex $a_y$ on the list $\lambda(y)$. We also maintain a heap of vertices in set $\set{a_y\mid y\in S}$, whose key is $\dist_G(T,y)+\ell_G(y,a_y)$. In every step, we select a vertex $a=a_y$ from the top of the heap, add it to $S$, and then advance the pointer $p_y$ until the first vertex that does not lie in $S$ is encountered, and we initialize pointer $p_a$. If vertex $a$ that was added to $S$ serves as vertex $a_y$ for several vertices $y\in S$, we advance the pointer $p_y$ for each such vertex $y$.  To summarize, if $S$ is the set of vertices that the algorithm discovers at the time it is terminated, then the running time of the algorithm is $O(|E(S)|\cdot \log n)$, where $n=|V(G)|$.

\paragraph{Ball-Growing Technique.}
We will repeatedly use the following simple lemma, whose proof uses the standard ball-growing technique of \cite{LR,GVY}, and is provided in Section \ref{sec: ball-growing} of Appendix for completeness.

\begin{lemma}\label{lem: ball growing}
	There is a deterministic algorithm, whose input consists of an $n$-vertex graph $G$ with lengths $\ell(e)\geq 1$ on edges $e\in E$, a vertex $v\in V(G)$ that is not an isolated vertex in $G$, a distance parameter $D\geq 1$, a precision parameter $0<\eps<1/4$, and $k\geq 0$ subsets $T_1,\ldots,T_k$ of vertices of $G$ (that need not be disjoint). The algorithm computes an integer $1<i\leq \frac{2k+2}{\eps}$, such that, if we denote by $S=B_G(v,2(i-1)D)$, and $S'=B_G(v,2iD)$, then:
	
	\begin{itemize}
		\item $|E_G(S')|\leq |E_G(S)|\cdot |E(G)|^{\eps}$; and
		\item for all $1\leq j\leq k$, $|T_j\cap S'|\leq |T_{j}\cap S|\cdot |T_j|^{\eps}$.
	\end{itemize}
The running time of the algorithm is bounded by $O(|E_G(S')|\cdot \log n)\leq O(|E_G(S)|\cdot |E(G)|^{\eps}\cdot \log n)$.
\end{lemma}

\paragraph{Neighborhood Covers.}
Neighborhood Cover is a central notion that we use throughout the paper. We use both a strong and a weak notion of neighborhood covers, that are defined as follows.

\begin{definition}[Neighborhood Cover]
	Let $G$ be a graph with lengths $\ell(e)>0$ on edges $e\in E(G)$, let $S\subseteq V(G)$ be a subset of its vertices, and let $D\leq D'$ be two distance parameters. A \emph{weak $(D,D')$-neighborhood cover} for the set $S$ of vertices in $G$ is a collection $\cset=\set{C_1,\ldots,C_r}$ of clusters of $G$, such that:
	
	\begin{itemize}
%		\item for all $1\leq i\leq r$, $C_i$ is a vertex-induced subgraph of $H$;
		\item for every vertex $v\in S$, there is some index $1\leq i\leq r$ with $B_G(v,D)\subseteq V(C_i)$; and
		\item for all $1\leq i\leq r$, for every pair $s,s'\in S\cap V(C_i)$ of vertices, $\dist_{G}(s,s')\leq D'$.
	\end{itemize} 
	A set $\cset$ of clusters of $G$ is a \emph{strong $(D,D')$-neighborhood cover} for vertex set $S$ if it is a weak $(D,D')$-neighborhood cover for $S$, and, additionally, for every cluster $C\in \cset$, for every pair $s,s'\in S\cap V(C)$ of vertices, $\dist_{C}(s,s')\leq D'$.
	
	If the set $S$ of vertices is not specified, then we assume that $S=V(G)$.
\end{definition}

\paragraph{Computing Initial Sparse Neighborhood Cover}
We will employ the following theorem, that allows us to efficiently compute a sparse neighborhood cover in a given input graph $G$.
The theorem extends the ball-growing technique, building on some ideas from  \cite{APSP-previous}.

\begin{theorem}\label{claim: cutting G}
	There is a deterministic algorithm whose input consists of an $n$-vertex graph $G$ with integral lengths $\ell(e)\geq 1$ on edges $e\in E(G)$, set $T\subseteq V(G)$ of special vertices called terminals with $|T|=k$, a precision parameter $0<\eps<1/16$, and two distance parameters $D\geq 1$ and $D'\geq \frac{2^{13}\cdot D}{\eps^2}$. The algorithm computes a collection $\cset$ of vertex-induced subgraphs (clusters) of graph $G$ with the following properties:
	
	\begin{itemize}
		\item for every cluster $C\in \cset$, there is some terminal $t\in T$, with $V(C)\subseteq B_G(t,D')$;
		
		\item every terminal $t\in T$ belongs to at most $\frac{64k^{\eps/16}}{\eps}$ clusters in $\cset$; and
		
		\item for every terminal $t\in T$, there is at least one cluster $C\in \cset$, with $B_{G}(t,D)\subseteq V(C)$.
	\end{itemize}
	
	Additionally, the algorithm computes, for every terminal $t\in T$, a cluster $C(t)\in \cset$ with $B_{G}(t,D)\subseteq V(C(t))$.
	The running time of the algorithm is $O\left (k+|E(G)|^{1+\eps/32}\cdot \log n\right )$.
\end{theorem}

\begin{proof}
Throughout the proof, we denote $V=V(G)$, $E=E(G)$, and $m=|E|$. We also let $\eps'=\eps/32$.
	The proof repeatedly uses the ball-growing algorithm from \Cref{lem: ball growing}. 
	Throughout the proof, we maintain a vertex-induced subgraph $G'\subseteq G$, and a collection $\cset$ of clusters of $G$. We ensure that, throughout the algorithm, the following two invariants hold.
	
	\begin{properties}{I}
		\item for every terminal $t\in T$, there is a cluster $C(t)\in \cset\cup\set{G'}$, such that $B_{G}(t,D)\subseteq V(C(t))$; and \label{inv: ball coverage}
		\item for every cluster $C\in \cset$, there is some terminal $t\in T$, with $V(C)\subseteq B_G(t,D')$. \label{inv: few reg vertices}
	\end{properties}

	For every  terminal $t\in T$, we also maintain a cluster 
	$C(t)\in \cset\cup\set{G'}$, such that $B_{G}(t,D)\subseteq V(C(t))$.

	For every  terminal $t\in T$, we maintain a counter $n_t$, that counts the number of clusters $C\in \cset\cup\set{G'}$ that contain $t$. Let $r=\ceil{1/\eps'}+1$. We also maintain a partition $(S_0,S_1,\ldots,S_{r})$ of the set  $T$ of terminals into classes, that is defined as follows. A terminal $t\in T$ belongs to class $S_0$ if $n_t\leq k^{2\eps'}$. For $1\leq j<r$, terminal $t$ belongs to class $S_j$ if $j\cdot k^{2\eps'}<n_t\leq (j+1)\cdot k^{2\eps'}$. If $n_t>r\cdot k^{2\eps'}$, then terminal $t$ belongs to class $S_{r}$. In fact, as we show later, our algorithm will ensure that $S_r=\emptyset$ always holds.
	
	At the beginning of the algorithm, we let $G'=G$ and $\cset=\emptyset$. Clearly, both invariants hold at this time. We also initialize, for every  terminal $t\in T$, $n_t=1$, and $C(t)=G$. We then let $S_0$ contain all  terminals in $T$, and set $S_1=S_2=\cdots=S_r=\emptyset$.
	We  perform iterations, as long as there is at least one terminal $t\in T$ with $C(t)=G'$. In every iteration, a single cluster $C\subseteq G$ will be added to set $\cset$, and some vertices and edges will be removed from $G'$.

	We now describe the execution of a single iteration. Let $t\in T$ be any terminal with $C(t)=G'$. If $t$ is an isolated vertex of $G'$, then we add a new cluster $C=\set{t}$ to $\cset$, and set $C(t)=C$. It is immediate to verify that all invariants continue to hold. We assume from now on that $t$ is not isolated in $G'$. Let $T'=T\cap V(G')$.

	We apply the algorithm from \Cref{lem: ball growing} to graph $G'$, vertex $t$, distance parameter $D$, precision parameter $\eps'$, and subsets $T',S_0,\ldots,S_r$ of vertices of $G'$ instead of $T_1,\ldots,T_k$. Let $1<i\leq \frac{2r+6}{\eps'}$ be the integer that the algorithm returns.
	Denote $A=B_{G'}(t,2(i-1)D)$, and $A'=B_{G'}(t,2iD)$. Recall that the algorithm guarantees that
	$|T'\cap A'|\leq |T'\cap A|\cdot |T'|^{\eps'}\leq |T\cap A|\cdot k^{\eps'}$, and for all $0\leq j\leq r$, $|S_j\cap A'|\leq |S_j\cap A|\cdot |S_j|^{\eps'}\leq |S_j\cap A|\cdot k^{\eps'}$. Recall also that the running time of the algorithm is bounded by $O(|E_{G'}(A)|\cdot |E(G')|^{\eps}\log m)\leq O(|E_{G'}(A)|\cdot m^{\eps'}\log m)$.

	Let $C=G'[A']$. We add cluster $C$ to set $\cset$, and we delete from $G'$ all vertices of $A$. We also increase the counter $n_{t'}$ for every  terminal $t'\in A'\setminus A$, and update the class $S_j$ to which the terminal belongs if needed. 
	Note that terminals of $A'\setminus A$ are the only terminals for which new copies are created; terminals of $A$ are added to cluster $C$ but they are removed from graph $G'$.
	Notice also that the edges of $E_{G'}(A)$ are deleted from graph $G'$; we will charge the running time of the current iteration to these vertices.
	
	Observe that $V(C)=A'=B_{G'}(t,2iD)$. Since $i\leq \frac{2r+6}{\eps'}\leq \frac{4}{(\eps')^2}\leq \frac{2^{12}}{\eps^2}$ 
	(since $r=\ceil{1/\eps'}+1$ and $\eps'=\eps/32$), and  $D'\geq \frac{2^{13}\cdot D}{\eps^2}$, we get that $V(C)\subseteq B_{G'}(t,D')\subseteq B_{G}(t,D')$. 
	This establishes Invariant  \ref{inv: few reg vertices}. We now prove that Invariant \ref{inv: ball coverage} also continues to hold by computing, for every terminal $t'\in T$, a cluster $C(t')\in \cset\cup\set{G'}$ with $B_G(t',D)\subseteq V(C(t'))$. Consider any terminal $t'\in T$. If, at the beginning of the iteration, $C(t')\in \cset$ held, then $C(t')$ remains unchanged, and we are guaranteed that $B_G(t',D)\subseteq V(C(t'))$ continues to hold. Consider now some terminal $t'\in T$, for which $C(t')=G'$ held at the beginning of the iteration,  so $B_{G}(t',D)\subseteq V(G')$ held at the beginning of the iteration. If $t'\not\in B_G(t,(2i-1)D)$, then it is easy to verify that $B_{G}(t',D)\subseteq V(G')$ continues to hold at the end of the iteration. Otherwise, $B_{G}(t',D)\subseteq V(C)$ must hold. For every  terminal $t'\in B_{G'}(t,(2i-1)D)$, we set $C(t')=C$, and for all other terminals $t'$, $C(t')$ remains unchanged. From the above discussion, for every terminal $t'\in T$, $B_G(t',D)\subseteq V(C(t'))$ now holds, establishing Invariant \ref{inv: ball coverage}. The updates to clusters $C(t')$ for terminals $t'\in T$ can be performed without increasing the asymptotic running time of the iteration.

	Note that, in the current iteration, the vertices of $A$ are deleted from $G'$. For every terminal $t'\in (A'\setminus A)\cap T$, we create a new copy of $t'$, that is added to cluster $C$ (in addition to the original copy of $t'$ that continues to lie in $G'$). 
	Recall that the algorithm from 
	\Cref{lem: ball growing} guarantees that
	$|T\cap A'|\leq |T\cap A|\cdot k^{\eps'}$. We assign a \emph{charge} of $k^{\eps'}$ units to every  terminal of $A\cap T$, that is responsible for ``paying'' for the  terminals of $(A'\setminus A)\cap T$. Notice that the number of the  terminals of $(A'\setminus A)\cap T$ is bounded by the total charge to the  terminals of $A\cap T$. Therefore, the number of newly created copies of  terminals of $T$ that are added to cluster $C$ is bounded by the charge to the terminals of $A\cap T$. Since the terminals of $A$ are deleted from $G'$, we will never charge them again.
	
	The algorithm terminates when  for every terminal $t\in T$, $C(t)\neq G'$ holds. We then return the final collection $\cset$ of clusters of $G$. Our invariants ensure that, for every cluster $C\in \cset$,  there is some terminal $t\in T$, with $V(C)\subseteq B_G(t,D')$. They also ensure that, for every terminal $t\in T$, there is a cluster $C(t)\in \cset$, with $B_{G}(t,D)\subseteq V(C(t))$, and our algorithm maintains such a cluster $C(t)$ for every  vertex $t$. 
	
	Notice that a  terminal $t\in T$ may be charged at most once by our algorithm, since, when terminal $t$ is charged, it is deleted from $G'$. The charge to every terminal $t\in T$ is bounded by $k^{\eps'}$. For every  terminal $t'\in T$, every new copy of $t'$ that is created by our algorithm, is charged to other  terminals. Since the total number of  terminals is $k$, and the charge to every  terminal is at most $k^{\eps'}$, we get that $\sum_{C\in \cset}|T\cap V(C)|\leq k^{1+\eps'}$. We will use this fact later, in order to bound the maximum number of copies of a  vertex that the algorithm creates.
	
	Next, we bound the running time of the algorithm. From our discussion, if a cluster $C$ was added to $\cset$ during some iteration $i$ of our algorithm, and the set $A\subseteq V(C)$ of vertices was deleted from $G'$ during the iteration, then the running time of that iteration is bounded by $O\left (|E_G(A)|\cdot m^{\eps'}\cdot \log n\right )$. It is then easy to verify that the total running time of the algorithm is bounded by: 
	$$O\left(|E(G)|\cdot m^{\eps'}\cdot \log m+k\right )\leq O\left (m^{1+\eps'}\cdot \log m+k\right )\leq O\left (k+m^{1+\eps/32}\cdot \log n\right ).$$
	
	Lastly, we prove that every  terminal of $T$ belongs to at most $\frac{64k^{\eps/16}}{\eps}$ clusters of $\cset$. 
	Recall that $r=\ceil{1/\eps'}+1$, $\eps'=\eps/32$, and set $S_r$ contains all  terminals that appear in at least $k^{2\eps'}\cdot r=k^{\eps/16}\cdot \left(\ceil{32/\eps}+1\right )$ clusters of $\cset\cup \set{G'}$.
	We show below that $S_r=\emptyset$ holds throughout the algorithm. It will then follow that every terminal appears in at most $k^{\eps/16}\cdot \left(\ceil{32/\eps}+1\right )\leq \frac{64k^{\eps/16}}{\eps}$ clusters of $\cset$. Therefore, in order to complete the proof of \Cref{claim: cutting G}, it is now enough to prove that $S_r=\emptyset$.

	For all $1\leq j\leq r$, we denote by $S_{\geq j}=S_j\cup S_{j+1}\cup\cdots\cup S_{r}$. Note that, once a  terminal $t$ is added to set $S_{\geq j}$, it remains in this set until the end of the algorithm. 
	We prove the following observation.
	
	\begin{observation}\label{obs: terminal growth}
		For all $1\leq j\leq r$, $|S_{\geq j}|\leq k^{1-j\eps'}$ holds throughout the algorithm.
	\end{observation}
	\begin{proof}
		Since, for all $1\leq j\leq r$, terminals may join set $S_{\geq j}$ over the course of the algorithm, but they may never leave it,
		it is enough to prove that, at the end of the algorithm, $|S_{\geq j}|\leq k^{1-j\eps'}$ holds for all $1\leq j\leq r$. The proof is by induction on $j$.
		
		The base is $j=1$. From the charging scheme that we have described above, every  terminal of $T$ receives a charge of at most $k^{\eps'}$, and $\sum_{C\in \cset}|V(C)\cap T|\leq k^{1+\eps'}$. A terminal may belong to set $S_1$ only if at least $k^{2\eps'}$ copies of the terminal lie in the clusters of $\cset$. Therefore, $|S_1|\leq \frac{k^{1+\eps'}}{k^{2\eps'}}\leq k^{1-\eps'}$.

		Consider now some integer $j>1$, and assume that the claim holds for $j-1$. Consider some terminal $t\in T$, that was added to set $S_{\geq j}$ at some time $\tau$ during the algorithm's execution. Let $\tau'<\tau$ be the time when terminal $t$ was added to set $S_{\geq (j-1)}$. Then at time $\tau'$, $n_t=(j-1)\cdot k^{2\eps'}+1$ held, and at time $\tau$, $n_t=j\cdot k^{2\eps'}+1$ held. Therefore, between time $\tau'$ and $\tau$, terminal $t$ belonged to class $S_{j-1}$, and during that time, $k^{2\eps'}$ new copies of this terminal were created.
		
		Consider now some iteration of the algorithm, when a new cluster $C$ was created by applying the algorithm from \Cref{lem: ball growing} to graph $G'$ and some terminal $t$. Let $i$ be the integer that the algorithm returned, let  $A=B_G(v,2(i-1)D)$, and let $A'=B_G(v,2iD)$. Recall that the only terminals for which new copies were created during this iteration are the terminals of $(A'\setminus A)\cap T$, and $|S_{j-1}\cap A'|\leq |S_{j-1}\cap A|\cdot k^{\eps'}$ held. We think of the new copies of the terminals as being added to the new cluster $C$, while graph $G'$ contains their original copies. Vertices of $A$ are deleted from graph $G'$ in the current iteration. We assign, to every terminal of $S_{j-1}\cap A$ a \emph{charge} of $k^{\eps'}$. Notice that the total charge assigned to all terminals of $S_{j-1}\cap A$ is at least as large as the number of terminals of $S_{j-1}$ that lie in $A'\setminus A$, so the charge is at least as large as the number of new copies of terminals of $S_{j-1}$ that were created in the current iteration. As the terminals of $A$ are deleted from graph $G'$ in the current iteration, they will never be charged again for any terminals of $S_{j-1}$. Therefore, a terminal that ever belonged to set $S_{j-1}$ may only be charged at most once for creating new copies of terminals of $S_{j-1}$, and the amount of the charge is $k^{\eps'}$. Since, from the induction hypothesis, at the end of the algorithm, $|S_{\geq (j-1)}|\leq k^{1-(j-1)\eps'}$ holds, the total number of copies of terminals of $S_{j-1}$ that were ever created during the algorithm is bounded by $|S_{\geq (j-1)}|\cdot k^{\eps'}\leq k^{1-j\eps'+2\eps'}$. As discussed already, in order for a terminal of $S_{j-1}$ to join set $S_{j}$, we need to create at least $k^{2\eps'}$ new copies of that terminal. We conclude that the total number of terminals that ever belonged to set $S_j$ over the course of the algorithm is bounded by $k^{1-j\eps'}$. If terminal $t$ belongs to $S_{\geq j}$ at the end of the algorithm, then it must have belonged ot $S_j$ at some time during the algorithm. Therefore, at the end of the algorithm, $|S_{\geq j}|\leq k^{1-j\eps'}$ holds.
	\end{proof}

	Since $r= \ceil{1/\eps'}+1$, we conclude that set $S_r$ of terminals remains empty throughout the algorithm, and so every  terminal of $T$ lies in at most $\frac{64k^{\eps/16}}{\eps}$ clusters of $\cset$.
\end{proof}

\iffalse
\mynote{probably not needed}
\paragraph{Decremental Connectivity/Spanning Forest.}
We use the results of~\cite{dynamic-connectivity}, who provide a deterministic data structure, that we denote by $\CONNSF(G)$, that, given an $n$-vertex unweighted undirected graph $G'$, that is subject to edge deletions, maintains a spanning forest of $G$, with total update time $O((m+n)\log^2n)$, where $m$ is the number of edges in the initial graph $G$. Moreover, the data structure supports connectivity queries $\conn(G,u,v)$: given a pair  $u,v$ of vertices of $G$, return ``yes'' if $u$ and $v$ are connected in $G$, and ``no'' otherwise. The running time to respond to each such query is  $O(\log n/\log\log n)$. 
%Since the data structure maintains a spanning forest of $G$, we can also use it to respond to a query $\path(G,u,v)$: given two vertices $u$ and $v$ in $G$, return any simple path connecting $u$ to $v$ in $G$ if such a path exists, and return $\emptyset$ otherwise. If $u$ and $v$ belong to the same connected component $C$ of $G$, then the running time of the query is $O(|V(C)|)$.
\fi

\subsection{Dynamic Graphs and Other Dynamic Objects}
Throughout the paper, we will consider graphs $G$ that undergo an online sequence $\Sigma=(\sigma_1,\sigma_2,\ldots,\sigma_q)$ of update operations. For now it may be convenient to think of the update operations  as being edge deletions, though we will consider additional update operations later. After each update operation (e.g. edge deletion), our algorithm will perform some updates to the data structures that it maintains. 
We refer to different ``times'' during the algorithm's execution. We refer to time 0 as the time at which the data structures of the algorithm have been initialized, but no updates from $\Sigma$ have yet occurred. For each integer $1\leq \tau\leq q$, we refer to ``time $\tau$ in the algorithm's execution'' as the time immediately after all updates to the data structures maintained by the algorithm following the $\tau$th update operation $\sigma_{\tau}\in \Sigma$ are completed. 
The \emph{time horizon} $\tset=\set{0,1,\ldots,|\Sigma|}$ is the time interval from the beginning of the algorithm, and until all updates in $\Sigma$ have been processed. Note that we view the time horizon as a collection of discreet time points.

When we say that some property holds at every time during the time horizon (or throughout an algorithm's execution), we mean that the property holds at each time $\tau\in \tset$. The property may not hold, for example, during the procedure that updates the data structures maintained by the algorithm, following some input update operation $\sigma_{\tau}\in \Sigma$.
For $\tau\geq 0$, we denote by $G\attime$ the graph $G$ at time $\tau$; that is, $G\attime[0]$ is the original graph, and for $\tau\geq 0$, $G\attime$ is the graph obtained from $G$ after the first $\tau$ update operations $\sigma_1,\ldots,\sigma_{\tau}$. 

Assume now that we are given a dynamic graph $G$, that undergoes a sequence of edge-deletion and edge-insertion updates. Whenever an edge $(u,v)$ is deleted from $G$, and then inserted into $G$ again, we view it as two different edges. For example, if we start with $E(G)=\emptyset$, and then iteratively insert edge $(u,v)$ into $G$ and then delete it $r$ times, then the total number of edges that ever belonged to $G$ over the course of this update sequence is $r$.

\paragraph{Other Dynamic Objects.}
Consider an algorithm that is applied to some dynamic graph $G$, with time horizon $\tset$. Suppose we define a dynamic set $S\subseteq V(G)$ of vertices, that may change over time. We assume that the set $S$ of vertices is initialized at some time $\tau\in \tset$, and undergoes changes afterwards, during which vertices may be added to or removed from $S$. We say that the set $S$ of vertices is \emph{decremental}, if, after the set $S$ of vertices is initialized at time $\tau$, vertices may leave it, but no new vertices may join it. We say that the set $S$ of vertices is \emph{incremental}, if, after $S$ is initialized at time $\tau$, vertices may join it, but they may not leave it. We may also consider other dynamic collections of objects, for which the notion of decremental or incremental set is defined similarly.

\subsection{Fully Dynamic Graphs and Vertex-Splitting}
Suppose we are given a  graph $G$ with lengths $\ell(e)$ on its edges $e\in E(G)$, that undergoes an online sequence $\Sigma$ of updates.
Recall that we say that $G$ is a fully-dynamic graph, if each of the updates in $\Sigma$ is either an edge deletion or an edge insertion. 
We will sometimes consider two other types of update operations. The first type is \emph{isolated vertex deletion}: given a vertex $x$ that is an isolated vertex in the current graph $G$, delete $x$ from $G$. The second type of update is \emph{vertex-splitting}: given a vertex $v\in V(G)$, and a non-empty subset $E'\subseteq \delta_G(v)$ of its adjacent edges, insert a new vertex $v'$ into $G$, and, for every edge $e=(v,u)\in E'$, insert an edge $(v',u)$ of length $\ell(e)$ into graph $G$. We will sometimes consider graphs that only undergo edge-deletion, isolated vertex-deletion, and vertex-splitting update operations. Intuitively, vertex-splitting operations insert edges into $G$, but this type of edge-insertions is relatively easy to deal with. On the other hand, if we are given a bound on the total number of vertices present in a dynamic graph $G$,  that undergoes edge-deletion, edge-insertion, isolated vertex-deletion, and vertex-splitting operations, then we can view $G$ as a standard fully-dynamic graph, that only undergoes edge-deletions and edge-insertions. In other words, vertex-splitting operations can be implemented via edge-insertions, and we will ignore isolated vertices. This is formally summarized in the next lemma.

\begin{lemma}\label{lem: vertex splitting to fully dynamic}
	Let $G$ be a graph that undergoes a sequence $\Sigma$ of online update operations of four types: edge-deletion, edge-insertion, isolated vertex-deletion, and vertex-splitting. Let $\tset$ be the time horizon associated with $\Sigma$, and assume that, for all $\tau\in \tset$, $|V(G\attime)|\leq N$ holds. Then there is a dynamic graph $H$ that, after initialization, undergoes an online sequence $\Sigma'$ of edge-insertions and edge-deletions, such that $|V(H)|=N$, and, for all $\tau\in \tset$, if we let $\tilde G\attime$ be the graph obtained from $G\attime$ by deleting all isolated vertices from it, and we let $\tilde H\attime$ be obtained similarly from $H\attime$, then $\tilde H\attime=\tilde G\attime$ holds. Moreover, there is a deterministic algorithm, that, given $G\attime[0]$, $N$, and the online update sequence $\Sigma$ for $G$, initializes the graph $H=H\attime[0]$, and computes the online update sequence $\Sigma'$ for $H$. The total update time of the algorithm is $O(|E(G)|+N+L(\Sigma))$, where $L(\Sigma)$ is the number of bits in the description of the update sequence $\Sigma$.
\end{lemma}

\begin{proof}
	The dynamic graph $H$ is defined as follows: at all times $\tau\in \tset$, graph $H\attime$ is obtained from graph $G\attime$ by adding a collection $S$ of $N-|V(G\attime)|$ isolated vertices to it, that we refer to as \emph{spare} vertices.
	Clearly, if we let $\tilde G\attime$ and $\tilde H\attime$ be the graphs obtained from $G\attime$ and $H\attime$, respectively, by deleting all isolated vertices from them, then $\tilde H\attime=\tilde G\attime$ holds.
	
	Given the initial graph $G\attime[0]$, it is immediate to compute the corresponding initial graph $H\attime[0]$, in time $O(|E(G)|+N)$. We now provide an algorithm for computing the update sequence $\Sigma'$ of edge-insertions and edge-deletions that allows us to maintain graph $H$ correctly.
	
	Let $\sigma\in \Sigma$ be an update to graph $G$. If $\sigma$ is the deletion of an edge $e$ from $G$, then we delete edge $e$ from $H$. If $\sigma$ is the insertion of an edge $e'$ into $G$, then we insert edge $e'$ into $H$. If $\sigma$ is the deletion of an isolated vertex $x$ from graph $G$, then we do not perform any updates to graph $H$, but we add vertex $x$ to the set $S$ of spare vertices.
	
	Lastly, assume that $\sigma$ is a vertex-splitting operation, that is applied to some vertex $v\in V(G)$, and a set $E'\subseteq \delta_G(v)$ of its incident edges. Since we are guaranteed that $|V(G)|\leq N$ holds at all times, it must be the case that $S\neq \emptyset$ currently holds. We let $x\in S$ be any spare vertex, that we will identify with $v'$ from now on, until $v'$ is deleted from $G$. We remove vertex $v'$ from the set $S$ of spare vertices. For every edge $e=(v,u)\in E'$, we then perform the insertion of the edge $(v',u)$ into graph $H$.
	
	It is easy to verify that, at all times $\tau\in \tset$, the graph that is obtained from $H\attime[0]$ by applying the sequence of edge-deletions and edge-insertions from $\Sigma'$ up to time $\tau$ is indeed $H\attime$. It is also easy to verify that the running time of the algorithm is  $O(|E(G)|+N+L(\Sigma))$.	
\end{proof}

\subsection{Basic and Modified Even-Shiloach Tree}
Suppose we are given a graph $G=(V,E)$ with integral lengths $\ell(e)\geq 1$ on its edges $e\in E$, a source $s$, and a distance bound $D\geq 1$. The Even-Shiloach Tree (\EST) algorithm of~\cite{EvenS,Dinitz,HenzingerKing}  maintains, for every vertex $v$ with $\dist_G(s,v)\leq D$, the distance $\dist_G(s,v)$, under the deletion of edges from $G$. It also maintains a shortest-path tree $\tau$ rooted at  $s$, that includes all vertices $v$ with $\dist_G(s,v)\leq D$. We denote the corresponding data structure by $\EST(G,s,D)$. %When the distance bound $D$ is unbounded (that is, $D=nL$, where $L=\max_{e\in E}\set{\ell(e)}$), then we denote the data structure by $\EST(G,s)$.
The total update time of the algorithm  is $O(m\cdot D\log n)$, where $m$ is the initial number of edges in $G$ and $n=|V|$. Throughout this paper, we refer to the corresponding data structure as \emph{basic \EST}. 

Note that the \EST data structure only supports decremental graphs. While we do not currently have similar data structures for fully dynamic graphs, in some cases \EST can be maintained under some limited types of edge insertions. Different data structures supporting different restricted kinds of edge insertions were considered in the past. In this paper, we need to extend the \EST data structure to support dynamic graphs that undergo three types of updates:
edge-deletion; isolated vertex-deletion; and vertex-splitting. The latter operation may not be applied to the source vertex $s$.

\iffalse
 of the following types:

\begin{itemize}
	\item Edge deletion: given an edge $e\in E(G)$, delete $e$ from $G$;
	\item Isolated vertex deletion: given a vertex $v\in V(G)$ that is currently an isolated vertex, delete $v$ from $G$; and
	\item Vertex splitting: given a vertex $v\in V(G)\setminus\set{s}$, and a non-empty subset $E'\subseteq \delta_G(v)$ of its incident edges, insert a new vertex $v'$ into $G$, and, for every edge $e=(v,u)\in E'$, insert an edge $(v',u)$ of length $\ell(e)$ into $G$. 
\end{itemize}
\fi

In the following theorem we extend the \EST data structure so that it can handle all update operations described above. The proof of the theorem is standard, and it is almost identical to a similar theorem that was proved in \cite{APSP-previous}. A similar data structure was used, either explicity or implicitly, in numerous other previous papers. For completeness, we provide the proof of the theorem in Section \ref{sec: appx-ES-tree} of Appendix.

\begin{theorem}\label{thm: ES-tree}
	There is a deterministic algorithm, that we refer to as \emph{modified \EST}, whose input is a connected graph $G=(V,E)$  with integral lengths $\ell(e)\geq 1$ on its edges $e\in E$, a source $s$, and a distance bound $D\geq 1$, such that the length of every edge in $G$ is at most $D$, with graph $G$ undergoing an online sequence of edge-deletion, isolated vertex-deletion and vertex-splitting updates (but vertex-splitting may not be applied to $s$). The algorithm  supports  $\shortestpath$  queries: given a vertex $x\in V(G)$, either correctly establish, in time $O(1)$, that $\dist_H(s,x)>D$, or return a shortest $s$-$x$ path $P$ in $G$, in time $O(|E(P)|)$. The algorithm also maintains a collection $S^*=\set{v\in V(G)\mid \dist_G(v,s)>D}$ of vertices. The total update time of the algorithm is $O(m^*\cdot D\cdot \log m^*)$, where $m^*$ is the total number of edges that ever belonged to graph $G$.
\end{theorem}

\subsection{Well-Connected Graphs}

We will employ well-connected graphs, that were introduced in
\cite{DMG}, together with some related algorithmic tools. Intuitively, we will replace expander graphs, that were used in the algorithm of \cite{APSP-previous} for \APSP with well-connected graphs. The main motivation for replacing expanders with well-connected graphs is that, typically, the use of expander graphs in distance-based problems  leads to a super-logarithmic loss in the approximation, while well-connected graphs and the algorithmic tools associated with them were explicitly designed to overcome this difficulty.
We start by defining well-connected graphs. The definition is identical to that in \cite{DMG}.

\begin{definition}[Well-Connected Graph] 
	Given an $n$-vertex graph $G$,  a set $S$ of its vertices called \emph{supported vertices}, and parameters $\eta,d>0$, we say that graph $G$ is $(\eta,d)$-well-connected with respect to $S$, if, for every pair $A,B\subseteq S$ of disjoint equal-cardinality subsets of supported vertices, there is a collection $\pset$ of paths in  $G$, that connect every vertex of $A$ to a distinct vertex of $B$, such that the length of each path in $\pset$ is at most $d$, and every edge of $G$ participates in at most $\eta$ paths in $\pset$.
\end{definition}	
	
For intuition, it would be convenient to think of $d=2^{\poly(1/\eps)}$, $\eta=n^{O(\eps)}$, and $|S|\geq |V(G)|-n^{1-\eps}$, for some precision parameter $0<\eps<1$.

The work of \cite{DMG} provides a fast algorithm, that, given a graph $G$ and a set $T$ of its vertices called terminals, either computes a well-connected graph $X$ that is defined over a large subset of $T$, together with its embedding into $G$, or returns two large sets $T_1,T_2\subseteq T$ of terminals, together with a relatively small set $E'$ of edges, such that $\dist_{G\setminus E'}(T_1,T_2)$ is large. Another algorithm that  \cite{DMG} provides, and that we will exploit, is for decremental \APSP in well-connected graphs. Given a graph $X$ that undergoes an online sequence of edge deletions, and a set $S$ of its vertices, such that $X$ is well-connected with respect to $S$, the algorithm maintains a large subset $S'\subseteq S$ of vertices, and supports \shortpath\ queries between vertices of $S'$: given a pair $x,y\in V(S')$ of such vertices, it returns a path $P$ connecting $x$ to $y$ in $X$, such that the length of $P$ is small, and the time required to respond to the query is $O(|E(P)|)$. However, the algorithm for decremental \APSP in well-connected graphs requires one additional input, called a \emph{\HSS}, for graph $X$. Intuitively, \HSS is a hierarchy of well-connected graphs that are embedded into each other, with graph $X$ being the topmost graph in the hierarchy. The algorithm for embedding a well-connected graph into a given graph $G$, that we mentioned above, in case it produces a well-connected graph $X$ and its embedding into $G$, also returns the required \HSS for graph $X$, that can then be used by the algorithm for the \APSP problem on the well-connected graph $X$. Therefore, the specifics of the definition of the \HSS are not important for us: the algorithm for embedding a well-connected graph will produce exactly the kind of \HSS that the algorithm for \APSP needs to use, and we will only apply the algorithm for \APSP in well-connected graphs to graphs $X$ that were obtained via the  embedding algorithm. But for completeness, we provide the definition of the \HSS from \cite{DMG} here.

\paragraph{\HSS.}
The \HSS uses two main parameters: the base parameter $N>0$, and a level parameter $j>0$. We also assume that we are given a precision parameter $0<\eps<1$. The notion of Hierarchical Support Structure is defined inductively, using the level parameter $j$. If $X$ is a graph containing $N$ vertices, then a level-1 \HSS for $X$ simply consists of a set $S(X)$ of vertices of $X$, with $|V(X)\setminus S(X)|\leq N^{1-\eps^4}$. Assume now that we are given a graph $X$ containing exactly $N^j$ vertices. A level-$j$ Hierarchical Support Structure for $X$ consists of a collection $\hset=\set{X_1,\ldots,X_r}$ of $r=N-\ceil{2N^{1-\eps^4}}$ graphs, such that for all $1\leq i\leq r$, $V(X_i)\subseteq V(X)$; $|V(X_i)|=N^{j-1}$; and $|E(X_i)|\leq N^{j-1+32\eps^2}$. We also require that $V(X_1),\ldots,V(X_r)$ are all mutually disjoint. Additionally, it must contain, for all $1\leq i\leq r$, a level-$(j-1)$ Hierarchical Support Structure for $X_i$, which in turn must define the set $S(X_i)$ of supported vertices for graph $X_i$. We require that each such graph $X_i$ is $(\eta_{j-1},\td_{j-1})$-well-connected with respect to $S(X_i)$, where $\tilde d_{j-1}=2^{\tilde c(j-1)/\eps^4)}$ for some constant $\tilde c$, and $\eta_{j-1}=N^{6+256(j-1)\eps^2}$. Lastly, the Hierarchical Support Structure for graph $X$ must contain an embedding of graph $X'=\bigcup_{i=1}^rX_i$ into $X$, via path of length at most $2^{O(1/\eps^4)}$, that cause congestion at most $N^{O(\eps^2)}$.  We then set $S(X)=\bigcup_{i=1}^rS(X_i)$, and we view $S(X)$ as the set of supported vertices for graph $X$, that is defined by the \HSS.

\paragraph{Embedding of a well-connected graph.}
We will use the following theorem from \cite{DMG}.

\begin{theorem}[Corollary 5.3 from \cite{DMG}]\label{cor: HSS witness}
	There is a deterministic algorithm, whose input consists of an $n$-vertex graph $G$, a set $T$ of $k$ vertices of $G$ called terminals, and parameters $\frac{2}{(\log k)^{1/12}}< \eps<\frac{1}{400}$, $d>1$ and $\eta>1$, such that $1/\eps$ is an integer. The algorithm computes one of the following:

	\begin{itemize}
		\item either a pair $T_1,T_2\subseteq T$ of disjoint subsets of terminals, and a set $E'$ of edges of $G$, such that:
		
		\begin{itemize}
			\item $|T_1|=|T_2|$ and $|T_1|\geq \frac{k^{1-4\eps^3}}{4}$;
			\item $|E'|\leq \frac{d\cdot |T_1|}{\eta}$; and
			\item for every pair $t\in T_1,t'\in T_2$ of terminals, $\dist_{G\setminus E'}(t,t')>d$;
		\end{itemize} 
		
		\item or a graph $X$ with $V(X)\subseteq T$, $|V(X)|=N^{1/\eps}$, where $N=\floor{k^{\eps}}$,  and maximum vertex degree at most   $k^{32\eps^3}$, together with an embedding $\pset$ of $X$ into $G$ via paths of length at most $d$ that cause congestion at most $\eta\cdot k^{32\eps^3}$, and a level-$(1/\eps)$ \HSS for $X$, such that $X$ is $(\eta',\td)$-well-connected with respect to the set $S(X)$ of vertices defined by the support structure, where $\eta'=N^{6+256\eps}$, and $\td=2^{\tilde c/ \eps^5}$, where $\tilde c $ is the constant used in the definition of the \HSS.
	\end{itemize}
	The running time of the algorithm is  $O\left (k^{1+O(\eps)}+|E(G)|\cdot k^{O(\eps^3)}\cdot(\eta+d\log n)\right )$.
\end{theorem}

\iffalse
\paragraph{Algorithm for the Distancing Player.}
We use the algorithm for the Distancing Player of the \DMG from \cite{DMG}. The algorithm either produces the desired $(\delta,d)$-distancing in the current graph $X$, or constructs a level-$\ceil{1/\eps}$ \HSS for $X$, together with a large set $S(X)$ of supported vertices, such that $X$ is well-connected with respect to $S(X)$.

\begin{theorem}[Theorem 2.2 in \cite{DMG}]\label{thm: construct HSS last level}
	There is a deterministic algorithm, whose input consists of a parameter $0<\eps<1/4$, such that $1/\eps$ is an integer, an integer $N>0$, and a graph $X$ with $|V(X)|=N^{1/\eps}$, such that $N$ is sufficiently large, so that $\frac{N^{\eps^4}}{\log N}\geq 2^{128/\eps^5}$ holds. The algorithm computes one of the following:
	
	\begin{itemize}
		\item either a $(\delta,d)$-distancing $(A,B,E')$ in  $X$, where $\delta=4\eps^3$, $d=2^{32/\eps^4}$ and $|E'|\leq \frac{|A|}{N^{\eps^3}}$; or
		\item a level-$(1/\eps)$ \HSS for $X$, such that  graph $X$ is $(\eta,\td)$-well-connected with respect to the set $S(X)$ of vertices defined by the support structure, where $\eta=N^{6+O(\eps)}$ and $\td=2^{O(1/\eps^5)}$.
	\end{itemize}
	The running time of the algorithm is bounded by: $O(|E(X)|^{1+O(\eps)})$.
\end{theorem}
\fi

\paragraph{\APSP in Well-Connected Graphs.}

Lastly, we need an algorithm for decremental \APSP in well-connected graphs from \cite{DMG}. Assume that we are given a graph $X$ that is obtained from \Cref{cor: HSS witness}. In other words, we are given a level-$(1/\eps)$ \HSS for $X$, together with a large set $S(X)$ of its vertices, so that $X$ is well-connected with respect to $S(X)$. We then assume that graph $X$ undergoes a sequence of edge deletions. As edges are deleted from $X$, the well-connectedness property may no longer hold, and the \HSS may be partially destroyed. Therefore, we only require that the algorithm maintains a large enough subset $S'(X)\subseteq S(X)$ of supported vertices, and that it can respond to \shortpath\ queries between pairs of vertices in $S'(X)$: given a pair $x,y$ of such vertices, the algorithm needs to return a path of length at most $2^{O(1/\eps^6)}$ in the current graph $X$ connecting them. We also require that the set $S'(X)$ is \emph{decremental}, so vertices can leave this set but they may not join it. The algorithm is summarized in the following theorem.

\begin{theorem}[Theorem 2.3 in \cite{DMG}]\label{thm: APSP in HSS full}
	There is a deterministic algorithm, whose input consists of:
	
	\begin{itemize}
		\item a parameter $0<\eps<1/400$, so that $1/\eps$ is an integer;
		\item an integral parameter $N$ that is sufficiently large, so that $\frac{N^{\eps^4}}{\log N}\geq 2^{128/\eps^6}$ holds;
		\item  a graph $X$ with $|V(X)|=N^{1/\eps}$; and
		\item a level-$(1/\eps)$ \HSS for  $X$, such that $X$ is $(\eta',\td)$-well-connected with respect to the set $S(X)$ of vertices defined by the \HSS, where $\eta'$ and $\td$ are parameters from \Cref{cor: HSS witness}.  
	\end{itemize} 
	Further, we assume that graph $X$ undergoes an online sequence of at most $\Lambda=|V(X)|^{1-10\eps}$ edge deletions. The algorithm maintains a set $S'(X)\subseteq S(X)$ of vertices of $X$, such that, at the beginning of the algorithm, $S'(X)=S(X)$, and over the course of the algorithm, vertices can leave $S'(X)$ but they may not join it. The algorithm ensures that $|S'(X)|\geq \frac{|V(X)|}{2^{4/\eps}}$ holds at all times, and it supports short-path queries between vertices of $S'(X)$: given a pair $x,y\in S'(X)$ of vertices, return a path $P$ connecting $x$ to $y$ in the current graph $X$, whose length is at most $2^{O(1/\eps^6)}$, in time $O(|E(P)|)$. The total update time of the algorithm is $O( |E(X)|^{1+O(\eps)})$. 
\end{theorem}

\subsection{Useful Inequalities}
\label{subsec: useful inequalities}

In many of our algorithms, we will be given a pair $W>0$, $\eps$ of parameters, such that $\frac{1}{(\log W)^{1/24}}\leq \eps\leq 1/400$ holds. It will be useful for us to establish several simple bounds that muat hold for these parameters. Observe first that:

\begin{equation}\label{eq: large W}
W>2^{1/\eps^{24}}\geq 2^{400^{24}},
\end{equation}

so in particular:

\begin{equation}\label{eq: large W 2}
W^{\eps}>2^{1/\eps^{23}}.
\end{equation}

Since, from \Cref{eq: large W}, parameter $W$ must be sufficiently large, we get that:

\[ \frac{1}{\eps^{12}}\leq (\log W)^{1/2}<\frac{\log W}{2\log\log W}, \]

and so:

\begin{equation}\label{eq: large W3}
W^{\eps^{12}}>\log^2 W.
\end{equation}

\section{Valid Input Structure, Valid Update Operations, and the Recursive Dynamic Recursive Neighborhood Cover Problem}
\label{sec: valid inputs}

Throughout this paper, we will work with inputs that have a specific structure. This structure is identical to the one defined in \cite{APSP-previous}, and it is designed in a way that will allow us to naturally compose different instances recursively, by exploiting the notion of neighborhood covers. In order to avoid repeatedly defining such inputs, we provide a definition here, and then refer to it throughout the paper. We also define the types of update operations that we allow for such inputs. After that, we formally define the Recursive Dynamic Neighborhood Cover problem (\recdynnc), and provide several useful observations and simple algorithmic tools for the problem.  In this section we also state our algorithm for the \recdynnc problem with slightly weaker guarantees, that allows us to prove 
\Cref{thm: main final dynamic NC algorithm}.

\subsection{Valid Input Structure and Valid Update Operations}

We start by defining a valid input structure; the definition is used throughout the paper and is intended as a shorthand for the types of inputs that many of our subroutines use. The definition is identical to the one from \cite{APSP-previous}.

\begin{definition}[Valid Input Structure]
	A valid input structure consists of a bipartite graph $H=(V,U,E)$, a distance threshold $D>0$ and integral lengths $1\leq \ell(e)\leq D$ for edges $e\in E$. 
	The vertices in set $V$ are called \emph{regular vertices} and the vertices in set $U$ are called \emph{supernodes}.
	We denote a valid input structure by $\iset=\left(H=(V,U,E),\set{\ell(e)}_{e\in E(H)}, D\right )$. If the distance threshold $D$ is not explicitly defined, then we set it to $\infty$.
\end{definition}

Intuitively, supernodes in set $U$ may represent clusters in a Neighborhood Cover $\cset$ of the vertices in $V$ with some (smaller) distance threshold, that is computed and maintained recursively. % 
Given a valid input structure $\iset=\left(H,\set{\ell(e)}_{e\in E(H)}, D\right )$, we allow the following types of update operations:

\begin{itemize}

\item {\bf Edge Deletion.} Given an edge $e\in E(H)$, delete $e$ from $H$.

\item {\bf Isolated Vertex Deletion.} Given a vertex $x\in V(H)$ that is an isolated vertex, delete $x$ from $H$; and

%\paragraph{Special Node Splitting.} Given a super-node $u\in U$ that is not an isolated vertex, and a value $0<\ell\leq D$, add a new regular vertex $v$ to graph $H$ where it joins the set $V$. Add a new edge $e=(u,v)$ to $H$, of length $\ell$.  Set $w(v)=1$ and decrease $w(u)$ by $1$. %. If $u\in U_i$, then set $\ell(e)=2^i$ (in other words, $\ell(e)$ is set to be equal to the lengths of all other edges incident to $u$). Set $w(v)=1$ and decrease $w(u)$ by $2$. 
%We will sometimes say that $v$ is a \emph{child node} of $u$. The operation may only be performed if, after the update, $w(u)\geq |\delta_H(u)|+|\notdelta(u)|+1$ holds.

%\item {\bf Edge Length Increase.} Given an edge $e\in E(H)$, and a value $2\ell(e)\leq \ell'\leq D$, set the length of edge $e$ to be $\ell'$.

\item {\bf Supernode Splitting.} The input to this update operation is a supernode $u\in U$ and a {\bf non-empty} subset $E'\subseteq \delta_H(u)$ of edges incident to $u$.
The update operation creates a new supernode $u'$, and, for every edge $e=(u,v)\in E'$, it adds a new edge $e'=(u',v)$ of length $\ell(e)$ to the graph $H$.  We will sometimes refer to $e'$ as a \emph{copy of edge $e$}.
\end{itemize}

For brevity of notation, we will refer to edge-deletion, isolated vertex-deletion, and supernode-splitting operations as \emph{valid update operations}.
Notice that valid update operations may not create new regular vertices, so vertices may be deleted from the vertex set $V$, but never added to it. A supernode splitting operation, however, adds a new supernode to graph $H$, and also inserts edges into $H$.
Unfortunately, this means that the number of edges in $H$ may grow as the result of the update operations, making it challenging to analyze the running times of various algorithms that we run on subgraphs $C\subseteq H$ in terms of $|E(C)|$. 
In order to overcome this difficulty, we use the notion of the \emph{dynamic degree bound}, which was also defined in \cite{APSP-previous}.

\begin{definition}[Dynamic Degree Bound]
We say that a valid input structure \newline  $\iset=\left(H=(V,U,E),\set{\ell(e)}_{e\in E(H)}, D\right )$, undergoing an online sequence $\Sigma$ of valid update operations has \emph{dynamic degree bound} $\mu$ if, for every regular vertex $v\in V$, the total number of edges incident to $v$ that are ever present in $H$ over the course of the time horizon $\tset$ is at most $\mu$. 
\end{definition}

We will usually denote by $N^0(H)$ the  number of regular vertices in the initial graph $H$. If $(\iset,\Sigma)$ have dynamic degree bound $\mu$, then we are guaranteed that the number of edges that are ever present in $H$ over the course of the update sequence $\Sigma$ is bounded by $N^0(H)\cdot \mu$.

In general, we will always ensure that the dynamic degree bound $\mu$ is quite low. It may be convenient to think of it as $m^{\poly(\eps)}$, where $m$ is the initial number of edges in the input graph $G$ for the \APSP problem, and $\eps$ is a precision parameter. Intuitively, every supernode of graph $H$ represents some cluster $C$ in a  $(\hat D,\hat D')$-neighborhood cover $\cset$ of $G$, for some parameters $\hat D,\hat D'\ll D$. Typically, each regular vertex of $H$ represents some actual vertex of graph $G$,  and an edge $(v,u)$ is present in $H$ iff vertex $v$ belongs to the cluster $C$ that supernode $u$ represents. Intuitively, we will ensure that the neighborhood cover $\cset$ of $G$ is constructed and maintained in such a way that the total number of clusters of $\cset$ to which a given regular vertex $v$ ever belongs over the course of the algorithm is small. This, in turn, will ensure that the dynamic degree bound for graph $H$ is small as well. %As an example of a recursive composition of such instances, we may be interested in computing a neighborhood cover of the resulting graph $H$, which will in turn define a neighborhood cover of $G$, for distance parameters $D,D'$, where $D$ is much larger than $\hat D'$. %This intuitive explanation shows one way in which the valid input structure and the corresponding edge replication bounds are defined and exploited. For this scenario, as already mentioned, we could just bound the total number of edges of $E'$ incident to a single vertex. But we will also use an additional different method for composing instances recursively, for which the edge-replication bound as defined above is necesary.

Note that we can assume without loss of generality that every vertex in the original graph $H\attime[0]$ has at least one edge incident to it, as otherwise it is an isolated vertex, and will remain so as long as it lies in $H$. Moreover, from the definition of the supernode-splitting operation, it may not be applied to an isolated vertex (as we require that the edge set $E'$ is non-empty). Therefore, any isolated vertex of $H\attime[0]$ can be ignored. We will therefore assume from now on that every supernode in the original graph $H\attime[0]$ has degree at least $1$. (This assumption is only used for convenience, so that we can bound the total number of vertices in $H\attime[0]$ by $O(|E(H\attime[0])|)$.)

The following simple observation, that was proved in \cite{APSP-previous}, shows that the distances in the graph $H$ may not decrease as the result of a valid update operation.

\begin{observation}[Observation 3.1 in full version of \cite{APSP-previous}]\label{obs: no dist increase}
	Consider the graph $H$ at any time during the execution of the sequence $\Sigma$ of valid update operations, and let $x,x'$ be any two vertices of $H$. Let $H'$ be the graph obtained after a single valid update operation on $H$. Then, if $x,x'\in V(H')$, then $\dist_{H'}(x,x')\geq \dist_{H}(x,x')$.
\end{observation}

\paragraph{Ancestors of Supernodes.}
Since supernodes may be inserted into graph $H$ as part of the update sequence that it undergoes, it will be convenient for us to track this process via the notion of \emph{ancestors}, which is defined in a natural way.

Let $\iset=\left(H,\set{\ell(e)}_{e\in E(H)},D\right )$ be a valid input structure, that undergoes a sequence $\Sigma$ of valid update operations, and let $\tset$ be the time horizon corresponding to $\Sigma$.
Let $u$ be a supernode that lies in $H$ at any time during time interval $\tset$, and let $\hat \tau(u)\in \tset$ be the last time when $u\in V(H)$ holds. For every time $\tau\in \tset$ with $\tau\leq \hat \tau(u)$, we define  a supernode $\anc\attime(u)$, that lies in $H\attime$, and is called \emph{an ancestor-supernode, or just an ancestor, of $u$ at time $\tau$}. 

The definition of the ancestor-supernode is inductive over the time when supernode $u$ was first added to graph $H$.
Assume first that supernode $u$ lies in the initial graph $H\attime[0]$. Then for every time $\tau\in \tset$ with $\tau\leq \hat \tau(u)$, we set $\anc\attime(u)=u$, so this supernode is an ancestor of itself.  Assume now that supernode $u$ does not belong to the initial graph $H\attime[0]$, and let $\tau'\in \tset$ be the time when $u$ was added to graph $H$. Then $u$ was added via the supernode-splitting operation, applied to some supernode $u'$. For every time $\tau\in \tset$ with $\tau\leq \hat \tau(u)$, if $\tau<\tau'$, we set $\anc\attime(u)=\anc\attime(u')$, and otherwise we set $\anc\attime(u)=u$.

We also need the following simple observation.

\begin{observation}\label{obs: supernode edge tracking}
	Let $\iset=\left(H,\set{\ell(e)}_{e\in E(H)},D\right )$ be a valid input structure, that undergoes a sequence $\Sigma$ of valid update operations, and let $\tset$ be the corresponding time horizon.
	Let $e=(u,v)$ be an edge that is present in graph $H$ at some time $\tau\in \tset$, with $u\in U$ and $v\in V$. Then for all $\tau'\in \tset$ with $\tau'<\tau$, there is an edge $e'=(\anc\attime[\tau'](u),v)$ in graph $H\attime[\tau']$, whose length is $\ell(e)$.
\end{observation}
\begin{proof}
The proof is by induction on $\tau-\tau'$. The base of the induction is when $\tau=\tau'+1$. Assume that edge $e=(u,v)$ is present in graph $H$ at time $\tau$. If supernode $u$ is present in graph $H$ at time $\tau-1$, then $\anc\attime[\tau'](u)=u$, and, since edges may only be inserted into $H$ via supernode splitting operation, edge $(u,v)$ must be present in $H$ at time $\tau-1$, and its length may not change over the course of the algorithm.

Otherwise, there must be a supernode $u'$ that lies in graph $H$ at time $\tau-1$, so that $\anc\attime[\tau'](u)=u'$. Then supernode $u$ was added to graph $H$ at time $\tau$ via a supernode splitting operation applied to $u'$. But then, from the definition of the supernode splitting operation, there must be an edge $e'=(u',v)$ in graph $H$ at time $\tau-1$, whose length is $\ell(e)$.

Assume now that we are given some integer $i>1$, and that the claim holds for all $0\leq \tau'<\tau$ with $\tau-\tau'<i$. Consider now some edge $e=(u,v)$ that was present in graph $H$ at some time $\tau\in \tset$, and consider the time $\tau'=\tau-i$. Let $u''=\anc\attime[\tau-1](u)$ and $u'=\anc\attime[\tau'](u)$. From the definition of ancestor-supernodes, $u'=\anc\attime[\tau'](u'')$ must also hold. We apply the induction hypothesis to indices $\tau$ and $\tau-1$, to conclude that edge $e'=(u'',v)$ of length $\ell(e')=\ell(e)$ is present in graph $H\attime[\tau-1]$. We then apply the induction hypothesis again to indices $\tau-1$ and $\tau'$, together with supernode $u''$ and edge $e'$ to conclude that edge $e''=(u',v)$ of length $\ell(e'')=\ell(e')=\ell(e)$ is present in graph $H\attime[\tau']$.
\end{proof}

\iffalse
\subsection{Extended Even-Shiloach Tree}
The following theorem is a slight extension of the \EST data structure of \cite{EvenS,Dinitz,HenzingerKing}, that accommodates all types of valid update operatins. The proof of the theorem is standard, and can be found in \cite{APSP-previous}. We note that a similar data structure was used, either explicity or implicitly, in numerous other previous papers.

\begin{theorem}\label{thm: ES-tree}
	There is a deterministic algorithm, that we refer to as \emph{generalized \EST}, that, given a valid input structure $\iset=\left(H,\set{\ell(e)}_{e\in E(H)},D\right )$, where graph $H$ undergoes a sequence $\Sigma$ of valid update operations with dynamic degree bound $\mu$, and given additionally a source vertex $s\in V(H)$, and a distance threshold $D^*>0$, such that the length of every edge in $H$ is bounded by $D^*$, supports  $\shortestpath$  queries: given a vertex $x\in V(H)$, either correctly establish, in time $O(1)$, that $\dist_H(s,x)>D^*$, or return a shortest $s$-$x$ path $P$ in $H$, in time $O(|E(P)|)$. The total update time of the algorithm is $\otilde(N^0\cdot \mu \cdot D^*)$, where $N^0$ is the number of regular vertices in the initial graph $H$.
\end{theorem}
\fi

\subsection{The Recursive Dynamic Neighborhood Cover (\recdynNC) Problem}

In this subsection we provide a formal definition of the Recursive Dynamic Neighborhood Cover problem from \cite{APSP-previous}.

\paragraph{Problem Definition.}
The input to the Recursive Dynamic Neighborhood Cover (\recdynNC) problem is a valid input structure $\iset=\left(H=(V,U,E),\set{\ell(e)}_{e\in E},D \right )$, where graph $H$ undergoes an online sequence $\Sigma$ of valid update operations with some given dynamic degree bound $\mu$. Additionally, we are given a desired approximation factor $\alpha$. We assume that we are also given some arbitrary fixed ordering $\oset$ of the vertices of $H$, and that any new vertex that is inserted into $H$ as the result of supernode-splitting updates is added at the end of the current ordering.
The goal is to maintain the following data structures:

\begin{itemize}
	\item   a collection $\uset$ of subsets of vertices of graph $H$, together with a collection $\cset=\set{H[S]\mid S\in\uset}$ of clusters in $H$, such that $\cset$ is a weak $(D,\alpha \cdot D)$ neighborhood cover for the set $V$ of regular vertices in graph $H$.
		For every set $S\in \uset$, the vertices of $S$ must be maintained in a list, sorted according to the ordering $\oset$;
	\item for every regular vertex $v\in V$, a cluster  $C=\clustercover(v)$, with $B_H(v,D)\subseteq V(C)$;
	\item for every vertex $x\in V(H)$, a list $\clusterlist(x)\subseteq \cset$ of all clusters containing $x$, and for every edge $e\in E(H)$, a list $\clusterlist(e)\subseteq \cset$ of all clusters containing $e$.
\end{itemize} 

The set $\uset$ of vertex subsets must be maintained as follows. Initially, $\uset=\set{V(H\attime[0])}$, where $H\attime[0]$ is the initial input graph $H$. After that, the only allowed changes to vertex sets in $\uset$ are:

\begin{itemize}
	\item $\delvertex(S,x)$: given a vertex set $S\in \uset$, and a vertex $x\in S$, delete $x$ from $S$; 
	\item $\addsupernode(S,u)$: if $u$ is a supernode that is lying in $S$, that just underwent a supernode splitting update, add the newly created supernode $u'$ to $S$; and
	\item $\csplit(S,S')$: given a vertex set $S\in \uset$, and a subset $S'\subseteq S$ of its vertices, add $S'$ to $\uset$.
\end{itemize}

We refer to the above operations as \emph{allowed changes to $\uset$}.
In other words, if we consider the sequence of changes that clusters in $\cset$ undergo over the course of the algorithm, the corresponding sequence of changes to vertex sets in $\set{U(C)\mid C\in \cset}$ must obey the above rules.

We note that, while we require that, at the beginning of the algorithm, $\uset=\set{V(H\attime[0])}$ holds, we allow the data structure to update this initial collection of vertex subsets via allowed operations,  before processing any updates to graph $H$. 
We sometimes refer to the resulting collection $\cset$ of clusters, that is obtained before any update from $\Sigma$ is processed, as \emph{initial collection of clusters}, or \emph{collection of clusters at time $0$}.

While it was convenient for us to define the allowed operations using the collection $\uset$ of subsets of vertices of $H$, in our algorithms we will usually directly work with the corresponding collection $\cset=\set{H[S]\mid S\in \uset}$ of clusters. Therefore it may be convenient for us to say that the allowed operations are applied to the clusters of $\cset$ directly: $\delvertex(C,x)$ deletes a vertex from cluster $C$; $\addsupernode(C,u)$ inserts a supernode into cluster $C$, together with the corresponding collection of edges; and $\csplit(C,C')$ creates a new cluster $C'\subseteq C$, where $C$ is an existing cluster that lies in $\cset$. In the latter case, we say that $C'$ was \emph{split off} from cluster $C$. We note that, in addition to the allowed operations, whenever an edge is deleted from graph $H$, we will usually also delete it from every cluster that contains it.

%We note that whenever an edge $e$ is deleted from graph $H$ as part of input update sequence $\Sigma$, we assume that $e$ is also deleted from every cluster $C\in \cset $ containing $e$, as $C\subseteq H$ must hold at all times. We can assume that this is implemented by automatically triggering $\deledge(C,e)$ operation for every cluster $C\in \cset$ containing $e$ whenever $e$ is deleted from $H$. Similarly, whenever the length of edge $e$ is increased, or an isolated vertex $v$ is deleted from $H$ via the input update sequence $\Sigma$, we mirror these operations in every cluster containing $e$ or $v$. Note that our definition of neighborhood cover requires that every cluster in $\cset$ is a vertex-induced subgraph of $H$. We will ensure that this is the case at every time $t$ of the algorithm (that is, after the $t$th element of the input update sequence $\Sigma$ is processed). In other words, updates to clusters in $\cset$ will come in batches, and we will ensure that this property holds between the batches. We did not include the edge length increase operation explicitly as a cluster-update operation, since the clusters are defined as subgraphs of $H$, so they inherit the lengths of their edges from $H$.

\paragraph{Ancestor Clusters.}
It will be convenient for us to define the notion of \emph{ancestors} of clusters in $\cset$, that is somewhat similar to that of ancestor-supernodes. Let $\tset$ be the time horizon of the update sequence $\Sigma$, and let $C$ be a cluster that ever belonged to $\cset$ over the course of the algorithm. For every time $\tau\in \tset$, we will define an \emph{ancestor} of cluster $C$ at time $\tau$, denoted by $\anc\attime(C)$. The definition is inductive over the time when cluster $C$ was first added to $\cset$.

Consider first the initial set $\cset$ of clusters, that the algorithm constructs prior to processing the first update in $\Sigma$. For every cluster $C\in \cset$, for every time $\tau\in \tset$, we set $\anc\attime(C)=C$, so each such cluster is an ancestor of itself. Consider now some time $\tau'\in \tset$ with $\tau'>0$, when a new cluster $C'$ is added to set $\cset$. Then there is some cluster $C\in \cset$, so that cluster $C'$ was split off from cluster $C$ at time $\tau'$. For every time $\tau\in \tset$, if $\tau<\tau'$, we set $\anc\attime(C')=\anc\attime(C)$, and otherwise we set $\anc\attime(C')=C'$.

\iffalse
The following observation is immediate from our definitions.

\begin{observation}\label{consistent covering property}
	Suppose we are given an algorithm that maintains a data structure for the \recdynnc problem, that ensures the Consistent Covering property. Assume that, at time $\tau$ during the time horizon $\tset$, for a regular vertex $x\in V(H\attime)$, $\coveringcluster(x)=C$ held. Then for all $\tau'<\tau$, at time $\tau'$, $\coveringcluster(x)=\anc\attime(C)$ holds.
\end{observation}
\fi

\paragraph{Consistent Covering Property.}
We require that the data structure for the \recdynnc problem obeys the \emph{Consistent Covering property}, that is defined as follows.

\begin{definition}[Consistent Covering Property]
	We say that a data structure for the \recdynnc problem maintains the \emph{Consistent Covering property}, if the following holds. 
	Consider any times $\tau'<\tau$ during the time horizon, and a regular vertex $x\in V(H\attime)$. Assume that, at time $\tau$, $\coveringcluster(x)=C$ held, and that $\anc\attime[\tau'](C)=C'$. Then, at time $\tau'$, $B_H(x,D)\subseteq V(C')$ held. Here, $D$ is he distance parameter in the input to the \recdynnc problem.
\end{definition}

\iffalse
\begin{definition}[Consistent Covering Property]
	We say that a data structure for the \recdynnc problem maintains the \emph{Consistent Covering property}, if the following holds. 
	Consider any time $\tau$ during the time horizon, and a regular vertex $x\in V(H\attime)$. Assume that, at time $\tau-1$, $\coveringcluster(x)=C$ held, and at time $\tau$, $\coveringcluster(x)=C'$ holds. Then either $C=C'$, or cluster $C'$ was created at time $\tau$ via the cluster splitting operation that was applied to cluster $C$.
\end{definition}
\fi

The Consistent Covering property was not explicitly defined in \cite{APSP-previous}, but the data structures for the \recdynnc problem provided in that work obey this property. We need this property in order to reduce fully-dynamic \APSP to \recdynnc.

In addition to maintaining the above data structures, an algorithm for the \recdynnc problem needs to support queries $\spquery(C,v,v')$: given two {\bf regular} vertices $v,v'\in V$, and a cluster $C\in \cset$ with $v,v'\in C$, return a path $P$ in the current graph $H$, of length at most $\alpha\cdot D$ connecting $v$ to $v'$ in $H$, in time $O(|E(P)|)$. %, for the given approximation factor $\alpha$. We note that path $P$ may not be contained in $C$.
This completes the definition of the \recdynNC problem.
The \emph{size} of an instance $\iset=\left(H=(V,U,E),\set{\ell(e)}_{e\in E},D \right )$ of the \recdynnc instance, that we denote by $N^0(H)$, is the number of regular vertices in the initial graph $H$.

In the remainder of the paper, we will always assume that a data structure that an algorithm for the \recdynnc problem maintains must obey the Consistent Covering property.

We also need the following two simple observations.

\begin{observation}\label{obs: supernode in cluster ancestor}
Suppose we are given an algorithm that maintains a data structure for the \recdynnc problem, and let $\cset$ be the collection of clusters that the algorithm maintains. Let $u$ be a supernode, and let $C\in \cset$ be a cluster, such that, at some time $\tau>0$ during the time horizon, $u\in V(C)$ held. Then for all $0\leq \tau'<\tau$, at time $\tau'$, $\anc\attime[\tau'](u)\in \anc\attime[\tau'](C)$ held.
\end{observation}

\begin{proof}
	It is enough to prove the observation for $\tau'=\tau-1$, since we can then apply it iteratively in order to extend it to all values $0\leq \tau'<\tau$. Therefore, we only prove the observation for $\tau'=\tau-1$.
	
	Let $u'=\anc\attime[\tau'](u)$ and let $C'=\anc\attime[\tau'](C)$. 
	Assume for contradiction that, at time $\tau'$, $u'\not\in V(C')$.
	We consider four cases.
	
	 The first case is when $u'=u$ and $C'=C$. Then supernode $u$ existed at time $\tau-1$, and it joined cluster $C$ at time $\tau$. This is impossible, since a supernode may only join a cluster if it was just created via the supernode-splitting operation.
	
	The second case is when $u'=u$ and $C'\neq C$. Then at time $\tau$, cluster $C$ was split off from cluster $C'$. But then $C\subseteq C'$ must hold, and so $u\in V(C')$ holds at time $\tau-1$.
	
	The third case is when $u'\neq u$ and $C'=C$. In this case, at time $\tau$, supernode $u$ was created via the supernode-splitting operation that was applied to supernode $u'$. Since supernode $u$ was then added to cluster $C$, it must be the case that $u'\in V(C)$ held at time $\tau-1$.
	
	Lastly, from our definition of time slots, it is possible that $u'\neq u$ and $C'\neq C$ both hold: that is, at time $\tau$,  cluster $C$ was created by splitting it off from cluster $C'$, and supernode $u$ was created via supernode-splitting operation applied to $u'$. These updates were performed one after another, and, from the analysis of Case 2 and Case 3, at time $\tau'$, $u'\in V(C')$ must have held.
\end{proof}

\subsubsection{Bounding the Distance Parameter $D$}
\label{subsubsec: bounding D}

Suppose we are given a valid update structure   $\iset=\left(H=(V,U,E),\set{\ell(e)}_{e\in E},D \right )$ that  undergoes an online sequence $\Sigma$ of valid update operations, with dynamic degree bound $\mu$, and let $N$ be the number of regular vertices in $H$ at the beginning of the algorithm. 
Assume that $(\iset,\Sigma)$ is an instance of the \recdynnc problem. 
We show, using standard techniques, that, at the cost of losing a factor $3$ in the approximation ratio, we can assume that $D=3N$, and that all edge lengths are integers between $1$ and $D$. We will use this simple observation multiple times.

Recall that, since $\iset$ is a valid input structure, all edges in $H$ have lengths at most $D$. We set the length of each  edge $e$ to be $\ell'(e)=\ceil{N\cdot \ell(e)/D}$.

For every pair $x,y$ of vertices, let $\dist'(x,y)$ denote the distance
between $x$ and $y$ with respect to the new edge length values.
Notice that for every pair $x,y$ of vertices, $\frac{N}{D}\cdot \dist(x,y)\leq\dist'(x,y)\leq\frac{N}{D}\cdot \dist(x,y)+2N$,
since the shortest $x$-$y$ path contains at most $2N$ edges. 

Therefore, if $\dist(x,y)\leq D$, then $\dist'(x,y)\leq 3N$. Moreover, if $P$ is an $x$-$y$ path with $\ell'(P)\leq \alpha N$, then $\ell(P)\leq \alpha D$ must hold.
It is now enough to solve the \recdynnc problem on graph $H$ with the new edge weights $\ell'(e)$ for $e\in E(H)$, and distance bound $D'=3N$.

\subsubsection{Updating Clusters of a Graph $H$}
\label{subsubsec: updating clusters}

Let  $\iset=\left(H=(V,U,E),\set{\ell(e)}_{e\in E},D \right )$ be a valid input structure, and assume that it undergoes an online sequence $\Sigma$ of valid update operations with associated time horizon $\tset$. Let $\tau\in \tset$ be any time point, and let $C$ be a vertex-induced subgraph of $H\attime[\tau]$. As graph $H$ undergoes updates (after time $\tau$), we will typically need to update cluster $C$ accordingly. Specifically, let $\sigma\in \Sigma$ be any update operation for graph $H$ that occurred after time $\tau$. If $\sigma$ is the deletion of an edge $e$, then, if $e\in E(C)$, we delete $e$ from $C$ as well. If $\sigma$ is the deletion of an isolated vertex $x$ from $H$, then, if $x\in V(C)$, we delete $x$ from $C$ as well. Assume now that $\sigma$ is a supernode-splitting operation, applied to supernode $u$ and a subset $E'\subseteq \delta_H(u)$ of edges. If $u\in V(C)$, and $E'\cap E(C)\neq\emptyset$, then we apply supernode-splitting operation in cluster $C$, to vertex $u$ and the set $E''=E'\cap E(C)$ of edges. Therefore, if we denote by $\Sigma'\subseteq \Sigma$ the sequence of update operations that graph $H$ undergoes since time $\tau$, then $\Sigma'$ naturally defines the corresponding sequence $\Sigma_C$ of valid update operations for cluster $C$. It is easy to verify that, if the dynamic degree bound of $(H,\Sigma)$ is $\mu$, then the dynamic degree bound of $(C,\Sigma_C)$ is at most $\mu$. We need the following simple observation.

\begin{observation}\label{obs: maintain ball covering property}
	Let $\iset=\left(H=(V,U,E),\set{\ell(e)}_{e\in E},D \right )$ be a valid input structure, and assume that it undergoes an online sequence $\Sigma$ of valid update operations, with associated time horizon $\tset$. Let $\tau\in \tset$ be any time point, and let $C$ be a vertex-induced subgraph of $H\attime$. Denote by $\Sigma'\subseteq\Sigma$ the sequence of updates that graph $H$ undergoes from time $\tau$ onwards, and let $\Sigma_C$ be the corresponding update sequence for $C$. For all $\tau'\in \tset$ with $\tau'>\tau$, let $C\attime[\tau']$ be the graph obtained from $C$ after applying the sequence $\Sigma_C$ of updates to it up to time $\tau'$. Then for all $\tau'\in \tset$ with $\tau'>\tau$, $C\attime[\tau']$ is a vertex-induced subgraph of $H\attime[\tau']$. Furthermore, for every regular vertex $v\in V$ and distance parameter $D'$, if $B_{H\attime}(v,D')\subseteq C\attime$, then for all  $\tau'\in \tset$ with $\tau'>\tau$, 
$B_{H\attime[\tau']}(v,D')\subseteq C\attime[\tau']$ holds.
\end{observation}
\begin{proof}
	It is immediate to verify that $C$ remains a vertex-induced subgraph of $H$ at all times $\tau'\in \tset$ with $\tau'>\tau$ by inspecting the changes to graphs $H$ and $C$ as the result of a single update operation.
	
	Consider now some regular vertex $v\in V$ and distance parameter $D'$, such that $B_{H\attime}(v,D')\subseteq C\attime$. We define two dynamic graphs: graph $G$ is the subgraph of $H$ induced by the set $B_{H}(v,D')$ of vertices, and graph $G'$ is the subgraph of $C$ induced by the set $B_C(v,D')$ of vertices. At time $\tau$, $G=G'$ must hold. It is easy to verify that, for every update operation $\sigma\in \Sigma$ that is applied to graph $H$ after time $\tau$, if $\sigma$ modifies graph $G$, then the corresponding update $\sigma'\in \Sigma_C$ leads to an identical modification of graph $G'$, and if $\sigma$ does not affect $G$, then neither does $\sigma'$. Therefore, for all $\tau'\in \tset$ with $\tau'>\tau$, $G\attime[\tau']=(G')\attime[\tau']$ holds. It is then immediate to see that  $B_{H\attime[\tau']}(v,D')\subseteq C\attime[\tau']$ holds for all such time points $\tau'$.
\end{proof}

\subsection{Main Technical Result for the \recdynnc Problem and Proof of \Cref{thm: main final dynamic NC algorithm}}

As one of our main technical results, we will prove the following theorem. 
The proof of the theorem appears in \Cref{sec: proof of recdynnc inner}.

\begin{theorem}\label{thm: main final dynamic NC algorithm inner}
	There is a deterministic algorithm for the \recdynNC problem, that,  given a valid input structure $\iset=\left(H=(V,U,E),\set{\ell(e)}_{e\in E},D \right )$ undergoing a sequence of valid update operations, with dynamic degree bound $\mu$, together with parameters $\hat W$ and  $1/(\log \hat W)^{1/100}\leq \eps<1/400$,  such that, if we denote by $N^0(H)$ the number of regular vertices in $H$ at the beginning of the algorithm, then $N^0(H)\cdot \mu\leq \hat W$ holds, achieves approximation factor $\alpha=(\log\log \hat W)^{2^{O(1/\eps^2)}}$,  with total update time $O((N^0(H))^{1+O(\eps)}\cdot \mu^{O(1/\eps)}\cdot D^3)$. Moreover, the algorithm ensures that for every regular vertex $v\in V$, the total number of clusters in the weak neighborhood cover $\cset$ that the algorithm maintains, to which vertex $v$ ever belongs over the course of the algorithm, is bounded by $\hat W^{4\eps^4}$.
\end{theorem}

Note that the guarantees provided by \Cref{thm: main final dynamic NC algorithm inner} are somewhat weaker than those required by \Cref{thm: main final dynamic NC algorithm}, in that the total update time of the algorithm depends polynomially on $D$. We can remove this polynomial dependence on $D$ using standard techniques; a similar idea was used in \cite{APSP-previous}, who also initially provided an algorithm for \recdynnc whose running time depended polynomially on $D$, and then removed this dependence. We complete the proof of \Cref{thm: main final dynamic NC algorithm} from \Cref{thm: main final dynamic NC algorithm inner} in Section \ref{subsec: reduce dependence on D} of Appendix. This part is almost identical to a similar proof from \cite{APSP-previous} (see the proof of Theorem 3.4 in \cite{APSP-previous}). The main difference is that the proof from \cite{APSP-previous} only dealt with a special case where the dynamic degree bound $\mu=2$, so no supernode-splitting updates were allowed; and the proof from  \cite{APSP-previous} did not need to explicitly establish the Consistent Covering property, though their algorithm ensured it.

עד כאן

\section{From \recdynnc to Fully Dynamic APSP -- Proof of \Cref{thm: main fully-dynamic APSP single scale}}
\label{sec: fully APSP inner main}

\iffalse
The majority of this section is dedicated to the proof of \Cref{thm: main fully-dynamic APSP single scale}. At the end of this section, we also provide the proof of \Cref{cor: T-emulator}, that builds on the data structures that the algorithm from  \Cref{thm: main fully-dynamic APSP single scale} maintains. We start with the proof of \Cref{thm: main fully-dynamic APSP single scale}.
\fi

This section is dedicated to the proof of \Cref{thm: main fully-dynamic APSP single scale}.
We assume that \Cref{assumption: alg for recdynnc2} holds, that is, 
	there is a deterministic algorithm for the \recdynNC problem, that,  given a valid input structure $\iset=\left(H=(V,U,E),\set{\ell(e)}_{e\in E},D \right )$ undergoing a sequence of valid update operations, with dynamic degree bound $\mu$, together with parameters $\hat W$ and  $1/(\log \hat W)^{1/100}\leq \eps<1/400$,  such that, if we denote by $N$ the number of regular vertices in $H$ at the beginning of the algorithm, then $N\cdot \mu\leq  \hat W$ holds, achieves approximation factor $\alpha(\hat W)$,  with total update time $O(N^{1+O(\eps)}\cdot \mu^{O(1/\eps)})$. The algorithm also ensures that, for every regular vertex $v\in V$, the total number of clusters in the weak neighborhood cover $\cset$ that the algorithm maintains, to which vertex $v$ ever belongs over the course of the algorithm, is bounded by $\hat W^{4\eps^3}$. Here, $\alpha(\cdot)$ is a non-decreasing function.
We denote the algorithm for \recdynnc problem with the above properties by $\aset$.

We assume that we are given an instance of the $D^*$-restricted \APSP problem, that consists of
an $n$-vertex graph $G$ with integral lengths $\ell(e)\geq 1$ on its edges $e\in E(G)$, together with a precision parameter $\frac{1}{(\log n)^{1/200}}<\eps<1/400$, and a distance parameter $D^*>0$, where graph $G$ undergoes an online sequence of edge insertions and deletions.
For convenience, we denote $\alpha=\alpha(n^3)$, where $\alpha(\cdot)$ is the approximation factor from \Cref{assumption: alg for recdynnc2}.

Let $m^*$ be the total number of edges that are ever present in $G$.  
We can assume w.l.o.g. that $n/2\leq m^*\leq n^2$ holds. In order to do so, we partition our algorithm into phases. At the beginning of each phase, we consider the current graph $G$, and denote by $n'$ the number of vertices of $G$ that are not isolated, and by $m'$ the number of edges currently in graph $G$, so $m'\geq n'/2$. Let $G'$ be the graph obtained from $G$ by deleting all isolated vertices, and then inserting a new set $S$ of $m'$ isolated vertices, that we refer to as spare vertices. 
We now consider the update sequence to graph $G$. Whenever an edge $e$ is deleted from $G$, we delete the same edge from $G'$. If an edge $e=(x,y)$ is inserted into $G$, and both $x$ and $y$ are vertices of $V(G')\setminus S$, we insert edge $e$ into $G'$. Otherwise, it must be the case that either $x$ or $y$ (or both) were isolated vertices in $G$ prior to the edge insertion. If $x\not\in V(G')\setminus S$, and $S\neq \emptyset$, we select an arbitrary vertex $v$ from $S$, that is identified with $x$ from now on, and delete it from $S$. Similarly, if $y\not\in V(G')\setminus S$, and $S\neq \emptyset$, we select an arbitrary vertex $v'$ from $S$, that is identified with $y$ from now on, and delete it from $S$. We then insert edge $e$ into $G'$. Once $m'/2$ edges have been inserted into $G$ since the beginning of the phase, the phase terminates. Notice that $S\neq\emptyset$ must hold throughout the phase.  We can view each of these phases as a separate instance of the \APSP problem (for the first phase, we consider the graph $G$ after the first edge insertion). It is now sufficient to design an algorithm for a single phase. Therefore, we assume from now on that $n/2\leq m^*\leq n^2$ holds.
Note that, from our discussion, if $m$ is the number of edges in the initial graph $G$, then the total number of edges that may be inserted over the course of the algorithm is bounded by $m$. Therefore, we can assume that $|V(G)|+|E(G\attime[0])|+|\Sigma|\leq 4m$.
We note that it may no longer be the case that
$\eps>\frac{1}{(\log m)^{1/200}}$, if $m$ is significantly smaller than the number of vertices in graph $G$ that served as input to \Cref{thm: main fully-dynamic APSP single scale}. We use another parameter $\hat m$, that is the maximum between $4m$ and the number of vertices in the original graph $G$, so that $\eps>\frac{1}{(\log \hat m)^{1/200}}$ holds. This will allow us to use Inequalities \ref{eq: large W}--\ref{eq: large W3} from  \Cref{subsec: useful inequalities}, with parameter $W$ replaced by $\hat m$.

From now on we denote by $m$ the number of edges in  graph $G$ that we obtain after the above transformation. %, and by $n$ the number of vertices in $G$.  

To summarize, from now on we assume that we are given an initial  $m$-edge graph $G$, that undergoes a sequence $\Sigma$ of online edge insertions and deletions, during which at most $m$ edges may be inserted into $G$, and $|V(G)|+|E(G\attime[0])|+|\Sigma|\leq 4m$.
We are also given a parameter $\hat m\geq 4m$, and a precision parameter $\frac{1}{(\log \hat m)^{1/200}}<\eps<1/400$. 
It is sufficient to design a deterministic algorithm for the $D^*$-restricted $\APSP$ problem on $G$,  that achieves approximation factor $\alpha'=(\alpha)^{O(1/\eps)}$, and has amortized update time at most $\hat m^{O(\eps)}$ per operation.
Equivalently, it is sufficient that the total update time of the algorithm is bounded by $O(m\cdot \hat m^{O(\eps)})$.
Query time for  \shortpath\ queries should be bounded by $O\left(2^{O(1/\eps)}\cdot \log \hat m\right )$ (unless we are required to return a path connecting the queried vertices, in which case the additional time required to respond to the query should be bounded by $O\left(|E(P)|\right )$, where $P$ is the returned path). Using the arguments described in \Cref{subsubsec: bounding D}, at the cost of losing a factor $4$ in the approximation ratio, we can assume that $D^*\le 4\hat m$, and that $D^*$ is an integral power of $2$.
 
% We now assume that we are given a graph $G$, with integral length $\ell(e)\geq 1$ on its edges, that udergoes a sequence $\Sigma$ of edge deletion and edge insertion operations, together with a precision parameter $\frac{2}{(\log m)^{1/100}}<\eps<1/400$, where 
% $|\Sigma|+|V(G)|+|E(G\attime[0])|=m$, and a distance parameter $D^*>0$. Using the arguments from \Cref{subsubsec: bounding D}, at the cost of losing a factor $4$ in the approximation, we can assume that $D^*\leq 3m$, and that $D^*$ is an integral power of $2$.
% Our goal is to design a data structure that supports \shortpath\ queries, as given in the definition of the $D^*$-restricted \APSP problem.

% For convenience, we denote $\alpha(m^{1+2\eps})$ by $\alpha$. 
 
 Throughout, we use the parameters $q=\ceil{1/\eps}$  and $M=\ceil{m^{\eps}}$.
Notice that $m\leq M^q\leq m^{1+2\eps}$. 
%It will be convenient for us to ensure that the sequence $\Sigma$  of updates to graph $G$ contains exactly $M^q$ edge insertions: otherwise, we can add, at the end of sequence $\Sigma$, a sequence of update operations that repeatedly insert and delete the same edge, to ensure that the above property holds. Note that $|V(G)|+|E(G\attime[0])|\leq M^q$.
% For convenience, we will denote $m=M^q$; since   $M^q\leq m^{1+2\eps}$, this update will not change any asymptotic bounds in the theorem statement, and we are still guaranteed that $\frac{1}{(\log m)^{1/100}}<\eps<1/400$ holds.
% Since every edge that is deleted from $G$ either belonged to $G\attime[0]$, or was inserted into $G$, we get that the number of edge deletions in $\Sigma$ is at most $2m$, and $|\Sigma|+|V(G)|+|E(G\attime[0])|\leq 4m$.

As usual, we let $\tset$ be the time horizon associated with the update sequence $\Sigma$.
Consider now graph $G$ at some time $\tau\in \tset$, and let $e\in E(G\attime)$ be any edge of $G$. We say that an edge $e\in E(G)$ is \emph{original}, if it was present in $G$ at the beginning of the algorithm, and was never deleted or inserted. If edge $e$ is not an original edge, then we say that it is an \emph{inserted} edge.

\paragraph{Distance Scales.}
Throughout, we use a parameter $\hat D=D^*\cdot 2^{10q+10}=D^*\cdot 2^{O(1/\eps)}$.
Note that, from Inequality \ref{eq: large W 2}, and since we have assumed that  $D^*\le 4\hat m$, we get that

\begin{equation}\label{eq: bound on D}
\hat D\leq D^*\cdot (\hat m)^{\eps^{20}}\leq 4\hat m^{1+\eps^{20}},
\end{equation}

and, from Inequality \ref{eq: large W3}:

\begin{equation} \label{eq: bound 2 on D}
\log\hat D\leq O(\log \hat m)\leq \hat m^{\eps^{12}}.
\end{equation}

For all $0\leq i\leq \log \hat D$, we define a distance scale $D_i=2^i$.

The data structure that our algorithm maintains is partitioned into $(q+1)$ levels. In order to describe the purpose of each level, we first need to define a partition of the time horizon into phases.

\paragraph{Hierarchical Partition the Time Horizon into Phases.}
We define a hierarchical partition of the time horizon into phases. Specifically, for every level $0\leq L\leq q$, we define a partition of the time horizon $\tset$ into \emph{level-$L$ phases}. For all $0\leq L'< L\leq q$, we will ensure that every level-$L$ phase is completely contained in some level-$L'$ phase.

There is a single level-$0$ phase, that spans the whole time horizon $\tset$. For all $0< L\leq q$, we partition the time horizon into at most $M^L$ level-$L$ phases, each of which spans a consecutive sequence $\Sigma'\subseteq\Sigma$ of updates, that contains exactly $M^{q-L}$ edge insertions (except for the last phase, that may contain fewer insertions).
In other words, if the $k$th level-$L$ phase ends at time $\tau$, then the $\tau$th update operation in $\Sigma$ is edge-insertion, and, since the beginning of the current level-$L$ phase, exactly $M^{q-L}$ edges have been inserted into $G$ via sequence $\Sigma$.
It will be convenient for us to ensure that the number of level-$L$ phases is exactly $M^L$. If this is not the case, then we add empty phases at the end of the last phase.

 For $1\leq k\leq M^{L}$, we denote the $k$th level-$L$ phase by $\Phi^L_k$, and the subsequence of $\Sigma$ containing all update operations that occur during Phase $\Phi^L_k$ by $\Sigma^L_k$.
We also associate the time interval $\tset^L_k$, corresponding to the update sequence $\Sigma^L_k$, with the level-$L$ phase $\Phi^L_k$. For all $0\leq L\leq q$, we will initialize the level-$L$ data structure from scratch at the beginning of each level-$L$ phase. Note that each level-$q$ phase only spans a single edge insertion. In other words, every time a new edge is inserted into $G$, we start a new level-$q$ phase, and recompute the level-$q$ data structure from scratch. Notice that our definition of phases indeed ensures that,  for all $0\leq L'< L\leq q$, every level-$L$ phase is completely contained in some level-$L'$ phase.  Intuitively, for all $1\leq L\leq q$, during each level-$L$ phase $\Phi^L_k$, the level-$L$ data structure will be ``responsible'' for all edges that were inserted into $G$ before the beginning of Phase $\Phi^L_k$, but after the beginning of the current level-$(L-1)$ phase. We now formalize this intuition.

\paragraph{Edge and Path Classification.}
Consider a level $0< L\leq q$, and some level-$L$ phase $\Phi^L_k$. Let $\Phi^{L-1}_{k'}$ be the unique level-$(L-1)$ phase that contains Phase $\Phi^L_k$. Let $\tau\in \tset$ be the beginning of Phase $\Phi^L_k$, and let $\tau'\in \tset$ be the beginning of Phase $\Phi^{L-1}_{k'}$ (note that it is possible that $\tau=\tau'$). We define the set $A^L_k$ of edges of graph $G$ that is associated with Phase $\Phi^L_k$. An edge $e$ belongs to set $A^L_k$ if and only if it was inserted into $G$ between time $\tau'$ and time $\tau$ (including time $\tau'$ and excluding time $\tau$). Notice that the cardinality of set $A^L_k$ is bounded by the  number of edges that may be inserted into $G$ during a single level-$(L-1)$ phase, so $|A^L_k|\leq M^{q-L+1}$. The set $A^L_k$ of edges does not change over the course of Phase $\Phi^L_k$. We also denote by $S^L_k$ the collection of vertices of $G$ that serve as endpoints to the edges of $A^L_k$. Intuitively, level-$L$ data structure is responsible for keeping track of the edges in set $A^L_k$, over the course of each level-$L$ phase $\Phi^L_k$. We will construct and maintain a level-$L$ graph $H^L$, that is initialized from scratch at the beginning of each level-$L$ phase $\Phi^L_k$, whose set of regular vertices contains a vertex representing every edge in $A^L_k$, and a vertex representing every vertex in $S^L_k$. Observe that, as the level $L$ increases, the cardinalities of the corresponding sets $A^L_k$ of edges decrease, so the graphs that we maintain are smaller. At the same time, as $L$ grows, the number of level-$L$ phases also grows. We will ensure that the time that is required to maintain a level-$L$ data structure over a course of each level-$L$ phase $\Phi^L_k$ is almost linear in $|A^L_k|$, allowing us to bound the total update time of the data structure maintained at each level by a function that is almost linear in $m$. 

For cosistency of notation, we let $\Phi^0_1$ denote the single level-$0$ phase, we let $A^0_1$ be the set of all edges that belonged to $G$ at the beginning of the algorithm. 

Consider again some time $\tau\in \tset$. For all $0\leq L\leq q$, we let $k_L$ be the integer, such that $\tau\in \tset^L_{k_L}$ holds. We partition all edges of the current graph $G\attime$ into $q+1$ \emph{levels}. For $0\leq L\leq q$, edge $e$ belongs to level $L$, if and only if $e\in A^L_{k_L}$. It is easy to see that, for every edge $e$ that lies in graph $G$ at time $\tau$, there is precisely one level in $\set{0,\ldots,q}$, to which edge $e$ belongs. We denote the level of edge $e$ by $\level(e)$. Note that, as the algorithm progresses, the level of a given edge may only decrease.

Consider again graph $G$ at time $\tau$, and let $P$ be any path that is contained in $G\attime$, with $|E(P)|\geq 1$. The \emph{level} of path $P$, denoted by $\level(P)$, is the largest level of any of its edges, $\level(P)=\max_{e\in E(P)}\set{\level(e)}$. 

For all $0\leq L\leq q$, the purpose of the level-$L$ data structure is to support \shortpath\ queries between pairs of vertices $x,y\in V(G)$, such that there exists a level-$L$ path in the current graph $G$ connecting $x$ to $y$, whose length is at most $D^*$. Since every path connecting $x$ to $y$ in $G$ belongs to one of the levels in $\set{0,\ldots,q}$, this will allow us to support \shortpath\ queries as required from the definition of $D^*$-restricted \APSP.

\paragraph{High-Level Description of the Construction.}
Consider some level $0\leq L\leq q$, and some level-$L$ phase $\Phi^L_k$. As noted already, at the beginning of Phase $\Phi^L_k$, we initialize the level-$L$ data structure from scratch. Let $\tau \in\tset$ denote the time when Phase $\Phi^L_k$ begins. Note that $\tau$ may also be a starting time of phases from other levels. In such cases, we assume that, when we execute the algorithm for initializing the level-$L$ data structure, then for all $0\leq L'<L$, the level-$L'$ data structure is already initialized.

Over the course of the level-$L$ phase $\Phi^L_k$, we will maintain a dynamic graph $H^L$. We will also initialize the corresponding valid input structure $\iset^L$, associated with graph $H^L$, that will undergo a sequence of valid update operations. The set of regular vertices of graph $H^L$ consists of two subsets: set $\set{v^L(x)\mid x\in S^L_k}$ of vertices, that represent the endpoints of the edges of $A^L_k$, and set 
$\set{v^L(e)\mid e\in A^L_k}$ of vertices, representing the edges of $A^L_k$. We refer to the former as \emph{type-1 regular vertices} and to the latter as \emph{type-2 regular vertices}. We describe the collection of supernodes of $H^L$ later.

For all $0\leq i\leq \log \hat D$, we will define and maintain a subgraph $H^L_i$, which is identical to $H^L$, but it excludes all edges whose length is above $D_i$. We will also define the corresponding valid input structure $\iset^L_i$. We will view $\iset^L_i$ as the input to the \recdynnc problem, with distance scale $D_i$, and we will apply Algorithm $\aset$ from \Cref{assumption: alg for recdynnc2} to it. We denote by $\cset^L_i$ the collection of clusters that this algorithm maintains. For every cluster $C\in \cset^L_i$, we say that the \emph{scale} of cluster $C$ is $i$, and we denote $\scale(C)=i$. We also denote $\cset^L=\bigcup_{i=0}^{\log \hat D}\cset^L_i$ and $\cset^{<L}=\bigcup_{L'<L}\cset^{L'}$.

We now provide additional details on the structure of the graph $H^L$, and specifically its supernodes and its edges. The collection of the supernodes of $H^L$ consists of two subsets. The first subset contains, for every vertex $x\in S^L_k$, the corresponding supernode $u^L(x)$, that connects, with an edge of length $1$, to the type-1 regular vertex $v^L(x)$. Additionally, for every edge $e\in A^L_k$, such that $x$ is an endpoint of $e$, we add an edge $(v^L(e),u^L(x))$ of length $\ell(e)$ to graph $H^L$. We refer to all supernodes we have defined so far as \emph{type-1 supernodes}. The second set of supernodes, called \emph{type-2 supernodes}, contains, 
for {\bf some} clusters $C\in \cset^{<L}$, the corresponding supernode $u^L(C)$. 

In order to decide which clusters of $\cset^{<L}$ have the corresponding supernode included in graph $H^L$, and in order to define the edges that are incident to such supernodes, we will define, for every cluster $C\in \cset^{<q}$, a decremental set $V^F(C)$ of vertices of $G$, which we call a  \emph{flattened set of vertices}. The specific definition of this set of vertices is somewhat technical and is deferred for later. For a cluster $C\in \cset^{<L}$, we add a supernode $u^L(C)$ to graph $H^L$ if and only if $V^F(C)$ contains at least one vertex of $S^L_k$. If supernode $u^L(C)$ is included in graph $H^L$, then we connect it with an edge to every type-1 regular vertex $v^L(x)$, for which $x\in V^F(C)$ holds. The length of the edge is $2^{\scale(C)}$. We now proceed to provide intuition on the flattened sets of vertices.

\paragraph{Flattened Sets of Vertices.}
Consider some level $0\leq L\leq q$, and some cluster $C\in \cset^L$. Intuitively, our layered constructions has created a hierarchical containment structure for the clusters: if, for some cluster $C'\in \cset^{<L}$, the correspoinding supernode $u^L(C')$ belongs to cluster $C$, then we can think of cluster $C$ as ``containing'' cluster $C'$, in some sense. A natural and intuitive way to define the flattened sets $V^F(C)$ of vertices, would then be the following.

If $C\in \cset^0$ is a cluster from level $0$, then we let $V^F(C)$ contain every vertex $x\in V(G)$, whose corresponding type-1 regular vertex $v^0(x)$ lies in $C$. Consider now some level $0<L\leq q$, and let $C\in \cset^L$ be any cluster. As before, for every vertex $x\in V(G)$ with $v^L(x)\in V(C)$, we add vertex $x$ to set $V^F(C)$. But additionally, for every supernode $u^L(C')$ that belongs to cluster $C$, we add all vertices of $V^F(C')$ to set $V^F(C)$, provided that $\scale(C')\leq \scale(C)$.

This simple intuitive definition of the flattened sets of vertices would serve our purpose in the sense that it would allow us to support the short-path queries as required. But unfortunately, due to the specifics of how the \recdynnc data structure is defined, we cannot control the cardinalities of the resulting flattened sets $V^F(C)$ of vertices, which could in turn lead to a running time that is too high.

In order to overcome this difficulty, we slightly modify the above definition of the flattened set of vertices. Specifically, for every level $0\leq L\leq q$, and every cluster $C\in \cset^L$, we will mark every supernode $u^L(C')\in V(C)$ as either \emph{important} or \emph{unimportant} for cluster $C$. We only include the vertices of $V^F(C')$ in set $V^F(C)$ if supernode $u^L(C')$ is marked as important for $C$. A status of a supernode $u^L(C')$ with respect to a cluster $C$ may switch from important to unimportant over the course of the algorithm, but it may never switch in the opposite direction. This allows us to guarantee that the set $V^F(C)$ of vertices remains decremental, which is crucial since the \recdynnc data structure does not support edge insertions, except in the case of supernode splitting. We defer the specific definition of important supernodes for later, but they are defined so that, on the one hand, we can control the cardinalities of the sets $V^F(C)\cap S^L_k$ of vertices (which is sufficient in order to make our construction efficient), while, on the other hand, still allowing us to support \shortpath\ queries.

We now proceed to formally define the level-$0$ data structure. We then define the data structures for levels $1\leq L\leq q$. Finally, we provide an algorithm for responding to \shortpath\ queries.

Throughout, we use the following parameters. 
We let $\Delta=\hat m^{16\eps^3}$, and, for all $0\leq L\leq q$, we let $\mu_L=(4\Delta)^{L+1}$.

%Let $\mu_0=2$, and, for $1\leq L\leq q$, we let $\mu_L=\left (16\cdot \Delta(8m\cdot \mu_{L-1})\right )^L$. We claim that, for all $0\leq L\leq q$, $\mu_L\leq m^{\eps}$ holds. Clearly, the claim is true for $\mu_0$. Assume now that the claim is true for some integer $0\leq L<q$. Recall that, $\Delta(\cdot)$ is a non-decreasing function, and, for some large enough constant $c$, $\Delta(x)\leq x^{\eps^2}$ holds for all $x\geq 0$. Therefore:

%\[\mu_{L+1}= \left (16\cdot \Delta(8m\cdot \mu_{L-1})\right )^{L+1}\leq 
%\left(16\Delta(8m^{1+\eps})\right )^L\leq \left(16m^{2\eps^2}\right )^L
% \]
%First, we denote $\Delta'=2\Delta(W)\cdot \log \hat D \leq O(\Delta(W)(q+\log D^*)\leq O\left(\frac{\Delta(W)\log W}{\eps}\right )$ (since $D^*\leq W$). 

We will ensure that, for every level $0\leq L\leq q$, the dynamic degree bound of graph $H^L$ is at most $\mu_L$, and every regular vertex of $H^L$ may lie in at most $\Delta$ clusters of $\cset^L$ over the course of a single level-$L$ phase.

We denote $\mu'=\mu_q=(4\Delta)^{q+1}$, so the dynamic degree bounds of all graphs $H^L$ that we maintain are bounded by $\mu'$. Notice that:

\begin{equation}\label{eq: bound mu}
\mu'\leq (4\hat m^{16\eps^3})^{\ceil{1/\eps}+1}\leq \hat m^{64\eps^2}. 
\end{equation}

%We also let $\mu_0=2$, and, for $1\leq L\leq q$, we let $\mu_L=(8\Delta')^L$. We will ensure that, for all $0\leq L\leq q$, the dynamic degree bound in graph $H^L$ is bounded by $\mu_L$. We let $\mu'=(8\Delta')^q\leq \left(\frac{\Delta(W)\log W}{\eps}\right )^{O(1/\eps)}$, so the dynamic degree bounds of all graphs $H^L$ that we maintain are bounded by $\mu'$.

\subsection{Level-$0$ Data Structure}
\label{subsec: fully level 0}

Recall that there is a single level-$0$ phase, that lasts for the whole duration of the time interval $\tset$ -- the time horizon for $\Sigma$.
The level-$0$ data structure is initialized once at the beginning of the algorithm. We construct a valid input structure $\iset^0=\left(H^0=(V^0,U^0,\tilde E^0),\set{\ell(e)}_{e\in \tilde E^0},\hat D \right )$, that, over the course of the time interval $\tset$,  undergoes a sequence of valid update operations.  Additionally, for all $0\leq i\leq \log \hat D$, we will maintain data structure $DS^0_i$, that, intuitively, solves the \recdynnc problem on $\iset^0$ with distance parameter $D_i$.
We now define the data structures that the algorithm maintains more formally.
We start by providing an algorithm for initializing the level-$0$ data structure, and then provide an algorithm for updating it. Lastly, we show that the resulting level-$0$ data structure can support approximate short-path queries between pairs of vertices $x,y\in V(G)$, such that there is a level-$0$ path of length at most $D^*$ connecting $x$ to $y$ in the current graph $G$.

\paragraph{Initialization.}
We initialize a bipartite level-$0$ graph $H^0=(V^0,U^0,\tilde E^0)$, where vertices of $V^0$ are called regular vertices, and vertices of $U^0$ are called supernodes. 
The initial graph $H^0$ is constructed as follows. For every vertex $x\in V(G)$, we introduce a regular vertex $v^0(x)$, that is added to $V^0$, and a supernode $u^0(x)$, that is added to $U^0$. We also add an edge $(v^0(x),u^0(x))$, whose length is $1$. We refer to vertices in $\set{v^0(x)\mid x\in V(G)}$ as \emph{type-1 regular vertices}, and to supernodes in set $\set{u^0(x)\mid x\in V(G)}$ as \emph{type-1 supernodes}.

For every edge $e=(x,y)\in E(G)$, we include a regular vertex $v^0(e)$ in $V^0$. The vertex connects to supernodes $u^0(x)$ and $u^0(y)$ with edges, whose lengths are $\ell(e)$ each. We refer to the set $\set{v^0(e)\mid e\in E(G)}$ of vertices as \emph{type-2 regular vertices}. Level-$0$ graph $H^0$ does not contain type-2 supernodes, unlike graphs from higher levels.
This completes the definition of the initial graph $H^0$.

For all $0\leq i\leq \log \hat D$, we also define a graph $H^0_i$, that is defined exactly like $H^0$, except that it does not include edges of $H^0$ whose length is greater than $D_i$. 

For all $0\leq i\leq \log \hat D$, we have  now defined a valid input structure $\iset^0_i=\left(H^0_i,\set{\ell(e)}_{e\in E(H^0_i)},D_i \right )$.
Below, we define an online sequence of valid update operations for each such graph $H^0_i$. We can therefore view $\iset^0_i$ as an instance of the \recdynnc problem. The number of regular vertices in graph $H^0_i$ is bounded by $|V(G^0)|+|E(G^0)|\leq 4m$. We let $\hat W=\max\set{4m\cdot \mu_0,\hat m}\leq \max\set{m\cdot \hat m^{64\eps^2},\hat m}\leq n^3$.
We initialize the data structure associated with algorithm $\aset$ for the \recdynnc problem  from \Cref{assumption: alg for recdynnc2}  
on this instance of \recdynNC with the parameter $\hat W$ that we defined. We denote the corresponding data structure by $\DS^0_i$, and we denote by $\cset^0_i$ the collection of clusters (the weak neighborhood cover) that the algorithm maintains. For every cluster $C\in \cset^0_i$, we say that the \emph{scale} of $C$ is $i$, and we denote $\scale(C)=i$. Recall that the approximation factor that Algorithm $\aset$ achieves is $\alpha(\hat W)\leq \alpha$.

\paragraph{Updates.}
As graph $G$ undergoes a sequence $\Sigma$ of edge insertions and deletions, we update graph $H^0$ as follows. Consider any update to graph $G$ that appears in $\Sigma$. If the update is the insertion of an edge into $G$, then we ignore it. Assume now that the update is the deletion of some edge $e$ from $G$. If edge $e$ is not an original edge of $G$ (that is, it was inserted at some point into $G$), then we also ignore it. Otherwise, we delete both edges that are incident to supernode $v^0(e)$ from $H^0$. 
This concludes the description of an algorithm that, given the online  sequence $\Sigma$ of edge insertions and deletions for graph $G$, produces an update sequence for graph $H^0$. Notice that we only employ edge-deletion operations to graph $H^0$. We never perform vertex-deletion or supernode-splitting. It is therefore immediate to verify that the dynamic degree bound of graph $H^0$ is at most $2\leq \mu_0$. It is immediate to verify that the time required to initialize graph $H^0$, and to produce the update sequence for it, given $\Sigma$, is bounded by $O(m)$. Notice also that the number of regular vertices in graph $H_0$ is bounded by $4m$. Therefore, for all $0\leq i\leq \log \hat D$, the time that is required in order to maintain data structure $\DS^0_i$ is bounded by: 

$$O\left (m^{1+O(\eps)}\cdot \mu_0^{O(1/\eps)}\right ) \leq  O\left (m^{1+O(\eps)}\cdot \hat m^{O(\eps)}\right )\leq O\left (m\cdot \hat m^{O(\eps)}\right ).$$ 

The total update time that is required in order to maintain the whole level-$0$ data structure, that we denote by $\DS^0$, is then bounded by:

\[T^0\leq O\left (m\cdot \hat m^{O(\eps)}\cdot \log \hat D\right )\leq O\left (m\cdot \hat m^{O(\eps)}\right ),  \]

since $\log \hat D\leq \hat m^{\eps^{12}}$ from Inequality  \ref{eq: bound 2 on D}.

Throughout the algorithm, we denote $\cset^0=\bigcup_{i=0}^{\log \hat D}\cset^0_i$. Since, for all $0\leq i\leq \log \hat D$, a regular vertex of $H^0$ may belong to at most $\hat W^{4\eps^3}\leq \hat m^{8\eps^3}$ clusters over the course of the algorithm, 
we get that every regular vertex of $H^0$ may belong to at most $2  \hat m^{8\eps^3}\cdot \log \hat D\leq \hat m^{16\eps^3}=\Delta$ clusters of $\cset^0$ over the course of the algorithm.

Recall that we denote by $\DS^0$ all level-$0$ data structures that we maintain. For convenience, for all $0\leq i\leq \log \hat D$, for every regular vertex $v\in V(H^0)$, we denote the cluster $\coveringcluster(v)$ that data structure $\DS^0_i$ maintains by  $\coveringcluster^0_i(v)$. Similarly, we denote the list $\clusterlist(v)$ of all clusters of $\cset^0_i$ containing $v$ that data structure $\DS^0_i$ maintains by $\clusterlist^0_i(v)$.  We then let $\clusterlist^0(v)=\bigcup_{i=0}^{\log \hat D}\clusterlist^0_i(v)$. Lastly, for all $0\le i\leq \log \hat D$, if $C$ is a cluster that currently lies in $\cset^0_i$, and $v,v'$ are two regular vertices of $H^0$ that currently lie in $C$, we denote by $\shortpath^0_i(C,v,v')$ a query $\shortpath(C,v,v')$ to data structure $\DS^0_i$. Recall that the data structure must return a path $P\subseteq H^0$ of length at most $\alpha\cdot D_i$ connecting $v$ to $v'$ in $H^0_i$, in time $O(|E(P)|)$. For convenience, we will think of $\shortpath^0_i(C,v,v')$ as a query that is supported by data structure $\DS^0$.

%We have now obtained a valid input structure  $\iset^0=\left(H^0=(V^0,U^0,\tilde E^0),\set{\ell(e)}_{e\in \tilde E^0},D^0 \right )$, that undergoes a sequence $\Sigma^0$ of valid update operations with dynamic degree bound $\mu$. For all $1\leq i\leq \log D^0$, we define an instance of the \recdynnc problem with distance scale $D_i$, as follows. The corresponding dynamic graph $H^0_i$ is identical to $H^0$, except that it excludes all edges whose length is greater than $D_i$. Therefore, we obtain a valid input structure $\iset^0_i=\left(H^0_i,\set{\ell(e)}_{e\in E(H^0_i)},D_i \right )$, together with a sequence $\Sigma^0_i$ of valid update operations, that is identical to $\Sigma^0$, except that it does not contain edges whose length is greater than $D_i$. We view $\iset^0_i$, together with update sequence $\Sigma^0_i$, as  an instance of the \recdynnc problem, with distance parameter $D_i$. 

\iffalse
\paragraph{Total update time.}
 Recall that we are guaranteed that, for all $1\leq i\leq \log D^0$, the total update time required for maintaining data structure $\DS^0_i$ is at most $(W')^{1+\delta}(\mu\cdot \log W)^c$. Since $D^*\leq W$, we get that the total update time that is needed in order to maintain all level-$0$ data structures is at most $O\left((W')^{1+\delta}\mu^c\cdot (\log W)^{c+1}\right )$. 
We denote the total update time for maintaining the level-$0$ data structure over the course of the algorithm by $T^0\leq c'\cdot (W')^{1+\delta}\cdot \mu^c(\log W)^{c+1}$, where $c'$ is a large enough constant. 
\fi

\paragraph{Supporting short-path queries.}
Next, we show an algorithm that supports short-path queries between pairs of vertices $x,y\in V(G)$, provided that there is a path $P$ connecting $x$ to $y$ in the current graph $G$, whose length is at most $D^*$, and level is $0$.
We start with the following simple claim.

\begin{claim}\label{claim: responding to queries level-0}
	Let $x,y$ be any pair of vertices of $G$, and assume that, at some time $\tau\in \tset$, there is a level-$0$ path $P$ in graph $G$, that connects $x$ to $y$, such that the length of $P$ is at most $D\leq D^*$. Then there is a path $P'$ in graph $H^0$, connecting $v^0(x)$ to $v^0(y)$, whose length is at most $2D+2$. 
\end{claim}
\begin{proof}
	Let $x=z_1,\ldots,z_r=y$ be the sequence of vertices on path $P$. For $1\leq j\leq r-1$, denote by $e_j=(z_j,z_{j+1})$ the $j$th edge on the path. Consider the following sequence of vertices in graph $H^0$:
	
	\[\left (v^0(z_1),u^0(z_1),v^0(e_1),u^0(z_2),v^0(e_2),
\ldots,v^0(e_{r-1}),u^0(z_r),v^0(z_r)\right ).  \]

It is immediate to verify that this sequence defines a valid path in graph $H^0$, and the length of the path is $2+\sum_{j=1}^{r-1}2\ell(e_j)\leq 2+2D$.
\end{proof}

We then obtain the following easy corollary.

\begin{corollary}\label{cor: level 0 queries}
	There is a deterministic algorithm, that, at any time $\tau\in \tset$, given a pair $x,y$ of vertices of $G$, and a distance parameter $D\leq D^*$ that is an integral power of $2$, responds "YES" or "NO", in time $O(\log \hat m)$. If the algorithm responds "NO", then $G$ does not contain a path connecting $x$ to $y$ of length at most $D$ that belongs to level $0$. If the algorithm responds "YES", then it can, additionally, compute a path $P'$ in graph $G$, connecting $x$ to $y$, whose length is at most $4D\cdot \alpha$, in time $O(|E(P')|)$.
\end{corollary}

We note that, once the algorithm from \Cref{cor: level 0 queries} responds ``YES'' or ''NO'', we may choose to terminate it. Alternatively, if it responds ``YES'', we may choose to continue the algorithm to obtain the desired path $P'$.

\begin{proof}
Suppose we are given a pair $x,y$ of vertices of $G$ and a 	distance parameter $D$, that is an integral power of $2$. We let $i=\log D$.  Notice that, from \Cref{claim: responding to queries level-0}, if there exists a level-$0$ path $P$ connecting $x$ to $y$ in $G$, whose length is at most $D$, such that all edges of $P$ are original edges, then there must be a path $\tilde P$  in graph $H^0$, of length at most $4D$, connecting $v^0(x)$ to $v^0(y)$.  We use data structure $DS^0_{i+2}$ to compute a cluster $C=\coveringcluster^0_{i+2}(v^0(x))$. We then check whether $v^0(y)\in v^0(C)$. If this is not the case, then we are guaranteed that there is no level-$0$ path in $G$ of length at most $D$ that connects $x$ to $y$. Therefore, if $v^0(y)\not\in V(C)$, we return ``NO'', and otherwise we return ``YES''. Since the vertices of $V(C)$ are maintained in a sorted list, and $|V(H^0)|\leq \hat m$, the running time of the algorithm so far is bounded by $O(\log \hat m)$. 

If the algorithm responds ``YES'', then we are guaranteed that $v^0(x),v^0(y)\in V(C)$. We can then execute query  $\spquery^0_{i+2}(C,v^0(x),v^0(y))$ in data structure $\DS^0_{i+2}$. The data structure must return  a path $\tilde P'$ in the current graph $H^0$, of length at most $\alpha\cdot 4D_i$, connecting $v^0(x)$ to $v^0(y)$, in time $O(|E(\tilde P')|)$. 
Since each type-1 regular vertex in $H^0$ has degree $1$, we can assume that no such vertex serves as an inner vertex on path $\tilde P'$.

We now transform path $\tilde P'$ in graph $H^0$ into a path $P'$, connecting $x$ to $y$, in graph $G$. In order to do so, we delete the first and the last vertices of $\tilde P'$, suppress all type-2 regular vertices, and replace every supernode $u^0(z)$ on the path with the corresponding vertex $z\in V(G)$. It is easy to verify that  the resulting path $P'$ connects $x$ to $y$ in the current graph $G$, and its length remains bounded by $4\alpha\cdot D_i$. The running time of this part of the algorithm is $O(|E(P')|)$.
\end{proof}

\paragraph{Flattened Sets of Vertices.}
For every cluster $C\in \cset^0$, we now define a flattened set $V^F(C)\subseteq V(G)$ of vertices. We do not maintain these sets of vertices explicitly, but we will use this definition in order to maintain data structures from higher levels. Consider some time $\tau\in \tset$, and a cluster $C$ that belonged to set $\cset^0$, at time $\tau$. We let the flattened set $V^F(C)\subseteq V(G)$ of vertices contain every vertex $x\in V(G)$, such that the corresponding regular vertex $v^0(x)$ lies in $V(C)$. Note that, from the definition of allowed updates to the set $\cset^0$ of clusters in the \recdynnc problem, regular vertices may not join a cluster after it is created. Therefore, for every cluster $C\in\cset^0$, the corresponding flattened set $V^F(C)\subseteq V(G)$ of vertices is decremental: once cluster $C$ is added to set $\cset^0$ and set $V^F(C)$ is initialized, vertices may leave it but they cannot join it.

\subsection{Level-$L$ Data Structure}
\label{subsec: Level L DS}

We now consider an integer $0<L\leq q$, and provide the description of level-$L$ data structure. Recall that the purpose of the data structure is to support approximate \shortpath\ queries between pairs of vertices $x,y\in V(G)$, such that there is a level-$L$ path in $G$ connecting $x$ to $y$, whose length is at most $D^*$. 

We assume that for all $0\leq L'<L$, the level-$L'$ data structure is defined already, and in particular for every cluster $C\in \cset^{<L}$, the flattened set $V^F(C)\subseteq V(G)$ of vertices is already defined (recall that we do not maintain these vertex sets explicitly).

Recall that the timeline is partitioned into $M^{L}$ level-$L$ phases, each of which spans a contiguous sequence $\Sigma'\subseteq \Sigma$ of updates to graph $G$, that contains at most $M^{q-L}$ edge insertions. For all $1\leq k\leq M^L$, the $k$th level-$L$ phase is denoted by $\Phi^L_k$,  the time interval corresponding to phase $\Phi^L_k$ is denoted by $\tset^L_k$, and the sequence of updates that graph $G$ undergoes during Phase $\Phi^L_k$ is denoted by $\Sigma^L_k\subseteq \Sigma$. From the definition of the hierarchical partition of the time line into phases, for all $0\leq L'<L$, phase $\Phi^L_k$ is completely contained in some level-$L'$ phase.
Let $\Phi^{L-1}_{k'}$ be the level-$(L-1)$ phase that contains $\Phi^L_k$. 
 Recall that we have denoted by $A^L_k$ the collection of all edges that were inserted into graph $G$ between time $\tau'$ -- the beginning of level-$(L-1)$ phase $\Phi^{L-1}_{k'}$, and time $\tau$ -- the beginning of level-$L$ phase $\Phi^L_k$ (including $\tau'$ and excluding $\tau$). Clearly, $|A^L_k|\leq M^{q-L+1}$ holds.
We also denoted by $S^L_k$ the collection of vertices of $G$ that serve as endpoints of the edges of $A^L_k$. The set $A^L_k$ of edges remains unchanged over the course of Phase $\Phi^L_k$, even if some edges of $A^L_k$ are deleted from graph $G$. Similarly, the set $S^L_k$ of vertices does not change over the course of Phase $\Phi^L_k$.

At the beginning of every level-$L$ phase, we initialize the level-$L$ data structures from scratch. Notice that, if $\tau$ is the time when some level-$L$ phase $\Phi^L_k$ starts, it is possible that $\tau$ is also the starting time of some level-$L'$ phase, for $1\leq L'<L$. However, in this case, time $\tau$ must also be the start of a level-$(L-1)$ phase, and so $A^L_k=\emptyset$ must hold. In this case, the level-$L$ data structure, graph $H^L$, and the set $\cset^L$ of clusters remain empty over the course of Phase $\Phi^L_k$. From now on we assume that, when a level-$L$ data structure is being initialized at the beginning of some level-$L$ phase $\Phi^L_k$, data structures from levels $0,\ldots,L-1$ have already been initialized.

We now consider some integer $1\leq k\leq M^{L}$, and provide a  description of the level-$L$ data structure $\DS^L$ that is maintained over the course of level-$L$ phase $\Phi^L_k$.
As mentioned already, we assume that data structures for levels $0,\ldots,L-1$ are already defined and initialized. 

Recall that, for all $0\leq L'<L$, for each cluster $C\in \cset^{L'}$, the level-$L'$ data structure defines a flattened set $V^F(C)\subseteq V(G)$ of vertices. This set is initialized once cluster $C$ joins set $\cset^{L'}$. 
Notice that, once a cluster $C$ joins set $\cset^{L'}$, it remains in this set for the duration of the current level-$L'$ phase (though the cluster may eventually become empty), and hence for the duration of Phase $\Phi^L_k$.
For all $0\leq L'<L$, for every level-$L'$ phase $\Phi^{L'}_{k'}$, we assume that the following properties hold:

\begin{properties}{P}
	\item For every cluster $C$ that lied in $\cset^{L'}$ at any time during Phase $\Phi^{L'}_{k'}$, the flattened set $V^F(C)\subseteq V(G)$ of vertices is initialized when cluster $C$ joins set $\cset^{L'}$, and after that vertices may leave $V^F(C)$ but not join it; \label{prop: flat set decremental}

	\item For every vertex $x\in V(G)$, there are at most  $(4\Delta)^{L'+1}$ clusters $C\in  \cset^{L'}$, such that $x$ ever belonged to $V^F(C)$ over the course of Phase $\Phi^{L'}_{k'}$; and \label{prop: every vertex in few flat sets}

	\item If a new cluster $C$ is added to set $\cset^{L'}$ at some time $\tau$ during Phase $\Phi^{L'}_{k'}$, due to a cluster splitting operation that is applied to a cluster $C'\in \cset^{L'}$ with $C\subseteq C'$, then, at time $\tau$, $V^F(C)\subseteq V^F(C')$ holds. \label{prop: cluster splitting flat sets}
\end{properties}

Notice that all these properties hold for the level-$0$ data structure. We assume that they hold for data structures from levels $0,\ldots,L-1$, and we describe a construction of the level-$L$ data structure that will ensure these properties for level $L$.

The next observation will be useful for us later.

\begin{observation}\label{obs: flat vertex sets ancestry}
Let $0\leq L'<L$ be a level, and assume that the level-$L'$ data structure ensures properties \ref{prop: flat set decremental} and \ref{prop: cluster splitting flat sets}. Consider any level-$L'$ phase $\Phi^{L'}_{k'}$, and a cluster $C$ that belonged to set $\cset^{L'}$ at time $\tau\in \tset^{L'}_{k'}$. Consider some time $\tau'<\tau$ with $\tau'\in \tset^{L'}_{k'}$, and let $C'=\anc\attime[\tau'](C)$. Denote by $S$ the set $V^F(C)$ at time $\tau$, and by $S'$ the set $V^F(C')$ at time $\tau'$. Then $S\subseteq S'$ must hold.
\end{observation}
\begin{proof}
	Consider a level $0\leq L'<L$, a level-$L'$ phase $\Phi^{L'}_{k'}$, and a cluster $C$ that lied in set $\cset^{L'}$ at some time $\tau\in \tset^{L'}_{k'}$. We define a dynamic set of vertices $Q\subseteq V(G)$, for all time points $\tau''\in \tset^{L'}_{k'}$ with $\tau''\leq \tau$. For each such time point $\tau''$, set $Q\attime[\tau'']$ is equal to the set $V^F(\anc\attime[\tau''](C))$ at time $\tau''$. From properties \ref{prop: flat set decremental} and \ref{prop: cluster splitting flat sets} it is immediate to verify that set $Q$ of vertices is decremental. The observation then follows since $Q\attime[\tau]=S$ and $Q\attime[\tau']=S'$.
\end{proof}

We now consider the current level-$L$ phase $\Phi^L_k$.
For convenience, for all $0<L'\leq L-1$, we denote by $\tilde \cset^{L'}$ the collection of all clusters $C$ that ever belonged to set $\cset^{L'}$ over the course of the current level-$L$ phase $\Phi^L_k$. Notice that, once cluster $C$ is added to set $\cset^{L'}$, it remains in $\cset^{L'}$ until the end of the current level-$L'$ phase, so in particular, it remains in $\cset^{L'}$ until the end of Phase $\Phi^L_k$.  We also denote $\tilde \cset^{<L}=\bigcup_{L'=0}^{L-1}\tilde \cset^{L'}$.

 As mentioned already, the level-$L$ data structure maintains a graph $H^L$ over the course of Phase $\Phi^L_k$. The definition of the graph depends on the flattened sets $V^F(C)\subseteq V(G)$ of vertices, for clusters $C\in \cset^{<L}$. As mentioned already, the sets $V^F(C)$ of vertices are not maintained explicitly, since this may be too costly, as their cardinalities may be quite large. However, in order to maintain graph $H^L$, we only need to maintain, for every cluster $C\in \cset^{<L}$, the set $V^F(C)\cap S^L_k$ of vertices, whose cardinality is significantly smaller. For convenience, for every cluster $C\in \cset^{<L}$, we denote by $Z^L_k(C)=V^F(C)\cap S^L_k$. The set $Z^L_k(C)$ of vertices is only maintained over the course of the current level-$L$ phase, by our level-$L$ data structure. Once the level-$L$ phase terminates, we recompute all such sets of vertices from scratch. Observe however that vertex sets $\set{Z^L_k(C)}_{C\in \cset^{<L}}$ are completely determined by the data structures from levels $0,\ldots,L-1$, and by the set $S^L_k$ of vertices, which is fixed throughout Phase $\Phi^L_k$. In particular, the definitions of the sets  $\set{Z^L_k(C)}_{C\in \cset^{<L}}$ of vertices do not depend on the data structures that we will maintain at level $L$.

The following properties  of the sets $\set{Z^L_k(C)}_{C\in \cset^{<L}}$ of vertices follow immediately from Properties \ref{prop: flat set decremental}--\ref{prop: cluster splitting flat sets}

\begin{properties}{P'}
	\item For a cluster $C\in \tilde \cset^{<L}$, vertex set $Z^L_k(C)\subseteq V(G)$ is initialized when cluster $C$ joins set $\cset^{<L}$ or at the start of phase $\Phi^L_k$ -- whatever happens later, and after that, over the course of Phase $\Phi^L_k$, vertices may leave $V^F(C)$ but not join it; \label{prop: flat set decremental for L}

	\item For every vertex $x\in S^L_k$, there are at most  $(4\Delta)^{L+1}\leq \mu_L$ clusters $C\in \tilde \cset^{<L}$, such that $x$ ever belongs to $Z^L_k(C)$ over the course of Phase $\Phi^L_k$; and \label{prop: every vertex in few flat sets for L}

	\item If a new cluster $C$ is added to set $\cset^{<L}$ at some time $\tau$ during Phase $\Phi^L_k$ because of a cluster splitting operation applied to a cluster $C'\in \cset^{<L}$, then, at time $\tau$, $Z^L_k(C)\subseteq Z_k^L(C')$ holds. \label{prop: cluster splitting flat sets for L}
\end{properties}

The data structure at level $L$ consists of three components. The first component, that we refer to as $\DSFS$, maintains the sets $\set{Z^L_k(C)}_{C\in \cset^{<L}}$  of vertices. The second data structure, that we refer to as $\DSbasic$, maintains the level-$L$ graph $H^L$, and the corresponding neighborhood cover data structures, that include the set $\cset^L$ of level-$L$ clusters. The third data structure, that we refer to as $\DSimp$ maintains, for every cluster $C\in \cset^L$ and supernode $u\in V(C)$, an \emph{importance status} of the supernode: whether the supernode is marked as important for cluster $C$. The importance status of the supernodes will in turn be used in order to define the flattened sets $V^F(C)$ of vertices for clusters $C\in \cset^L$, and data structure $\DSimp$ will be exploited by data structures $\DSFS$ from higher levels $L''>L$, in order to maintain the sets $Z^{L''}_{k''}(C)$ of vertices for clusters $C\in \cset^{<L''}$. We start by providing a high-level definition of each of these three data structures, and state the invariants that the data structures maintain. We then provide a more detailed description of the implementation of each of the data structures.

\paragraph{Data Structure $\DSFS$.}
Intuitively, the purpose of the $\DSFS$ data structure is to maintain the sets $\set{Z^L_k(C)}_{C\in \cset^{<L}}$ of vertices over the course of level-$L$ phase $\Phi^L_k$. However, the number of clusters in set $\cset^{<L}$ may be prohibitively large, and for many of these clusters, $Z^L_k(C)=\emptyset$ may hold. Instead, the data structure will maintain a set $\cset^*\subseteq \cset^{<L}$ of clusters, and it will only explicitly maintain the sets $Z^L_k(C)$ of vertices for clusters $C\in \cset^*$. The data structure ensures that the set $\cset^*$ of clusters has the following properties:

\begin{properties}{F}
	\item the set $\cset^*$ of clusters is initialized at the beginning of Phase $\Phi^L_k$, and, over the course of Phase $\Phi^L_k$, clusters may join it but they may not leave it; \label{propflat1: incremental set}
	
	\item at the beginning of Phase $\Phi^L_k$, for every cluster $C\in \cset^*$, $Z^L_k(C)\neq \emptyset$ holds, and whenever a new cluster $C$ joins set $\cset^*$, $Z^L_k(C)\neq \emptyset$ holds at that time;
	\label{propflat1.5: joining Cstar}
	
	\item throughout Phase $\Phi^L_k$, if a cluster $C\in \cset^{<L}$ does not lie in $\cset^*$, then $Z^L_k(C)=\emptyset$ holds; and \label{propflat2: not in Cstar emptyset}

	\item if $\tau\in \tset^{L}_k$ is a time that is not the beginning of Phase $\Phi^L_k$, then a cluster $C$ may be added to set $\cset^*$ at time $\tau$ only if cluster $C$ was added to set $\cset^{<L}$ at time $\tau$, due to a cluster splitting operation that was applied to some cluster $C'$, and $C'\in \cset^*$ at time $\tau$. \label{propflat3: new cluster in set}
\end{properties}

We note that, if a new cluster $C$ is added to set $\cset^{<L}$ at some time $\tau\in \tset^{L}_k$ due to a cluster splitting operation that was applied to cluster $C'$, and $Z^L_k(C)\neq \emptyset$, then, from Property \ref{prop: cluster splitting flat sets for L}, $Z^L_k(C')\neq \emptyset$ holds at time $\tau$, and so $C'\in \cset^*$ must hold.

We ensure that the set $\cset^*$ of clusters that data structure $\DSFS$ maintains over the course of Phase $\Phi^L_k$ obeys the above invariants. For every cluster $C\in \cset^*$, the data structure also maintains  the corresponding set $Z^L_k(C)\subseteq V(G)$ of vertices. We denote by $\TFS$ the total update time that data structure $\DSFS$ requires over the course of a single phase. We emphasize that data structure $\DSFS$ only depends on data structures maintained at levels $0,\ldots,L-1$, and the set $S^L_k$ of vertices that is fixed throughout Phase $\Phi^L_k$, so it does not depend on any other data structures maintained at level $L$.

\paragraph{Data Structure \DSbasic.}
Data structure $\DSbasic$ is responsible for maintaining the level-$L$ a bipartite graph $H^L=(V^L,U^L,\tilde E^L)$, over the course of Phase $\Phi^L_k$. The graph is initialized at the beginning of Phase $\Phi^L_k$, and after that it undergoes a sequence of online valid update operations, that is computed by data structure \DSbasic, based on the sequence $\Sigma$ of updates that graph $G$ undergoes, changes to the set $\cset^*$ of clusters, and the sets $Z^L_k(C)$ of vertices for clusters $C\in \cset^*$ that data structure $\DSFS$ maintains. For all $0\leq i\leq \log \hat D$, we also maintain a graph $H^L_i$, that is obtained from $H^L$ by deleting from it all edges whose length is greater than $D_i$. We then obtain a corresponding valid input structure $\iset^L_i$, that undergoes a sequence of valid update operations. We apply the algorithm $\aset$ for the \recdynnc problem from \Cref{assumption: alg for recdynnc2} to this input structure, and parameter $\hat W=\max\set{4m\cdot \mu_L,\hat m}$, and obtain a set $\cset^L_i$ of clusters that the algorithm maintains over the course of Phase $\Phi^L_k$.  We then denote $\cset^L=\bigcup_{i=0}^{\log \hat D}\cset^L_i$. We ensure that every regular vertex of $H^L$ may belong to at most $\Delta$ clusters of $\cset^L$ over the course of Phase $\Phi^L_k$. 
We now formally define dynamic graph $H^L$.

\begin{definition}[Graph $H^L$]\label{def: graph HL}
Dynamic graph $H^L=(V^L,U^L,\tilde E^L)$ is defined as follows.
\begin{itemize}
\item The set $V^L$ of regular vertices of graph $H^L$  is the union of two subsets: the set of type-1 regular vertices $\set{v^L(x)\mid x\in S^L_k}$, and the set of type-2 regular vertices $\set{v^L(e)\mid e\in A^L_k}$. 

\item
The set $U^L$ of supernodes of graph $H^L$ is the union of two subsets -- the set of type-1 supernodes $\set{u^L(x)\mid x\in S^L_k}$, and the set of type-2 supernodes $\set{u^L(C)\mid C\in \cset^*}$.

\item The edges of graph $H^L$ are defined as follows. Let $e=(x,y)\in A^L_k$ be a an edge. If edge $e$ was already deleted from $G$, then vertex $v^L(e)$ is isolated in graph $H^L$. Otherwise, it is connected to supernodes $u^L(x)$ and $u^L(y)$ with edges of length $\ell(e)$. Consider now a vertex $x\in S^L_k$. We add an edge $(v^L(x),u^L(x))$ to graph $H^L$ of length $1$. Additionally, for every cluster $C\in \cset^*$ with $x\in Z^L_k(C)$, we add an edge $(v^L(x),u^L(C))$, whose length is $2^{\scale(C)}$.
\end{itemize} 
\end{definition}

Note that, from the above definition, the set $V^L$ of regular vertices, and the set of type-1 supernodes remains unchanged over the course of Phase $\Phi^L_k$.

Notice also that, for every type-2 supernode $u^L(C)$, all edges that are incident to the supernode have the same length -- $2^{\scale(C)}$, and all neighbors of $u^L(C)$ are type-1 regular vertices, that represent the vertices of $Z^L_k(C)$. We denote by $N(u^L(C))$ the collection of all vertices of $H^L$ that are neighbors of supernode $u^L(C)$. Since the set $Z^L_k(C)$ of vertices is decremental, once supernode $u^L(C)$ is added to graph $H^L$, the set $N(u^L(C))$ of its neighbors is also decremental.

We show below an algorithm that maintains the dynamic graph $H^L$. After the graph $H^L$ is initialized at the beginning of Phase $\Phi^L_k$, it will only undergo a sequence of valid update operations. We show that the resulting dynamic graph is consistent with the definition of graph $H^L$ listed above. We need one more detail regarding graph $H^L$ that will be useful for us. As mentioned already, once we initialize graph $H^L$, it will only undergo valid update operations: edge-deletion, isolated vertex deletion, and supernode-splitting. The supernode splitting operation will only be applied to type-2 supernodes, and it will mirror cluster splitting from levels $0,\ldots, L-1$. In other words, supernode $u^L(C)$ may only undergo a supernode splitting operation at time $\tau$, if the corresponding cluster $C\in \cset^{<L}$ underwent a cluster-splitting operation at time $\tau$. If $C'\subseteq C$ is the new resulting cluster, then the new supernode that is obtained by splitting $u^L(C)$ is $u^L(C')$ -- that is, it represents cluster $C'$. It is then easy to verify that, for every type-2 supernode $u^L(C)$ that ever lies in graph $H^L$, for every time $\tau\in \tset^L_k$, if we denote $\anc\attime(u^L(C))$ by $u^L(C')$, then $C'=\anc\attime(C)$ (if $C\in \cset^{L'}_i$, then the ancestor-cluster of $C$ is defined with respect to $\cset^{L'}_i$). In particular, both $C$ and $C'$ must have the same scale $i$.

For every cluster $C\in \cset^{<L}$, if supernode $u^L(C)$ belongs to graph $H^L$, and it is not an isolated vertex of $H^L$, then we will designate some vertex $x\in V(G)$ to be the \emph{representative} vertex of cluster $C$ at level $L$, and denote $\sigma_L(C)=x$. Vertex $x$ may only be a representative of cluster $C$ only if $x\in V^F(C)$, and $v^L(x)$ is a vertex of $H^L$. Note that, if $x$ is a representative of $u^L(C)$, then edge $(v^L(x),u^L(C))$ belongs to graph $H^L$. For convenience, we summarize these requirements in the following rules.

\begin{properties}{B}
	\item for every cluster $C\in \cset^{<L}$, if $u^L(C)$ is a vertex of $H^L$, and it is not an isolated vertex, then a \emph{level-$L$ representative} $\sigma_L(C)\in V(G)$ of cluster $C$ must be maintained by the data structure. \label{prop: representatives 1}
	
	\item if $C\in \cset^{<L}$, and $x=\sigma^L(C)$, then $x\in V^F(C)$ and $v^L(x)\in V(H^L)$ must hold. \label{prop: representatives 2}
\end{properties}

 Generally, we will let $\sigma_L(C)$ be any vertex that satisfies these properties. If $x=\sigma_L(C)$, and, after some time, vertex $x$ is deleted from set $V^F(C)$, then we select any other vertex $y\in V^F(C)$ with $v^L(y)\in V(H^L)$ to be a new representative of cluster $C$ at level $L$.

\paragraph{Data Structure \DSimp.}

The purpose of data structure $\DSimp$ is to maintain, for every cluster $C\in \cset^L$, for every type-2 supernode $u\in V(C)$, the \emph{importance status} of supernode $u$. In other words, supernode $u$ is marked as either important or unimportant for cluster $C$. If a type-2  supernode $u$ does not belong to cluster $C$, then we will say that $u$ is \emph{unimportant} for cluster $C$. However, data structure $\DSimp$ only  maintains explicitly the importance status of a type-2 supernode $u$ for a cluster $C\in \cset^L$ if $u\in V(C)$. In addition to maintaining the importance status of type-2 supernodes for clusters, the data structure maintains, for every type-2 supernode $u\in V(H^L)$, the list of all clusters $C\in \cset^L$, for which $u$ is marked as important supernode, together with pointers to these clusters.

The definition of important supernodes completely determines the definition of the flattened sets $V^F(C)$ of vertices for clusters $C\in \cset^L$. For a cluster $C\in \cset^L$, the set $V^F(C)$ of vertices is the union of two subsets. The first subset is $\set{x\in S^L_k\mid v^L(x)\in V(C)}$. The second subset contains, for every type-2 supernode $u^L(C')$ that is marked as important for cluster $C$, all vertices of $V^F(C')$ (recall that $C'\in \cset^{<L}$ must hold, so set $V^F(C')$ is already defined).

On the one hand, we would like to mark supernodes of clusters as important quite generously, in order to ensure that we can support short-path queries. For example, marking all type-2 supernodes of every cluster $C$ as important for $C$ would accomplish this. But on the other hand, we need to limit the sets $V^F(C)$ of vertices, in order to ensure that the cardinalities of the sets $Z^{L''}_{k''}(C)$ of vertices, for levels $L''>L$ are not too large, as we need to maintain them explicitly.

A reasonable compromise between these two goals is the following: we would like to ensure that a type-2 supernode $u=u^L(C')$ is marked as important for cluster $C$ if and only if all these conditions hold: (i) $u\in V(C)$; (ii) $\scale(C')\leq \scale(C)$; and (iii) $N(u)\subseteq V(C)$. Defining the importance status of supernodes $u$ in this way would indeed allow us to accomplish the goals stated above, but it would introduce a significant problem: the status of a supernode $u$ for a cluster $C$ may flip from being unimportant to being important. We cannot allow this to happen, since this may cause vertices to join the set $V^F(C)$, which is unacceptable (see Condition \ref{prop: flat set decremental}, which is crucial in order to ensure that no edges are inserted into graph $H^L$ except via supernode splitting operations). Therefore, we modify the above definition to ensure that the status of a supernode may never switch from unimportant to important.

For every cluster $C\in \cset^{L'}$, for every type-2 supernode $u=u^{L'}(C')\in V(C)$, data structure $\DSimp$ will maintain the \emph{importance status} of supernode $u$, that is, the supernode will be marked as either being \emph{important} or \emph{unimportant} for $C$. This marking will obey the following rule. 

\begin{properties}{R}
	\item	Consider some time $\tau\in \tset^L_k$, a cluster $C$ that belongs to $\cset^{L}$ at time $\tau$, and a type-2 supernode $u=u^{L}(C')$ that belongs to $V(C)$ at time $\tau$ (so $C\in \cset^{<L}$ holds). Supernode $u$  is marked as \emph{important} for $C$ at time $\tau$ if and only if all of the following hold:
	
	\begin{itemize}
		\item $\scale(C')\leq \scale(C)$;
		\item at time $\tau$, $N(u)\neq \emptyset$ and $N(u)\subseteq V(C)$ hold;
		 and
		\item if $\tau$ is not the beginning of Phase $\Phi^L_k$, then for every time $\tau'\in \tset^L_k$ with $\tau'< \tau$, if we denote $\tilde C=\anc\attime[\tau'](C)$, and $\tilde u=\anc\attime[\tau'](u)$, then supernode $\tilde u$ was marked as important for cluster $\tilde C$ at time $\tau'$.
	\end{itemize}
	\label{rule for importance}
\end{properties}

It is easy to see that the following rule is equivalent to Rule \ref{rule for importance}. 
\begin{properties}{R'}
	\item	Consider some time $\tau\in \tset^L_k$, a cluster $C$ that belongs to $\cset^{L}$ at time $\tau$, and a type-2 supernode $u=u^{L}(C')$ that belongs to $V(C)$ at time $\tau$. Supernode $u$  is marked as \emph{important} for $C$ at time $\tau$ if and only if $\scale(C')\leq \scale(C)$, and  for every time $\tau'\in \tset^L_k$ with $\tau'\leq \tau$, if we denote $\tilde C=\anc\attime[\tau'](C)$ and $\tilde u=\anc\attime[\tau'](u)$, then, at time $\tau'$, $N(\tilde u)\neq \emptyset$ and $N(\tilde u)\subseteq V(\tilde C)$ held.
	\label{rule2 for importance}
\end{properties}

The equivalence easily follows since, if we denote by $\tilde C=\anc\attime[\tau'](C)$ and $u^{L}(\tilde C')=\anc\attime[\tau'](u^{L}(C'))$, then $\scale(\tilde C)=\scale(C)$, and $\scale(\tilde C')=\scale(C')$.
Additionally, from \Cref{obs: supernode in cluster ancestor}, $u^{L}(\tilde C')\in V(\tilde C)$ holds at time $\tau'$.
We will use the two rules interchangeably, as in some proofs one of them is more convenient to use than the other.

We now proceed to describe algorithms that maintain level-$L$ data structures in more detail. We note that, while data structure \DSFS only depends on data structures maintained at levels $0,\ldots,L-1$, it exploits data structure \DSimp maintained at those levels. Because of this, we describe the data structures in a somewhat different order. First, we assume that data structure \DSFS with properties that are defined above exists, and describe data structure \DSbasic that relies on it. Next, we describe data structure \DSimp, that uses data structure \DSbasic. Only then we describe data structure \DSFS, that relies on data structures \DSimp from lower levels.

%We note that the flat sets $V^F(C)$ of vertices for clusters $C\in \cset^{L'}$ are not maintained explicitly by level-$L'$ data structure. Instead, it maintains something we call \emph{cluster-supernode imporatance status}: For every cluster $C\in \cset^{L'}$ and supernode $u\in V(C)$, it maintains a bit $\imp(C,u)$, indicating whether supernode $u$ is considered \emph{important} for cluster $C$. For every supernode $u\in V(H^{L'})$, it also maintains a list $\impclusters(u)$ of all clusters $C\in \cset^{L'}$, such that $u$ is marked as an important supernode for $C$. If supernode $u$ does not lie in cluster $C$, then we will implicitly consider it to be unimportant for cluster $C$. This importance status of supernodes in clusters of $\cset^{<L}$, together with the sets of vertices contained in these clusters completely determine the corresponding sets $V^F(C)$ of vertices of $G$.

\subsubsection{Data Structure $\DSbasic$}
\label{subsubsec: DSbasic}

We fix a level-$L$ phase $\Phi^L_k$. We assume that data structures from levels $0,\ldots,L-1$ are now defined, and we assume that there exists data structure \DSFS, that, over the course of Phase $\Phi^L_k$, maintains a collection $\cset^*\subseteq \cset^{<L}$ of clusters, for which Properties \ref{propflat1: incremental set}--\ref{propflat3: new cluster in set} hold. The data structure additionally maintains the sets $Z^L_k(C)$ of vertices, for all clusters $C\in \cset^*$. 
From Property \ref{prop: flat set decremental for L}, once cluster $C$ joins set $\cset^*$, vertices may leave set $Z^L_k$, but they may not join it.
Recall that we have denoted by $\TFS$ the total update time of data structure $\DSFS$ over the course of Phase $\Phi^L_k$. 

At the beginning of Phase $\Phi^L_k$, we are given a set $A^L_k$ of edges of $G$, with $|A^L_k|\leq M^{q-L+1}$, and the set $S^L_k$ of vertices of $G$, that serve as endpoints of the edges of $A^L_k$. Both sets remain unchanged over the course of the phase. At the beginning of the phase, data structure $\DSFS$ produces the initial set $\cset^*$ of clusters that contains, from Property \ref{propflat2: not in Cstar emptyset}, all clusters $C\in \cset^{<L}$ with $Z^L_k\neq \emptyset$, and only such clusters. For each such cluster, the data structure also produces the initial set $Z^L_k\subseteq V(G)$ of vertices. It is now immediate to initialize graph $H^L$, so that it matches Definition \ref{def: graph HL}. From Property \ref{prop: every vertex in few flat sets for L},  for every vertex $x\in S^L_k$, there are at most  $\mu_L=(4\Delta)^{L+1}$ clusters $C\in  \cset^{*}$ with $x\in Z^L_k(C)$, so the degree of every regular vertex in $H^L$ is at most $\mu_L$. 
For every cluster $C\in \cset^*$, we let $\sigma_L(C)\in V(G)$ be an arbitrary vertex $x\in Z^L_k(C)$, that becomes the representative of $C$ at level $L$.
Clearly, graph $H^L$ can be initialized in time $O(|A^L_k|\cdot \mu_L)\leq O(M^{q-L+1}\cdot \mu_L)\leq O(M^{q-L+1}\cdot\hat m^{O(\eps)})$, since $\mu_L\leq \mu'\leq \hat m^{O(\eps^2)}$ from Inequality \ref{eq: bound mu}.

Next, we describe an algorithm that correctly maintains graph $H^L$, by only applying valid update operations to it. Updates to graph $H^L$ are triggered by the arrival of new updates from sequence $\Sigma$ to graph $G$. Consider any such update $\sigma_{\tau}$, which must be either the deletion of an edge from $G$, or the insertion of an edge into $G$.

If an edge $e$ is inserted into $G$, then we simply ignore this update. Assume now that an edge $e$ is deleted from $G$. If edge $e$ was inserted into $G$ during Phase $\Phi^L_k$, then we ignore this update. Otherwise, if $e\in A^L_k$, then we delete the two edges that are incident to vertex $v^L(e)$ from graph $H^L$. No further updates to $H^L$ or data structures from lower levels are required. Otherwise, we update data structures from levels $0,\ldots,L-1$, and data structure $\DSFS$, with the deletion of the edge $e$.
As the result of this update, the set $\cset^*$ of clusters may change, and additionally, for some clusters $C\in \cset^*$, the set $Z^L_k(C)$ of vertices may change. Each of such changes may require updating graph $H^L$.

Assume first that, for some cluster $C\in \cset^*$, set $Z^L_k$ of vertices has changed. From Property \ref{prop: flat set decremental for L}, vertices may leave set $Z^L_k$, but they may not join it. Therefore, the change must be the deletion of some vertices from $Z^L_k$. For each such vertex $x$, we delete the edge $(v^L(x),u^L(C))$ from graph $H^L$.
If vertex $x$ is the representative vertex of cluster $C$ at level $L$, then, if vertex $u^L(C)$ is not an isolated vertex in $H^L$, there must be another vertex $y\in V(G)$ with $v^L(y)\in Z^L_k(C)$. We let any such vertex $y$ become the new representative vertex for $C$ at level $L$.

Assume now that a new cluster $C$ just jointed the set $\cset^*$. From Property \ref{propflat3: new cluster in set}, this can only happen if  cluster $C$ was just added to set $\cset^{<L}$, due to a cluster splitting operation that was applied to some cluster $C'$, that currently belongs to $\cset^*$.
Notice that, if $C'\in \cset^{L'}_i$ holds for some $0\leq L'<L$ and $0\leq i\leq \log \hat D$, then $C\in \cset^{L'}_i$ must also hold, since the cluster splitting operation was executed by data structure $\DS^{L'}_i$. So in particular, $\scale(C)=\scale(C')$. From Property \ref{prop: cluster splitting flat sets for L}, $Z^L_k(C)\subseteq Z_k^L(C')$ currently holds. Therefore, for every vertex $x\in Z^L_k(C)$, edge $(v^L(x),u^L(C'))$ currently lies in graph $H^L$. We let $E'=\set{(v^L(x),u^L(C'))\mid x\in Z^L_k(C)}$ be the collection of all such edges. We apply the supernode splitting operation to supernode $u^L(C')$, with the set $E'$ of edges, to obtain a new supernode $u^L(C)$, that represents the new cluster $C$. Note that, from Property \ref{propflat1.5: joining Cstar}, $Z^L_k(C)\neq \emptyset$ must hold, so $E'\neq \emptyset$ as well. It is easy to verify that, at the end of this supernode splitting operation, the new supernode $u^L(C)$ is connected with an edge of length $2^{\scale(C)}=2^{\scale(C')}$ to every regular vertex $v^L(x)$ with $x\in Z^L_k(C)$, and only to such vertices. We select any such vertex $x\in V^L_k(C)$ to be the representative vertex of cluster $C$ for level $L$, setting $\sigma_L(C)=x$.

This completes the algorithm for maintaining graph $H^L$ over the course of Phase $\Phi^L_k$. The running time required for updating graph $H^L_k$ is asymptotically bounded by the total update time of data structure $\DSFS$. Therefore, the running time of the algorithm so far is bounded by $O(M^{q-L+1}\cdot \hat m^{O(1)}+\TFS)$.
Notice that, from Property \ref{prop: every vertex in few flat sets for L}, for every vertex $x\in S^L_k$, there are at most  $\mu_L$ clusters $C\in \tilde \cset^{<L}$, such that $x$ ever belongs to $Z^L_k(C)$ over the course of Phase $\Phi^L_k$. It is then easy to verify that the dynamic degree bound of graph $H^L$ is bounded by $\mu_L$. It is also easy to see that the data structure ensures Properties \ref{prop: representatives 1} and \ref{prop: representatives 2} for representative vertices  of clusters of $\cset^{<L}$ at level $L$.

For all $0\leq i\leq \log \hat D$, we also obtain dynamic graph $H^L_i$, that is identical to $H^L$, except that it excludes all edges whose length is greater than $D_i$. Once graph $H^L$ is initialized, it is easy to initialize each such graph $H^L_i$. When an update operation is applied to graph $H^L$, then we apply the same update operation to graph $H^L_i$, except that we ignore edges whose length is greater than $D_i$. In particular, supernodes $u^L(C)$ with $\scale(C)>i$ are never split, as all their incident edges have length greater than $D_i$. The running time that is required in order to maintain all graphs $H^L_i$, for $0\leq i\leq \log \hat D$ is bounded by $O\left ((M^{q-L+1}\cdot \hat m^{O(\eps)}+\TFS)\cdot \log \hat D\right )\leq O\left (M^{q-L+1}\cdot \hat m^{O(\eps)}+\TFS\cdot \log \hat D\right )$, since $\log \hat D\leq \hat m^{\eps^{12}}$ from Inequality  \ref{eq: bound 2 on D}.

Consider now some scale $0\le i\leq \log \hat D$. Notice that the number of regular vertices in graph $H^L_i$ is $|A^L_k|+|S^L_k|\leq 3M^{q-L+1}<m$.

We have now defined a valid input structure  $\iset^L_i=\left(H^L_i,\set{\ell(e)}_{e\in E(H^L_i)},D_i \right )$, that undergoes a sequence of valid update operations. We view it as an input to the \recdynnc problem, with distance parameter $D_i$ and parameter $\hat W=\max\set{4m\mu_L,\hat m}\leq \hat m^2$. We apply Algorithm $\aset$ for the \recdynnc problem from \Cref{assumption: alg for recdynnc2}
to this instance of \recdynNC. We denote the corresponding data structure by $\DS^L_i$, and we denote by $\cset^L_i$ the collection of clusters (the weak neighborhood cover) that the algorithm maintains. For every cluster $C\in \cset^L_i$, we say that the \emph{scale} of $C$ is $i$, and we denote $\scale(C)=i$. Recall that the algorithm achieves approximation factor $\alpha(\hat W)\leq \alpha(n^3)=\alpha$.
Since the number of regular vertices in $H^L_i$ is at most $3M^{q-L+1}$, and the dynamic degree bound of graph $H^L_i$ is $\mu_L\leq \mu'\leq \hat m^{64\eps^2}$, from 
\Cref{assumption: alg for recdynnc2}, the total update time of data structure $\DS^L_i$ is bounded by:

\[O\left (M^{(q-L+1)(1+O(\eps))}\cdot (\mu_L)^{O(1/\eps)}\right )\leq O\left (M^{q-L+1}\cdot\hat m^{O(\eps)}\right ).
\]

Recall that algorithm $\aset$ also guarantees that every regular vertex of $H^L_i$ may belong to at most $\hat W^{4\eps^3}\leq \hat m^{8\eps^3}$ different clusters of $\cset^L_i$ over the course of Phase $\Phi^L_k$.

Lastly, we denote $\cset^L=\bigcup_{i=0}^{\log \hat D}\cset^L_i$. From the above discussion, every regular vertex of $H^L$ may belong to at most $2\hat m^{8\eps^3}\log \hat D\leq \hat m^{16\eps^3}=\Delta$ clusters of $\cset^L$ over the course of Phase $\Phi^L_k$ (we have used the fact that $\log \hat D\leq \hat m^{\eps^{12}}$ from Inequality  \ref{eq: bound 2 on D}.). 

This completes the description of data structure $\DSbasic$. We denote by $\Tbasic$ that total update time of this data structure over the course of Phase $\Phi^L_k$. From our discussion:

\[\Tbasic\leq  O\left (M^{q-L+1}\cdot \hat m^{O(\eps)}+ \TFS\cdot \log \hat D\right )\leq O\left (M^{q-L+1}\cdot \hat m^{O(\eps)}+ \TFS\cdot \hat m^{O(\eps^3)}\right ). \]

For convenience, for all $0\leq i\leq \log \hat D$, for every regular vertex $v\in V(H^L)$, we denote the cluster $\coveringcluster(v)$ that data structure $\DS^L_i$ maintains by  $\coveringcluster^L_i(v)$. Similarly, we denote the list $\clusterlist(v)$ of all clusters of $\cset^L_i$ containing $v$ that data structure $\DS^L_i$ maintains by $\clusterlist^L_i(v)$.  We then let $\clusterlist^L(v)=\bigcup_{i=0}^{\log \hat D}\clusterlist^L_i(v)$. Lastly, for all $0\le i\leq \log \hat D$, if $C$ is a cluster that currently lies in $\cset^L_i$, and $v,v'$ are two regular vertices of $H^L$ that currently lie in $C$, we denote by $\spquery^L_i(C,v,v')$ a query $\spquery(C,v,v')$ to data structure $\DS^L_i$. Recall that the data structure must return a path $P\subseteq H^L$ of length at most $\alpha\cdot D_i$ connecting $v$ to $v'$ in $H^L_i$, in time $O(|E(P)|)$. For convenience, we will think of $\spquery^L_i(C,v,v')$ as a query that is supported by data structure $\DS^L$.

\subsubsection{Data Structure $\DSimp$}
\label{subsubsection: DSimp}

In this subsection we describe data structure $\DSimp$. Recall that, over the course of Phase $\Phi^L_k$, the data structure must maintain, for every cluster $C\in \cset^L$, for every type-2 supernode $u\in V(C)$, the importance status of supernode $u$, which is a single bit indicating whether supernode $u$ is important for cluster $C$. For brevity, if a type-2 supernode $u$ does not belong to cluster $C$, then we say that it is unimportant for $C$, though there is no need to maintain the importance status in such a case explicitly. The importance status of every type-2 supernode in a cluster $C\in \cset^L$ is completely determined by Rule \ref{rule for importance} (or its equivalent rule \ref{rule2 for importance}). Our goal is to implement a data structure that maintains the importance status of every type-2 supernode in every cluster of $\cset^L$ efficiently. In addition to maintaining the importance status of supernodes in clusters of $\cset^L$, we maintain, for every type-2 supernode $u\in V(H^L)$, an ordered list of all clusters $C\in \cset^L$, such that $u$ is important for $C$, together with pointers to each such cluster. We now describe initialization of data structure $\DSimp$, and the algorithm for updating it.

\paragraph{Initialization.}
Consider the initial collection $\cset^{L}$ of clusters that is constructed at the beginning of Phase $\Phi^{L}_{k}$. 
According to Rule \ref{rule for importance}, 
for every cluster $C\in \cset^{L}$, and every type-2 supernode $u=u^L(C')\in V(C)$, supernode $u$ is important for cluster $C$ if and only if the following three conditions hold: (i) $\scale(C')\leq \scale(C)$; (ii) $N(u)\neq \emptyset$; and (iii) $N(u)\subseteq V(C)$. Here, $N(u)$ is the set of all vertices that are neighbors of $u$ in graph $H^L$. The most natural way to implement the initialization is then to consider every cluster $C\in \cset^L$, and every type-2 supernode $u\in V(C)$ one by one, and then establish whether $u$ is important for cluster $C$. Unfortunately, this algorithm may be inefficient, since, unlike regular vertices, supernodes may belong to a large number of clusters of $\cset^L$, and spending $N(u)$ time on each such occurrence of a supernode in a cluster may be too expensive.

In order to initialize the data structure efficiently, for every cluster $C\in \cset^{L}$ and type-2 supernode $u=u^{L}(C')\in V(C)$, we create a counter $\rho(C,C')$. The counter will count the number of vertices of $N(u)$ that lie in $V(C)$. Initially, we set the counter $\rho(C,C')$ to $0$.  

Next, we consider every type-1 regular vertex $v^{L}(x)\in V^{L}$ one by one. Recall that data structure $\DSbasic$ ensures that vertex $v^{L}(x)$ may belong to at most $\Delta$ clusters of $\cset^{L}$, and it may be a neighbor of at most $\mu_{L}$ supernodes of $H^{L}$. When vertex $v^{L}(x)\in V^{L}$ is processed,  we consider every cluster $C\in \cset^{L}$ with $v^{L}(x)\in V(C)$. For each such cluster, we consider every type-2 supernode $u^{L}(C')$ that is a neighbor of $v^{L}(x)$. If $u^{L}(C')\in V(C)$, then we increase the counter $\rho(C,C')$ by $1$.

 From the above discussion, the time required to process a single regular type-1 vertex of $V^{L}$ is $O(\mu_{L}\cdot \Delta)$, and the time to process all type-1 vertices of $V^{L}$ is bounded by $O(|A^L_k|\cdot \mu_{L}\cdot \Delta)\leq O(M^{q-L+1}\cdot \mu_{L}\cdot \Delta)$. 

Once all type-1 regular vertices of $H^{L}$ are processed, we consider every type-2 supernode $u^{L}(C')\in V(H^{L})$, and we compute $|N(u)|$ -- the number of neighbors of $u$ in $H^L$.  This calculation can be performed for all supernodes of $H^{L}$ in time $O(|E(H^{L})|)\leq O(|A^L_k|\cdot \mu_L)\leq O(M^{q-L+1}\cdot \mu_{L})$.

Lastly, we consider every cluster $C\in \cset^{L}$, and every type-2 supernode $u^{L}(C')\in V(C)$. If $\scale(C')\leq \scale(C)$, $N(u^L(C'))\neq\emptyset$, and  $\rho(C,C')=|N(u^L(C'))|$, then we mark supernode $u^{L}(C')$ as  important for $C$. We then add cluster $C$ to the list of clusters that supernode $u^{L}(C')$ maintains, together with a pointer to cluster $C$. Otherwise, we mark $u^L(C')$ it as unimportant for $C$.
The time that is required in order to perform this last step is bounded by $O\left (\sum_{C\in \cset^L}|V(C)|\right )$, which, in turn, is bounded by the time that is required in order to initialize the set $\cset^L$ of clusters, that is bounded by $O(\Tbasic)$.

Overall, the running time of the algorithm for initializing the data structure is bounded by:

\[O(M^{q-L+1}\cdot \mu_{L}\cdot \Delta+\Tbasic)\leq O(M^{q-L+1}\cdot \hat m^{O(\eps^2)}+\Tbasic)  \]

\paragraph{Updates.}
We now provide an algorithm that maintains, for every cluster $C\in \cset^L$, the importance status of every type-2 supernode  $u\in V(C)$, over the course of Phase $\Phi^{L}_{k}$, so that Rule \ref{rule for importance} is obeyed.
The importance status of a supernode may change either due to the valid update operations that graph $H^L$ undergoes (that includes an insertion of a new type-2 supernode into a cluster $C$ due to the supernode-splitting operation), or due to the allowed updates that clusters of $\cset^L$ undergo. We now consider each of these possible changes in turn, starting with the valid update operations that are applied to graph $H^L$.

First, if an isolated vertex is deleted from graph $H^{L}$, then no further updates to data structure $\DSimp$ are necessary. Assume now that an edge $(u,v)$ is deleted from graph $H^L$, where $u$ is a supernode. If $u$ becomes an isolated vertex in $H^L$, then we consider every cluster $C\in \cset^L$ for which $u$ is marked as an important supernode (using the list maintained by vertex $u$), and we mark $u$ as an unimportant supernode for each such cluster, setting the list of clusters maintained by vertex $u$ to $\emptyset$.

Assume now that graph $H^L$ undergoes a supernode-splitting operation, applied to some supernode $u^{L}(C')$, that creates a new supernode $u^{L}(C'')$. We consider every cluster $C\in \cset^{L}$, to which the new supernode $u^{L}(C'')$ was added. From the definition of allowed changes to the clusters in the definition of the \recdynnc problem, it must be the case that $u^{L}(C')\in V(C)$ holds. If $u^{L}(C')$ is marked as unimportant for $C$, then we mark $u^{L}(C'')$ as unimportant for $C$ as well. Assume now that $u^{L}(C')$ is marked as important for $C$. In this case, from Rule \ref{rule for importance}, every vertex in $N(u^{L}(C'))$ lies in $C$. From the definition of the supernode-splitting operation,  $N(u^{L}(C''))\neq \emptyset$ and $N(u^{L}(C''))\subseteq N(u^{L}(C'))$ must hold. Furthermore, since such a supernode splitting operation may only occur if cluster $C'$ just underwent a cluster-splitting operation that created a new cluster $C''$, it must be the case that $\scale(C'')=\scale(C')\leq \scale(C)$.  Therefore, if $u^{L}(C')$ is marked as important for $C$, we mark $u^{L}(C'')$ as important for $C$ as well, and add cluster $C$ to the list of clusters that supernode $u^{L}(C'')$ maintains. %Oherwise, we mark $u^{L}(C'')$ as unimportant for $C$.

As graph $H^{L}$ undergoes valid update operations, the collection $\cset^{L}$ of clusters also undergoes a sequence of allowed changes. We have already discussed how the insertion of a new supernode into a cluster is handled (recall that a new supernode $u'$ may only be inserted into a cluster $C$ if $u'$ was just created via a supernode-splitting operation applied another supernode $u\in V(C)$). It remains to consider the $\delvertex$ and $\csplit$ operations.

Consider first a $\delvertex$ update, when some vertex $z$ is deleted from some cluster $C\in \cset$. If the vertex is a supernode, and it is marked as important for $C$, then we delete $C$ from the list of clusters maintained by supernode $z$. Assume now that $z$ is a regular vertex. We only need to update the data structure if it is a type-1 regular vertex, that is, $z=v^{L}(x)$ for some vertex $x\in V(G)$. In this case, we consider every supernode $u^{L}(C')$ that is a neighbor of $z$ in $H^{L}$ one by one. For each such supernode $u^{L}(C')$, if it is currently marked as important for $C$, we mark it as unimportant for $C$. Note that, since the degree of $v^{L}(x)$ is bounded by $\mu_{L}$, the time required to update the data structure due to the deletion of $v^{L}(x)$ from $C$ is bounded by $O(\mu_{L})$.

Lastly, consider a $\csplit$ update, where a new cluster $C'$ is created, by splitting a cluster $C\in \cset^{L}$. We consider every supernode $u\in V(C')$ one by one. If $u$ is marked as unimportant for $C$, then we mark it as unimportant for $C'$. Otherwise, we check whether every  neighbor vertex of $u$ in $H^L$ also lies in $V(C')$. This can be done in time $O(|\delta_{C'}(u)|)$. If this is the case, we mark $u$ as important for $C'$, and otherwise we mark it as unimportant for $C'$. Note that the total time required to process this update is bounded by $O(|E(C')|)$.

This concludes our algorithm for maintaining the importance status of the supernodes in the clusters of $\cset^{L}$. It is immediate to verify that the resulting data structure obeys Rule \ref{rule for importance}.

Recall that the total time required to initialize data structure \DSimp is bounded by $O(M^{q-L+1}\cdot \hat m^{O(\eps^2)}+\Tbasic)$. 
Since data structure $\DSbasic$ needs to maintain both the graph $H^L$ and all clusters in $\cset^L$ explicitly, 
from the above discussion, it is easy to verify that the additional time that is required in order to maintain the importance status of the supernodes in each cluster of $\cset^{L}$ is bounded by $O(\Tbasic\cdot \mu_{L})\le O(\Tbasic\cdot \hat m^{O(\eps^2)})$. Overall, the total update time of data structure $\DSimp$ over the course of a single level-$L$ phase is bounded by:

\[O\left (M^{q-L+1}\cdot\hat m^{O(\eps^2)}+\Tbasic\cdot \hat m^{O(\eps^2)}\right ).\]

We will need the following observation regarding the importance status of supernodes.

\begin{observation}\label{obs: supernode important in few clusters}
	Let $u$ be a type-2 supernode that belonged to graph $H^L$ at some time during Phase $\Phi^L_k$. Then the total number of clusters $C\in \cset^L$, such that $u$ was ever marked as an important supernode for $C$ over the course of Phase $\Phi^L_k$, is at most $\Delta$.
\end{observation}
\begin{proof}
	Let $u=u^L(C')$ be a type-2 supernode that belonged to graph $H^L$ at some time during Phase $\Phi^L_k$.  Assume that $u$ was added to graph $H^{L}$ at time $\tau$.  If $\tau$ is not the beginning of Phase $\Phi^L_k$, then node $u$ was added to $H^L$ due to the supernode-splitting operation, and, from the definition of the operation, $N(u)\neq \emptyset$ holds at time $\tau$. Otherwise, from the definition of the initial graph $H^L$, it is easy to verify that $N(u)\neq \emptyset$ must hold at time $\tau$ as well. Note that, from the definition of valid update operations, the set $N(u)$ of the neighbors of vertex $u$ in $H^L$ is decremental.
	
	Among all vertices that lie in $N(u)$ at time $\tau$, we choose a single vertex $v^{L}(x)$, as follows. If, at the end of Phase $\Phi^{L}_{k}$, $N(u)\neq \emptyset$ holds, then we let $v^{L}(x)$ be any vertex lying in set $N(u)$ at the end of the phase. Otherwise, we let $v^{L}(x)$ be the vertex of $H^{L}$, that belonged to set $N(u)$ at time $\tau$, and left this set last.  
	
	Consider now some cluster $C$, such that, at some time $\tau'\geq\tau$ during Phase $\Phi^L_k$, supernode $u$ was marked as an important supernode for $C$. Then at time $\tau'$, it must be the case that $N(u)\subseteq V(C)$, and, in particular, $v^{L}(x)\in V(C)$. Since every regular vertex of $H^{L}$ may belong to at most $\Delta$ clusters of $\cset^{L}$ over the course of Phase $\Phi^{L}_{k}$, we conclude that supernode $u$ may be marked as an important supernode for at most  $\Delta$ clusters of $\cset^{L}$ over the course of Phase $\Phi^{L}_{k}$.
\end{proof}

We can now define the flattened sets $V^F(C)$ of vertices for clusters $C\in \cset^L$. Consider a cluster $C$ that belonged to set $\cset^L$ at some time $\tau\in \tset^L_k$. Then, at time $\tau$, the flattened set $V^F(C)\subseteq V(G)$ of vertices  is the union of two subsets. The first subset is  $\set{x\in V(G)\mid v^L(x)\in V(C)}$, and the second subset is the union of sets $V^F(C')$ for clusters $C'\in \cset^{< L}$, for which supernode $u^{L}(C')$ lies in graph $H^{L}$, and is marked as an important supernode for $C$ at time $\tau$:

\[V^F(C)= \set{x\in V(G)\mid v^L(x)\in V(C)}\cup \left(\bigcup_{u^{L}(C')\mbox{ important for } C}V^F(C')\right ).\]

In the next subsection, we analyze the flattened sets $V^F(C)$ of vertices for clusters $C\in \cset^L$, and in particular we show that Properties \ref{prop: flat set decremental}--\ref{prop: cluster splitting flat sets} hold for them.

\subsubsection{Flattened Vertex Sets -- Analysis}

We start by establishing that Property \ref{prop: cluster splitting flat sets} holds for the flattened sets $V^F(C)$ of vertices for clusters $C\in \cset^L$.

\begin{observation}\label{obs: cluster splitting flat sets}
	If a new cluster $C$ is added to set $\cset^{L}$ at some time $\tau$ during Phase $\Phi^{L}_{k}$, due to a cluster splitting operation that is applied to a cluster $C'\in \cset^{L}$ with $C\subseteq C'$, then, at time $\tau$, $V^F(C)\subseteq V^F(C')$ holds. 
\end{observation}
\begin{proof}
	Since, at time $\tau$, $C\subseteq C'$ holds, every type-1 regular vertex $v^{L}(x)$ that belongs to $V(C)$ must also lie in $V(C')$.  Furthermore, from Rule \ref{rule for importance}, a supernode $u\in V(C')$ may only be marked as important for $C$ at time $\tau$, if  $u$ is marked as important for $C'$ at time $\tau$. From the definition of of the flattened set of vertices, it is now immediate that $V^F(C)\subseteq V^F(C')$ holds at time $\tau$.
\end{proof}

Next, we establish Property \ref{prop: flat set decremental} in the following observation.
\begin{observation}\label{obs: flat set decremental}
For every cluster $C$ that lied in $\cset^{L}$ during Phase $\Phi^{L}_{k}$, the set $V^F(C)$ of vertices is decremental. In other words, after $V^F(C)$ is initialized (when cluster $C$ joins $\cset^{L}$), vertices may leave $V^F(C)$ but not join it.
\end{observation}
\begin{proof}
Consider a cluster $C$ that was added to set $\cset^L$ at some time $\tau$ during Phase $\Phi^L_k$.	Recall that:

\[V^F(C)= \set{x\in V(G)\mid v^L(x)\in V(C)}\cup \left(\bigcup_{u^{L}(C')\mbox{ important for } C}V^F(C')\right ).\] 

Consider first the collection $\set{x\in V(G)\mid v^L(x)\in V(C)}$ of vertices of $G$. Once cluster $C$ is created, it may only undergo allowed updates. From the definition of the allowed updates, no new regular vertices may join $C$. Therefore, vertices may leave set $\set{x\in V(G)\mid v^L(x)\in V(C)}$, but they may not join it.

Consider now some type-2 supernode $u^L(C')$. Recall that $C'\in \cset^{<L}$ must hold, and, from our assumption, Property \ref{prop: flat set decremental} holds for all levels $L'<L$. Since level-$L$ phase $\Phi^L_k$ is contained in some level-$L'$ phase, we conclude that the set $V^F(C')$ of vertices is decremental over the course of Phase $\Phi^L_k$.

Assume first that supernode $u^L(C')$ belonged to cluster $C$ at time $\tau$ when cluster $C$ was created. Note that, from Rule \ref{rule for importance}, the importance status of $u^L(C')$ for cluster $C$ may switch from important to unimportant, but not in the other direction.

Therefore, from the above discussion, the only way that new vertices may join set $V^F(C)$ is when a new supernode $u^L(C')$ is inserted into cluster $C$, that is marked as important for $C$, and so the vertices of $V^F(C')$ are added to set $V^F(C)$. Recall that a new supernode $u^L(C')$ may only be added to cluster $C$ if some type-2 supernode $u^L(C'')$, that currently lies in cluster $C$, just underwent supernode-splitting operation, and supernode $u^L(C')$ was created as the result of this operation. From Rule \ref{rule for importance}, supernode $u^L(C')$ may only be marked as important for $C$, if supernode $u^L(C'')$ is currently marked as such. We now show that $V^F(C')\subseteq V^F(C'')$ holds, and so the vertices of $V^F(C')$ already lie in set $V^F(C)$.

Indeed, let $L'<L$ be the level for which $C''\in \cset^{L'}$. Recall that supernode-splitting operations at level $L$ mirror cluster-splitting operations at lower levels. In other words, a supernode-splitting operation that creates a new supernode $u^L(C')$ may only be applied to supernode $u^L(C'')$ in graph $H^L$, if cluster $C''\in \cset^{L'}$ just underwent a cluster-splitting operation, that resulted in the creation of a new cluster $C'\subseteq C''$. However, in this case, since we have assumed that  Property \ref{prop: cluster splitting flat sets} holds for all levels below $L$, we get that, when cluster $C'$ was created, $V^F(C')\subseteq V^F(C'')$ held. Therefore, no new vertices may be added to set $V^F(C)$ due to the supernode-splitting operation.
\end{proof}

Lastly, we establish Property \ref{prop: every vertex in few flat sets}.
\begin{observation}\label{obs: vertex in few flat sets}
For every vertex $x\in V(G)$, there are at most  $(4\Delta)^{L+1}$ clusters $C\in  \cset^{L}$, such that $x$ ever belonged to set $V^F(C)$ over the course of Phase $\Phi^{L}_{k}$.
\end{observation}
\begin{proof}
%	The proof is by induction on $L'$. The base of the induction is $L'=0$. Recall that there is only one level-$0$ phase. Consider any vertex $x\in V(G)$. If $C$ is a cluster that ever belonged to set $\cset^0$, and $x$ ever belonged to $V^F(C)$, then at some time during the algorithm, $v^1(x)\in V(C)$ must have held. Recall that we are guaranteed that, for every regular vertex of $H^1$, there are at most $O(\Delta(W)\cdot \log W)\leq \Delta$ clusters of $\cset^0$ to which it every belongs over the course of the algorithm. Therefore, there may be at most $\Delta$ clusters $C$ that belonged to $\cset^0$ over the course of the algorithm, such that $x\in V^F(C)$ ever held.
%	Consider now some level $0<L'<L$, and assume that the claim holds for levels $0,\ldots,L'-1$. 
Consider a level-$L$ phase $\Phi^L_k$, and some vertex $x\in V(G)$. 

For convenience, for all $0\leq L'\leq L$, we denote by $\tilde \cset^{L'}$ the collection that contains all clusters $C$ that ever belonged to $\cset^{L'}$ over the course of the level-$L$ phase $\Phi^{L}_{k}$. Recall that, for all $0\leq L'<L$,  current level-$L$ phase $\Phi^{L}_{k'}$ is contained in the current level-$L'$ phase. 

Recall first data structure $\DSbasic$ ensures that every regular vertex may belong to at most $\Delta$ clusters in $\tilde \cset^L$ over the course of Phase $\Phi^L_k$. Therefore, the number of clusters $C\in \tilde \cset^L$, for which $v^L(x)\in V(C)$ ever held over the course of the phase is at most $\Delta$.

Consider now some level $0\leq L'<L$.
Since we have assumed that  Property \ref{prop: every vertex in few flat sets} holds for all levels below $L$, there are at most $(4\Delta)^{L'+1}$ clusters $C\in \tilde \cset^{L'}$, such that $x\in V^F(C)$ ever held over the course of Phase $\Phi^{L}_{k}$. Let $\cset^*(x)$ denote the set of all clusters $C\in \bigcup_{L'=0}^{L-1}\tilde \cset^{L'}$, such that $x\in V^F(C)$ held at any time during Phase $\Phi^{L}_{k}$. From the above discussion, $|\cset^*(x)|\leq \sum_{L'=0}^{L-1}(4\Delta)^{L'+1}\leq 2\cdot (4\Delta)^{L}$. 
From \Cref{obs: supernode important in few clusters}, every supernode of $H^{L}$ may be marked as important for at most $\Delta$ clusters over the course of Phase $\Phi^L_k$. Therefore, for every cluster $C\in \cset^*(x)$, there are at most $\Delta$ clusters $C'\in \tilde \cset^L$, for which $u^L(C)$ was ever marked as important for $C'$ over the course of Phase $\Phi^L_k$. We conclude that the number of clusters $C'\in \tilde \cset^L$, for which $x\in V^F(C)$  ever held over the course of Phase $\Phi^L_k$ is bounded by:

\[\Delta+|\cset^*(x)|\cdot \Delta\leq \Delta+ 2\cdot (4\Delta)^{L}\cdot \Delta\leq (4\Delta)^{L+1}.\]

%	Let $\tilde \cset^{L'}$ be the collection of all clusters that ever belonged to $\cset^{L'}$ over the course of Phase $\Phi^{L'}_{k'}$. Let $\cset(x)\subseteq \tilde \cset^{L'}$ denote the subset of clusters $C$ such that $x\in V^F(C)$ ever held over the course of Phase $\Phi^{L'}_{k'}$. We partition $\cset(x)$ into two subsets.
%	Set $\cset_1(x)$ contains all clusters $C\in \tilde \cset(x)$, such that $v^{L'}(x)$ ever belonged to $V(C)$ over the course of Phase $\Phi^{L'}_{k'}$, and $\cset_2(x)$ contains all remaining clusters. Since every regular vertex may lie in at most $\Delta$ clusters of $\tilde \cset^{L'}$, we get that $|\cset_1(x)|\leq \mu_{L'}$. Next, we bound $\cset_2(x)$. For every cluser $C\in \cset_2(x)$, there must be a cluster $C'\in \cset^*(x)$, such that supernode $u(C')$ was marked as important supernode for cluster $C$. Recall that $|\cset^*(x)|\leq 2\cdot (4\Delta)^{L'-1}$, and, from 
\end{proof}

To summarize, we have now established that Properties \ref{prop: flat set decremental}--\ref{prop: cluster splitting flat sets} hold for level $L$ and the sets $\set{V^F(C)}_{C\in \cset^L}$ of vertices. 

\subsubsection{Data Structure $\DSFS$}

In this subsection we describe data structure $\DSFS$, that is maintained as part of the level-$L$ data structure, over the course of a single level-$L$ phase $\Phi^L_k$. The purpose of the data structure is to efficiently maintain the sets $Z^L_k(C)=V^F(C)\cap S^L_k$ of vertices for clusters $C\in \cset^{<L}$. This data structure only depends on data structures maintained at levels $0,\ldots,(L-1)$, and is independent of other level-$L$ data structures, except that it uses the set $S^L_k$ of vertices (that is computed at the beginning of Phase $\Phi^L_k$, and does not change over the course of the phase).

We now fix a level-$L$ phase $\Phi^L_k$. For convenience, we denote by $\tilde T^L_k$ the total update time of data structures from levels $0,\ldots,(L-1)$ during Phase $\Phi^L_k$. We will use $\tilde T^L_k$ in order to bound the total update time of $\DSFS$ over the course of Phase $\Phi^L_k$.
We start by describing the algorithm for initializing data structure $\DSFS$, and then provide an algorithm for updating it.

\subsubsection*{Initialization}
At the beginning of Phase $\Phi^L_k$, we construct a collection $\cset^*\subseteq \cset^{<L}$ of clusters $C$, that contains every cluster $C\in \cset^{<L}$, for which $Z^L_k(C)\neq \emptyset$ holds, and only such clusters. For every cluster $C\in \cset^*$, we also construct the corresponding set $Z^L_k(C)=V^F(C)\cap S^L_k$ of vertices. We emphasize that the sets $V^F(C)$ of vertices are not maintained explicitly.

Consider now some cluster $C\in \cset^{<L}$. In order to efficiently maintain the sets $Z^L_k(C)$ of vertices as the phase progresses, we will  store, together with every vertex $x\in Z^L_k(C)$, some additional information. Specifically, assume that $C\in \cset^{L'}$ for some level $0\leq L'<L$. We will store, together with every vertex $x\in Z^L_k(C)$, a list $\supernodes(C,x)$, that contains every type-2 supernode $u=u^{L'}(C')$, such that $u$ is marked as important for $C$, and $x\in Z^L_k(C')$ holds. Intuitively, these are the supernodes that are responsible for including vertex $x$ in set $Z^L_k(C)$, and this additional data structure will allow us to maintain the set $Z^L_k(C)$ efficiently, as these supernodes may become unimportant for $C$ over the course of the phase.

For every vertex $x\in S^L_k$, we also maintain pointers from $x$ to every cluster $C\in \cset^{<L}$ with $x\in Z^L_k(C)$.

In order to initialize the \DSFS data structure, we start by setting $\cset^*=\emptyset$, and then process all levels $L'=0,\ldots,L-1$ one by one.

\paragraph{Processing Level $0$.}
We first describe an algorithm for processing level $0$. For every vertex $x\in S^L_k$, we use data structure $\clusterlist^0(v^0(x))$ to compute all clusters $C\in \cset^0$ with $v^0(x)\in V(C)$. For each such cluster $C$, if $C$ does not belong to set $\cset^*$ yet, we add $C$ to $\cset^*$ and initialize $Z^L_k(C)=\set{x}$. Otherwise, if $C$ already lies in $\cset^*$, we add vertex $x$ to set $Z^L_k$. Since every regular vertex may lie in at most $\Delta$ clusters of $\cset^0$, the running time required to process level $0$ is bounded by $O(\Delta\cdot |S^L_k|)$.

\paragraph{Processing Level $L'$.}
Consider now some integer $0<L'\leq L-1$, and assume that levels $0,\ldots,L'-1$ were already processed. We now show an algorithm to process level $L'$.
The algorithm for processing level $L'$ consists of two steps.

In the first step, we process every cluster $C$ that lies in $\cset^*$ one by one. Each such cluster $C$ must belong to $\cset^{<L'}$. In order to process cluster $C$, we check whether $u^{L'}(C)$ is a vertex of $H^{L'}$. If this is not the case, then no further processing of $C$ is required. Assume now that $u^{L'}(C)$ is a vertex of $H^{L'}$. Then we consider every cluster $C'\in \cset^{L'}$, such that $u^{L'}(C)$ was marked as an important supernode for $C'$ (in which case $u^{L'}(C)\in V(C')$ must hold). Recall that data structure $\DSimp$ for level $L'$ maintains a pointer from $u^{L'}(C)$ to every such cluster $C'$.
For each such cluster $C'$, if
 $C'$ does not currently lie in $\cset^*$, then we add it to $\cset^*$, and initialize $Z^L_k(C')=\emptyset$. If $C'$ currently lies in $\cset^*$, then set $Z^L_k(C')$ is already initialized.
Next, we consider every vertex $x\in Z^L_k(C)$ one by one. For each such vertex $x$, if $x$ already lies in set $Z^L_k(C')$, then we add supernode $u^{L'}(C)$ to the list $\supernodes(C,x)$. Otherwise, we add $x$ to $Z^L_k(C')$, and we initialize the list $\supernodes(C,x)$ to contain supernode $u^{L'}(C)$. This completes the algorithm for processing a cluster $C\in \cset^*$, and the algoritm for the first step of processing level $L'$.

Recall that, from Property \ref{prop: every vertex in few flat sets}, for every vertex $x\in S^L_k$, there are at most $(4\Delta)^{L'+1}$ clusters $C\in \cset^*$ with $x\in V^F(C)$ before level $L'$ is processed. In particular, $|\cset^*|\leq |S^L_k|\cdot (4\Delta)^{L'+1}$ must hold then. Moreover, from \Cref{obs: supernode important in few clusters}, every supernode of $H^{L'}$ may be marked as important for at most $\Delta$ clusters of $\cset^{L'}$. It is then easy to verify that the running time of the first step is bounded by $O\left(|S^L_k|\cdot (4\Delta)^{L'+2}\right )$.

In the second step, we consider every vertex $x\in S^{L}_k$ one by one.  
For each such vertex $x\in S^L_k$, if $v^{L'}(x)\in V(H^{L'})$, then we use data structure $\clusterlist^{L'}(v^{L'}(x))$ to compute all clusters $C\in \cset^{L'}$ with $v^{L'}(x)\in V(C)$. For each such cluster $C$, if $C$ does not belong to $\cset^*$ yet, we add $C$ to $\cset^*$ and initialize $Z^L_k(C)=\set{x}$ and $\supernodes(C,x)=\emptyset$. Otherwise, we add vertex $x$ to $Z^L_k(C)$ if it does not lie in this set already. Since every regular vertex may lie in at most $\Delta$ clusters of $\cset^{L'}$, the running time of this second step is $O(|S^L_k|\cdot \Delta)$.

This completes the algorithm for processing level $L'$. Once all levels $0\leq L'<L$ are processed, we add, for every vertex $x\in S^L_k$, pointers from $x$ to every cluster $C\in \cset^*$ with $x\in Z^L_k(C)$. This completes the algorithm for initializing the $\DSFS$ data structure.
From the discussion so far, the running time required to initialize the data structure is bounded by $O\left(|S^L_k|\cdot (4\Delta)^{L+1}\right )\leq O\left(|S^L_k|\cdot \hat m^{O(\eps^2)}\right )$. Clearly, set $\cset^*$ contains all clusters $C\in \cset^{<L}$ with $Z^L_k(C)\neq\emptyset$, and only such clusters. Therefore, Properties \ref{propflat1.5: joining Cstar} and \ref{propflat2: not in Cstar emptyset} hold after the initialization of data structure $\DSFS$.

\subsubsection*{Updating the $\DSFS$ Data Structure.}

We now describe an algorithm for  maintaining the set $\cset^*$ of clusters over the course of Phase $\Phi^L_k$, so that Properties \ref{propflat1: incremental set} -- \ref{propflat3: new cluster in set}  hold. For every cluster $C\in \cset^*$, we will also maintain the set $Z^L_k$ of vertices, and, for every vertex $x\in Z^L_k(C)$, the list $\supernodes(C,x)$, containing every supernode $u^L(C')$ that is marked as important for $C$, for which $x\in V^F(C')$ holds.

It is easy to verify that we only need to update the set $\cset^*$ of clusters if a new cluster $C'$ is created via the cluster splitting operation of some cluster $C\in \cset^*$ (if a cluster splitting operation is applied to a cluster $C\not\in \cset^*$, then $Z^L_k(C)=\emptyset$ must hold, and, from Property \ref{prop: cluster splitting flat sets}, $Z^L_k(C')=\emptyset$ holds as well, so we do not need to add $C'$ to $\cset^*$). Additionally, we may need to update the sets $Z^L_k(C)$ of vertices for clusters $C\in \cset^*$, if one of the following happen: (i) a regular vertex is deleted from a cluster $C\in \cset^*$; or (ii) a supernode that was marked as important for a cluster $C\in \cset^*$ becomes unimportant for $C$. Notice also that, if some vertex $x$ is deleted from a set $Z^L_k(C)$ for some cluster $C\in \cset^*$ with $C\in \cset^{L'}$, then this change may propagate to higher levels: if, for some level $L''>L'$, supernode $u^{L''}(C)$ lies in $H^{L''}$, and it is an important supernode for some cluster $C'\in \cset^{L''}\cap \cset^*$, then we may need to delete $x$ from $Z^L_k(C')$ as well. We also need to update the data structure if a supernode splitting operation occurs, for some supernode $u\in V(C)$, where $C\in \cset^*$. In this case, the set $Z^L_k(C)$ may not change, but we may need to update the lists $\supernodes(C,x)$ of some vertices $x\in Z^L_k(C)$. We now consider each of these updates one by one.

\paragraph{Deletion of a regular vertex from a cluster.}
Let $C\in \cset^{L'}$ be a cluster that currently lies in set $\cset^*$, for some $0\leq L'<L$, and assume that some type-1 regular vertex $v^{L'}(x)$ was just deleted from $C$ (if the deleted vertex is a type-2 regular vertex, then no update is needed). If $x\not\in S^L_k$, then no other updates are needed. Assume now that $x\in S^L_k$.  If the list $\supernodes(C,x)$ is empty, then we delete $x$ from $Z^L_k(C)$. Otherwise, no update is needed to the data structure. 
Note that the running time required to update the data structure is asymptotically bounded by the number of clusters containing vertices $v^{L'}(x)$, for $0\leq L'<L$. Therefore, the total time required in order to update data structure $\DSFS$ due to the deletions of regular vertices from clusters is asymptotically bounded by the total update time of data structures from levels $0,\ldots,L-1$ over the course of Phase $\Phi^L_k$.
%If, after this deletion, $Z^L_k(C)=\emptyset$ holds, then we delete $C$ from $\cset^*$.

\paragraph{A supernode becomes unimportant for a cluster.}
Let $C\in \cset^{L'}$ be a cluster that currently lies in set $\cset^*$, for some $0\leq L'<L$, and assume that some type-2 supernode $u^{L'}(C')\in V(C)$ that was previously marked as important for $C$, is now marked as unimportant (this includes the case when $u^{L'}(C')$ is deleted from cluster $C$). We consider every vertex $x\in Z^L_k(C')$ one by one. For each such vertex $x$, we delete supernode $u^{L'}(C')$ from the list $\supernodes(C,x)$. If the list becomes empty, and there is no type-1 regular vertex $v^{L'}(x)$ in $V(C)$, then we delete $x$ from set $Z^L_k(C')$. %If $Z^L_k(C)$ becomes empty, then we delete $C$ from $\cset^*$.
The total time required in order to update the data structure $\DSFS$  is asymptotically bounded by the number of deletions performed in the lists $\supernodes(C,x)$ for vertices $x\in S^L_k$.

\paragraph{A new cluster is created.}
Let $C\in \cset^{*}$ be a cluster, with $C\in \cset^{L'}$, for some $0\leq L'<L$, and assume that a new cluster $C'$ was just created added to $\cset^{L'}$ via a cluster-splitting operation applied to cluster $C$.

 We initialize $Z^L_k(C')=\emptyset$, and then consider every type-1 regular vertex $v^{L'}(x)\in V(C')$ one by one. For each such vertex, if $x\in S^L_k$, then we add $x$ to $Z^L_k(C')$. Next, we process every type-2 supernode $u^{L'}(C'')\in V(C')$. For each such supernode $u^{L'}(C'')$, if it is marked as important for $C'$, then we consider every vertex $y\in Z^L_k(C'')$. If $y\in Z^L_k(C')$ holds, then we add supernode $u^{L'}(C'')$ to the list $\supernodes(C',y)$. Otherwise, we add $y$ to $Z^L_k(C')$, and we initialize the list $\supernodes(C',y)$ to contain the supernode $u^{L'}(C'')$. 
  Once all regular vertices and type-2 supernodes of $C'$ are processed, if $Z^L_k(C')\neq \emptyset$, then we add $C'$ to $\cset^*$.
 The time required in order to update data structure \DSFS due to the creation of a new cluster $C'$ is asymptotically bounded by $|V(C')|$ plus the total number of entries that were added to lists $\supernodes(C',x)$ for vertices $x\in S^L_k$.
  
Note that the splitting of cluster $C$ may lead to supernode splitting of the supernode $u^{L''}(C)$ in graphs $H^{L''}$ with $L''\geq L'$, to which such a supernode belongs. We now provide an algorithm that updates the data structure due to supernode splitting.

\paragraph{Supernode splitting.}
Consider some level $0\leq L'<L$, and assume that some supernode $u^{L'}(C')$ just underwent a supernode-splitting operation, through which a new supernode $u^{L'}(C'')$ was created, where $C''\subseteq C'$ is a cluster that was just split off from cluster $C'$. As a result, the newly created supernode $u^{L'}(C'')$ may have been added into some clusters $C\in \cset^{L'}$. Let $C$ be any such cluster, and recall that $u^{L'}(C')\in V(C)$ must hold (due to the definition of allowed updates to clusters maintained by the \recdynnc data structure). If supernode $u^{L'}(C'')$ is not marked as an important cluster for $C$, then no further updates are necessary. Assume now that $u^{L'}(C'')$ is marked as important for $C$. In this case, $u^{L'}(C')$ must also be important for $C$, and so $Z^{L}_k(C')\subseteq Z^L_k(C)$ currently holds. Since cluster $C''$ was just split off from cluster $C'$, from 
Property \ref{prop: cluster splitting flat sets},
  $V^F(C'')\subseteq V^F(C')$ currently holds, and so $Z^L_k(C'')\subseteq Z^L_k(C')$. In particular, $Z^L_k(C'')\subseteq Z^L_k(C)$ must hold. For every vertex $x\in Z^L_k(C'')$, we add supernode $u^{L'}(C'')$ to list $\supernodes(C,x)$.
 The time required in order to update data structure \DSFS due to the supernode splitting operation is asymptotically bounded by the total number of entries that were added to lists $\supernodes(C,x)$ for vertices $x\in S^L_k$ and clusters $C\in \cset^*$, plus the number of clusters in $\cset^{<L}$, into which the new supernode was inserted. The latter can be charged to the total update time of data structures from levels $0,\ldots,L-1$.

\paragraph{Propagated updates.}
Assume that some vertex $x$ was just deleted from set $Z^L_k(C)$ of any cluster $C\in \cset^*$, and assume that $C\in \cset^{L'}$. For every level $L'<L''<L$, we check whether $u^{L''}(C)$ is a supernode of $H^{L''}$, and if so, we consider every cluster $C'\in \cset^{L''}$ for which $u^{L''}(C)$ is an important supernode, using the list that the supernode maintains. Consider any such cluster $C'\in \cset^{L''}$, and notice that $C'\in \cset^*$ and $x\in V^F(C')$ must hold. We delete $u^{L''}(C)$ from the list $\supernodes(C',x)$. If the list becomes empty, and $v^{L''}(x)\not\in V(C)$, then we delete $x$ from $Z^{L}_k(C')$ (which, in turn, may trigger updates in higher levels). Notice however that the running time of all such updates can be charged to the number of supernodes deleted from lists $\supernodes(C',x)$, for vertices $x\in S^L_k$ and clusters $C'\in \cset^*$.

This completes the algorithm for maintaining the collection $\cset^*$ of clusters, and sets $Z^L_k(C)$ of vertices for clusters in $\cset^*$, over the course of Phase $\Phi^L_k$.  
We claim that the algorithm guarantees Properties \ref{propflat1: incremental set} -- \ref{propflat3: new cluster in set}. Indeed, it is immediate to see that, once the set $\cset^*$ of clusters is initialized, we never delete clusters from it, so Property \ref{propflat1: incremental set} holds. It is also easy to verify that, at the beginning of Phase $\Phi^L_k$, we correctly initialize the set $\cset^*$ of clusters to contain every cluster $C\in \cset^{<L}$ with $Z^L_k(C)\neq \emptyset$, and only such clusters. As the phase progresses, we only add new clusters $C$ to $\cset^*$ if $Z^L_k(C)\neq \emptyset$ holds. Therefore, Property \ref{propflat1.5: joining Cstar} is maintained.

As noted already, Property \ref{propflat2: not in Cstar emptyset} holds at the beginning of the phase. Since vertex sets $Z^L_k(C)$ are decremental, the only way that the set $Z^L_k(C)$ of vertices is non-empty for a cluster $C\not\in \cset^*$, is if $C$ is a newly created cluster. However, in this case, cluster $C$ must have been created via a cluster-splitting operation of some cluster $C'$, and, from Property \ref{prop: cluster splitting flat sets}, $V^F(C)\subseteq V^F(C')$ must hold. But then $Z^L_k(C)\subseteq Z^L_k(C')$, and so $Z^L_k(C')\neq\emptyset$. Therefore, $C'\in \cset^*$ holds, and our algorithm will add cluster $C$ to set $\cset^*$ the moment it is created. This establishes Property \ref{propflat2: not in Cstar emptyset}.  Property \ref{propflat3: new cluster in set} follows immediately from the algorithm.

We conclude that the set $\cset^*$ of clusters that we maintain satisfies Properties \ref{propflat1: incremental set} -- \ref{propflat3: new cluster in set}. It is then easy to verify that the sets $Z^L_k(C)$ of vertices for clusters $C\in \cset^*$ are maintained correctly by our algorithm.

We now analyze the total update time of data structure $\DSFS$ over the course of Phase $\Phi^L_k$. 
Recall that we have already bounded the running time of the algorithm that initializes the data structure by $O\left(|S^L_k|\cdot (4\Delta)^{L+1}\right )$.

Let $\tilde \cset^*$ denote the collection $\cset^*$ at the end of the phase. Since set $\cset^*$ is incremental, set $\tilde \cset^*$ maintains every cluster that ever belonged to $\cset^*$ over the course of the phase.

For every vertex $x\in S^L_k$, we denote by $\tilde \cset(x)$ the collection of all clusters $C\in \tilde \cset^*$, for which $x\in Z^L_k(C)$ ever held during Phase $\Phi^L_k$. Let $\Pi(x)$ be the collection of all pairs $(C,u)$, where $C\in \tilde \cset(x)$, and supernode $u$ ever belonged to list $\supernodes(C,x)$. We bound the total update time of the algorithm for maintaining data structure $\DSFS$ (excluding the initialization) as follows. For every vertex $x\in S^L_k$, we let the budget of $x$ be $\beta(x)=|\tilde \cset(x)|+|\Pi(x)|$. Whenever vertex $x$ is added to a set $Z^L_k(C)$ of a newly created cluster $C$, or whenever the list $\supernodes(C,x)$ is modified, we can charge the time required to process these changes to the budget of $x$. This is because, for every cluster $C$ with $x\in Z^L_k(C)$, every supernode can be added to the list $\supernodes(C,x)$ at most once, and then deleted from the list at most once (because the status of a supernode $u$ for cluster $C$ may never switch from unimportant to important, and because set $V^F(C')$ of vertices is decremental for all clusters $C'\in \cset^{<L}$). Furthermore, vertex $x$ may only be added to set $Z^L_k(C)$ once when cluster $C$ is created, and after that it may only be deleted from the set. The additional running time that is required by the algorithm for maintaining data structure $\DSFS$ is asymptotically bounded by the time that is required in order to maintained data structures from levels $0,\ldots,L-1$ over the course of Phase $\Phi^L_k$, that we denoted by $\tilde T^L_k$. Overall, the total update time that is required in order to maintain data structure $\DSFS$ (excluding initialization) is bounded by:

$$O\left(\tilde T^L_k+\sum_{x\in S^L_k}\beta(x)\right )$$.

Next, we bound $\sum_{x\in S^L_k}\beta(x)$. First, from Property \ref{prop: every vertex in few flat sets}, $|\tilde \cset(x)|\leq \sum_{L'=0}^{L-1} (4\Delta)^{L'+1}\leq (4\Delta)^{L+1}$. Next, we bound $|\Pi(x)|$. Consider a pair $(C,u)\in \Pi(x)$, and assume that $C\in \cset^{L'}$, for some $0<L'<L$, and $u=u^{L'}(C')$. Then $C'\in \tilde\cset(x)$ must hold, and supernode $u^{L'}(C')$ is important for cluster $C$.

Consider now some level $0\leq L''<L$, and a cluster $C'\in\tilde\cset(x)\cap \tilde \cset^{L''}$. Let $L''<L'<L$ be any level, such that supernode $u^{L'}(C')$ belongs to graph $H^{L'}$ at any time during Phase $\Phi^L_k$. From \Cref{obs: supernode important in few clusters}, the total number of clusters $C\in \tilde \cset^{L'}$, such that supernode $u^{L'}(C')$ was ever marked as important for $C$ over the course of Phase $\Phi^L_k$ is at most $\Delta$. Therefore, there are at most $\Delta$ pairs $(C,u^{L'}(C'))\in \Pi(x)$, where $C\in  \cset^{L'}$ ever held. We conclude that for every cluster $C'\in \tilde \cset(x)$, there are at most $L\cdot \Delta$ pairs $(C,u^{L'}(C'))\in \Pi(x)$, where $0\leq L'<L$. Therefore, $|\Pi(x)|\leq |\tilde \cset(x)|\cdot L\cdot \Delta\leq  L\cdot (4\Delta)^{L+1}\cdot \Delta$. Overall, we get that:

\[\sum_{x\in S^L_k}\beta(x)\leq |S^L_k|\cdot \left( (4\Delta)^{L+1}+L\cdot (4\Delta)^{L+1}\cdot \Delta\right )\leq |S^L_k|\cdot (4\Delta)^{2L}.\]

Overall, the total update time that is required in order to maintain the flat vertex sets (including initialization) is bounded by $O\left(\tilde T^L_k+|S^L_k|\cdot (4\Delta)^{2L}\right )\leq O\left(\tilde T^L_k+M^{q-L+1}\cdot \hat m^{O(\eps^2)}\right )$, since $|S^L_k|\leq 2|A^L_k|\leq 2M^{q-L+1}$, $\Delta= \hat m^{16\eps^3}$, and $L\leq \ceil{1/\eps}$.

\subsubsection{Analysis of Total Update Time}

We now bound the total update time of the level-$L$ data structure. We first fix a single level-$L$ phase $\Phi^L_k$, and analyze the total update time of the level-$L$ data structure over the course of the phase. Recall that the total update time required by data structure $\DSFS$ is bounded by:

\[\TFS\leq O\left(\tilde T^L_k+M^{q-L+1}\cdot \hat m^{O(\eps^2)}\right ),\]

where $\tilde T^L_k$ is total update time of data structures from levels $0,\ldots,L-1$ during phase $\Phi^L_k$. 

The total update time of data structure $\DSbasic$ is:

\[\Tbasic\leq  O\left (M^{q-L+1}\cdot \hat m^{O(\eps)}+ \TFS\cdot \hat m^{O(\eps^3)}\right ). \]
%\[\Tbasic\leq  O\left (\left (M^{q(1+\delta)-L+1}\cdot (\mu_L\cdot \log W)^c+ \TFS\right )\cdot \log \hat D\right ), \]

and the total update time of data structure $\DSimp$ is bounded by:

\[O\left (M^{q-L+1}\cdot\hat m^{O(\eps^2)}+\Tbasic\cdot \hat m^{O(\eps^2)}\right ).\]

Overall, we get that the total update time of the level-$L$ data structure over the course of Phase $\Phi^L_k$ is bounded by:

\[ 
\begin{split}
&O\left (M^{q-L+1}\cdot \hat m^{O(\eps^2)}+\Tbasic\cdot \hat m^{O(\eps^2)}\right )\\
&\quad\quad\quad\quad\leq O\left (M^{q-L+1}\cdot \hat m^{O(\eps^2)}+M^{q-L+1}\cdot \hat m^{O(\eps)}+ \TFS\cdot\hat m^{O(\eps^2)}\right )\\
&\quad\quad\quad\quad\leq O\left (M^{q-L+1}\cdot \hat m^{O(\eps)}+\tilde T^L_k\cdot \hat m^{O(\eps^2)} \right ).
%&\quad\quad\quad\quad\leq O\left (M^{q(1+\delta)-L+1}\cdot (\mu_L\cdot \log W)^{c+1}\cdot (4\Delta')^{2L}\cdot  \log \hat D    +\tilde T^L_k\cdot\mu_L\cdot  \log \hat D\right ).
\end{split}  \]

%\mynote{add to thm statement that $2^{O(1/\eps)}\leq W$. Also we assume that $\Delta'\geq 2$}
%Since $\mu_L=(8\Delta')^L\leq (8\Delta')^{2/\eps}\leq (\Delta')^{4/\eps}$ (as $\Delta'\geq \Delta(W)\geq 2$), and $\hat D=D^*\cdot 2^{10q+10}=D^*\cdot 2^{O(1/\eps)}\leq W\cdot 2^{O(1/\eps)}\leq \poly(W)$ (since $D^*\leq W$ and $\eps>1/(\log W)^{1/4}$ from the definition of $W$-bounded \APSP (see \Cref{def: bounded APSP}), we get that the total update time of the level-$L$ data structure over the course of Phase $\Phi^L_k$ is at most:

%\[
%O\left (M^{q(1+\delta)-L+1}\cdot (\Delta'\cdot  \log W)^{O(1/\eps)}    +\tilde T^L_k\cdot(\Delta')^{O(c/\eps)}  \log W\right ).
%\]
%
%(we have also used the fact that $L\leq q=\ceil{1/\eps}$).

Recall that the number of level-$L$ phases is $M^L$. Let $T^{<L}$ denote the total update time of data structures for levels $0,\ldots,L-1$ over the course of the whole algorithm. Then the total update time of the level-$L$ data structure over the course of the whole algorithm is bounded by:

\[
\begin{split}
O\left (M^{q+1}\cdot \hat m^{O(\eps)}+ T^{<L}\cdot \hat m^{O(\eps^2)}\right )
&\leq O\left (m\cdot \hat m^{O(\eps)}+  T^{<L}\cdot \hat m^{O(\eps^2)}\right ).
\end{split}\]

(We have used the fact that $M^q<m^{1+\eps}\leq m\cdot \hat m^{\eps}$.)

Let $T^0$ denote the total update time of the level-$0$ data structure over the course of the algorithm. We then get the following recursion:

\[
T^{<1}=T^0\leq O\left(m\cdot \hat m^{O(\eps)}\right ),
\]

and for $1<L\leq q+1$:

\[T^{<L} \leq O\left (m\cdot \hat m^{O(\eps)}+  T^{<(L-1)}\cdot \hat m^{O(\eps^2)}\right ).\]

Since $q=\ceil{1/\eps}$, it is then easy to see that:

\[T^{<q+1}\leq T^0\cdot \left(\hat m^{O(\eps^2)}\right )^{q+1}\leq T^0\cdot \hat m^{O(\eps)}\leq  O\left(m\cdot \hat m^{O(\eps)}\right ).\]

The amortized update time of the algorithm is bounded by $\hat m^{O(\eps)}$ per operation, as required.

\subsection{Responding to Short-Path Queries}
\label{sec: short path query support}

In this section we provide an algorithm to support   \shortpath\  queries. We start by showing that, for every cluster $C\in \cset^{<q+1}$, all vertices lying in the flattened set $V^F(C)$ are close to each other in the current graph $G$. We also show an algorithm that, for all $0\leq L\leq q$, given a path $P$ in graph $H^L$ between a pair $v^L(x),v^L(y)$ of type-1 regular vertices, transforms it into a path connecting $x$ to $y$ in $G$, while approximately preserving its length. We will exploit this path-transformation algorithm in order to support  \shortpath\  queries.

\subsubsection{Paths between Vertices in Flattened Sets, and Path Transformation}

Consider some level $0\leq L\leq q$, and some time $\tau\in\tset$ during the time horizon. Let $C$ be a cluster that lies in set $\cset^L$ at time $\tau$. Recall that we have defined a dynamic set $V^F(C)$ of vertices associated with cluster $C$, that is used in order to construct and maintain graphs at higher levels. We show that every pair of vertices in set $V^F(C)$ is connected by a short path in the current graph $G$, and we further show an efficient algorithm that computes such a path. Recall that $\alpha=\alpha(n^3)$ is the approximation factor that Algorithm $\aset$ from \Cref{assumption: alg for recdynnc} achieves.

\begin{claim}\label{claim: short paths flat sets}
	There is a large enough constant $\tilde c$, and a deterministic algorithm that, at any time $\tau\in \tset$, given a level $0\leq L\leq q$, a cluster $C\in\cset^L$, and a pair of vertices $x,y\in V^F(C)$, returns a path $P$ connecting $x$ to $y$ in the current graph $G$, whose length is at most $(\tilde c\alpha)^{L+1}\cdot D_i$, where $i=\scale(C)$. The running time of the algorithm is bounded by $O(|E(P)|)$.
\end{claim}

\begin{proof}
	Throughout, we denote $i=\scale(C)$.
	The proof is by induction on the level $L$.  The base case is when $L=0$. From the definition of the flattened sets of vertices, if $x,y\in V^F(C)$, then $v^0(x),v^0(y)\in V(C)$ must hold. We execute query $\spquery^0_i(C,x,y)$ in data structure $\DS^0_i$, that must return a path $Q$ connecting vertex $v^0(x)$ to vertex $v^0(y)$ in graph $H^0$, whose length is at most $\alpha\cdot D_i$. We can assume w.l.o.g. that the path contains no type-1 regular vertices except for its endpoints, since every type-1 regular vertex has degree $1$ in $H^0$. By suppressing all type-2 regular vertices on path $P$, deleting the first and the last vertex on the path, and replacing each supernode $u(z)$ with the corresponding vertex $z\in V(G)$, we obtain a path $P$ connecting $x$ and $y$ in the current graph $G$, whose length is at most $\alpha \cdot D_i$. The running time of the algorithm is bounded by $O(|E(P)|)$.
	
For the step of the induction, we consider some level $0<L\leq q$, and assume that the claim holds for all levels $0\leq L'<L$. In the remainder of the proof, whenever we refer to dynamic graphs or data structures that our algorithm maintains, we refer to them at time $\tau$, unless stated otherwise.
Let $C\in \cset^L_i$ be the given cluster, and let $x,y\in V^F(C)$ be a pair of vertices.
	
From the definition of the flattened set $V^F(C)$ of vertices, either (i) $v^L(x)\in V(H^L)$ and $v^L(x)\in V(C)$; or (ii) there is a supernode $u^L(C')$, that is marked as important for $C$, and $v^L(x)\in V^F(C')$. Assume first that $v^L(x)\in V(H^L)$. Then we define a vertex $z=x$, and we let $Q$ be a path in graph $G$, that only consists of vertex $x$. 
Assume now that the latter holds, that is, there is a supernode $u^L(C')$, which is marked as important for $C$, and $v^L(x)\in V^F(C')$.
Let $0\leq L'<L$ be the level with $C'\in \cset^{L'}$.
Since supernode $u^L(C')$ is marked as important for $C$, $N(u^L(C'))\neq \emptyset$ (from Rule \ref{rule for importance}), and so $u^L(C')$ is not an isolated vertex in $H^L$. Therefore, our data structure maintains a representative vertex from cluster $C'$ at level $L$, that we denote by $z=\sigma_L(C')$. From Property \ref{prop: representatives 2}, $z\in V^F(C')$ must hold, and vertex $v^L(z)$ lies in graph $H^L$. From the definition of graph $H^L$, edge $(v^L(z),u^L(C'))$ belongs to $H^L$. From the definition of important supernodes, since supernode $u^{L}(C')$ is marked as important for cluster $C$, all regular vertices that are neighbors of $u^L(C')$ in $H^L$ belong to $C$, so in particular $v^L(z)\in V(C)$. Lastly, since supernode $u^L(C')$ is marked as important for cluster $C$, $\scale(C')\leq \scale(C)$ must hold. We denote $\scale(C')=i'$; recall that $\scale(C)$ is denoted by $i$.
We apply the algorithm from the induction hypothesis to level $L'$, cluster $C'$, and vertices $x$ and $z$, that both belong to $V^F(C')$. We then obtain a path $Q$ in graph $G$, that connects $x$ to $z$, whose length is bounded by $(\tilde c\alpha)^{L'+1}\cdot D_{i'}\leq (\tilde c\alpha)^{L}\cdot D_i$. The time required to compute the path is bounded by $O(|E(Q)|)$.
	 
	 To summarize, so far we have defined a vertex $z\in V(G)$, with $v^L(z)\in V(C)$, and we have computed a path $Q$ in graph $G$, connecting $x$ to $z$, whose length is at most $(\tilde c\alpha)^{L}\cdot D_i$.
	 
	 We repeat the same procedure with vertex $y$, to obtain a vertex $z'\in V(G)$, with $v^L(z')\in V(H^L)$ and $v^L(z')\in V(C)$.
	We also compute a path $Q'$ connecting $z'$ to $y$  in graph $G$, whose length is at most $(\tilde c\alpha)^{L}\cdot D_i$, in time $O(|E(Q')|)$.

	Next, we perform query $\spquery^L_i(C,v^L(z),v^L(z'))$ in data structure $\DS^L_i$, to obtain a path $\tilde P$ in graph $H^L$, that connects vertices $v^L(z)$ and $v^L(z')$, whose length is at most $\alpha\cdot D_i$. The query time is bounded by $O(|E(\tilde P)|)$.

	Let $u^L(C_1),u^L(C_2),\ldots,u^L(C_r)$ denote the sequence of type-2 supernodes on path $\tilde P$, as we traverse it  from vertex $v^L(z)$ to vertex $v^L(z')$. For all $1\leq j\leq r$, let $i_j=\scale(C_j)$. %Since the length of every edge incident to supernode $u^L(C_j)$ in graph $H^L$ is $D_{i_j}$, the length of the path $\tilde P$ in $H^L$ is at least $\sum_{j=1}^rD_{i_j}$, and so $\sum_{j=1}^rD_{i_j}\leq \alpha\cdot D_i$. From our construction of graph $H^L$, all neighbor-vertices of a type-2 supernode are type-1 regular vertices. 
	For $1\leq j\leq r$, we denote by $v^L(z_j),v^L(z'_j)$ the two vertices that immediately precede and immediately follow supernode $u^L(C_j)$ on path $\tilde P$.
		We will use the following observation in order to transform path $\tilde P$ in graph $H^L$ into a path in graph $G$ that connects $z$ to $z'$, while approximately preserving the path length.

	\begin{observation}\label{obs: path transformation}
		There is a deterministic algorithm, that, given an index $1\leq j\leq r$, computes a path  $\tilde Q_j$ in graph $G$, that connects $z_j$ to $z'_j$, whose length is at most $(\tilde c\alpha)^{L}\cdot D_{i_j}$.  The running time of the algorithm is $O(|E(\tilde Q_j)|)$. 
	\end{observation}
\begin{proof}
	We fix an index $1\leq j\leq r$. From the construction of graph $H^L$, since edges $(v^L(z_j),u^L(C_j))$, $(v^L(z'_j),u^L(C_j))$ are present in graph $H^L$, it must be the case that $z_j\in V^F(C_j)$ and $z'_j\in V^F(C_j)$. Let $L'<L$ be the level with $C_j\in \cset^{L'}$. We  apply the algorithm from induction hypothesis to level $L'$, cluster $C_j$, and vertices $z_j,z'_j$, to obtain a path $\tilde Q_j$ connecting $z_j$ to $z'_j$ in graph $G$, whose length is at most $(\tilde c\alpha)^{L'+1}\cdot D_{i_j}\leq (\tilde c\alpha)^{L}\cdot D_{i_j}$. The time required to compute the path is bounded by $O(|E(\tilde Q_j)|)$.
\end{proof}

Consider now some index $1\leq j<r$, and let $P_j$ be the subpath of path $\tilde P$ between $v^L(z'_j)$ and $v^L(z_{j+1})$. We denote the length of $P_j$ by $\ell'_j$. We now define a path $\tilde Q'_j$ connecting $z'_j$ to $z_{j+1}$ in graph $G$, such that the length of $\tilde Q'_j$ is at most $\ell'_j$. If $z'_j=z_{j+1}$, then $\tilde Q_j=(z'_j)$. Otherwise, we denote the sequence of supernodes on path $\tilde Q_j$ (which must all be type-1 supernodes) by $u^L(a_1),u^L(a_2),\ldots,u^L(a_{r_j})$. From the definition of graph $H^L$, $a_1=z'_j$, and $a_{r_j}=z_{j+1}$ must hold. We can also assume that every consecutive pair of supernodes on the path are distinct (since otherwise the path traverses the same edge back and forth). It is then easy to verify that, for all $1\leq b<r_j$, the regular vertex that appears between $u^L(a_b)$ and $u^L(a_{b+1})$ on path $P_j$ must be a type-2 regular vertex, that we denote by $v^L(e_b)$, and moreover, $a_b,a_{b+1}$ are the endpoints of edge $e_b$. Therefore, the sequence $(e_1,e_2,\ldots,e_{r_j})$ of edges in graph $G$ define a path that connects $z'_j$ to $z_{j+1}$. It is immediate to verify that the length of the path is bounded by $\ell'_j$.

We also consider the subpath $P_0$ of $\tilde P$ between $v^L(z)$ and $v^L(z_1)$, whose length is denoted by $\ell'_0$, and the subpath $P_r$ of $\tilde P$ between $v^L(z_r)$ and $v^L(z')$, whose length is denoted by $\ell'_r$. Using exactly the same algorithm, we compute a path $\tilde Q'_0$ connecting $z$ to $z_1$ in $G$ of length at most $\ell'_0$, and a path $\tilde Q'_r$ connecting $z_r$ to $z'$ in $G$, whose length is at most $\ell'_r$.

It is immediate to verify, from the construction of graph $H^L$, that the length of path $\tilde P$ in $H^L$ is at least $\sum_{j=0}^r\ell'_j+\sum_{j=1}^{r}D_{i_j}$. Therefore, $\sum_{j=0}^r\ell'_j+\sum_{j=1}^{r}D_{i_j}\leq \alpha\cdot D_i$. 

Consider now a path $P'$ in graph $G$, that is obtained by concatenating the paths $\tilde Q'_0,\tilde Q_1,\tilde Q'_1,\ldots,\tilde Q_r,\tilde Q'_r$. Then path $P'$ connects $z$ to $z'$ in $G$, and its length is bounded by $\sum_{j=0}^r\ell'_j+\sum_{j=1}^{r}(\tilde c\alpha)^L\cdot D_{i_j}\leq \tilde c^L \alpha^{L+1}\cdot D_i$. Lastly, we let $P$ be the path obtained by concatenating paths $Q,P'$ and $Q'$. Then $P$ is a path in graph $G$ that connects $x$ to $y$, and its length is bounded by $2(\tilde c\alpha)^L\cdot D_i+\tilde c^L \alpha^{L+1}\cdot D_i\leq (\tilde c\cdot\alpha)^{L+1}\cdot D_i$. It is easy to verify that the running time of the algorithm is bounded by $O(|E(P)|)$.
\end{proof}

The following corollary of \Cref{claim: short paths flat sets} allows us to efficiently transform paths in graph $H^L$, for all $1\leq L\leq q$, into paths in graph $G$, while approximately preserving the path length.

\begin{corollary}\label{cor: path transform}
	There is a large enough constant $c'$, and a deterministic algorithm, that, at any time $\tau\in \tset$, given a level $0\leq L\leq q$, a pair $v^L(z),v^L(z')\in H^L$ of type-1 regular vertices, and a path $P$ connecting $v^L(z)$ to $v^L(z')$ in $H^L$, whose length is denoted by $D'$, computes a path $P'$ in graph $G$, connecting $z$ to $z'$, so that the length of $P'$ is bounded by $(c'\alpha)^{L+1}\cdot D'$. The running time of the algorithm is bounded by $O(|E(P')|)$.
\end{corollary}
\begin{proof}
	The proof of the corollary uses arguments similar to those that appeared in the proof of \Cref{claim: short paths flat sets}.
	
	Let $u^L(C_1),u^L(C_2),\ldots,u^L(C_r)$ denote the sequence of type-2 supernodes on path $P$, as we traverse it  from $v^L(z)$ to $v^L(z')$. For all $1\leq j\leq r$, let $i_j=\scale(C_j)$. As in the proof of \Cref{claim: short paths flat sets}, we denote, for all $1\leq j\leq r$, by $v^L(z_j)$ and $v^L(z'_j)$ the two vertices that immediately precede and immediately follow supernode $u^L(C_j)$ on path $P$. For $1\leq j<r$, we also let $P_j$ be the subpath of $P$ between vertices $v^L(z'_j)$ and $v^L(z_{j+1})$, We also let $P_0$ be the subpath of $P$ between $v^L(z)$ and $v^L(z_1)$, and $P_r$ the subpath of $P$ between $v^L(z'_r)$ and $v^L(z')$. For $0\leq j\leq r$, let $\ell_j$ be the length of path $P_j$ in graph $H^L$. For convenience, we denote $z=z'_0$ and $z'=z_{r+1}$. Notice that the length of path $P$ in $H^L$ is  $D'\geq \sum_{j=0}^r\ell_j+\sum_{j=1}^rD_{i_j}$.
	
	Using the same algorithm as in the proof of  \Cref{claim: short paths flat sets}, for all $0\leq j\leq r$, we compute a path $P'_j$ in graph $G$, that connects vertices $z'_j$ and $z_{j+1}$, whose length is at most $\ell_j$, in time $O(|E(P'_j)|)$.
	
	Next, we consider each index $1\leq j\leq r$ one by one. From the definition of graph $H^L$, $z_j,z'_j\in V^F(C_j)$ must hold. We now use the algorithm from  \Cref{claim: short paths flat sets} to compute a path $P''_j$ in graph $G$, connecting vertices $z_j$ and $z'_j$, whose length is at most $(\tilde c\alpha)^{L+1}\cdot D_{i_j}$, in time $O(|E(P''_j)|)$.
	
	Lastly, we obtain the final path $P'$ connecting $z$ to $z'$ in graph $G$ by concatenating the paths $P'_0,P''_1,P'_1,\ldots,P''_r,P'_r$. The length of the resulting path is bounded by:

	\[ \sum_{j=0}^r\ell_j+\sum_{j=1}^r(\tilde c\alpha)^{L+1}D_{i_j}\leq (\tilde c\alpha)^{L+1}\cdot D'. \]
\end{proof}

Next, we define and analyze the central notion that we will use in order to support short-path queries, namely, covering chains.

\subsubsection{Covering Chains}
\label{subsubsec: covering chains}

The notion of covering chains is central to our algorithm for supporting \shortpath\ queries.
In order to define covering chains, we first need to define the notion of a \emph{covering quadruple}.

\begin{definition}[Covering Quadruple]
	Let $\tau\in \tset$ be any time during the time horizon, let $0\leq L\leq q$ be a level, $0\leq i\leq \log \hat D$ a scale, $C$ a cluster that lies in $\cset^{L}_i$ at time $\tau$, and $x\in V(G)$ a vertex. We say that $(L,i,C,x)$ is a \emph{valid covering quadruple} at time $\tau$ if vertex $v^L(x)$ lies in graph $H^L$, and moreover, $C=\coveringcluster^L_i(v^L(x))$ at time $\tau$.
\end{definition}

Note that, if  $(L,i,C,x)$ is a valid covering quadruple, then it must be the case that $v^L(x)\in V(C)$, and, from the definition of the flattened vertex sets, $x\in V^F(C)$ must also hold.
We are now ready to define the notion of covering chains.

\begin{definition}[Covering Chain]\label{def: covering chain}
Let $\tau\in \tset$ be any time during the time horizon, and let $J=((L_1,i_1,C_1,x_1),\ldots,(L_r,i_r,C_r,x_r))$ be a sequence of valid covering quadruples. We say that $J$ is a \emph{valid covering chain for vertex $x\in V(G)$ at time $\tau$}, if the following hold:

\begin{itemize}
	\item $0=L_1<L_2<\cdots<L_r\leq q$;
	\item for all $1<j\leq r$, $i_{j}=i_1+10(L_j+1)$;
	\item $x_1=x$; and
	\item for all $1\leq j< r$, supernode $u^{L_{j+1}}(C_{j})$ lies in graph $H^{L_{j+1}}$ at time $\tau$, and moreover, $x_{j+1}$ is the representative vertex of cluster $C_j$ at level $L_{j+1}$ at time $\tau$. 
\end{itemize}

We say that the length of the chain $J$ is $r$. We also say that the level of chain $J$ is $L_r$, and its scale is $i_1$. We may sometimes say that $J$ lies at level $L_r$ and scale $i_1$. Lastly, we say that vertex $x$ \emph{owns} chain $J$.
\end{definition}

Notice that, if $J=((L_1,i_1,C_1,x_1),\ldots,(L_r,i_r,C_r,x_r))$ is a valid covering chain for some vertex $x\in V(G)$, then for all $0\leq j<r$, it must be the case that $x_j,x_{j+1}\in V^F(C_j)$. Indeed, vertex $x_{j+1}$ may only be a representative of cluster $C_j$ at level $L_{j+1}$ if $x_{j+1}\in V^F(C_j)$ holds, and,  since $(L_j,i_j,C_j,x_j)$ is a valid covering quadruple, as established above,  $x_j\in V^F(C_j)$ must also hold.

For a vertex $x\in V(G)$, a level $0\leq L\leq q$, and a scale $0\leq i\leq \log D^*$, we denote by $\jset^L_i(x)$ the set of all valid level-$L$ chains at scale $i$ that vertex $x$ owns. Note that the set $\jset^L_i(x)$ of covering chains may change over time. We also denote
$\jset^L(x)=\bigcup_{i=0}^{\log  D^*}\jset^L_i(x)$, and
 $\jset(x)=\bigcup_{L=0}^q\jset^L(x)$ -- the set of all covering chains owned by $x$. 
%When the time $\tau$ is clear from context, or unimportant, we omit the superscript $(\tau)$. We will also refer to the \emph{current set $\jset(x,L,i)$ of covering chains} to mean the set $\jset\attime(x,L,i)$, where $\tau$ is the current time in the algorithm's execution.

Covering chains will be used in order to respond to \shortpath\ queries. Specifically, given a \shortpath\ query between a pair $x,y$ of vertices of $G$, we start by computing, for every level $0\leq L\leq q$, the collections $\jset^L(x)$ and $\jset^L(y)$ of covering chains. For every pair $J\in \jset^L(x)$, $J'=\jset^L(y)$ of such chains, we then perform a distance query between vertices $v^L(x_r)$, $v^L(x'_{r'})$ in $\DS^L$, where $x_r$ is the vertex that appears in the last quadruple of $J$, and $x'_{r'}$ is the vertex that appears in the last quadruple of $J'$. The responses to these distance queries will then guide the algorithm for computing a short path between the input vertices $x$ and $y$.

We start by showing that, for every vertex $x$ and level $L$, we can compute the set $\jset^L(x)$ of covering chains efficiently. We then explore several properties of covering chains that will be useful for us later.
Recall that we have denoted $\mu'=(4\Delta)^{q+1}$, and that the dynamic degree bounds in all graphs $\set{H^L}_{L=0}^q$ are at most $\mu'$.

\begin{claim}\label{claim: compute covering chain}
	There is a large enough constant $\hat c$ and a deterministic algorithm that, at any time $\tau\in \tset$ during the time horizon, given a vertex $x\in V(G)$, a level $0\leq L\leq q$, and a scale $0\leq i\leq \log D^*$, computes the current set $\jset^L_i(x)$ of covering chains at level $L$ and scale $i$ that $x$ owns. The running time of the algorithm is at most $O(\hat c^{L+1}\cdot L^2)$. Moreover, at any time $\tau$ during the time horizon $\tset$, $|\jset^L_i(x)|\leq \hat c^{L+1}$ holds.
\end{claim}
\begin{proof}
	The proof is by induction on the level $L$. The base of the induction is when $L=0$.  Given a scale $0\leq i\leq \log D^*$ and a vertex $x\in V(G)$, we compute cluster $C_1=\coveringcluster^0_i(v^0(x))$ in time $O(1)$ using data structure $\DS^0_i$. We then obtain a covering chain that consists of a single covering quadruple $J=(0,i,C_1,x)$. It is easy to verify that $J$ is the only level-$0$ covering chain at scale $i$ that vertex $x$ owns. We then set $\jset^L_i(x)=\set{J}$. The running time of the algorithm is $O(1)\leq \hat c$, if $\hat c$ is a large enough constant.
	
Consider now some integer $0<L\leq q$, and assume that the claim holds for all levels $0\leq L'<L$. 
We also assume that, for all $0\leq L'<L$, the time that is required in order to compute set $\jset^{L'}_i(x)$, given the sets $\jset^{0}_i(x),\ldots,\jset^{L'-1}_i(x)$, is bounded by $L'\cdot (\hat c)^{L'+1}$.

 We start by computing, for all $0\leq L'<L$, the corresponding collection $\jset^{L'}_i(x)$ of covering chains, and we denote $\jset'=\bigcup_{L'=0}^{L-1}\jset^{L'}_i(x)$. We then gradually construct the collection $\jset''=\jset^L_i(x)$ of covering chains. Notice that, if $J=((L_1,i_1,C_1,x_1),\ldots,(L_r,i_r,C_r,x_r))$  is a covering chain that $x$ owns at level $L$ and scale $i$, then $r>1$ must hold, and furthermore, $J'=((L_1,i_1,C_1,x_1),\ldots,(L_{r-1},i_{r-1},C_{r-1},x_{r-1}))$ is a valid covering chain that $x$ owns at level $L_{r-1}$ and scale $i$. In particular, $J'\in \jset'$ must hold. we say that $J'$ is the \emph{prefix} of chain $J$.
Clearly, for every chain $J\in \jset^L_i(x)$, there is a chain $J'\in \jset'$ that is the prefix of $J$.
Therefore, we will consider every chain in $\jset'$ one by one, and for each such chain $J'$, we will attempt to extend it by a single quadruple in order to obtain a chain in $\jset^L_i(x)$.
 
 Consider any chain $J'=((L_1,i_1,C_1,x_1),\ldots,(L_{r-1},i_{r-1},C_{r-1},x_{r-1}))\in \jset'$, and recall that $L_1=0$, $i_1=i$, and $x_1=x$ must hold. Let $i_r=i+10L+10$. We check, in time $O(1)$, whether supernode $u^{L}(C_{r-1})$ lies in graph $H^L$. If so, and if it is not an isolated vertex of $H^L$, then we let $x_r$ be the unique representative of cluster $C_{r-1}$ at level $L$, and we let $C_r=\coveringcluster^L_{i_r}(x_r)$. 
 Vertex $x_r$ and cluster $C_r$ can be computed in time $O(1)$ using data structure $\DS^L$.
 Let $J$ be a chain that is obtained from $J'$ by appending the quadruple $(L,i_r,C_r,x_r)$ at the end of $J'$. Clearly, $J\in \jset^L_i(x)$ holds, and moreover, it is easy to verify that $J$ is the only covering chain in $\jset^L_i(x)$, such that $J'$ is a prefix of $J$. We then add $J$ to $\jset''$ and continue to the next iteration. Notice that processing chain $J'\in \jset'$ takes time $O(r)\leq O(L)$, where $r$ is the length of the chain.
 
 It is easy to verify that $|\jset^L_i(x)|\leq |\jset'|=\bigcup_{L'=0}^{L-1}|\jset^{L'}_i(x)|$. From the induction hypothesis, for all $0\leq L'<L$, $|\jset^{L'}_i(x)|\leq \hat c^{L'+1}$ holds. Therefore, $|\jset^L_i(x)|\leq \sum_{L'=0}^{L-1}\hat c^{L'+1}\leq \hat c^{L+1}$.

 The running time that is required in order to compute set $\jset^L_i(x)$, once the sets $\set{\jset^{L'}_i(x)}_{L'=1}^{L-1}$ have been computed, is bounded by $O(L\cdot |\jset'|)\leq O(L\cdot \hat c^{L+1})$. From the induction hypothesis, the time required to process levels $L'=0,\ldots,L-1$ is bounded by $\sum_{L'=0}^{L-1} O(L'\cdot \hat c^{L'+1})\leq  O(L^2\cdot (\hat c^{L+1}))$. Therefore, the total time that is required in order to construct the set $\jset^L_i$ of covering chains is bounded by $O(L^2\cdot (\hat c^{L+1}))$.
\end{proof}

In the next crucial claim we show that, if $J=((L_1,i_1,C_1,x_1),\ldots,(L_r,i_r,C_r,x_r))$ is a covering chain that vertex $x$ owns at time $\tau$, then $x\in V^F(C_r)$ must hold.

\begin{claim}\label{claim: covering chain to flat set}
	Let $x\in V(G)$ be any vertex, and let $J=((L_1,i_1,C_1,x_1),\ldots,(L_r,i_r,C_r,x_r))$ be a covering chain that $x$ owns at some time $\tau\in \tset$. Then at time $\tau$, $x\in V^F(C_r)$ holds.
\end{claim}
\begin{proof}
	The proof is by induction on the length $r$ of the chain $J$. The base of the induction is when the length of $J$ is $1$, so $J=(L_1,i_1,C_1,x_1)$. Recall that $L_1=0$ and $x_1=x$ must hold.  Furthermore, at time $\tau$, $C_1=\coveringcluster^0_{i_1}(v^0(x))$. In particular, regular vertex $v^0(x)$ must lie in cluster $C_1$ at time $\tau$. From the definition of set $V^F(C_1)$, it must be the case that $x\in V^F(C_1)$ at time $\tau$.
	
	Assume now that we are given an integer $0<r\leq q$, and that the claim holds for all integers $0\leq r'<r$. Let $x\in V(G)$ be a vertex, and let $J=((L_1,i_1,C_1,x_1),\ldots,(L_r,i_r,C_r,x_r))$ be a covering chain that $x$ owns at some time $\tau\in \tset$. Denote $J'=((L_1,i_1,C_1,x_1),\ldots,(L_{r-1},i_{r-1},C_{r-1},x_{r-1}))$. Clearly, $J'$ is a valid covering chain of length $r-1$ that $x$ owns at time $\tau$, so, from the induction hypothesis, at time $\tau$, $x\in V^F(C_{r-1})$ holds. For convenience, we denote $C'=C_{r-1}$, $L'=L_{r-1}$, $C=C_r$, and $L=L_r$. We also denote $i'=i_{r-1}$ and $i=i_r$. Recall that $i\geq i'+10$, from the definition of a covering chain. Our goal is to prove that, at time $\tau$, $x\in V^F(C)$ holds. 
	
	From the definition of a covering chain, supernode $u^L(C')$ lies in graph $H^L$, and $x_r$ is the representative vertex of cluster $C'$ at level $L$. In particular, it must be the case that $x_r\in V^F(C')$ (see Property \ref{prop: representatives 2} of representative vertices). 
	Therefore, graph $H^L$ contains an edge $(v^L(x_r),u^L(C'))$, whose length is $2^{\scale(C')}=D_{i'}$. 
	For convenience, denote $u=u^L(C')$.
	Recall that we denoted by $N(u)$ the collection of neighbors of vertex $u$. From the definition of graph $H^L$, for every vertex $v\in N(u)$, the length of the edge $(u,v)$ in $H^L$ is $D_{i'}$, and so $\dist_{H^L}(v,v^L(x_r))\leq 2D_{i'}$. Recall that $C=\coveringcluster^L_i(v^L(x_r))$. Therefore, $u\in V(C)$, and $N(u)\subseteq V(C)$ holds at time $\tau$. 
	
	Notice that, if supernode $u$ is marked as important for cluster $C$ at time $\tau$, then $V^F(C')\subseteq V^F(C)$ holds at time $\tau$, and so $x\in V^F(C)$ must hold. Therefore, it is now enough to prove that supernode $u$ is marked as important for cluster $C$ at time $\tau$. The following observation will then finish the proof of \Cref{claim: covering chain to flat set}.
	
	\begin{observation}\label{obs: supernode is important}
		Supernode $u=u^L(C')$ is marked as important for cluster $C$ at time $\tau$.
	\end{observation}
\begin{proof}
	Since vertex $x_r$ is a representative vertex of cluster $C'$ at time $\tau$, $N(u)\neq \emptyset$ at time $\tau$, and, as we have established already, $N(u)\subseteq V(C)$ holds at time $\tau$. Moreover, $\scale(C')<\scale(C)$. Therefore, according to Rule \ref{rule2 for importance}, in order to prove that $u$ is marked as an important for $C$ at time $\tau$, it is enough to prove that, 
	 for every time $\tau'< \tau$ during the current level-$L$ phase, if $C^*=\anc\attime[\tau'](C)$ and $u^*=\anc\attime[\tau'](u)$, then $N\attime[\tau'](u^*)\neq \emptyset$ and $N\attime[\tau'](u^*)\subseteq V(C^*)$ holds at time $\tau'$ (here, $N\attime[\tau'](u^*)$ is the set of all vertices that are neighbors of $u^*$ at time $\tau'$).
	 
We now fix some time $\tau'<\tau$ during the current level-$L$ phase.
Denote  	$C^*=\anc\attime[\tau'](C)$ and $u^*=\anc\attime[\tau'](u)$. 
Since $C_r=\coveringcluster^L_i(v^L(x_r))$ holds at time $\tau$, and $C^*=\anc_{\tau'}(C)$, from the Consistent Covering property of the \recdynnc data structure,  at time $\tau'$, it must be the case that $B_{H^L}(v^L(x_r),2^i)\subseteq V(C^*)$.

%Assume now that $u^*=u^L(\tilde C)$. Then, in the level-$L'$ data structure, $\tilde C=\anc\attime[\tau'](C')$ holds. 

Since, at time $\tau$, edge $(v^L(x_r),u)$ lies in graph $H^L$ (as $x_r\in V^F(C'))$, from \Cref{obs: supernode edge tracking},  
edge $(v^L(x_r),u^*)$ belongs to graph $H^L$ at time $\tau'$, so in particular, $N\attime[\tau'](u^*)\neq \emptyset$. Furthermore, from the same observation, the lengths of the edges incident to $u^*$ at time $\tau'$ are equal to the lengths of the edges incident to $u$ at time $\tau$, which, in turn, are all equal to $D_{i'}$.  Therefore, 
the length of every edge that connects a vertex of $N\attime[\tau'](u^*)$ to $u^*$ at time $\tau'$ is $D_{i'}$.  %Let $N_{\tau'}(u^*)$ be the set of all type-1 regular vertices that are neighbors of $u^*$ at time $\tau'$. Then 

We conclude that, for each vertex $v\in N\attime[\tau'](u^*)$, at time $\tau'$, $\dist_{H^L}(v,v^L(x_r))\leq 2D_{i'}$ must hold. Since, at time $\tau'$, $B_{H^L}(v^L(x_r),2^i)\subseteq V(C^*)$, and $i\geq i'+10$, we get that, at time $\tau'$, $N_{\tau'}(u^*)\subseteq V(C^*)$ held.

From Rule \ref{rule2 for importance}, we conclude that supernode $u$ is marked as important for cluster $C$ at time $\tau$.
\end{proof}
\end{proof}

\subsubsection{A Central Claim}

In this subsection we state and prove the central claim that allows us to support \shortpath\ queries. 
The section uses the notion of levels of edges and of paths of graph $G$, that was defined at the beginning of \Cref{sec: fully APSP inner main}.
We also need one additional definition.

Consider some time $\tau\in \tset$, a vertex $x\in V(G)$, a scale $0\leq i\leq \log D^*$, and a level $0\leq L\leq q$. Recall that we have defined a collection $\jset^L_i(x)$ of covering chains at level $L$ and scale $i$ that $x$ owns at time $\tau$. We define a set $R^L_i(x)$ of vertices of $G$, as follows. Vertex $x'$ lies in $R^L_i(x)$ if and only if there exists a covering chain $J=((L_1,i_1,C_1,x_1),\ldots,(L_r,i_r,C_r,x_r))\in \jset^L_i(x)$, such that $x'=x_r$. Recall that chain $J=((L_1,i_1,C_1,x_1),\ldots,(L_r,i_r,C_r,x_r))$ has level $L$ if $L_r=L$, and it has scale $i$ if $i_1=i$.
We are now ready to state the central claim.

\begin{claim}\label{claim: short path existence}
	Consider any time $\tau\in\tset$, level $0\leq L\leq q$, and scale $0\leq i\leq \log  D^*$. Let $x,y\in V(G)$ be any pair of vertices, such that, at time $\tau$, there is a path $P$ connecting $x$ to $y$ in $G$, whose length is at most $D_i\leq D^*$, and $\level(P)=L$. Then $R^L_i(x),R^L_i(y)\neq \emptyset$, and moreover, at time $\tau$, there is a path $P'$ in graph $H^L$, connecting a vertex $v^L(x')$, with $x'\in R_i^L(x)$ to a vertex $v^L(y')$, with $y'\in R_i^L(y)$, whose length is at most $2^{10L+4}\cdot D_i$.
\end{claim}

\begin{proof}
	The proof is by induction on the level $L$.

\paragraph{Induction Base.}
The base case is when $L=0$. Consider a pair $x,y$ of vertices of $G$, and a path $P\subseteq G$ connecting $x$ to $y$, whose length is at most $D_i$, such that path $P$ lies at level $0$. In other words, every edge of $P$ is an original edge of graph $G$.
	
From the definition of covering chains, the set $\jset^0_i(x)$ of level-$0$ scale-$i$ covering chains that vertex $x$ owns at time $\tau$ only contains a single chain $J=(0,i,C,x)$, where $C=\coveringcluster^0_i(v^0(x))$. Therefore, $R^0_i(x)=\set{x}$. Similarly, the set $\jset^0_i(y)$ of level-$0$ scale-$i$ covering chains that vertex $y$ owns at time $\tau$ only contains a single chain $J'=(0,i,C',y)$, where $C'=\coveringcluster^0_i(v^0(y))$, and so $R^0_i(y)=\set{y}$.
	
We now show that there exists a path $P'$, connecting vertices $v^0(x)$ and $v^0(y)$ in graph $H^0$, whose length is at most $4D_i$.
	We denote the sequence of the sequence of vertices on path $P$ by $(x=z_1,z_2,
\ldots,z_r=y)$. For $1\leq j<r$, we denote $e_j=(z_j,z_{j+1})$. Consider the following sequence of vertices in graph $H^0$: 

\[v^0(z_1),u^0(z_1),v^0(e_1),u^0(z_2),v^0(e_2),\ldots,v^0(e_{r-1}),u^0(z_r),v^0(z_r).\]

Recall that graph $H^0$ contains edges  $(v^0(z_1),u^0(z_1))$ and $(u^0(z_r),v^0(z_r))$ of length $1$ each. Additionally, for all $1\leq j<r$, since $e_j=(z_j,z_{j+1})$, edges $(v^0(e_j),u^0(z_j))$ and $(v^0(e_j),u^0(z_{j+1}))$ are both present in $H^0$, and have length $\ell(e_j)$ each. It is the easy to see that the above sequence of vertices defines a path in graph $H^0$, that connects $v^0(x)$ to $v^0(y)$, and that the length of the path is at most $2+2\sum_{j=1}^{r-1}\ell_G(e_j)\leq 4\ell_G(P)\leq 4D_i$.

\paragraph{Induction Step.}
We now consider a level $0<L\leq q$, and we assume that the claim holds for all levels $0\leq L'<L$. Let $x,y\in V(G)$ be a pair of vertices, and let $P$ be a level-$L$ path in graph $G\attime$ that connects $x$ to $y$, so that the length of the path is at most $D_i$. Recall that $\hat D=D^*\cdot 2^{10q+10}$,
Since $D_i\leq D^*$, we get that $D_i\cdot 2^{10q+10}\leq \hat D$.

We denote by $\Phi^L_k$ the current level-$L$ phase. From the definition of a level-$L$ path, there must be at least one edge $e\in E(P)$, so that $e\in A^L_k$. We denote $E(P)\cap A^L_k=\set{e_1,e_2,\ldots,e_r}$, and we assume that the edges are indexed in the order of their appearance on path $P$, as we traverse it from $x$ to $y$. For all $1\leq j\leq r$, we denote $e_j=(x_j,y_j)$, and we assume that $x_j$ appears before $y_j$ on path $P$, as we traverse it from $x$ to $y$. Note that it is possible that $x=x_1$ or $y=y_r$ (or both).

Let $\pset=\set{P_0,P_1,\ldots,P_{r}}$ be the collection of paths that is obtained from $P$ once we delete the edges of $E(P)\cap A^L_k$ from it. For convenience, we denote $x=y_0$ and $y=x_{r+1}$. Then for all $0\leq j \leq r$, path $P_j$ has endpoints $y_j$ and $x_{j+1}$. Some of the paths in $\pset$ may consist of a single vertex. For example, if $x=x_1$, then $P_0=(x)$, and if, for some $0<j\leq r$, $y_j=x_{j+1}$, then $P_j=(y_j)$.
For all $0\leq j\leq r$, we denote the length of path $P_j$ in graph $G$ by $\ell_j$. We also let $i_j$ be the smallest integer, such that $\ell_j\leq 2^{i_j}$ holds, so, if $\ell_j\neq 0$, then $\ell_j\leq 2^{i_j}<2\ell_j$.

Observe first that every vertex of $\set{x_1,y_1,\ldots,x_r,y_r}$ lies in $S^L_k$, and so for each such vertex $z$, there is a corresponding type-1 regular vertex $v^L(z)$ in graph $H^L$. We start by showing that, for all $1\leq j< r$, there is a path $Q_j$ in graph $H^L$, that connects $v^L(y_{j})$ to $v^L(x_{j+1})$, whose length is comparable to $\ell_j$. In fact this path will only contain two edges.

\begin{claim}\label{claim: 2-hop paths}
	For all $1\leq j< r$,  there is a path $Q_j$ in graph $H^L$, connecting $v^L(y_{j})$ to $v^L(x_{j+1})$, such that the length of the path is at most $2^{i_j+10L+1}$.
\end{claim}
\begin{proof}
	We fix an index $1\leq j< r$. 
	If $y_j=x_{j+1}$, then we let $Q_j$ be a path that consists of a single vertex -- vertex $v^L(y_j)$ (that must belong to $H^L$ from our discussion above). From now on we assume that $y_j\neq x_{j+1}$.
	
	For ease of notation, we denote $z=y_j$ and $z'=x_{j+1}$. We also denote $P'=P_j$, and $i'=i_j$. Therefore, $P'$ is a path in graph $G$, connecting vertex $z$ to vertex $z'$. The length of $P'$ is at most $2^{i'}$, and $\level(P')<L$.
	
	Throughout the proof, whenever we refer to graph $G$, data structures that we maintain, or any other dynamic objects, by default we refer to all these objects at time $\tau$ (unless stated otherwise). %Recall that we assume that path $P'$ lies in graph $G$, connecting $z$ to $z'$, and that its length is at most $2^{i'}$. 
	The proof of \Cref{claim: 2-hop paths} easily follows from the following observation.
	
\begin{observation}\label{obs 2-hop paths}
	There is a cluster $C\in \cset^{<L}$, with $\scale(C)\leq i'+10L$, such that $z,z'\in V^F(C)$.
\end{observation}

Indeed, assume that \Cref{obs 2-hop paths} holds. Then, from the definition of graph $H^L$, it must be the case that $u^L(C)$ is a vertex of $H^L$ (as $z,z'\in S^L_k$), and moreover, edges $(v^L(z),u^L(C))$ and $(u^L(C),v^L(z'))$ are present in $H^L$. Since the length of each such edge is $2^{\scale(C)}\leq 2^{i_j+10L}$, we obtain the desired path $Q_j$ connecting $v^L(z)$ to $v^L(z')$ in $H^L$ by simply concatenating these two edges.  The length of the path is bounded by  $2^{i_j+10L+1}$. In order to complete the proof of \Cref{claim: 2-hop paths}, it is now enough to prove \Cref{obs 2-hop paths}.

\begin{proofof}{\Cref{obs 2-hop paths}}
	Let $L'=\level(P')$, so $L'<L$. From the induction hypothesis, there is a path $\tilde Q$ in graph $H^{L'}$, of length at most $2^{i'+10L'+4}$, connecting some vertices $v^{L'}(\tilde z),v^{L'}(\tilde z')$, where $\tilde z\in R^{L'}_{i'}(z)$ and $\tilde z'\in R^{L'}_{i'}(z')$. 
	
	From the definition of set $R^{L'}_{i'}(z)$ of vertices, there is a covering chain $J\in \jset^{L'}_{i'}(z)$, whose last quadruple is $(L',i'', C,\tilde z)$, where $i''= {i'+10L'+10}$.  From the definition of a covering quadruple, $C\in \cset^{L'}_{i''}$ holds, and $ C=\coveringcluster_{i''}^{L'}(v^{L'}(\tilde z))$. Note that $\scale(C)=i''={i'+10L'+10}\leq i'+10L$.
	From \Cref{claim: covering chain to flat set}, $z\in V^F(C)$ holds. From now on, it remains to show that $z'\in V^F(C)$ holds as well.

	As before, from the definition of the set $R^{L'}_{i'}(z')$ of vertices, there is a covering chain $J'\in \jset^{L'}_{i'}(z')$, whose last quadruple is $(L',i'',C',\tilde z')$, where $i''={i'+10L'+10}$ as before. We consider the penultimate quadruple of the chain $J'$, that we denote by $(L'',\tilde i,\tilde C,\tilde z'')$. Recall that $\tilde i=\scale(\tilde C)={i'+10L''+10}\leq {i'+10L'}$, since $L''<L'$ must hold.
	
	From the definition of a covering chain, supernode $u^{L'}(\tilde C)$ lies in graph $H^{L'}$, and vertex $\tilde z'$ is its representative in graph $H^{L'}$.
	
	Note that, if we let $J''$ be the chain obtained from $J'$ by deleting its last quadruple, then we obtain a valid containment chain that vertex $z'$ owns. The scale of the chain remains $i'$, and its level is $L''$. Therefore, from \Cref{claim: covering chain to flat set}, $z'\in V^F(\tilde C)$ holds. We need the following observation to complete the proof of \Cref{obs 2-hop paths}
	
	\begin{observation}\label{obs: other vertex in}
		At time $\tau$, supernode $u^{L'}(\tilde C)$ lies in cluster $C$, and it is marked as an important supernode for cluster $C$.
	\end{observation}

Assume first that the observation is correct. We now get that, at time $\tau$, $V^F(\tilde C)\subseteq V^F(C)$ holds, and so $z'\in V^F(C)$ must hold, completing the proof of \Cref{obs 2-hop paths}. It now remains to prove \Cref{obs: other vertex in}, which we do next.

\begin{proofof}{\Cref{obs: other vertex in}}
Recall that there is a path $\tilde Q$ in graph $H^{L'}$, whose length is at most
 $2^{i'+10L'+4}$, connecting $v^{L'}(\tilde z)$ to $v^{L'}(\tilde z')$. On the other hand, edge $(u^L(\tilde C),v^{L'}(\tilde z'))$ lies in graph $H^{L'}$, and its length is $2^{\scale(\tilde C)}\leq 2^{i'+10L'}$. Therefore, the distance between vertices $v^{L'}(\tilde z)$ and $u^{L'}(\tilde C)$  in graph $H^{L'}$ at time $\tau$ is bounded by $2^{i'+10L'+4}+2^{i'+10L'}\leq 2^{i'+10L'+5}$. 
 
Recall that  $C=\coveringcluster_{i''}^{L'}(v^{L'}(\tilde z))$, and that
 $i''={i'+10L'+10}$. Therefore, cluster $C$ contains all vertices of  $B_{H^{L'}}(u^{L'}(\tilde C),2^{i'+10L'})$. Since every edge incident to $u^{L'}(\tilde C)$ has length $2^{\scale(\tilde C)}\leq 2^{i'+10L'}$, we get that, at time $\tau$, $N(u^{L'}(\tilde C))\subseteq V(C)$.
 Notice that, from the above discussion, at time $\tau$, supernode $u^{L'}(\tilde C)$ had at least one neighbor, and $\scale(\tilde C)<\scale(C)$ holds.
 %From Rule \ref{rule for importance}, supenode $u^{L'}(\tilde C)$ is marked as important for cluster $C$ at time $\tau$ if and only if, for every time $\tau'<\tau$ during the current level-$L'$ phase, supernode $\anc\attime[\tau'](u^{L'}(\tilde C))$ was marked as important for cluster $\anc\attime[\tau'](C)$.

From Rule \ref{rule2 for importance} regarding important supernodes, it is now enough to prove the following: for every time $\tau'<\tau$ during the current level-$L'$ phase, if we denote by $N_{\tau'}$ the set of all neighbors of supernode $\anc\attime[\tau'](u^{L'}(\tilde C))$ at time $\tau'$, then, at time $\tau'$, $N_{\tau'}\neq \emptyset$ held, and every vertex of $N_{\tau'}$ belonged to cluster $\anc\attime[\tau'](C)$. If this condition holds, then, from  Rule \ref{rule2 for importance}, supernode $u^{L'}(\tilde C)$ is marked as important for $C$ at time $\tau$.
The following observation will then complete the proof of  \Cref{obs: other vertex in}.

\begin{observation}\label{obs: important supernode inner inner}
	Let $\tau'<\tau$ be any time during the current level-$L'$ phase. Denote $u^{L'}(\tilde C')=\anc\attime[\tau'](u^{L'}(\tilde C))$, $C'=\anc\attime[\tau'](C)$, and let $N_{\tau'}$ be the set of all vertices of $H^{L'}$ that were neighbors of supernode $u^{L'}(\tilde C')$ at time $\tau'$. Then, at time $\tau'$,  $N_{\tau'}\neq\emptyset$, and $N_{\tau'}\subseteq V(C')$ hold.
\end{observation}
\begin{proof}
	We fix some time $\tau'<\tau$ during the current level-$L'$ phase. In this proof, whenever we refer to graphs, data structures, and other dynamic objects that our algorithm maintains, we refer to these objects at time $\tau'$, unless stated otherwise.
	From the Consistent Covering Property, at time $\tau'$, $B_{H^{L'}}(v^{L'}(\tilde z),2^{i''})\subseteq V(C')$ held. 
	Since  $i''={i'+10L'+10}$, cluster $C'$ contains all vertices of $B_{H^{L'}}(v^{L'}(\tilde z),2^{i'+10L'+10})$.

Recall that we have established that, at time $\tau$, the distance between 
vertices $v^{L'}(\tilde z)$ and $v^{L'}(\tilde z')$ in $H^{L'}$ was at most $2^{i'+10L'+4}$ (due to path $\tilde Q$). 
Since, from \Cref{obs: no dist increase}, distances between regular vertices of $H^{L'}$ may only grow overtime, at time $\tau'$, the distance between 
vertices $v^{L'}(\tilde z)$ and $v^{L'}(\tilde z')$ was also at most $2^{i'+10L'+4}$.

From \Cref{obs: supernode edge tracking}, since edge $(v^{L'}(\tilde z'),u^{L'}(\tilde C))$ lies in graph $H^{L'}$ at time $\tau$, it must be the case that edge $(v^{L'}(\tilde z'),u^{L'}(\tilde C'))$ lies in graph $H^{L'}$ at time $\tau'$. Moreover, the lengths of all edges incident to $u^{L'}(\tilde C')$ at time $\tau'$ are equal to the lengths of the edges incident to $u^{L'}(\tilde C)$ at time $\tau$; equivalently,  $\scale(\tilde C')=\scale(\tilde C)\leq i'+10L'$ must hold. We conclude that the length of every edge connecting vertex $u^{L'}(\tilde C')$ to a vertex of $N_{\tau'}$ is at most $2^{i'+10L'}$ at time $\tau'$, and that the distance between vertices $v^{L'}(\tilde z)$ and $u^{L'}(\tilde C')$ in graph $H^{L'}$ at time $\tau'$ is at most $2^{i'+10L'+4}+2^{i'+10L'}\leq 2^{i'+10L'+5}$. Since cluster $C'$ contains all vertices of $B_{H^{L'}}(v^{L'}(\tilde z),2^{i'+10L'+10})$, we get that, at time $\tau'$, every vertex in $B_{H^{L'}}(u^{L'}(\tilde C'),2^{i'+10L'})$ belonged to cluster $C'$. In particular, $N_{\tau'}\subseteq V(C')$ holds at time $\tau'$.
\end{proof}
\end{proofof} 
\end{proofof}

So far \Cref{claim: 2-hop paths} allows us to transform each path $P_j\in \pset$, for $0<j< r$ into a corresponding path in graph $H^{L'}$. We use the next claim to deal with path $P_0$.

\begin{claim}\label{claim: 2-hop paths first and last}
	There is a path $Q_0$ in graph $H^L$, connecting some vertex $v^L(y'_0)$ with $y'_0\in R^L_i(x)$ to vertex $v^L(x_{1})$, such that the length of the path is at most $2^{i+10L+1}$.
\end{claim}

\begin{proof}
	For convenience, we denote $y_0$ by $z$ and $x_1$ by $z'$. We also denote $P_0$ by $P'$. Recall that path $P'$ connects $z$ to $z'$ in graph $G$, and its length is bounded by $D_i=2^i$.

	Assume first that $z=z'$, and so $P'=(z)$. 
	Notice that in this case, $v^L(z)\in V(H^L)$ holds. We construct a level-$L$ scale-$i$ covering chain $J$ for $z$ as follows. The chain consists of two quadruples. The first quadruple is $(L_1,i,z,C)$, where $L_1=0$, and $C=\coveringcluster^0_i(v^0(z))$. Recall that $z\in V^F(C)$ must hold, and, since $v^L(z)\in V(H^L)$, supernode $u^L(C)$ is present in graph $H^L$. We let $\hat z$ be the representative vertex of cluster $C$ for level $L$, let $i'=i+10L+10$, and let $\hat C=\coveringcluster^L_{i'}(v^L(\hat z))$. The second quadruple of the chain $J$ is $(L,i',\hat z,\hat C)$. Note that, since, as we established already, $D_i\cdot 2^{q+10L+10}\leq \hat D$, $i'\leq \log \hat D$ holds. Since we obtain a valid level-$L$ scale-$i$ chain $J$ that vertex $z$ owns, we conclude that $\hat z\in R^L_i(z)$. Since $z=z'$, and $z\in V^F(C)$, edge $(v^L(z'),u^L(C))$ lies in graph $H^L$, as does edge $(v^L(\hat z),u^L(C))$. We denote $y'_0=\hat z$, and we return path $Q_0$ that is a concatenation of edges $(v^L(\hat z),u^L(C))$ and $(u^L(C),v^L(z'))$. Since the length of each of these edges is $2^{\scale(C)}=2^i$, the length of the resulting path is $2^{i+1}$. We assume from now on that $z\neq z'$.

	Let $L'=\level(P')$, so $L'<L$. 
	Whenever we refer to dynamic graphs, data structures that our algorithm maintains, and other dynamic objects, we refer to them at time $\tau$, unless stated otherwise.
	
	From the induction hypothesis, there is a path $\tilde Q$ in graph $H^{L'}$, of length at most $2^{i+10L'+4}$, connecting some vertices $v^{L'}(\tilde z),v^{L'}(\tilde z')$, where $\tilde z\in R^{L'}_{i}(z)$ and $\tilde z'\in R^{L'}_{i'}(z')$. 
	
	From the definition of set $R^{L'}_{i}(z)$, there is a covering chain $J\in \jset^{L'}_{i}(z)$, whose last quadruple is $(L',i', C,\tilde z)$, where $i'= {i+10L'+10}$.  From the definition of a covering quadruple, $C\in \cset^{L'}_{i'}$ holds, and $ C=\coveringcluster_{i'}^{L'}(v^{L'}(\tilde z))$. Note that $\scale(C)=i'={i+10L'+10}\leq i+10L$.
	From \Cref{claim: covering chain to flat set}, $z\in V^F(C)$ holds.
	We use the following observation.
	
\begin{observation}\label{obs: second vertex in cluster}
	At time $\tau$,  $z'\in V^F(C)$ holds.
\end{observation}

The proof of the observation uses arguments that are identical to those used in the proof of \Cref{obs 2-hop paths}, and we do not repeat them here.
Notice that, from the definition of graph $H^L$, since a vertex $z'\in S^L_k$ lies in the flattened set $V^F(C)$ of vertices, supernode $u^{L}(C)$ is present in graph $H^L$, and it connects to vertex $v^L(z')$ with an edge, whose length is $2^{\scale(C)}\leq 2^{i+10L'+10}\leq 2^{i+10L}$. Let $z''$ be the vertex that serves as the representative of cluster $C$ at level $L$. 

Denote $i''=i+10L+10$. 
Since we have established that $D_i\cdot 2^{10q+10}\leq \hat D$, we get that $i''\leq \log \hat D$ holds.
Let $J'$ be the covering chain that is obtained from $J$, by appending the covering quadruple $(L,i'',C^*,z'')$, where $C^*=\coveringcluster^L_{i''}(v^L(z''))$. Then $J'$ is a valid level-$L$ scale-$i$ covering chain that belongs to vertex $z$. Since $z''$ is the representative of cluster $C$ at level $L$, edge $(v^L(z''),u^L(C))$ is present in graph $H^L$, and it length is $2^{\scale(C)}\leq 2^{i+10L}$. If $z''=z'$, then we let path $Q_0$ consist of a single vertex $z''$. Otherwise, path $Q_0$ is a concatenation of the edges $(v^L(z''),u^L(C))$ and $(u^L(C),v^L(z'))$. The length of path $Q_0$ is then bounded by $2^{i+10L+1}$. Since $J'$ is a valid level-$L$ scale-$i$ covering chain owned by $z$, we get that $z''\in R^L_i(z)$ holds as required.
\end{proof}

Lastly, we need the following claim in order to deal with path $P_r$. Its proof is identical to the proof of \Cref{claim: 2-hop paths first and last}, with vertex $x_{r+1}$ playing the role of vertex $y_0$ and vertex $y_r$ playing the role of $x_1$.

\begin{claim}\label{claim: last}
There is a path $Q_r$ in graph $H^L$, connecting some vertex $v^L(x'_{r+1})$ with $x'_{r+1}\in R^L_i(x_{r+1})$ to vertex $v^L(y_{r})$, such that the length of the path is at most $2^{i+10L+1}$.
\end{claim}	

We are now ready to complete the proof of \Cref{claim: short path existence}. 
Recall that we have denoted $E(P)\cap A^L_k=\set{e_1,e_2,\ldots,e_r}$,
where the edges are indexed in the order of their appearance on $P$. For $1\leq j\leq r$, we denoted $e_j=(x_j,y_j)$, where vertex $x_j$ appears closer to $x$ on path $P$ than $y_j$. We also denoted $y_0=x$ and $x_{r+1}=y$. For all $1\leq j< r$, we have now defined a path $Q_j$ in graph $H^L$, that connects vertex $v^L(y_j)$ to vertex $v^L(x_{j+1})$, whose length is at most $2^{i_j+10L+1}$. From our definition of the paths $P_1,\ldots,P_{r-1}$, for all $1\leq j\leq r$, the length of path $P_j$ is $\ell_j\geq 2^{i_j-1}$. Lastly, we have defined a path $Q_0$ in graph $H^L$, that connects some vertex $v^L(y'_0)$ with $y'_0\in \rset^L_i(x)$ to vertex $v^L(x_1)$, and a path $Q_r$ in graph $H^L$, that connects vertex $v^L(y_{r})$ to some vertex $v^L(x'_{r+1})$, with $x'_{r+1}\in R^L_i(y)$. The lengths of both these paths are bounded by $2^{i+10L+1}$. 

For all $1\leq j<r$, we also construct a path $Q'_j$ in graph $H^L$, that
connects vertices $v^L(x_j)$ and $v^L(y_j)$: consider the following sequence of vertices in graph $H^L$: $(v^L(x_j),u^L(x_j),v^L(e_j),u^L(y_j),v^L(y_j))$. It is easy to verify from the definition of graph $H^L$ that this sequence defines a path in graph $H^L$, whose length is $2+2\ell_G(e_j)$.

By concatenating the paths $Q_0,Q'_1,Q_1,\ldots,Q'_r,Q_r$, we obtain the desired path $P'$ in graph $H^L$, that connects some vertex $v^L(x')$ with $x'\in \rset^L_i(x)$ to some vertex $v^L(y')$, with $y'\in R^L_i(y)$.

From the above discussion, the length of the path is bounded by:

\[2\cdot 2^{i+10L+1}+ \sum_{j=1}^{r-1}2^{i_j+10L+1}+\sum_{j=1}^r(2+2\ell_G(e_j))\leq 2^{i+10L+2}+2^{10L+3}\cdot \ell_G(P)\leq 2^{i+10L+4}.  \]
\end{proof}
\end{proof}

\subsubsection{Algorithm for Responding to Short-Path Queries}

We are now ready to describe our algorithm for supporting \shortpath\ queries. Suppose we are given a pair $x,y\in V(G)$ of vertices at some time $\tau\in \tset$. Recall that our goal is to either 
to respond either ``YES'' or ``NO'', in time $O\left(2^{O(1/\eps)}\cdot \log \hat m\right )$, so that, if the algorithm responds ``NO'', then  $\dist_G(x,y)>D^*$ holds. If the algorithm responds ``YES'', then it may be asked additionally to compute a path $P$ in the current graph $G$, connecting $x$ to $y$, of length at most $D^*\cdot \alpha^{O(1/\eps)}$, in time $O(|E(P)|)$. 

Let $x$ and $y$ be a pair of input vertices, and let $i=\log D^*$.
We consider every level $0\leq L\leq q$ one by one. When level $L$ is considered, we use the algorithm from \Cref{claim: compute covering chain}, in order to compute the collections $\jset^L_i(x)$ and $\jset^L_i(y)$ of covering chains. Recall that the running time of the algorithm is bounded by $O(2^{O(L)}\cdot L^2)\leq 2^{O(1/\eps)}$. Recall also that $|\jset^L_i(x)|,|\jset^L_i(y)|\leq 2^{O(L)}\leq 2^{O(1/\eps)}$. 
Using the collections $\jset^L_i(x)$ and $\jset^L_i(y)$ of covering chains, we can now compute the sets $R^L_i(x),R^L_i(y)$ of vertices of $G$, in time $2^{O(1/\eps)}$. Clearly, $|R^L_i(x)|,|R^L_i(y)|\leq 2^{O(1/\eps)}$. Next, we consider every pair $z\in R^L_i(x),z'\in R^L_i(y)$ of vertices. Let $i'=i+10L+4$, and recall that $i'\leq \log \hat D$. 
Recall that, for each pair  $z\in R^L_i(x),z'\in R^L_i(y)$ of vertices, $v^L(z),v^L(z')\in V(H^L)$ must hold. We let $C=\coveringcluster^L_{i'}(v^L(z))$, and we check, in time $O(\log \hat m)$, whether $v^L(z')\in V(C)$ holds. If so, we respond ``YES''. If the algorithm is then asked to compute a path connecting $x$ to $y$ in graph $G$, then we do so as follows.

First, we perform query $\spquery^L_{i'}(C,v^L(z),v^L(z'))$ in data stucture $\DS^L_{i'}$. Recall that the data structure must return a path $Q$ of length at most $2^{i'}\cdot \alpha$, connecting $v^L(z)$ to $v^L(z')$ in $H^L$, in time $O(|E(Q)|)$. We apply the algorithm from \Cref{cor: path transform} to path $Q$ in graph $H^L$ in order to compute a path $P'$ in graph $G$, connecting $z$ to $z'$, whose length is bounded by $\alpha^{O(L)}\cdot 2^{i'}\cdot \alpha\leq  \alpha^{O(1/\eps)}\cdot D^*$. The running time of the algorithm is $O(|E(P')|)$. Next, we compute a path $P_0$ connecting $x$ to $z$ in $G$, as follows. Since $z\in R^L_i(x)$, there must be some covering chain $J\in \jset^L_i(x)$, such that, if we denote by $(L_r,i_r,x_r,C_r)$ the last quadruple of the chain, then $z=x_r$, $L_r=L$, and $i_r=i+10L$ holds. As observed already, $z\in V^F(C_r)$ must hold (see the discussion immediately following \Cref{def: covering chain}). Moreover, from \Cref{claim: covering chain to flat set}, $x\in V^F(C_r)$ holds as well. We can now use the algorithm from \Cref{claim: short paths flat sets} to compute a path $P_0$ connecting $x$ to $y$ in graph $G$, whose length is at most $\alpha^{O(L)}\cdot 2^{i+10L}\leq  \alpha^{O(1/\eps)}\cdot D^*$. The time required to compute path $P_0$ is bounded by $O(|E(P_0)|)$. We compute a path $P_1$ connecting $z'$ to $y$ in $G$, of length at most $ \alpha^{O(1/\eps)}\cdot D^*$ similarly. By concatenating paths $P_0,P'$ and $P_1$, we obtain a path $P$ in graph $G$ connecting $x$ to $y$, whose length is at most  $\alpha^{O(1/\eps)}\cdot D^*$. The running time required to compute path $P$ is bounded by $O(|E(P)|)$.

Once every level $0\leq L\leq q$ is processed, if the algorithm never returned ``YES'', then we return ``NO''. We claim that graph $G$ may not contain a path connecting $x$ to $y$, whose length is at most $D^*$. Indeed, assume for contradiction that there is a path $P$ connecting $x$ to $y$ in $G$, and the length of $P$ is $D^*$. Denote $\level(P)=L$, and let $i'=i+10L+4$. From \Cref{claim: short path existence},  there is a path $P'$ in graph $H^L$, connecting a vertex $v^L(x')$, with $x'\in R_i^L(x)$ to a vertex $v^L(y')$, with $y'\in R_i^L(x)$, whose length is at most $2^{10L+4}\cdot D_i$. Consider an iteration of our algorithm for processing level $L$ when the pair $z=x'$, $z'=y'$ of vertices was considered. Since there is a path of length at most $2^{i+10L+4}=2^{i'}$ connecting $v^L(z)$ to $v^L(z')$, if we denote $C=\coveringcluster^L_{i'}(v^L(z))$, then $v^L(z')\in V(C)$ must hold, and so our algorithm must have returned ``YES'', a contradiction.

Excluding the time that is required in order to compute a path $P$ once the algorithm returns ``YES'', for every level $0\leq L\leq q$, the algorithm spends time $O(2^{O(1/\eps)})$ in order to compute the sets $\jset^L_i(x),\jset^L_i(y)$ of covering chains. The time required to process every pair $z\in R^L_i(x),z'\in R^L_i(y)$ of vertices is $O(\log \hat m)$, and the number of such vertex pairs is bounded by $2^{O(1/\eps)}$. Therefore, processing a single level requires time $O\left(2^{O(1/\eps)}\cdot \log \hat m\right )$. Since the number of levels is $q+1\leq 2/\eps$, overall, the time required to process a query (excluding the time to compute a path $P$ once the algorithm returns ``YES'') is bounded by $O\left(2^{O(1/\eps)}\cdot \log \hat m\right )$.

%\input{T-emulator}
%--------------------

\section{Algorithm for the \recdynnc Problem - Proof of \Cref{thm: main final dynamic NC algorithm inner}}
\label{sec: proof of recdynnc inner}

We start by providing a high-level overtview of our algorithm for the \recdynnc problem. The discussion here is somewhat over-simplified and is only intended in order to provide intuition. As in \cite{APSP-previous}, the structure of the proof is inductive. Assume that our goal is to solve the \recdynnc problem on an instance $\iset=\left(H,\set{\ell(e)}_{e\in E(H)},D \right )$, that contains $W$ regular vertices. Assume that we are given a precision parameter $\eps$, such that $1/\eps$ is an integer. We start by providing a rather straightforward algorithm for the \recdynnc problem, that, on instances of size at most $W^{3\eps}$, has total update time at most $W^{c\eps}\cdot\poly(D)$, for some fixed constant $c$, where $D$ is the distance parameter in the initial instance of the \recdynnc problem. This algorithm serves as the basis of the induction. Next, for all $1\leq z\leq 1/\eps$, we show that, if there is an algorithm for the \recdynnc problem, that, on instances  of size at most $W^{z\eps}$ has total update time at most $W^{c\eps+z\eps}\cdot \poly(D)$, then there is an algorithm for the \recdynnc problem that, on instances of size at most $W^{(z+1)\eps}$, has total update time at most $W^{c\eps+(z+1)\eps}\cdot \poly(D)$. By letting $z=\ceil{1/\eps}$, we can thus obtain the desired algorithm for \recdynnc. The most challenging part of the proof is, naturally, the inductive step, in which we assume the existence of an algorithm for the \recdynnc problem of a certain size, and prove the existance of an algorithm for the \recdynnc problem of a larger size. 
To this end, we will need to develop a number of subroutines that build on the fact that there exists an algorithm for the \recdynnc problem of some specific size. In order to simplify the notation, we state here a generic assumption that can be used in all these subroutines as needed, by substituting the correct parameters. 

\begin{assumption}\label{assumption: alg for recdynnc}
	For some parameters $W>0$ and $c'>0$, there is a deterministic algorithm $\aset$ for the \recdynNC problem, that,  given a valid input structure $\iset=\left(H=(V,U,E),\set{\ell(e)}_{e\in E},D \right )$ undergoing a sequence of valid update operations with dynamic degree bound $\mu$, such that the number of regular vertices in $H$ at the beginning of the algorithm is $W'\leq W$, achieves approximation factor $\alpha(W')$, where $\alpha(\cdot)\geq 2$ is a non-decreasing function, and  has total update time at most $W'\cdot W^{\delta}\cdot D^{c'}\cdot \mu^c$ for some large enough constant $c\geq 22$ that does not depend on $W$ or $W'$, and a parameter $0<\delta<1$ that may depend on $W$. Moreover, the algorithm ensures that, for every regular vertex $v$ of $H$, the total number of clusters in the neighborhood cover $\cset$ that the algorithm maintains, to which vertex $v$ ever belongs over the course of the algorithm is bounded by $\Delta(W')$, where $\Delta(\cdot)\geq 2$ is a non-decreasing function.
\end{assumption}

Observe that, if the assumption holds for some $W>0$, then it holds for all parameters $0<\tilde W<W$.
We emphasize that we do not make the above assumption right now. Instead we will develop a number of algorithms, that work provided the above assumption is correct.

We now turn to formally prove  \Cref{thm: main final dynamic NC algorithm inner}.
Assume that we are given a valid input structure $\iset=\left(H,\set{\ell(e)}_{e\in E},D \right )$ undergoing a sequence of valid update operations, with dynamic degree bound $\mu$, and parameters $0< \eps<1/400$, and let $N$ be the number of regular vertices in $H$. In order to prove \Cref{thm: main final dynamic NC algorithm inner}, we use induction. Let $q=\ceil{1/\eps}$. We prove by induction that, for all $3\leq i\leq q$, there is a deterministic algorithm for solving the \recdynnc problem on instances with at most $N^{i\eps}$ regular vertices. The base of the induction is a straightforward algorithm that simply creates a separate cluster for every regular vertex of $H$. For the step of the induction, we use a subroutine, that, given two parameters $W_1>W_2$, assuming that there exists an algorithm for the \recdynnc problem on graphs with $W_2$ regular vertices, provides an algorithm for the \recdynnc problem on graphs with $W_1$ regular vertices. The subroutine is summarized in the following theorem.

\begin{theorem}\label{recdynnc from assumption}
Assume that, for some parameter $W>0$ and $c'=3$ \Cref{assumption: alg for recdynnc} holds.
Then there is a deterministic algorithm for the \recdynnc problem, that,  on input $\iset=\left(H,\set{\ell(e)}_{e\in E(H)},D \right )$, that undergoes a sequence $\Sigma$ of valid update operations with dynamic degree bound $\mu$, a parameter $\hat W\geq N$, where $N$ is the number of regular vertices in $H$ at the beginning of the algorithm, such that $W<\hat W\leq W^{1.5}$, and a precision parameter $\frac{1}{(\log W)^{1/24}}\leq \eps\leq 1/400$, such that $1/\eps$ is an integer, achieves approximation factor

\[\alpha^*=\max\set{2^{O(1/\eps^6)}\cdot \log\log N,\frac{(\alpha( W))^2\cdot\log\log N}{\eps^{16}}}.\] 

The algorithm ensures that, for every regular vertex $v$ of $H$, the total number of clusters to which $v$ ever belongs over the course of the algorithm is bounded by $N^{4\eps}$.
The algorithm has total update time:

\[\begin{split}
O\left (N\cdot W^{\delta+c\eps}\cdot (\Delta(W))^c\cdot \mu^{c} \cdot D^{3}\cdot (\log D)^c\right )+O\left (N\cdot W^{O(\eps)}\cdot D^3\cdot \mu^{4}\cdot \left(\frac{\hat W}{W}\right )^8\cdot \Delta(W)\right ),
\end{split}
\] 

\end{theorem}

We also use a simpler analogue of \Cref{recdynnc from assumption}, for the regime where the dynamic degree bound $\mu$ is close to the number of regular vertices in the input graph $H$.

\begin{theorem}\label{recdynnc simple}
There is a deterministic algorithm for the \recdynnc problem, that,  on input $\iset=\left(H,\set{\ell(e)}_{e\in E(H)},D \right )$, that undergoes a sequence $\Sigma$ of valid update operations with dynamic degree bound $\mu$, such that $\mu\geq N^{1/10}$ holds, where $N$ is the number of regular vertices in $H$ at the beginning of the algorithm, and a precision parameter $\frac{1}{(\log N)^{1/24}}\leq \eps\leq 1/400$, such that $1/\eps$ is an integer, achieves approximation factor
$\alpha^*=2^{O(1/\eps^6)}\cdot \log\log N$.
The algorithm ensures that, for every regular vertex $v$ of $H$, the total number of clusters to which $v$ ever belongs over the course of the algorithm is bounded by $N^{4\eps}$.
The algorithm has total update time:  $O\left( N^{1+9\eps}\cdot \mu^{22}\cdot D \right )$.
\end{theorem}

We provide the proofs of \Cref{recdynnc from assumption} and \Cref{recdynnc simple} below, after we complete the proof of \Cref{thm: main final dynamic NC algorithm inner} using them. The proof easily follows from the following lemma.

\begin{lemma}\label{lemma: inductive}
	Let $N>0$, and $1/(\log N)^{1/50}\leq \eps<1/400$ be parameters, such that $1/\eps$ is an integer, and let $q=1/\eps$. For all $3\leq i\leq q$, there is a deterministic algorithm for the \recdynnc problem, that, given a valid input structure $\iset=\left(H,\set{\ell(e)}_{e\in E(H)},D \right )$ undergoing a sequence of valid update operations with dynamic degree bound $\mu$, such that, if we denote by $N^0(H)$ the number of regular vertices in $H$ at the beginning of the algorithm, then $N^0(H)\leq N^{i\eps}$, achieves approximation factor $\alpha_i=(\log\log (N^2))^{\tilde c\cdot 2^{2i}/\eps^{24}}$, and has total update time:

	\[\begin{split}
	N^0(H)\cdot N^{\tilde c\eps+\tilde ci\eps^2}\cdot \mu^{\tilde c i} \cdot D^{3},
	\end{split}
	\]
	
	where $\tilde c$ is a fixed large enough constant.
	The algorithm also ensures that, for every regular vertex $v$ of $H$, there are at most $\Delta_i$ clusters that ever contain $v$ over the course of the algorithm, where $\Delta_3=N^{3\eps}$, and, for all $i>3$, $\Delta_i=N^{4\eps^4}$.
\end{lemma}

\iffalse
\begin{lemma}\label{lemma: inductive}
	Let $N>0$, and $1/(\log N)^{1/50}\leq \eps<1/400$ be parameters, such that $1/\eps$ is an integer, and let $q=\ceil{1/\eps}$. For all $3\leq i\leq q$, there is a deterministic algorithm for the \recdynnc problem, that, given a valid input structure $\iset=\left(H,\set{\ell(e)}_{e\in E},D \right )$ undergoing a sequence of valid update operations with dynamic degree bound $\mu$, such that, if we denote by $N^0(H)$ the number of regular vertices in $H$ at the beginning of the algorithm, then $N^0(H)\leq N^{i\eps}$, achieves approximation factor $\alpha_i=(\log\log (N\mu))^{\tilde c\cdot 2^{2i}/\eps^{18}}$, and has total update time:

	\[\begin{split}
N^0(H)\cdot \hat W^{\tilde c\eps+\tilde ci\eps^2}\cdot \mu^{\tilde c i} \cdot D^{3},
	\end{split}
	\]
	
	where $\tilde c$ is a fixed large enough constant.
	 The algorithm also ensures that, for every regular vertex $v$ of $H$, there are at most $\Delta_i$ clusters that ever contain $v$ over the course of the algorithm, where $\Delta_3=\hat W^{3\eps}$, and, for all $i>3$, $\Delta_i=\hat W^{4\eps^4}$.
\end{lemma}
\fi

The proof of \Cref{lemma: inductive} is conceptually straightforward, but technically somewhat cumbersome. We prove it by induction on $i$. The base case, where $i=3$, is shown by a straightforward algorithm that creates a separater cluster $C_v$ for every regular vertex $v$, and maintains an \EST in $C_v$ rooted at vertex $v$, with depth bound $D$. In order to perform the induction step, we use the induction hypothesis to establish that \Cref{assumption: alg for recdynnc} holds for some setting of parameters, and then apply \Cref{recdynnc from assumption} (or \Cref{recdynnc simple} if the parameter $\mu$ is large). We defer the proof of the lemma to Section \ref{subsec: proof of inductive lemma 1} of Appendix.
\Cref{thm: main final dynamic NC algorithm inner} immediately follows from 
\Cref{lemma: inductive}: we let $\eps'=1/\ceil{1/\eps}$, so that $1/\eps'$ is an integer, and $\eps/2\leq \eps'\leq \eps$ holds. Since, from the statement of \Cref{thm: main final dynamic NC algorithm inner},  $1/(\log \hat W)^{1/100}\leq \eps<1/400$ holds, it is easy to verify that  $1/(\log \hat W)^{1/50}\leq \eps'<1/400$. \Cref{thm: main final dynamic NC algorithm inner} follows from applying the algorithm from \Cref{lemma: inductive} to the input instance, with parameter $\eps'$ and $i=\ceil{1/\eps'}$.

In the remainder of this subsection, we focus on the proofs of  \Cref{recdynnc from assumption} and \Cref{recdynnc simple}.
In order to prove both theorems, we define two new problems. One problem, called \maintaincluster, was already defined in \cite{APSP-previous}. The goal of the \maintaincluster problem is to manage a single cluster that is added to the neighborhood cover $\cset$. The algorithm must ensure that, at all times, the diameter of the cluster $C$ is not too large, and, if the diameter of $C$ becomes too large, it must provide a witness in the form of two regular vertices $x,y$ of $C$, for which $\dist_C(x,y)$ is large. The algorithm for the \maintaincluster problem is also responsible for supporting $\spquery$ queries between pairs of regular vertices in cluster $C$. In the second problem, that we call \maintainNC, the goal is to maintain the neighborhood cover $\cset$. The algorithm is not required to support $\spquery$ queries in clusters of $\cset$. However, given any such cluster $C$ and a pair $x,y$ of its regular vertices, such that $\dist_C(x,y)$ is large, it needs to provide a sequence of valid updates to cluster $C$, at the end of which either $x$ or $y$ are deleted from $C$. Below, we define each of these two problems, and provide algorithms for them, assuming \Cref{assumption: alg for recdynnc}. We then complete the proof of \Cref{recdynnc from assumption} and \Cref{recdynnc simple}, by carefully combining the two resulting algorithms.

\subsection{\maintaincluster Problem}
\label{subsec: maintain cluster problem}

In this subsection, we define a problem called \maintaincluster, and state our algorithm for solving this problem. The problem was initially defined in \cite{APSP-previous}. Our definition is essentially identical, except that we use slightly different parameters. 

Intuitively, the input to the \maintaincluster problem is a valid input structure $\iset=\left(C,\set{\ell(e)}_{e\in E(C)},D\right )$, where  $C$ is a connected subgraph of the original graph $H$.
Graph $C$ undergoes a sequence of valid update operations, with dynamic degree bound $\mu$. We assume that we are given as input a distance parameter $D^*\geq D$. The algorithm is required to support queries 
$\spquery(C,v,v')$, in which, given a pair $v,v'\in V(C)$ of regular vertices of $C$, it needs to return a path of length at most $\alpha D^*$ connecting them in $C$, where $\alpha$ is the approximation factor of the algorithm. At any time, the algorithm is allowed to raise a flag $F_C$, and  to supply a pair 
$x,y$ of regular vertices of $C$, with $\dist_C(x,y)>D^*$; we refer to $x,y$ as a \emph{witness pair}.  The algorithm then receives, as part of its input update sequence $\Sigma$, a sequence $\Sigma'$ of edge-deletions and isolated vertex-deletions, at the end of which either $x$ or $y$ are deleted from $C$. We call sequence $\Sigma'$ a \emph{flag lowering sequence}.
Once the flag lowering sequence $\Sigma'$ is processed, flag $F_C$ is lowered. However, the algorithm may raise the flag $F_C$ again immediately, as long  as it provides a new pair $x',y'$ of vertices with $\dist_C(x',y')>D^*$. We emphasize that we view the resulting flag lowering sequences $\Sigma'$ as part of the input sequence $\Sigma$ of valid update operations that cluster $C$ undergoes.
Queries $\spquery$ may only be asked when flag $F_C$ is down.
We also emphasize that the initial cluster $C$ that serves as input to the \maintaincluster problem may have an arbitrarily large diameter, and so the algorithm for the \maintaincluster problem  may repeatedly raise the flag $F_C$, until it is able to support $\spquery(C,v,v')$ (that intuitively means that the diameter of $C$ has fallen under $\alpha D^*$).
 We now provide a formal definition of the problem, starting from the definition of a valid flag-lowering sequence.
 
 \begin{definition}[Valid flag-lowering sequence]\label{def: flag lowering sequence}
Given a valid input structure \newline $\iset=\left(C,\set{\ell(e)}_{e\in E(C)},D\right )$ and a pair $x,y$ of regular vertices of $C$, a sequence $\Sigma'$ of valid update operations for $C$ is called a \emph{valid flag-lowering sequence}, if:

\begin{itemize}
	\item sequence $\Sigma'$ only contains edge-deletion and isolated vertex-deletion updates; 
	\item once the updates from $\Sigma'$ are applied to $C$, either $x$ or $y$ are deleted from $C$. %; and
	%\item either (i) none of the vertices and edges of $B_C(x, D^*/2)$ are deleted via sequence $\Sigma'$; or (ii) none of the vertices and edges of $B_C(y, D^*/2)$ are deleted via sequence $\Sigma'$.
\end{itemize}
\end{definition}

\begin{definition}[\maintaincluster problem] The input to the \maintaincluster problem is a valid input structure $\iset=\left(C,\set{\ell(e)}_{e\in E(C)},D\right )$, where $C$ is a connected graph, a distance parameter $D^*>D$ that is an integral power of $2$, and the desired approximation factor $\alpha$. 
Graph $C$ undergoes an online sequence $\Sigma$ of valid update operations, and we are given its dynamic degree bound $\mu$. %Additionally, we are given %the desired approximation factor $\alpha$ and 
% a parameter $\hat W\geq N^0(C)\cdot \mu$, where $N^0(C)$ is the number of regular vertices of $C$ at the beginning of the algorithm. 
The algorithm must  support queries $\spquery(C,v,v')$: given a pair $v,v'\in V(C)$ of regular vertices of $C$, return a path $P$ of length at most $\alpha\cdot D^*$ connecting them in $C$, in time $O(|E(P)|)$. The algorithm may, at any time, raise a flag $F_C$, at which time it must  supply a pair  $\hat v,\hat v'$ of regular vertices of $C$  (called a \emph{witness pair}), with $\dist_C(\hat v,\hat v')>D^*$.
 Once flag $F_C$ is raised, the algorithm will obtain, as part of its input update sequence $\Sigma$, a valid flag-lowering sequence $\Sigma'$. Flag $F_C$ is lowered after the updates from $\Sigma'$ are processed by the algorithm. Queries $\spquery$ may only be asked when flag $F_C$ is down.
\end{definition}

We note that each flag-lowering sequence $\Sigma'$ is viewed as part of the sequence $\Sigma$ of valid update operations that cluster $C$ undergoes.

We will use the following theorem, that provides a simple algorithm for the \maintaincluster problem, in the regime where the dynamic degree bound $\mu$ is very large compared to the initial number of regular vertices in $C$.

\begin{theorem}\label{thm: maintain cluster algorithm2}
	There is a deterministic algorithm for the \maintaincluster problem, that,  on input $\iset=\left(C,\set{\ell(e)}_{e\in E(C)},D \right )$, that undergoes a sequence $\Sigma$ of valid update operations with dynamic degree bound $\mu\geq (N^0(C))^{1/10}$, where $N^0(C)$ is the number of regular vertices in $C$ at the beginning of the algorithm, and a distance bound $D^*$, achieves approximation factor $1$, and  has total update time: $O(N^0(C)\cdot \mu^{22}\cdot D^*)$.
\end{theorem}

\begin{proof}
	For brevity, we denote $N=N^0(C)$, and $\hat W=N\cdot \mu$. Note that the total number of edges that are ever present in graph $C$ is bounded by $\hat W$. Let $V'$ be the set of all regular vertices of $C$ at the beginning of the algorithm. For every regular vertex $v\in V'$, we use the algorithm from \Cref{thm: ES-tree} in order to maintain a modified \EST $T_v$, rooted at vertex $v$, in graph $C$, with depth parameter $D^*$, as graph $C$ undergoes valid update operations. We denote by $S^*(v)$ the set $S^*=\set{v'\in V(C)\mid \dist_C(v,v')>D^*}$ of vertices that the algorithm maintains. Recall that the total update time of the algorithm from \Cref{thm: ES-tree} is bounded by $O(\hat W\cdot D^*\cdot \log \hat W)$. The total update time that is required in order to maintain all trees $T_v$, for all $v\in V'$, is bounded by:
	
	\[ O(\hat W\cdot N\cdot D^*\cdot \log \hat W)\leq O(N^3\cdot \mu^2\cdot D^*)\leq O(N\cdot \mu^{22}\cdot D^*), \]
	
	since $N\leq \mu^{10}$.
	Whenever, for any vertex $v\in V'$, a new regular vertex $v'$ is added to set $S^*(v)$, we raise the flag $F_C$, and supply $v,v'$ as a witness pair. If a regular vertex $v$ is deleted from $C$, we delete the corresponding \EST data structure $T_v$. 
	
	It remains to describe an algorithm to respond to queries $\spquery(C,v,v')$. Observe that the query may only be asked when flag $F_C$ is down, so $v'\not\in S^*(v)$ currently holds. We execute query $\shortestpath$ in data structure $T_v$, to compute a path $P$ connecting $v$ to $v'$ in $C$, whose length is at most $D^*$.
\end{proof}

%This completes the definition of the \maintaincluster problem.
Our main result for the \maintaincluster problem is summarized in the following theorem. The proof is deferred to \Cref{sec: balanced pseudocut}.

\begin{theorem}\label{thm: main maintain cluster algorithm}
	Assume that, for some parameter $W>0$ and $c'=3$ \Cref{assumption: alg for recdynnc} holds.
	Then there is a deterministic algorithm for the \maintaincluster problem, that,  on input $\iset=\left(C,\set{\ell(e)}_{e\in E(C)},D \right )$, that undergoes a sequence $\Sigma$ of valid update operations with dynamic degree bound $\mu$, a parameter $\hat W\geq N^0(C)$, where $N^0(C)$ is the number of regular vertices in $C$ at the beginning of the algorithm, such that $W<\hat W\leq W^{1.5}$, a distance parameter $D^*\geq D$, and a precision parameter $\frac{1}{(\log W)^{1/24}}\leq \eps\leq 1/400$, such that $1/\eps$ is an integer, achieves approximation factor $\alpha'=\max\set{2^{O(1/\eps^6)},(8\alpha(W))^{2}}$, and  has total update time: %$O\left (\hat W^{1+O(\eps)} \cdot \hat D^3\cdot \rho^8\right )$. %Moreover, the algorithm ensures that for every regular vertex $v\in V$, the total number of clusters in the neighborhood cover $\cset$ that the algorithm maintains, to which vertex $v$ ever belonged. is bounded by $W^{O(1/\log\log W)}$. It also ensures that the neighborhood cover $\cset$ that it maintains is a strong $(D,\alpha\cdot D)$ neighborhood cover.

\[O\left (\hat W\cdot W^{\delta+c\eps/8}\cdot(\Delta(W))^c\cdot \mu^c \cdot (D^*)^{3}\cdot (\log D^*)^c\right )+O\left (\hat W^{1+O(\eps)}\cdot (D^*)^3\cdot \mu^4\cdot \Delta(W)\cdot \left(\frac{\hat W}{W}\right )^8\right ).
\]

\iffalse

\[O\left (\hat W^{1+c\eps/8}\cdot W^{\delta}\cdot (\Delta(W))^c\cdot \mu^c \cdot (D^*)^{3}\cdot (\log D^*)^c\cdot (\log \hat W)^c\right )+O\left (\hat W^{1+O(\eps)}\cdot (D^*)^3\cdot \mu^2\cdot \left(\frac{\hat W}{W}\right )^7\right ).
\]\fi
\end{theorem}

\subsection{\maintainNC Problem}
\label{subsec: manageNC problem}

Intuitively, \maintainNC problem is vey similar to the \recdynnc problem: given a valid input structure $\iset=\left(H=(V,U,E),\set{\ell(e)}_{e\in E},D \right )$ that undergoes an online sequence of valid update operations, we need to maintain a collection $\cset$ of clusters of $H$, and, for every regular vertex $v$ of $H$, a cluster $\coveringcluster(v)\in \cset$ that contains $B_H(v,D)$, such that the Consistent Covering Property holds. The clusers in $\cset$ may only undergo allowed changes as before. However, we no longer require that queries $\spquery$ are supported; intuitively, this will be ensured by applying the algorithm for the \maintaincluster problem to every cluster $C\in \cset$. However, the algorithm for the \maintaincluster problem may, from time to time, raise flag $F_C$ for some cluster $C\in \cset$, and supply a pair $x,y$ of regular vertices of $C$, such that $\dist_C(x,y)$ is sufficiently large. In such cases, the algorithm for the \maintainNC problem is responsible for producing a valid Flag Lowering sequence $\Sigma'$ for $C$. It must then apply the updates from the flag-lowering sequence $\Sigma'$ to cluster $C$, deleting all corresponding edges and vertices from $C$. These updates may be interspersed with cluster-splitting updates applied to cluster $C$. We will partition the collection $\cset$ of clusters that the algorithm maintains into two subsets: set $\cset_1$ of \emph{primary} clusters, and set $\cset_2$ of \emph{secondary} clusters. Intuitively, every cluster $C\in \cset_2$ is significantly smaller than $H$, and we will eventually apply the algorithm for the \maintainNC probem to each such cluster recursively. Therefore, once a cluster $C$ is added to the set $\cset_2$ of secondary clusters, we no longer need to keep track of it, or to update it. The primary clusters will be managed via the $\maintaincluster$ problem. The main challenge is to ensure that every regular vertex of $H$ only belongs to a sufficiently small number of primary and secondary clusters over the course of the entire algorithm.
We now define the \maintainNC problem formally.

\paragraph{Problem Definition.}
The input to the \maintainNC problem is the same as the input to the \recdynnc problem: we are given a valid input structure $\iset=\left(H=(V,U,E),\set{\ell(e)}_{e\in E},D \right )$, where graph $H$ undergoes an online sequence $\Sigma$ of valid update operations with some given dynamic degree bound $\mu$. Additionally, we are given three parameters: $W_1$, $W_2$, and $\gamma>0$, such that $W_2< W_1\leq W_2^{1.5}$, and, if $N^0(H)$ denotes the number of regular vertices in $H$ at the beginning of the algorithm, then $N^0(H) \leq W_1$ holds. Additionally, we are given a precision parameter $\frac{1}{(\log W_2)^{1/4}}\leq \eps'\leq 1/400$.

As in the \recdynnc problem, the goal is to maintain a collection $\cset$ of clustes of $H$. At the beginning of the algorithm, $\cset=\set{H}$ must hold, and, as the algorithm progresses, clusters in $\cset$ may only undergo allowed changes: $\delvertex,\addsupernode$, and $\csplit$, which are defined exactly like in the $\recdynnc$ problem. 

The notion of ancestor-clusters is also defined exactly like in the \recdynnc problem. For every regular vertex $v\in V$, the algorithm must maintain a cluster $C=\clustercover(v)$, with $B_H(v,D)\subseteq V(C)$, and, like in the \recdynnc problem, we require that the Consistent Covering property holds: namely, if, at time $\tau$, $C=\coveringcluster(v)$ holds for a regular vertex $v$, then, for all $\tau'<\tau$,  at time $\tau'$, $B_{H}(v,D)\subseteq V(\anc\attime[\tau'](C))$ held.

The collection $\cset$ of clusters that the algorithm maintains is partitioned into two subsets: set $\cset_1$ of \emph{primary clusters}, and set $\cset_2$ of \emph{secondary clusters}. We require that the following properties hold:

\begin{properties}{R}
 \item at all times, $|\cset_1|\leq \frac{W_1^{1+\eps'}}{W_2}$ holds;	 \label{restriction: few primary clusters}
 \item if we denote, for every cluster $C\in \cset_1$, the number of regular vertices that were in $C$ when it was created by $N^0(C)$, then $\sum_{C\in \cset_1}N^0(C)\leq  W_1^{1+2\eps'}$ holds at all times; \label{restruction: few regular vertices in primary clusters}
 
 \item once a cluster $C$ joins set $\cset_2$, no further updates may be applied to $C$, except those corresponding to the updates that graph $H$ undergoes (e.g. if an edge or an isolated vertex are deleted from $H$, then the same edge or isolated vertex are deleted from $C$, and if a supernode $u\in U$ that lies in $C$ undergoes supernode-splitting, a similar supernode-splitting update is applied to $C$). In particular, cluster-splitting may only be applied to secondary clusters; \label{restriction: no updates for secondary clusters}
 
 \item for every regular vertex $v\in V$, if $\coveringcluster(v)\in \cset_2$ ever held, and if $\tau$ is the first time when $\coveringcluster(v)\in \cset_2$, then, at time $\tau$, cluster $C=\coveringcluster(v)$ just joined $\cset_2$, and, from time $\tau$ onwards, $\coveringcluster(v)=C$ always holds; \label{restricioin: secondary covering cluster}
 
 \item when a cluster $C$ joins set $\cset_2$, it may contain at most $W_2$ regular vertices; and \label{restriction: few vertices in secondary clusters}
 \item for every regular vertex $v$ of $H$, the total number of clusters of $\cset_2$ to which $v$ ever belongs is bounded by $W_2^{\eps'/4}$. \label{restriction: regualr vertex in few secondary clusters}
\end{properties} 

Unlike the \recdynnc problem, we no longer require that $\spquery$ queries are supported; this will eventually be ensured by applying the algorithm for the \maintaincluster problem to each primary cluster, and by solving the \maintainNC problem recursively on each secondary clusters. However, we require that the algorithm for the \maintainNC problem supports \emph{flag lowering} operations:

\paragraph{Flag-Lowering Operation.} At any time, the algorithm for the \maintaincluster problem may receive a primary cluster $C\in \cset_1$, and a pair $x,y$ of regular vertices of $C$, with $\dist_C(x,y)>\gamma\cdot D$. The algorithm is then required to compute a valid flag-lowering sequence $\Sigma'$ (see \Cref{def: flag lowering sequence}) for $(C,x,y)$. The algorithm then must apply the updates from the flag-lowering sequence $\Sigma'$ to cluster $C$, though these updates can be intersperesed with cluster-splitting updates, in which new clusters $C'\subseteq C$ are added to set $\cset$. Once all updates from $\Sigma'$ are processed, the Flag-Lowering operation terminates.

Our main result for the \maintainNC problem is summarized in the following theorem. This algorithm is one of our main technical contributions. 

\begin{theorem}\label{thm: alg for maintainNC}
	There is a deterministic algorithm for the \maintainNC problem, that, given a valid input structure $\iset=\left(H=(V,U,E),\set{\ell(e)}_{e\in E},D \right )$ that undergoes an online sequence $\Sigma$ of valid update operations with  dynamic degree bound $\mu$, and parameters $W_1,W_2$, and $\eps'$ as in the problem definition, together with a parameter $\gamma=\frac{2^{28}\log\log W_1}{(\eps')^4}$, has total update time $O\left(\frac{W_1^{2+3\eps'}\cdot \mu^{2}}{W_2}\right )$.
\end{theorem}

The proof of the theorem is provided in \Cref{sec: alg for maintainNC}.

\subsection{Completing the Proof of \Cref{recdynnc from assumption}}

We assume that, for some parameter $W>0$ and $c'=3$ \Cref{assumption: alg for recdynnc} holds. Assume that we are given a valid input structure $\iset=\left(H,\set{\ell(e)}_{e\in E(H)},D \right )$,
 that undergoes a sequence $\Sigma$ of valid update operations with dynamic degree bound $\mu$. Let $N$ denote the number of regular vertices of $H$ at the beginning of the algorithm, and let $\hat W\geq N$ be the parameter that is given as part of input. Recall that $W<\hat W\leq W^{1.5}$. Additionally, we are given a precision parameter $\frac{1}{(\log W)^{1/24}}\leq \eps\leq 1/400$, such that $1/\eps$ is an integer. We denote the parameters $\Delta(N)$ and $\alpha(W)$ from \Cref{assumption: alg for recdynnc} by $\Delta$ and $\alpha$, respectively. We will additionally use a parameter $\eps'=\eps^2$. Clearly, $\frac{1}{(\log W)^{1/12}}\leq \eps'\leq 1/400$ holds.
Let $\gamma=\frac{2^{30}\log\log N}{(\eps')^4}$, and let $D^*$ be the smallest itegral power of $2$ with $D^*\geq \gamma\cdot D$, so $D^*\leq \frac{2^{31}\log\log N}{(\eps')^4}\cdot D$.

Our algorithm maintains a collection $\rset$ of clusters of the input graph $H$, that we call \emph{basic clusters}. This collection of clusters is different and separate from the neighborhood cover $\cset$. At the beginning of the algorithm, we set $\rset=\set{H}$. As the algorithm progresses, clusters may be added to set $\rset$, but they may never leave it. For every cluster $R$ that ever belonged to $\rset$, we denote by $N^0(R)$ the number of regular vertices that $R$ contained when it first joined $\rset$. Notice that $N^0(R)\leq N$ must hold. We say that cluster $R$ is \emph{large}, if $N^0(R)\geq N^{3\eps}$, and we say that it is \emph{small} otherwise. 
Once a cluster $R$ joins the set $\rset$, as graph $H$ continues to undergo valid update operations, we update cluster $R$ accordingly, as described in \Cref{subsubsec: updating clusters}. 
Whenever a new cluster $R$ joins the collection $\rset$ of basic clusters, we will execute an algorithm $\aset(R)$ that will process this cluster. We now describe this algorithm.

\paragraph{Algorithm $\aset(R)$.}
Let $R$ be a cluster that was just added to the set $\rset$ of basic clusters. Assume first that cluster $R$ is large. Denote $W_1(R)=N^0(R)$ and $W_2(R)=W_1(R)/N^{\eps}$. Note that, since $R$ is a large cluster, $W_1(R)\geq N^{3\eps}$, and so $W_2(R)<W_1(R)\leq (W_2(R))^{1.5}$ holds. We initialize the algorithm for the \maintainNC problem from \Cref{thm: alg for maintainNC} on cluster $R$, with parameters $W_1=W_1(R), W_2=W_2(R)$ and $\eps'$. Recall that cluster $R$ undergoes a sequence of valid update operations, that correspond to the valid update operations performed in graph $H$. Therefore, we obtain a valid input to the \maintainNC problem. We denote the algorithm from \Cref{thm: alg for maintainNC} applied to this input by $\aset'(R)$, and the data structure that it maintains by $\DS'(R)$. Recall that the total update time of the algorithm is bounded by:

\[O\left(\frac{(W_1(R))^{2+3\eps'}\cdot \mu^{2}}{W_2(R)}\right )\leq O\left((N^0(R))^{1+3\eps'}\cdot N^{\eps}\right )\leq O\left(N^0(R)\cdot N^{2\eps}\right ).  \]

We denote the collection $\cset$ of clusters that Algorithm $\aset'(R)$ maintains by $\wset(R)$, and we denote the collections $\cset_1,\cset_2$ of clusters by $\wset_1(R)$ and $\wset_2(R)$, respectively.

Whenever a new cluster $C$ is added to the set $\wset_1(R)$ of primary clusters, we initialize the algorithm for the \maintaincluster problem 
from \Cref{thm: main maintain cluster algorithm} on cluster $C$. Recall that cluster $C$ undergoes a sequence of valid update operations that correspond to the valid update operations that cluster $R$ undergoes. We use the distance parameter $D^*$ that we defined above, and we let $\hat W(C)=N^0(C)$, where $N^0(C)\leq W_1(R)\leq N$ is the number of regular vertices in cluster $C$ when it joins the set $\wset_1(R)$. The  precision parameter $\eps$ remain unchaged. Recall that we have assumed that \Cref{assumption: alg for recdynnc} holds for a parameter $W$, and that $W<\hat W\leq W^{1.5}$ holds. 
If $\hat W(C)>W$, then we define $W'=W$, and otherwise we let $W'=\hat W(C)/2$. Clearly, $W'<\hat W(C)\leq W^{1.5}$, and $\frac{\hat W(C)}{W'}\leq \frac{\hat W}{W}$ hold. Since \Cref{assumption: alg for recdynnc} holds for parameter $W$ and for $c'=3$, it must also hold for $W'$ and for $c'=3$. 
We denote the algorithm from \Cref{thm: main maintain cluster algorithm} for the \maintaincluster problem, applied to the above input, by $\aset^*(C)$. Recall that the total update time of the algorithm is bounded by:

\[\begin{split}
&O\left (\hat W(C)\cdot W^{\delta+c\eps/8}\cdot \Delta^c\cdot \mu^c \cdot (D^*)^{3}\cdot (\log D^*)^c\right )+O\left ((\hat W(C))^{1+O(\eps)}\cdot (D^*)^3\cdot \mu^4\cdot \left(\frac{\hat W}{W}\right )^8\cdot \Delta\right )\\
&\quad\quad\leq O\left (N^0(C)\cdot W^{\delta+c\eps/8} \cdot \Delta^c\cdot \mu^{c} \cdot (D^*)^{3}\cdot (\log D^*)^c\right )\\ &\quad\quad\quad\quad\quad\quad\quad\quad\quad\quad\quad\quad+O\left (N^0(C)\cdot \hat W^{O(\eps)}\cdot (D^*)^3\cdot \mu^{4}\cdot \left(\frac{\hat W}{W}\right )^8\cdot\Delta\right ).
\end{split}
\]

Whenever Algorithm $\aset^*(C)$ raises flag $F_C$, and supplies a pair $x,y$ of regular vertices of $C$ with $\dist_C(x,y)>D^*$, we supply this pair of vertices to Algorithm $\aset'(R)$. Recal that $D^*\geq \frac{2^{30}\log\log N}{(\eps')^4}\cdot D\geq \frac{2^{28}\log\log  W_1(R)}{(\eps')^4}\cdot D$ holds, as required. Algorithm $\aset'(R)$ then provides a flag-lowering sequence $\Sigma'$, which is given to Algorithm $\aset^*(C)$ for processing.

Lastly, whenever a new cluster $C$ joins the set $\wset_2(R)$ of secondary clusters, we add $C$ to the set $\rset$ of basic clusters. This completes the description of Algorithm $\aset(R)$ for the case where $R$ is a large cluster. We now analyze its total update time. Recall that the total update time of Algorithm $\aset'(R)$ is bounded by $O\left(N^0(R)\cdot N^{2\eps}\right )$.
The total update time of Algorithm $\aset^*(C)$ for each primary cluster $C\in \wset_1$ is bounded by:

\[\begin{split}
& O\left (N^0(C)\cdot W^{\delta+c\eps/8}\cdot \Delta^c\cdot \mu^{c} \cdot (D^*)^{3}\cdot (\log D^*)^c\right )\\&\quad\quad\quad\quad\quad\quad+O\left(N^0(C)\cdot \hat W^{O(\eps)}\cdot (D^*)^3\cdot \mu^{4}\cdot \left(\frac{\hat W}{W}\right )^8\cdot\Delta\right ).
\end{split}
\]

Since, from Property \ref{restruction: few regular vertices in primary clusters}, $\sum_{C\in \wset_1(R)}N^0(C)\leq  (W_1(R))^{1+2\eps'}\leq N^0(R)\cdot N^{2\eps'}$, we get that the total update time of Algorithm $\aset(R)$ is bounded by:

\[\begin{split}
&O\left (N^0(R)\cdot W^{\delta+c\eps/8}\cdot \Delta^c\cdot \mu^{c} \cdot (D^*)^{3}\cdot (\log D^*)^c\right )\\
&\quad\quad\quad\quad+O\left(N^0(R)\cdot \hat W^{O(\eps)}\cdot (D^*)^3\cdot \mu^{4}\cdot \left(\frac{\hat W}{W}\right )^8\cdot\Delta\right ).
\end{split}
\]

Next, we describe Algorithm $\aset(R)$ for the case where the basic cluster $R\in \rset$ is small. Recall that, in this case, $N^0(R)\leq N^{3\eps}$ holds. We construct a collection $\wset_1(R)$ of clusters that contains, for every regular vertex $v\in V(R)$, a cluster $C_v$, that is initially a copy of the cluster $R$. As cluster $R$ undergoes valid update operations, the same valid update operations are applied to cluster $C_v$. If vertex $v$ is deleted from $R$, then we delete all edges and vertices from cluster $C_v$. We initialize the algorithm from \Cref{thm: ES-tree} for maintaining a modified \EST in graph $C_v$, with source vertex $v$, distance bound $D^*$, as the graph undergoes valid update operations; since supernode-splitting is a special case of vertex-splitting, every update that graph $C_v$  undergoes is either edge-deletion, or isolated vertex-deletion, or vertex-splitting.
Whenever some vertex $x\in V(C_v)$ is added to the set $S^*$ of vertices (in which case $\dist_{C_v}(x,v)>D^*$ must hold), we delete vertex $x$ with its all incident edges from $C_v$. We denote the algorithm for maintaining \EST in graph $C_v$ by $\aset^*(C_v)$. Recall that the total update time of the algorithm is bounded by  $O(m^*\cdot D^*\cdot \log m^*)$, where $m^*$ is the total number of edges that ever belonged to graph $C_v$. Since $m^*\leq N^0(R)\cdot \mu\leq N^{3\eps}\cdot \mu$, we get that the total update time of Algorithm $\aset^*(C_v)$ is bounded by $O(N^{4\eps}\cdot \mu^2 \cdot D^* )$, and the total update time of Algorithm $\aset(R)$ is bounded by $O(N^{7\eps}\cdot \mu^2 \cdot D^* )$.

For the remainder of the proof of \Cref{recdynnc from assumption}, it will be convenient for us to define a partitioning tree $\tset$ associated with the collection $\rset$ of basic clusters. The set of vertices of tree $\tset$ contains, for every cluster $R$ that ever belonged to $\rset$, a corresponding vertex $v(R)$. 
Let $C^0$ be the cluster that was added to $\rset$ first (so at the time when $C^0$ was added to $\rset$, $C^0=H$ held).
The root of the tree is vertex $v(C^0)$. consider now any cluster $R\neq C^0$ that ever belonged to $\rset$. Then there must be another cluster $R'\in \rset$, with $R\in \wset_2(R')$. We then make vertex $v(R)$ the child-vertex of $v(R')$ in the tree $R$, and we say that cluster $R$ is a \emph{child-cluster} of cluster $R'$.
Note that, from Property \ref{restriction: few vertices in secondary clusters}, and from our definition of parameters $W_1(R')=N^0(R')$ and $W_2(R')=W_1(R)'/N^{\eps}$, if cluster $R$ is a child-cluster of cluster $R'$, then $N^0(R)\leq N^0(R')/N^{\eps}$. From our algorithm, if vertex $v(R)$ is a leaf in the tree $\tset$, then $N^0(R)<N^{3\eps}$. Since $N^0(C^0)\leq N$, the height of the tree $\tset$ is bounded by $1/\eps$. For all $0\leq i\leq \ceil{1/\eps}$, we denote by $\lset_i$ the collection of all basic clusters $R$, such that the distance from vertex $v(R)$ to vertex $v(C_0)$ in the tree $\tset$ is exactly $i$, so $\lset_0=\set{C^0}$. For  $0\leq i\leq \ceil{1/\eps}$, we may sometimes refer to the clusters of $\lset_i$ as \emph{level-$i$ clusters}. Consider now any cluster $R\in \rset$, such that $v(R)$ is not a leaf vertex of the tree $\tset$. From Property \ref{restriction: regualr vertex in few secondary clusters}, for every regular vertex $x$ of $R$, the total number of clusters in $\wset_2(R)$ to which $x$ ever belongs is bounded by $N^{\eps'/4}$. Therefore:

\[ \sum_{R'\in \wset_2(R)}N^0(R')\leq N^0(R)\cdot N^{\eps'/4}. \]

We then get that, for all $0\leq i<\ceil{1/\eps}$:

\[ \sum_{R'\in \lset_{i+1}}N^0(R')\leq N^{\eps'/4}\cdot \sum_{R\in \lset_i}N^0(R). \]

Altogether, we get that:

\begin{equation}\label{eq: bound num of reg vertices}
\sum_{R\in \rset}N^0(R)\leq N^{\eps'/(2\eps)}N^0(H)\leq N^{1+\eps/2},
\end{equation}

sine $\eps'=\eps^2$.

We are now ready to complete the description of the algorithm for the \recdynnc problem. At the beginning of the algorithm, we let $\rset=\set{H}$, and we apply Algorithm $\aset(H)$ to cluster $H$. The collection $\cset$ of clusters that the algorithm maintains is $\cset=\bigcup_{R\in \rset}\wset_1(R)$. Notice that, from the definition of the \maintainNC problem, for every cluster $R\in \rset$, the clusters in set $\wset(R)$ only undergo allowed updates. Therefore, in order to show that all updates to clusters in $\cset$ are allowed updates, it is enough to show that, whenever a new cluster $C$ is added to $\cset$, there is some cluster $C'$ that is currently in $\cset$, with $C\subseteq C'$. Assume that some new cluster $C$ was added to set $\cset$. Then there must be some basic cluster $R\in \cset$, with $C\in \wset_1(R)$. If $C\neq R$, then cluster $C$ was created by splitting-off from some other cluster $C'\in \wset_1(R)$, so $C\subseteq C'$ and $C'\in \cset$ hold. Assume now that $C=R$. Then cluster $R$ was just added to the set $\rset$ of basic clusters, so there is another cluster $R'\in \rset$, such that $R$ was just added to the set $\wset_2(R')$ of secondary clusters. But then, from the definition of the \maintainNC problem, there is some primary cluster $C'\in \wset_1(R')$, such that cluster $R\subseteq C'$; in other words, cluster $R$ was created by splitting it off from $C'$. But then $C'\in \cset$ currently holds, and $C\subseteq C'$, as required.

Next, we define variables $\coveringcluster(v)$ for regular vertices of $H$. In order to do so, for every basic cluster $R\in \rset$, we define a collection $T(R)$ of regular vertices of $R$, that cluster $R$ is ``responsible'' for covering. We will ensure that, for every regular vertex $v\in T(R)$, either there is some primary cluster $C\in \wset_1(R)$ with $\coveringcluster(v)=C$; or there is some secondary cluster $R'\in \wset_2(R)$ with $v\in T(R')$.

Recall that, at the beginning of the algorithm, we set $\rset=\set{H}$. We then let $T(H)$ contain all regular vertices of $H$. For every regular vertex $v$ of $H$, we also set $\coveringcluster(v)=H$ at this time. Assume now that some cluster $R$ was added to set $\rset$. Then at this time, we set, for every regular vertex $v\in T(R)$, $\coveringcluster(v)=R$. Assume first that $R$ is a large basic cluster. Consider now any fixed regular vertex $v\in T(R)$. Recall that Algorithm $\aset'(R)$ maintains a cluster $\coveringcluster(v)\in \wset(R)$, that we denote, for convenience, by $\coveringcluster^R(v)$. As long as $\coveringcluster^R(v)\in \wset^1(R)$, we set $\coveringcluster(v)=\coveringcluster^R(v)$. Consider now the first time $\tau$ when $\coveringcluster^R(v)\in \wset_2(R)$, and denote $R'=\coveringcluster^R(v)$. Then, from Property \ref{restricioin: secondary covering cluster}, cluster $R'$ just jointed set $\wset_2(R)$, and hence it was just added to the set $\rset$ of basic clusters. Recall that, at this time, we set $\wset_1(R)=\set{R}$, and we add cluster $R$ to set $\cset$. We then add vertex $v$ to set $T(R')$, and set $\coveringcluster(v)=R'$. Assume now that cluster $R$ is a small cluster, and let $v$ be a regular vertex that lies in $T(R)$. Recall that set $\wset_1(R)$ of clusters contains a cluster $C_v$ associated with vertex $v$. We set $\coveringcluster(v)=C_v$, and it remains unchanged for the remainder of the algorithm.
We now argue that our algorithm obeys the Consistent Covering property.

\begin{claim}\label{claim: consistent covering property}
	The algorithm obeys the Consistent Covering property.
\end{claim}

\begin{proof}
	Let $v$ be any regular vertex of $H$, and assume that, at some time $\tau$, $\coveringcluster(v)=C$ held. From the definition of the set $\cset$ of clusters, there must be a basic cluster $R\in \rset$, with $C\in \wset_1(R)$.
	
	We denote by $(v(H)=v(R_0),v(R_1),\ldots,v(R_q)=v(R))$ the path connecting vertex $v(H)$ to vertex $v(R)$ in the tree $\tset$. For all $0\leq i\leq q$, we let $\tau_i$ be the time when cluster $R_i$ was added to set $\rset$. Since $\coveringcluster(v)\in \wset_1(R)$ holds at time $\tau$, it must be the case that, at time $\tau_q$, $v\in T(R)$ and $\coveringcluster(v)=R$ held. Moreover, for all $0\leq i<q$, at time $\tau_i$, $v\in T(R_i)$, and $\coveringcluster(v)=R_i$ held. Observe that, for all $0\leq i\leq q$, $\anc\attime[\tau_i](C)=R_i$.
	
	Consider now some time $\tau'<\tau$, and let $C'=\anc\attime[\tau'](C)$. Assume first that $\tau'\geq \tau_q$, so $C'\in \wset_1(R)$ holds. If $R$ is a large cluster, then, since Algorithm $\aset'(R)$ for the \maintainNC probem ensures the Consistent Covering property, at time $\tau'$, $B_R(v,D)\subseteq V(C')$ held. Moreover, since, at time $\tau_q$, $\coveringcluster(v)=R$ held, we get that, at time $\tau'$, $B_H(v,D)\subseteq V(R)$ holds. Therefore, at time $\tau'$, $B_H(v,D)\subseteq B_R(v,D)\subseteq V(C')$. If $R$ is a small cluster, then $C'=C_v=R=C$ holds at time $\tau_q$. As before, since, at time $\tau_q$, $\coveringcluster(v)=R$ held, we get that, at time $\tau'$, $B_H(v,D)=B_R(v,D)\subseteq V(C')$.
	
	Assume now that $\tau'<\tau_q$. Then there is an integer $0\leq i<q$, such that $\tau_i\leq \tau'<\tau_{i+1}$. Note that, at time $\tau_{i+1}$, $\coveringcluster(v)=R_{i+1}$ held. From our algorithm, it is easy to verify that $\anc\attime[\tau'](R_{i+1})=\anc\attime[\tau'](C)=C'$. Since Algorithm $\aset'(R_i)$ for the \maintainNC probem ensures the Consistent Covering property, at time $\tau'$, $B_{R_i}(v,D)\subseteq V(C')$ held. Moreover, since, at time $\tau_i$, $\coveringcluster(v)=R_i$ held, we get that, at time $\tau'$, $B_H(v,D)\subseteq V(R_i)$ holds. Therefore, at time $\tau'$, $B_H(v,D)\subseteq B_{R_i}(v,D)\subseteq V(C')$. 
\end{proof}

\subsubsection*{Bounding the Number of Clusters a Vertex May Belong to}
Consider any regular vertex $v\in V(H)$. We start by bounding the number of basic cluster $R\in \rset$, for which $v\in V(R)$ ever held. Recall that $v\in V(H)$ held at the beginning of the algorithm. Moreover, from Property \ref{restriction: regualr vertex in few secondary clusters} of the \maintainNC problem, if $R\in \rset$ is a large cluster for which $v\in V(R)$ held when cluster $R$ was added to $\rset$, then the total number of child-clusters $R'$ of $R$ for which $v\in V(R)$ ever held is bounded by $N^{\eps'/4}=N^{\eps^2/4}$. Therefore, for all $0\leq i\leq \ceil{1/\eps}$, the total number of clusters $R\in \lset_i$ for which $v\in V(R)$ ever held is bounded by $N^{i\eps^2/4}$. Since the height of the tree $\tset$ is bounded by $1/\eps$, the total number of clusters $R\in \rset$, for which $v\in V(R)$ ever held is bounded by $N^{\eps}$. Consider now some basic cluster $R\in \rset$, for which $v\in V(R)$ ever held. If $R$ is a large cluster, then, from Property \ref{restriction: few primary clusters}:

\[|\wset_1(R)|\leq  \frac{(W_1(R))^{1+\eps'}}{W_2(R)}\leq N^{2\eps}, \]

since $W_1(R)\leq N$, and $\frac{W_1(R)}{W_2(R)}= N^{\eps}$. If $R$ is a small cluster, then $|\wset_1(R)|=N^0(R)\leq N^{3\eps}$. In any case, if $v\in V(R)$ for a basic cluster $R\in \rset$, then $v$ may belong to at most $N^{3\eps}$ clusters of $\wset_1(R)$. Clearly, if $v\not\in V(R)$, then it may not belong to any cluster of $\wset_1(R)$. Overall, we get that, for a regular vertex $v$ of $H$, the total number of clusters in $\cset$ to which $v$ may ever belong is bounded by 
 $N^{\eps}\cdot N^{3\eps}\leq N^{4\eps}$.
Next, we bound the total update time of the algorithm.

\subsubsection*{Bounding the Total Update Time}

Recall that, for a single cluster $R\in \rset$, the total update time of Algorithm $\aset(R)$ is bounded by:

\[\begin{split}
&O\left (N^0(R)\cdot  W^{\delta+c\eps/4}\cdot \Delta^c\cdot \mu^{c} \cdot (D^*)^{3}\cdot (\log D^*)^c\right )\\&\quad\quad\quad\quad\quad\quad+O\left (N^0(R)\cdot \hat W^{O(\eps)}\cdot (D^*)^3\cdot \mu^{4}\cdot \left(\frac{\hat W}{W}\right )^8\cdot \Delta\right ),
\end{split}
\]

if $R$ is a large cluster, and by $O(N^{7\eps})\cdot \mu^2 \cdot D^* )$ if $R$ is a small cluster.

From Inequality \ref{eq: bound num of reg vertices},
$\sum_{R\in \rset}N^0(R)\leq N^{1+\eps/2}$, so in particular the total number of small clusters is bounded by $N^{1+\eps/2}$. Therefore, the total update time of all algorithms $\aset(R)$ for small clusters $R\in \rset$ is bounded by $O(N^{1+8\eps}\cdot \mu^2 \cdot D^* )$, and the total update time of all algorithms $\aset(R)$ for large clusters $R\in \rset$ is bounded by:

\[\begin{split}
O\left (N\cdot W^{\delta+c\eps/2}\cdot \Delta^c\cdot \mu^{c} \cdot (D^*)^{3}\cdot (\log D^*)^c+N\cdot W^{O(\eps)}\cdot (D^*)^3\cdot \mu^{4}\cdot \left(\frac{\hat W}{W}\right )^8\cdot \Delta\right ),
\end{split}
\]

The total update time of algorithms $\aset(R)$ for all $R\in \rset$ is then bounded by:

\[\begin{split}
O\left (N\cdot W^{\delta+c\eps}\cdot \Delta^c\cdot \mu^{c} \cdot D^{3}\cdot (\log D)^c+N\cdot W^{O(\eps)}\cdot D^3\cdot \mu^{4}\cdot \left(\frac{\hat W}{W}\right )^8\cdot \Delta\right ),
\end{split}
\]

\iffalse
\[\begin{split}
O\left (\hat W^{1+c\eps}\cdot W^{\delta}\cdot \Delta^c\cdot \mu^{c+1} \cdot D^{3}\cdot (\log D)^c\cdot (\log \hat W)^c+\hat W^{1+O(\eps)}\cdot D^3\cdot \mu^{3}\cdot \left(\frac{\hat W}{W}\right )^7\right ).
\end{split}
\]
\fi

(we have used the fact that $D^*\leq O\left(\frac{\log\log N}{(\eps')^4}\cdot D\right )\leq O\left(\frac{\log\log N}{\eps^8}\cdot D\right )\leq O( W^{\eps}\cdot D)$, since $\frac{1}{(\log W)^{1/24}}\leq \eps\leq 1/400$).

It is easy to verify that the total time that is required in order to maintain variables $\coveringcluster(v)$ for regular vertices $v$ of $H$ is asymptotically bounded by the above time. The time that is required to maintain the lists $\clusterlist(x)$ for vertices $x\in V(H)$ and $\clusterlist(e)$ for edges $e\in E(H)$ is asymptotically bounded by the time required to maintain the clusters of $\cset$. Overall, the total update time of the algorithm is bounded by:

\[\begin{split}
O\left (N\cdot W^{\delta+c\eps}\cdot \Delta^c\cdot \mu^{c} \cdot D^{3}\cdot (\log D)^c+N\cdot W^{O(\eps)}\cdot D^3\cdot \mu^{4}\cdot \left(\frac{\hat W}{W}\right )^8\cdot \Delta\right ),
\end{split}
\]

\iffalse
\[\begin{split}
O\left (\hat W^{1+c\eps}\cdot W^{\delta}\cdot \Delta^c\cdot \mu^{c+1} \cdot D^{3}\cdot (\log D)^c\cdot (\log \hat W)^c+\hat W^{1+O(\eps)}\cdot D^3\cdot \mu^{3}\cdot \left(\frac{\hat W}{W}\right )^7\right ).
\end{split}
\]
\fi

It now remains to provide an algorithm to respond to $\spquery$ queries.

\subsubsection*{Responding to Queries.}

We now provide an algorithm for responding to query $\spquery(C,x,y)$, where $C\in \cset$ is a cluster, and $x,y$ is a pair of regular vertices of $C$. Since $C\in \cset$, there is some basic cluster $R$ with $C\in \wset_1(R)$. If $R$ is a small cluster, then  $C=C_z$ holds for some regular vertex $z$ of $R$, and, from the definition of cluster $C_z$, both $x$ and $y$ currently lie in the \EST that Algorithm $\aset^*(C_z)$ maintains. The \EST is rooted at vertex $z$, and has depth $D^*$. We perform $\shortestpath$  query in this data structure, to obtain a path $P_1$ connecting $x$ to $z$ in $C$, whose length is at most $D^*$, in time $O(|E(P_1)|)$. Similarly, we obtain a path $P_2$ connecting $y$ to $z$ in $C$,  whose length is at most $D^*$, in time $O(|E(P_1)|)$. By concatenating the two paths, we obtain a path $P$ in cluster $C$, that connects $x$ to $y$, and has length at most $2D^*\leq \frac{2^{32}\log\log N}{(\eps')^4}\cdot D\leq \frac{2^{32}\log\log N}{\eps^8}\cdot D\leq \alpha^*\cdot D$.

Assume now that $R$ is a large cluster, so $C\in \wset_1(R)$ holds. Recall that we used Algorithm $\aset^*(C)$ for solving the \maintaincluster problem in graph $C$. We perform query $\spquery(C,x,y)$ in the corresponding data structure, and obtain a path $P$ connecting $x$ to $y$ in graph $C$, in time $O(|E(P)|)$, such that the length of the path is bounded by:

\[
\begin{split}
\max\set{2^{O(1/\eps^6)},(8\alpha( W))^{2}}\cdot D^*&\leq \max\set{2^{O(1/\eps^6)},(8\alpha( W))^{2}}\cdot \frac{2^{31}\log\log N}{(\eps')^4}\cdot D
\\&\leq  \max\set{2^{O(1/\eps^6)},(8\alpha( W))^{2}}\cdot \frac{2^{31}\log\log N}{\eps^8}\cdot D\\
&\leq \max\set{2^{O(1/\eps^6)}\cdot \log\log N,\frac{(\alpha( W))^2\cdot\log\log N}{\eps^{16}}}\cdot D\\
&\leq \alpha^*\cdot D.
\end{split}\]

\subsection{Completing the Proof of \Cref{recdynnc simple}}

The proof of \Cref{recdynnc simple} is essentially identical to that of \Cref{recdynnc from assumption}, with one difference: for every large basic cluster $R\in \rset$, for every primary cluster $C\in \wset_1(R)$, we apply the algorithm for the \maintaincluster problem from \Cref{thm: maintain cluster algorithm2} to it, instead of the algorithm from \Cref{thm: main maintain cluster algorithm}. Since we have assumed that $\mu\geq N^{1/10}$, we get that $\mu\geq (N^0(C))^{1/10}$, and so the total update time of the algorithm from \Cref{thm: maintain cluster algorithm2} on cluster $C$ is bounded by $O(N^0(C)\cdot \mu^{22}\cdot D^*)$.

As before, the total update time of Algorithm $\aset'(R)$ remains bounded by:

\[ O\left(N^0(R)\cdot N^{2\eps}\right ).  \]

As before, from Property \ref{restruction: few regular vertices in primary clusters}, $\sum_{C\in \wset_1(R)}N^0(C)\leq  N^0(R)\cdot N^{2\eps'}$. Altogether, the total update time of Algorithm $\aset(R)$ is bounded by:

\[\begin{split}
O\left(N^0(R)\cdot N^{2\eps}\right )+O\left(\sum_{C\in \wset_1(R)}N^0(C)\cdot \mu^{22}\cdot D^*\right )\leq O\left(N^0(R)\cdot N^{2\eps}\cdot \mu^{22}\cdot D^*\right ).
\end{split}
\]

The total update time of all algorithms $\aset(R)$ for small clusters $R\in \rset$ remains bounded by $O(N^{1+8\eps}\cdot \mu^2 \cdot D^* )$. Since, from Inequality
\ref{eq: bound num of reg vertices}, 
$\sum_{R\in \rset}N^0(R)\leq  N^{1+\eps/2}$, we get that the total update time of the algorithm is bounded by:

\[ O(N^{1+8\eps}\cdot \mu^2 \cdot D^* )+O\left (\sum_{R\in \rset} N^0(R)\cdot N^{2\eps}\cdot \mu^{22}\cdot D^*\right )\leq O\left( N^{1+8\eps}\cdot \mu^{22}\cdot D^* \right )\leq  O\left( N^{1+9\eps}\cdot \mu^{22}\cdot D \right ),\]

since $D^*\leq O(\hat W^{\eps}D)$.

Since the approximation factor of the algorithm from \Cref{thm: maintain cluster algorithm2}  is $1$, it is easy to see that our algorithm for the \recdynnc problem achieves approximation factor at most $\frac{2D^*}{D}\leq\frac{2^{32}\log\log N}{\eps^8}\leq 2^{O(1/\eps^6)}\cdot \log\log N$.

\section{Algorithm for the \maintainNC Problem -- Proof of \Cref{thm: alg for maintainNC}}
\label{sec: alg for maintainNC}

In this section we prove \Cref{thm: alg for maintainNC} by providing an algorithm for the \maintainNC problem. Recall that we are given as input 
a valid input structure $\iset=\left(H=(V,U,E),\set{\ell(e)}_{e\in E},D \right )$, where graph $H$ undergoes an online sequence $\Sigma$ of valid update operations with some given dynamic degree bound $\mu$. Additionally, we are given parameters: $W_1$, $W_2$,  $0<\eps'<1$, and $\gamma=\frac{2^{28}\log\log W_1}{(\eps')^4}$. We are guaranteed that $W_2\leq W_1\leq W_2^{1.5}$, and that, if $N^0(H)$ denotes the number of regular vertices in $H$ at the beginning of the algorithm, then $N^0(H) \leq W_1$ holds. 
We denote by $\tset$ the time horizon associated with $\iset$ and $\Sigma$, and we will use two additional distance parameters: $D'=\frac{2^{14}}{(\eps')^2}\cdot D$ and $ D^*=\frac{2^{14}\log\log W_1}{(\eps')^2}D'=\frac{2^{28}\log\log W_1}{(\eps')^4}D=\gamma \cdot D$.

Our algorithm will maintain a collection $\cset$ of clusters, that is partitioned into two subsets: set $\cset_1$ of primary clusters, and set $\cset_2$ of secondary clusters. Whenever a new cluster $C$ is added to set $\cset$, it immediately joins either $\cset_1$ or $\cset_2$, and it remains in that set until the end of the algorithm. 

 The algorithm will maintain a collection $T$ of regular vertices of $H$, that we refer to as \emph{terminals}, and a vertex-induced subgraph $H'$ of $H$. Intuitively, a regular vertex $v$ lies in $T$ if $\coveringcluster(v)\in \cset_1$. At the beginning of the algorithm, we let $T$ contain all regular vertices, and, as the algorithm progresses, vertices may leave the set $T$ but they may never join it. 
For every regular vertex $v$ of $H$, we will maintain a cluster $\coveringcluster(v)\in \cset$, such that Consistent Covering property holds. We will ensure that, throughout the algorithm, properties \ref{restriction: few primary clusters} -- \ref{restriction: regualr vertex in few secondary clusters} hold. 
Additionally, we will ensure that the following invariants always hold.

\begin{properties}{I}
	\item for every regular vertex $v$, if $v\not\in T$, then $\coveringcluster(v)\in \cset_2$; 
	%and $B_H(v,D)\subseteq V(\coveringcluster(v))$;
	\label{invariant: non terminal covered by secondary}
	
	\item for every regular vertex $v\in T$, $B_{H}(v,D)\subseteq V(H')$; \label{invariant: terminal covered by H'}
	
	\item for every regular vertex $v\in T$, there is some primary cluster $C'\in \cset_1$, such that $B_{H'}(v,D')\subseteq V(C')$. \label{invariant: special clusters for regular vertices}
\end{properties}

For every terminal $v\in T$, our algorithm will maintain a cluster $\specialcluster(v)\in \cset_1$, with $B_{H'}(v,D')\subseteq V(\specialcluster(v))$. It also ensures that $\coveringcluster(v)=\specialcluster(v)$ for all $v\in T$, and that the following Extended Consistent Covering property holds:

\begin{properties}[3]{I}
	\item for every time $\tau\in \tset$, if $t\in T$ and  $C=\specialcluster(t)$ at time $\tau$, and, for some $\tau'<\tau$, $C'=\anc\attime[\tau'](C)$, then, at time $\tau'$, $C'=\specialcluster(t)$ held. \label{inv: consistent covering for primary}
\end{properties}

\iffalse
Lastly, we also ensure the following invariant.

\begin{properties}[4]{I}
	\item At the beginning of the algorithm, the set $T$ of terminals contains all regular vertices of $H$. After that, vertices may leave $T$ but they may not join $T$. For every terminal $t\in T$, $\coveringcluster(t)=\specialcluster(t)$ always holds. When a regular vertex $v$ leaves the set $T$ for the first time, a new cluster $C\subseteq\specialcluster(v)$ is created via cluster-splitting of cluster $\specialcluster(v)$, that is added to set $\cset_2$. From that time onwards, $\coveringcluster(v)=C$ always holds.\label{inv: covering clusters}
\end{properties}

 Note that Invariants \ref{inv: extended consistent covering property} and \ref{inv: covering clusters} ensure that our algorithm has the Consistent Covering property.
 \fi

For every cluster $C\in \cset_1$, we maintain a list $\lambda(C)\subseteq T$ of terminals, that contains every terminal $t\in T$ with $\specialcluster(t)=C$.

Throughout, we denote by $V$ the set of all regular vertices that lie in $H$ at the beginning of the algorithm, and we let $N=|V|$. 
For every cluster $C\in \cset$, we denote by $N(C)$ the number of regular vertices that currently lie in $C$, and by $N^0(C)$ the number of regular vertices that lied in $C$ when the cluster was first added to $\cset$.

For every regular vertex $v\in V$, we maintain a counter $n_v$, whose value is the number of clusters $C\in \cset_2\cup \set{H'}$, with $v\in V(C)$. These counters will be used in order to ensure that every regular vertex of $V$ only lies in a small number of clusters of $\cset_2$, using techniques that are almost identical to those employed in the proof of \Cref{claim: cutting G}. The main difference is that the algorithm from \Cref{claim: cutting G} is only applied once to a static graph $G$, while our algorithm will gradually cut clusters off from graph $H'$ and add them to $\cset_2$, as graph $H$ undergoes a sequence of valid update operations. 

We use parameters $\hat \eps=\eps'/32$ and $r=\ceil{1/\hat \eps}+1$.
As in the proof of \Cref{claim: cutting G}, we maintain a partition of the set $V$ of vertices into $r+1$ classes $(S_0,S_1,\ldots,S_r)$, that are defined as follows.  A regular vertex $v\in V$ belongs to class $S_0$ if $n_v\leq N^{2\hat\eps}$. For $1\leq j<r$, vertex $v$ belongs to class $S_j$ if $j\cdot N^{2\hat\eps}<n_v\leq (j+1)\cdot N^{2\hat \eps}$. If $n_v>r\cdot N^{2\hat\eps'}$, then vertex $v$ belongs to class $S_{r}$. As in the proof of \Cref{claim: cutting G}, we will ensure that set $S_r$ remains empty throughout the algorithm. This will allow us to ensure requirement \ref{restriction: regualr vertex in few secondary clusters}.

In order to ensure that Requirements \ref{restriction: few primary clusters} and \ref{restruction: few regular vertices in primary clusters} are  satisfied, we will employ \emph{cluster budgets}, associated with clusters of $\cset_1$. We will assign, to every cluster $C\in \cset_1$, a budget $\beta(C)$. When cluster $C$ is first added to set $\cset_1$, we set the budget $\beta(C)=(N^0(C))^{1+\eps'}$. As the algorithm progresses, we may sometimes decrease the budget $\beta(C)$, but we will ensure that $\beta(C)\geq W_2$ and $\beta(C)\geq N(C)$ always holds. We then denote by $\beta=\sum_{C\in \cset_1}\beta(C)$. We will ensure that, at the beginning of the algorithm, $\beta\leq W_1^{1+\eps'}$ holds, and that the total budget $\beta$ does not increase over the course of the algorithm. This will ensure that, on the one hand, $\sum_{C\in \cset_1}N(C)\leq \sum_{C\in \cset_1}\beta(C)\leq W_1^{1+\eps'}$ always holds, while, on the other hand, $|\cset_1|\leq \frac{\beta}{W_2}\leq \frac{W_1^{1+\eps'}}{W_2}$. The algorithm will ensure the following invariant:

\begin{properties}[4]{I}
\item	For every cluster $C\in \cset_1$, $\beta(C)\geq W_2$ and $\beta(C)\geq (N(C))^{1+\eps'}$ holds at all times. Throughout the algorithm, the total budget $\beta=\sum_{C\in \cset_1}\beta(C)$ does not grow.\label{inv: budgets}
\end{properties}

At the beginning of the algorithm, we set $H'=H$, $\cset_1=\set{H}$, and $\cset_2=\emptyset$. We set the budget of cluster $H$ to be $\beta(H)=(N^0(H))^{1+\eps'}$. We also let $T$ contain all regular vertices of $H$. For every regular vertex $v$, we let $\coveringcluster(v)=H$, and $\specialcluster(v)=H$. We also initialize the list $\lambda(H)$ to contain all vertices of $T$, and, for every regular vertex $v\in V$, we set $n_v=1$. It is easy to verify that all invariants hold at the beginning of the algorithm. We also set $S_0=V$ and $S_1=S_2=\cdots=S_r=\emptyset$.

 As the algorithm progresses, and graph $H$ undergoes valid update operations, the clusters in set $\cset=\cset_1\cup \cset_2$ are updated accordingly, as described in \Cref{subsubsec: updating clusters}. The only additional changes to the clusters in set $\cset$, and to variables in $\set{\coveringcluster(v)}_{v\in V}$ and $\set{\specialcluster(t)}_{t\in T}$ will be done during Flag Lowering operations. Budgets of clusters in $\cset_1$ may also only be updated during Flag Lowering operations.
Therefore, if $\tset'\subseteq \tset$ is any time interval during which no Flag Lowering operations are performed, and all invariants held at the beginning of $\tset'$, then, from \Cref{obs: maintain ball covering property}, all invariants continue to hold at the end of $\tset'$.
It now remains to describe an algorithm for a Flag Lowering Operation.

\subsection{Flag-Lowering Operation}
\label{subsec: flag lowering}

We assume that we are given a collection $\cset=\cset_1\cup \cset_2$ of clusters, for which all invariants hold. We also assume that we are given a cluster $C\in \cset_1$, and a pair $v,v'$ of regular vertices that lie in $C$, such that $\dist_C(v,v')>D^*$. Our goal is to produce a valid flag-lowering sequence $\Sigma'$, which is then applied to cluster $C$, possibly interspersed with cluster-splitting updates, in which new clusters $C'\subseteq C$  are created. Our goal is to ensure that, at the end of the Flag-Lowering Operation, all invariants will continue to hold. The algorithm consists of two phases.
In Phase 1, we will either compute two terminals $t,t'\in \lambda(C)$, such that $\dist_C(t,t')\geq D^*/2$, and each of $B_C(t,D'),B_C(t',D')$ contains at least $W_2$ regular vertices (in which case we say that the phase is unsuccessful); or we will compute a flag lowering sequence $\Sigma'$, that will be applied to cluster $C$, possibly interspersed with cluster-splitting updates (in which case we say that the phase is successful). 
All clusters that are created during Phase 1 are added to set $\cset_2$. If Phase 1 is successful, then we do not continue to Phase 2. Otherwise, during Phase 2, we will create a new cluster $C'\subseteq C$, that is added to the set $\cset_1$ of primary clusters, and we will delete some edges and vertices from $C$, including at least one of the terminals $t,t'$. We now describe each of the phases in turn.

\subsubsection{Phase 1: Reaching Terminals}

In Phase 1, we will either compute two  terminals $t,t'\in \lambda(C)$ with $\dist_C(t,t')\geq D^*/2$, such that both $B_C(t,D')$ and $B_C(t',D')$ contain at least $W_2$ regular vertices, or we will delete some vertices and edges from cluster $C$ (the deletions may be interspersed with cluster-splitting updates), producing a valid flag-lowering update sequence $\Sigma'$. The main challenge is to execute this procedure efficiently, so that, if no pair of terminals with the above properties are found, then the running time of the algorithm can be charged to the edges that are deleted from $C$.

Assume first that $v$ is an isolated vertex in $C$. If $v\not\in \lambda(C)$, then we simply delete $v$ from $C$; it is easy to verify that all invariants continue to hold, since $C\neq\specialcluster(v)$. Assume now that $v\in \lambda(C)$, so $C=\specialcluster(v)$. In this case, $B_{H'}(v,D')\subseteq V(C)$ must hold, and, from Invariant \ref{invariant: terminal covered by H'}, $B_{H}(v,D)\subseteq V(H')$. Since $H'$ is a vertex-induced subgraph of $H$, it then follows that $B_H(v,D)\subseteq V(C)$. We create a new cluster $C'\subseteq C$ that only contains the vertex $v$, by spliting the cluster off from $C$. Cluster $C'$ is added to set $\cset_2$, and vertex $v$ is deleted from the set $T$ of terminals. We set $\coveringcluster(v)=C$. The flag-lowering sequence $\Sigma'$ then only consists of a single operation: deletion of vertex $v$ from $C$. Once vertex $v$ is deleted from $C$, the Flag-Lowering Operation terminates. It is easy to verify that all invariants continue to hold. If vertex $v'$ is an isolated vertex in $C$, then we proceed in exactly the same way. Therefore, we assume from now on that vertices $v$ and $v'$ are not isolated in $C$.

Our algorithm for Phase 1 will execute two procedures in parallel: procedure $\aset_1$, that processes vertex $v$, and procedure $\aset_2$, that processes vertex $v'$. The two procedures are executed in parallel in order to ensure the efficiency of the algorithm: we will consider the procedure that terminates first, and we may choose to not complete the other procedure. 
Each one of the two procedures may delete edges and vertices from $C$, and may create new clusters by splitting them off from $C$. However, we will ensure that all edges and vertices of $C$ that procedure $\aset_1$ deletes lie in the subgraph of $C$ induced by $B_C(v,D^*/3)$, and similarly, all edges and vertices of $C$ that procedure $\aset_2$ deletes lie in the subgraph of $C$ induced by $B_C(v',D^*/3)$. Additionally, when procedure $\aset_1$ creates a new cluster $C'\subseteq C$ by splitting it off from $C$, then $V(C')\subseteq B_C(v,D^*/3)$ holds, and similarly, when procedure $\aset_2$ creates a new cluster $C'\subseteq C$ by splitting it off from $C$, then $V(C')\subseteq B_C(v',D^*/3)$ holds. In fact Procedure $\aset_1$ will perform a number of BFS searches from vertex $v$ and its nearby vertices, and it will never explore any vertices that lie outside 
$B_C(v,D^*/3)$. Similarly, procedure $\aset_2$ will only explore vertices of $B_C(v',D^*/3)$. Therefore, the two procedures can be executed in parallel and independently, since they do not interfere with each other in any way. We focus on describing procedure $\aset_1$; the description of procedure $\aset_2$ is identical, except that vertex $v$ is replaced with $v'$.

\subsubsection*{Procedure $\aset_1$}
We start by applying the algorithm from \Cref{lem: ball growing} to graph $C$ and vertex $v$, with distance parameter $D'$, and parameter $\eps'$ remaining unchanged (we do not supply the sets $T_1,\ldots T_k$ of vertices, so $k=0$). Denote by $1<i\leq \frac{2}{\eps'}$ the integer that the algorithm returned, and denote $S_v=B_C(v,2(i-1)D')$, and $S'_v=B_C(v,2iD')$. Recall that the running time of the algorithm from \Cref{lem: ball growing} is bounded by $O(|E_C(S_v)|\cdot |E(C)|^{\eps'}\cdot \log (W_1\cdot \mu)\leq O(|E_C(S_v)|\cdot  (W_1\cdot \mu)^{\eps'}\log  (W_1\cdot \mu))$. We then check whether $S'_v\cap \lambda(C)=\emptyset$ holds. If this is the case, then we say that Procedure $\aset_1$ was successful. Note that, in this case, for every terminal $t\in \lambda(C)$, $B_C(t,D')\cap S_v=\emptyset$. Therefore, we can delete all edges incident to the vertices of $S_v$, and all vertices of $S_v$ from cluster $C$, without violating any invariants. 
Since $i\leq \frac{2}{\eps'}$, and $S'_v=B_C(v,2iD')$, while $D^*>\frac{64D'}{\eps'}$, all edges and vertices that we delete lie in $B_C(v,D^*/4)$. Moreover, the algorithm from \Cref{lem: ball growing} essentially performs a BFS from vertex $v$ in cluster $C$ up to depth $2iD'$, so it does not explore any vertices of $C$ that lie outside $S'_v$. We then terminate Procedure $\aset_1$.

From now on we assume that $S'_v\cap \lambda(C)\neq \emptyset$ holds, and we denote $T_1=S'_v\cap \lambda(C)$. We now perform iterations, as long as $T_1\neq \emptyset$.

\paragraph{Iteration Description.}
We now describe a single iteration. We let $t\in T_1$ be any terminal. We apply the algorithm from \Cref{lem: ball growing} to the current cluster $C$, vertex $t$, distance parameter $2D$, and precision parameter $\eps'$ remaining unchanged. We also set $k=r+1$, and we use sets $S_0\cap V(C),S_1\cap V(C),\ldots,S_r\cap V(C)$ instead of $T_1,\ldots,T_k$.

Let $1<i_t\leq\frac{2r+4}{\eps'}$ be the integer that the algorithm from \Cref{lem: ball growing} returned. We also let $B_t=B_C(t,4(i_t-1)D)$ and $B'_t=B_C(t,4i_tD)$. Since $i_t\leq \frac{2r+4}{\eps'}$, $r=\ceil{\frac 1{\hat \eps}}+1\leq \frac{4}{\hat \eps}$, and
$\hat \eps=\frac{\eps'}{32}$, we get that $4i_tD\leq \frac{16r}{\eps'}\cdot D\leq \frac{2^{11}}{(\eps')^2}\cdot D$. Since $D'=\frac{2^{14}}{(\eps')^2}\cdot D$ and $D^*>4D'$, we get that $B'_t\subseteq B_C(t,D'/2)\subseteq B_C(t,D^*/8)$. Note that the algorithm from \Cref{lem: ball growing} performs a BFS from vertex $t$ in cluster $C$ up to distance $2i_tD$, so all vertices of $C$ that it encounters are contained in $B_C(t,D^*/8)$, as required. 
Recall that the algorithm guarantees that, 
for all $0\leq j\leq r$, $|S_j\cap B'_t|\leq |S_{j}\cap B_t|\cdot N^{\eps'}$.

We now consider two cases. The first case happens if the number of regular vertices in $B'_t$ is at least $W_2$. In this case, we say that Procedure $\aset_1$ was \emph{unsuccessful}, and we return terminal $t$. We also say that the current iteration was a \emph{type-1} iteration. Notice that we are guaranteed that $B_C(t,D')$ contains at least $W_2$ regular vertices.

Assume now that the number of regular vertices in set $B'_t$ is less than $W_2$. We then say that the current iteration is a \emph{type-2} iteration.
Since $t\in \lambda(C)$, $\specialcluster(t)=C$ must hold. Therefore, from Invariant \ref{invariant: special clusters for regular vertices}, $B_{H'}(t,D')\subseteq V(C)$. Since $C$ is a vertex-induced subgraph of $H'$, we get that $B_{H'}(t,D')=B_C(t,D')$. We use the following simple observation.

\begin{observation}\label{obs: terminal coverage}
	Let $t'$ be any terminal in $T$. If $\dist_{H'}(t,t')\leq 4i_tD-2D$, then $B_H(t',D)\subseteq B'_t$. Otherwise, $B_H(t',D)\cap B_t=\emptyset$.
\end{observation}
\begin{proof}
	Let $t'\in T$ be any terminal. Assume fist that $\dist_{H'}(t,t')\leq 4i_tD-2D$. From Invariant \ref{invariant: terminal covered by H'}, $B_H(t',D)\subseteq H'$ holds, and, since $H'$ is a vertex-induced subgraph of $H$, $B_H(t',D)=B_{H'}(t',D)$. It is also easy to see that $B_{H'}(t',D)\subseteq B_{H'}(t,4i_tD)=B_C(t,4i_tD)=B'_t$. 
	
	Assume now that $\dist_{H'}(t,t')>4i_tD-2D$. Then for every vertex $y\in B_{H}(t',D)=B_{H'}(t',D)$, $\dist_{H'}(t,y)\geq 4i_tD-3D>4(i_t-1)D$ holds. Therefore, if $y\in V(C)$, then $\dist_C(t,y)>4(i_t-1)D$, and in any case $y\not\in B_t$. We conclude that $B_H(t',D)\cap B_t=\emptyset$.
\end{proof}

Note that, for a terminal $t'\in T$, if $\dist_{H'}(t,t')\leq 4i_tD-2D$ then $\dist_C(t,t')\leq 4i_tD-2D$ must hold and vice versa.

Let $C'$ be the subgraph of $C$ induced by the set $B'_t$ of vertices.  We apply the cluster splitting operation to cluster $C$, creating a new cluster $C'$, that is added to set $\cset_2$. Recall that the number of regular vertices in $C'$ is at most $W_2$, as required.
Next, we consider every terminal $t'\in T\cap B_{C}(t,4i_tD-2D)$. For each such terminal $t'$, we set $\coveringcluster(t')=C'$, and we delete $t'$ from the set $T$ of terminals, and from set $T_1$ if $t'\in T_1$ holds. From \Cref{obs: terminal coverage}, for each such terminal $t'$, $B_H(t',D)\subseteq V(C')$ indeed holds. From this point onward, $\coveringcluster(t')$ will not change, and it will remain equal to $C'$ for the remainder of the algorithm. We claim that this the assignment of $\coveringcluster(t')$ obeys the Consistent Covering property over the course of the entire algorithm. Indeed, let $\tau$ be the time when cluster $C'$ is created. Until time $\tau$, Consistent Covering property for terminal $t'$ followed from Invariant \ref{inv: consistent covering for primary}, as $\coveringcluster(t')=\specialcluster(t')$ always held. Consider now time $\tau$, when $\coveringcluster(C)$ was set to $C'$. Note that, at time $\tau$, $C=\specialcluster(t)$ and $\dist_H(t,t')<D'/2$ holds. Consider now any time $\tau'<\tau$, and let $\hat C=\anc\attime[\tau'](C)$. From Invariant \ref{inv: consistent covering for primary}, at time $\tau'$, $\hat C=\specialcluster(t)$ held, and so, at time $\tau'$, $B_{H'}(t,D')\subseteq V(\hat C)$ held. Since, at time $\tau$,  $\dist_H(t,t')\leq D'/2$, and since distances in $H$ may not shrink as the result of valid update operations, we get that, at time $\tau'$, $\dist_H(t,t')\leq D'/2$ also held, and so, at time $\tau'$, $B_H(t',D)\subseteq B_H(t,D')$. We conclude that, at time $\tau'$, $B_H(t',D)\subseteq V(\hat C)$ holds. Therefore, for every terminal $t'$ that we deleted from $T$ during the current iteration, we are guaranteed that the Consistent Covering property holds for $t'$ over the course of the entire algorithm.

Lastly, we delete all edges incident to the vertices of $B_t$ from cluster $C$, and then delete the vertices of $B_t$ from $C$. These deletions become part of the flag lowering sequence $\Sigma'$ that we construct. 
 We also delete the vertices of $B_t$ from graph $H'$. 

Notice that the regular vertices of $B_t$ now belong to cluster $C'$, and they were deleted from graph $H'$. Therefore, for a regular vertex $x\in B_t$, the counter $n_x$ does not need to be updated. However, if a regular vertex $x$ lies in set $B'_t\setminus B_t$, then  $x$ now lies in cluster $C'$, and it continues to lie in graph $H'$. We then say that a \emph{new copy of vertex $x$} was created -- the copy that lies in cluster $C'$. For each regular vertex $x\in B'_t\setminus B_t$, we increase the value of the counter $n_x$, and, if needed, we update the set $S_j$, for $0\leq j\leq r$, to which the vertex belongs.

We now show that all invariants continue to hold at the end of the iteration. Invariants \ref{invariant: non terminal covered by secondary} and \ref{invariant: terminal covered by H'}  continue to hold from \Cref{obs: terminal coverage}. It is immediate to verify that Invariant \ref{invariant: special clusters for regular vertices} continues to hold as well, since the changes to graph $H'$ and cluster $C$ over the course of the current iteration were identical. Invariant \ref{inv: consistent covering for primary} continues to hold, since we did not change clusters $\specialcluster(t'')$ for any terminal that remains in $T$ at the end of the iteration, and Invariant \ref{inv: budgets} continues to hold, since we did not make any changes to cluster budgets. 

This completes the description of the iteration.
If the current iteration was a type-1 iteration, then its running time can be bounded by $O(|E(C)|\log (W_1\cdot \mu))\leq O(W_1\cdot \mu\cdot \log (W_1\cdot \mu))$. Otherwise, if we let $E'=E_C(B_t)$ be the collection of edges that were deleted from cluster $C$ over the course of the iteration, then, from \Cref{lem: ball growing}, the running time of the iteration is bounded by $O(|E'|\cdot |E(C)|^{\eps'}\cdot \log (W_1\cdot \mu))\leq O(|E'|\cdot (W_1\cdot \mu)^{\eps'}\cdot \log (W_1\cdot \mu))$.

Procedure $\aset_1$ terminates either after a type-1 iteration, when a terminal $t$ with $B_{H'}(t,D')$ containing at least $W_2$ regular vertices is found (in which case we say that it is unsuccessful), or once $T_1=\emptyset$ holds (in which case we say that it is successful). In the latter case, we delete all edges that are incident to the vertices of $S_v$ from cluster $C$, and we delete all vertices of $S_v$ from cluster $C$. These deletions become a part of the flag-lowering update sequence $\Sigma'$. We will charge the running time of the algorithm from \Cref{lem: ball growing}, when it was used by Procedure $\aset_1$ for the first time, to the edges incident to vertices of $S_v$ that we just deleted from $C$. Since all terminals from $T_1$ have been deleted from the set $T$ of terminals, for each remaining terminal $t'\in \lambda(C)$, we are guaranteed that $B_{H'}(t',D')$ is disjoint from $S_v$, so we are guaranteed that $B_{H'}(t',D')\subseteq V(C)$ continues to hold. It is immediate to verify that all invariants continue to hold.

We now bound the running time of Procedure $\aset_1$. Let $E'$ be the set of all edges that were deleted over the course of the procedure from cluster $C$. If Procedure $\aset_1$ was successful, then its total running time is bounded by $O(|E'|\cdot (W_1\cdot \mu)^{\eps'}\cdot \log(W_1\cdot \mu))$. Otherwise, the running time of the first application of the algorithm from 
\Cref{lem: ball growing} and of the last iteration can be bounded by $O(|E(C)|\log (W_1\cdot \mu))\leq O(W_1\mu \log (W_1\cdot \mu))$, while the running time of the remainder of the algorithm is bounded by  $O(|E'|\cdot (W_1\cdot \mu)^{\eps'}\cdot \log (W_1\cdot \mu))$. Therefore, if Procedure $\aset_1$ is unsuccessful, its running time is bounded by $O((W_1\cdot \mu)^{1+\eps'}\cdot \log (W_1\cdot \mu))$.

This concludes the description of Procedure $\aset_1$. Procedure $\aset_2$ is similar, except that we replace vertex $v$ with vertex $v'$.

\subsubsection*{Completing the Description of Phase 1}

We execute Procedures $\aset_1$ and $\aset_2$ in parallel. Notice that the execution of each procedure can be partitioned into iterations. In every iteration, we first invoke the algorithm from \Cref{lem: ball growing}, that performs a weighted BFS from some given vertex. After that, we execute a clean-up step, during which some vertices and edges may be deleted from $C$ or from $H'$, and a new cluster may be created by performing cluster-splitting of cluster $C$. The running time of the lean-up step is asymptotically bounded by the running time of the algorithm from \Cref{lem: ball growing}. We assume w.l.o.g. that Procedure $\aset_1$ terminates before Procedure $\aset_2$ does, and we denote by $\tset(\aset_1)$ the running time of Procedure $\aset_1$. If, at the time when Procedure $\aset_1$ terminates, procedure $\aset_2$ performs a clean-up step, then we allow it to complete the clean-up step, and then halt it (possibly temporarily). The running time of both procedures so far is bounded by $O(\tset(\aset_1))$.

Assume first that Procedure $\aset_1$ was successful. In this case, by the end of the procedure, vertex $v$ is deleted from cluster $C$. Let $\Sigma'$ denote the sequence of all edge-deletions and isolated vertex-deletions that procedures $\aset_1$ and $\aset_2$ applied to cluster $C$, and let $E'$ be the set of all edges that both procedures deleted so far from $C$. Then $\Sigma'$ is valid flag lowering sequence for $C$. We say Phase 1 was succecssful, and we terminate the Flag Lowering operation. Notice that the running time of Phase 1 is bounded by 
$O(\tset(\aset_1))\leq O(|E'|\cdot (W_1\cdot \mu)^{\eps'}\cdot \log (W_1\cdot \mu))$ in this case.

Assume now that Procedure $\aset_1$ was unsuccessful, and let $t$ be the terminal that the procedure returned. Then $\dist_C(v,t)\leq D^*/4$, and $B_C(t,D')$ contains at least $W_2$ regular vertices. In this case, we resume the execution of Procedure $\aset_2$, until it terminates. We denote by $\tset(\aset_2)$ the running time of Procedure $\aset_2$. 
We consider again two cases.

The first case happens if Procedure $\aset_2$ was successful. In this case, by the end of the procedure, vertex $v'$ is deleted from cluster $C$. Let $\Sigma'$ denote the sequence of all edge-deletions and isolated vertex-deletions that procedures $\aset_1$ and $\aset_2$ applied to cluster $C$, and let $E'$ be the set of all edges that both procedures deleted from $C$. Then $\Sigma'$ is valid flag lowering sequence for $C$. We say Phase 1 was successful, and we terminate the Flag Lowering operation. Notice that the running time of Phase 1 is bounded by 
$O(\tset(\aset_2))\leq O(|E'|\cdot (W_1\cdot \mu)^{\eps'}\cdot \log (W_1\cdot \mu))$ in this case.

Lastly, we consider the second case, where Procedure $\aset_2$ was unsuccessful. In this case, we say that Phase 1 was unsuccessful, and we let $t'$ be the terminal that Procedure $\aset_2$ returned. We have now obtained two terminals $t,t'$, such that both $B_C(t,D')$ and $B_C(t',D')$ contain at least $W_2$ regular vertices each. Moreover, since, at the beginning of the algorithm, $\dist_C(v,t)\leq D^*/4$, $\dist_C(v',t')\leq D^*/4$, and $B_C(v,v')\geq D^*$ held, we get that at the beginning of the algorithm, $\dist_C(t,t')\geq D^*/2$ held. Since edge and vertex deletions from $C$ may not decrease distances between vertices, we are guaranteed that $\dist_C(t,t')\geq D^*/2$ currently holds. In this case, we will continue to Phase 2. Note that the running time of Phase 1 in this case is bounded by $O(\tset(\aset_2))\leq O((W_1\cdot \mu)^{1+\eps'}\log (W_1\cdot \mu))$. We also let $\Sigma'$ be the sequence of all edge-deletions and isolated vertex-deletions that Procedures $\aset_1$ and $\aset_2$ performed over the course of Phase 1. While $\Sigma'$ may not be a valid flag-lowering sequence (as it is possible that it did not delete either $v$ or $v'$ from $C$), we will extend it in the remainder of the algorithm, to ensure that it becomes a valid flag-lowering sequence.

We note that budgets of clusters in $\cset_1$ were not updated over the course of Phase 1, so Invariant \ref{inv: budgets} continues to hold at the end of the phase. We have already established that Invariants \ref{invariant: non terminal covered by secondary} --  \ref{inv: consistent covering for primary} continue to hold over the course of the phase, and that, for every vertex $x$ that is deleted from the set $T$ of the terminals during the current phase, the Consistent Covering property holds over the course of the entire algorithm.
Let $E'$ denote the collection of edges that were deleted over the course of Phase 1 from cluster $C$. If Phase 1 is successful, then its running time is bounded by $O(|E'|\cdot(W_1\cdot \mu)^{\eps'}\cdot \log (W_1\cdot \mu))$, and if the phase was unsuccessful, then its running time is bounded by  $O( (W_1\cdot \mu)^{1+\eps'}\cdot \log (W_1\cdot \mu))$.

\subsubsection{Phase 2: Separating Terminals}

The input to Phase 2 is a pair $t,t'\in \lambda(C)$ of terminals, such that $\dist_C(t,t')\geq D^*/2$, and each of $B_C(t,D')$ and $B_C(t',D')$ contains at least $W_2$ regular vertices. Over the course of Phase 2, we will construct a new cluster $C'\subseteq C$ that will be added to the set $\cset_1$ of primary clusters and the set $\cset$ of clusters that our algorithm maintains, by applying a cluster-splitting operation to cluster $C$. We will also delete some edges and vertices from cluster $C$, and we will update budget $\beta(C)$ and define budget $\beta(C')$, so that the total budget $\beta$ does not grow, and Invariant \ref{inv: budgets} is satisfied. Over the course of Phase 2, we will execute two procedures: procedure $\aset'_1$, which runs a BFS in cluster $C$ sarting from vertex $t$, and procedure $\aset'_2$, that runs a BFS in $C$ starting from $t'$. We now describe procedure $\aset'_1$; procedure $\aset'_2$ is identical, except that we replace terminal $t$ with $t'$.

\subsubsection*{Procedure $\aset_1'$}

Procedure $\aset_1'$ will perform a weighted BFS from vertex $t$ in graph $C$, up to a certain depth that will be determined during the procedure. We will again use the ball-growing technique, but in a manner that is slightly different from that in \Cref{lem: ball growing}. As in the proof of  \Cref{lem: ball growing}, we define \emph{layers} of the BFS, where for all $i> 0$, layer $L_i$ is defined as follows:

\[L_i=B_C(v,2iD')\setminus B_C(v,2(i-1)D').   \]

%For all $i\geq 0$, we denote by $m_i$ the total number of edges $e\in E(G)$, such that both endpoints of $e$ lie in $L_0\cup L_1\cup\cdots\cup L_i$.
 Let $N'$ denote the number of regular vertices currently in cluster $C$. For all $i> 0$, let $N_i$ be the number of regular vertices that lie in 
 $L_1\cup\cdots\cup L_i$, and let $\eta_i=\frac{N'}{N_i}$. Clearly, $N_1\leq N_2\leq\cdots$, and $\eta_1\geq\eta_2\geq\cdots$.
We now define the notion of eligible layer, which is somewhat different from that in the proof of \Cref{lem: ball growing}.

\begin{definition}[Eligible Layer]
	Let $i>1$ be an integer. We say that layer $L_i$ of the BFS is \emph{eligible}, if one of the following condition holds: 
	
	\begin{itemize}
			\item either $N_i\leq \frac{N'}{4}$ and $N_i\leq N_{i-1}\cdot \eta_{i-1}^{\eps'/4}$; or
			
			\item $\frac {N'} 4\leq N_i\leq \frac {N'} 2$, and $N_i\leq N_{i-1}\cdot \left(1+\frac{(\eps')^2}{16}\right )$.
	\end{itemize}
\end{definition}

We use the following simple observation, that follows from standard arguments.

\begin{observation}\label{obs: eligible layer2}
 If $B_C(t,D^*/4)$ contains at most $\frac{N'}{2}$ regular vertices, then there is an integer $1<i\leq \frac{2^{10}\log\log W_1}{(\eps')^2}$, such that layer $L_i$ is eligible.
\end{observation}
\begin{proof}
	Denote $z= \frac{2^{10}\log\log W_1}{(\eps')^2}$. Recall that $D^*=\frac{2^{14}\log\log W_1}{(\eps')^2}D'$. Since we have assumed that $B_C(t,D^*/4)$  contains at most $\frac{N'}{2}$ regular vertices, we get that $N_z\leq \frac{N'}{2}$.
	Assume for contradiction that the claim is false, so none of the layers $L_2,L_3,\ldots,L_z$ is eligible.

	Let $i^*$ be the smallest integer, for which $N_{i^*}\geq \frac{N'}{4}$ holds. Then for all $i^*<i\leq z$, $\frac{N'}{4}\leq N_i\leq \frac{N'}{2}$ holds, and $N_i>N_{i-1}\cdot \left(1+\frac{(\eps')^2}{16}\right )\geq N_{i-1}+\frac{N'\cdot (\eps')^2}{64}$. We claim that $i^*\geq z-\frac{64}{(\eps')^2}$ must hold. Indeed, otherwise, if we let $i^{**}=i^*+\frac{64}{(\eps')^2}$, then $i^{**}\leq z$ and $N_{i^{**}}>N'$ must hold, a contradiction. We conclude that $i^*\geq z-\frac{64}{(\eps')^2}\geq \frac{z}{2}$.
	
	Next, we will designate some layers $L_i$, for $1\leq i\leq i^*$ as \emph{special} layers. Instead of defining special layers explicitly, we describe an algorithm that computes special layers. The first special layer is layer $L_1$. Assume now that the largest index of a special layer that we have defined so far is $i$. We let $i'$ be the smallest integer, such that $i<i'<i^*$, and $\eta_{i'}\leq \sqrt{\eta_i}$. If no such integer exists, then we terminate the algorithm for computing special layers. Otherwise, we designate $L_{i'}$ as a special layer and continue the algorithm.

 Let $1=i_1<i_2<\cdots<i_q<i^*$ denote the indices of special layers.
Recall that $B_C(t,D')$ contains at least $W_2$ regular vertices, so $N_1\geq W_2$, and $\eta_{i_1}=\eta_1\leq \frac{N'}{W_2}$. Since $i_q<i^*$, we get that $N_{i_q}<\frac{N'}{4}$, and so $\eta_{i_q}>4$. From the definition of special layers, for all $1<j\leq q$, $\eta_{i_j+1}\leq \sqrt{\eta_{i_j}}$. It then follows that $q\leq \log\log(N'/W_2)\leq \log\log W_1$.
	
	Next, we consider a consecutive pair of special layers, that we denote by $L_i$ and $L_{i'}$, with $i'>i$. We claim that $i'-i\leq 16/\eps'$. Indeed, consider any index $i<i''< i'$. Since we have assumed that layer $L_{i''}$ is not eligible, we get that:
	
	\[ N_{i''}> N_{i''-1}\cdot \eta_{i''-1}^{\eps'/4}\geq N_{i}\cdot \eta_{i}^{\eps'/8}. \]
	
	Therefore, $N_{i''}>N_i\cdot \eta_i^{(i''-i)\eps'/8}$. If $i'>16/\eps'$, then $N_{i'}>N_i\cdot \eta_i=N'$, which is impossible.
Using a similar reasoning, $i^*-i_q\leq 16/\eps'$ must hold. Overall, we get that the number of special layers is $q\leq\log\log W_1$, and the number of layers lying between any pair of special layers is bounded by $16/\eps'$. Similarly, $i^*\leq i_q+16/\eps'$. We conclude that $i^*\leq \frac{32q}{\eps'}\leq \frac{32\log\log W_1}{\eps'}<\frac{z}{2}$, a contradiction.
\end{proof}

Algorithm $\aset'_1$ simply performs a weighted BFS in graph $C$ starting from vertex $t$, until it encounters the first layer $L_i$, such that either $N_i>\frac{N'}{2}$, or layer $L_i$ is eligible. Clearly, the running time of the algorithm is bounded by $O(|E(C)|\log (W_1\mu)\leq O(W_1\mu\log (W_1\mu))$. Algorithm $\aset_2$ is identical to $\aset_1$, except that we run the BFS starting from vertex $t'$ instead of $t$.

\subsubsection*{Completing the Description of Phase 2}

The algorithm for Phase 2 executes procedures $\aset_1$ and $\aset_2$ one after another. Recall that $\dist_C(t,t')\geq D^*/2$, and that $ D^*=\frac{2^{14}\log\log W_1}{(\eps')^2}D'$. From \Cref{obs: eligible layer2}, we are then guaranteed that at least one of the two procedures must terminate with an integer $1<i\leq \frac{2^{10}\log\log W_1}{(\eps')^2}$, such that layer $L_i$ is eligible for the corresponding BFS procedure. We assume w.l.o.g. that it is $\aset_1$, and we denote by $i$ the index of the eligible layer that Procedure $\aset_1$ computed. Let $C'$ be the subgraph of $C$ induced by the vertices of $L_1\cup \cdots L_i$. Notice that the number of regular vertices in $C'$ is $N(C')=N_i$. We add cluster $C'$ to set $\cset_1$, and we set its budget $\beta(C')$ to be $N_i^{1+\eps'}$. From our algorithm, $N_i\geq W_2$ must hold. Next, we delete from cluster $C$ all edges that are incident to vertices of $L_1\cup\cdots\cup L_{i-1}$, and the vertices of $L_1\cup\cdots\cup L_{i-1}$. These deletions are added to the flag lowering sequence $\Sigma'$. For ease of discussion, we denote by $C''$ cluster $C$ that is obtained after these deletions, and, when we refer to cluster $C$ itself in the remainder of the description of the phase, we refer to it at the beginning of the phase. We denote the number of regular vertices in $C''$ by $N''$. Our algorithm guarantees that $N''\geq \frac{N'}{2}$, and also that $N''\geq W_2$. We set the new budget of cluster $C''$ to be $\beta(C'')=(N'')^{1+\eps'}$. We will show below that these updates to budgets of clusters do not increase the total budget $\beta$.

Consider now some terminal $t''\in \lambda(C)$, and recall that $C=\specialcluster(t'')$, and so $B_{H'}(t'',D')\subseteq V(C)$. Notice that, if $\dist_C(t,t'')>(2i-1)D'$, then $B_{H'}(t'',D')=B_C(t'',D')$ is disjoint from vertices of $L_1\cup\cdots\cup L_{i-1}$, and so $B_{H'}(t'',D')\subseteq V(C')$ continues to hold. Otherwise, all vertices of $B_{H'}(t'',D')=B_C(t'',D')$ belong to $L_1\cup\cdots\cup L_i$, so 
$B_{H'}(t'',D')\subseteq V(C')$ holds. We consider every terminal $t''\in \lambda(C)$ with $\dist_C(t,t'')\leq (2i-1)D'$. For each such terminal, we set $\coveringcluster(t'')=C'$, and we add all such terminals to list $\lambda(C')$. From our discussion, after these updates, Invariants \ref{invariant: special clusters for regular vertices} and \ref{inv: consistent covering for primary} continue to hold. It is also easy to verify that Invariants \ref{invariant: non terminal covered by secondary} and \ref{invariant: terminal covered by H'} continue to hold at the end of the phase.

Recall that, from our discussion, the number of regular vertices in both $C'$ and $C''$ is at least $W_2$, and, from the definition of cluster budgets, we ensure that $\beta(C'),\beta(C'')\geq W_2$, and $\beta(C')\geq (N(C'))^{1+\eps'}$, while $\beta(C'')\geq (N(C''))^{1+\eps'}$. In order to establish Invariant \ref{inv: budgets}, it is enough to prove that the total budget $\beta$ does not increase.

Recall that Invariant \ref{inv: budgets} guaranteed that, at the beginning of the phase, $\beta(C)\geq (N')^{1+\eps'}$ held. Recall also that we have set $\beta(C')=N_i^{1+\eps'}$, and $\beta(C'')=(N'')^{1+\eps'}$. Therefore, it is enough to prove that:

\begin{equation}
\beta(C)\geq \beta(C')+\beta(C'').\label{ineq: budgets to prove}
%N_i^{1+\eps'}+(N'')^{1+\eps'}\leq (N')^{1+\eps'}. \label{ineq: budgets to prove}
\end{equation}

We consider two cases. The first case is when $N_i\leq \frac{N'}{4}$, so $N_i\leq N_{i-1}\cdot \eta_{i-1}^{\eps'/4}$ must hold. Notice that: 
$N''\leq N'-N_{i-1}$, and so:

 $$\beta(C'')\leq (N'-N_{i-1})^{1+\eps'}\leq (N'-N_{i-1})\cdot (N')^{\eps'}.$$ 
 
 At the same time, $N_i\leq N_{i-1}\cdot \eta_{i-1}^{\eps'/4}=N_{i-1}\cdot \frac{(N')^{\eps'/4}}{N_{i-1}^{\eps'/4}}$. Therefore:

\[\begin{split} 
\beta(C')& = N_i^{1+\eps'}\\
&\leq N_{i-1}^{(1-\eps'/4)\cdot (1+\eps')}\cdot (N')^{\eps'/4+(\eps')^2/4}\\
&\leq N_{i-1}^{1+3\eps'/4-(\eps')^2/4}\cdot (N')^{\eps'/4+(\eps')^2/4}\\
&\leq N_{i-1}\cdot (N')^{\eps'}.
 \end{split}\]

Therefore, altogether:

\[\begin{split}
\beta(C')+\beta(C'') &\leq (N'-N_{i-1})\cdot (N')^{\eps'}+N_{i-1}\cdot (N')^{\eps'}\leq (N')^{1+\eps'}\leq \beta(C). 
\end{split}\]

It remains to consider the second case, where $\frac {N'} 4\leq N_i\leq \frac {N'} 2$, and $N_i\leq N_{i-1}\cdot \left(1+\frac{(\eps')^2}{16}\right )$.

In this case:

 $$\beta(C'')\leq (N'-N_{i-1})^{1+\eps'}\leq (N'-N_{i-1})\cdot (N')^{\eps'},$$

as before. At the same time:

\[
\begin{split}
\beta(C')&=N_i^{1+\eps'}\\
&\leq N_{i-1}^{1+\eps'}\cdot  \left(1+\frac{(\eps')^2}{16}\right )^{1+\eps'}\\
&\leq N_{i-1}\left(\frac{N'}{2}\right )^{\eps'}\cdot \left(1+\frac{(\eps')^2}{16}\right )^{1+\eps'}.
%\\&\leq N_{i-1}\cdot (N')^{\eps'}.
\end{split}\]

(we have used the fact that $N_{i-1}\leq N'/2$). 
Recall that for all $a\geq 2$, $(1+1/a)^a\leq e$. Therefore:

%Recall that, for all  $\delta\in [0,1]$ and for all $0\leq a\leq 1$: $(1+\delta)^a\leq 1+\delta a$. Therefore:

\[\left(1+\frac{(\eps')^2}{16}\right )^{(1+\eps')/\eps'}\leq  \left(1+\frac{(\eps')^2}{16}\right )^{2/\eps'}\leq \left(1+\frac{(\eps')^2}{16}\right )^{(16/(\eps')^2)\cdot (\eps'/8)}\leq e^{\eps'/8}\leq 2.\]

We conclude that 
$\left(1+\frac{(\eps')^2}{16}\right )^{1+\eps'}\leq 2^{\eps'}$, and $\beta(C')\leq N_{i-1}\cdot (N')^{\eps'}$.

Altogether, we get that:

\[\beta(C')+\beta(C'')\leq N_{i-1}\cdot (N')^{\eps'}+(N'-N_{i-1})\cdot (N')^{\eps'}\leq (N')^{1+\eps'}\leq \beta(C). \]

We conclude that Invariant \ref{inv: budgets} continues to hold at the end of the phase.

This finishes the description of the second phase of the algorithm. We refer to the algorithm for the Flag Lowering Operation that we have described so far as \emph{basic algorithm}. Let $E'$ be the set of edges that were deleted from cluster $C$ over the course of Phase 1. If Phase 1 was successful, then the running time of the basic algorithm is bounded by $O(|E'|\cdot (W_1\cdot \mu)^{\eps'}\cdot \log (W_1\cdot \mu))$ (and we did not execute Phase 2). If Phase 1 was unsuccessful, then the running time of the basic algorithm is bounded by $O((W_1\cdot \mu)^{1+\eps'}\log (W_1\cdot \mu))$, and we have added a new cluster to set $\cset_1$. Recall that, if Phase 1 is successful, then the resulting sequence $\Sigma'$ of edge- and vertex-deletions is a valid flag-lowering sequence for $C$, that is, either $v$ or $v'$ are deleted from $C$ at the end of the sequence. If Phase 1 is unsuccessful, then the sequence $\Sigma'$ of edge- and vertex-deletions from cluster $C$ that we obtain at the end of the basic algorithm may not be a valid flag-lowering sequence, if $v$ and $v'$ both remain in cluster $C$ once the updates from $\Sigma'$ are applied to $C$. In this case, we repeat the basic algorithm, with the same cluster $C$ and pair $v,v'$ of vertices. We keep repeating the basic algorithm until either $v$ or $v'$ are deleted from $C$. If the basic algorithm is repeated $k+1$ times, and $E'$ is the set of edges that are eventually deleted from $C$ via the resulting flag-lowering sequence $\Sigma'$, then the total running time of the algorithm is bounded by 
$O(|E'|\cdot (W_1\cdot \mu)^{\eps'}\cdot \log (W_1\cdot \mu) + k(W_1\cdot \mu)^{1+\eps'}\log (W_1\cdot \mu))$. At the same time, at least $k$ new clusters were added to $\cset_1$ at the end of the Flag Lowering operation.

This completes the description of the Flag Lowering operation, and of the algorithm for the \maintainNC problem.

\subsubsection{Analysis}

Note that our algorithm maintains a collection $\cset$ of clusters of $H$. 
At the beginning of the algorithm, $\cset=\set{H}$ holds, and, as the algorithm progresses, clusters in $\cset$ only undergo allowed changes, as required. We argue that the algorithm has the Consistent Covering property. Indeed, if a regular vertex $v$ is ever deleted from the set $T$ of terminals, then we have shown already that the Consistent Covering property holds for that vertex throughout the entire algorithm. If vertex $v$ remains in set $T$ throughout the algorithm, then Consistent Covering property for that vertex follows from Invariants \ref{invariant: terminal covered by H'}, \ref{invariant: special clusters for regular vertices} and \ref{inv: consistent covering for primary}.

Next, we show that Properties \ref{restriction: few primary clusters}--\ref{restriction: regualr vertex in few secondary clusters} hold throughout the algorithm. Properties \ref{restriction: no updates for secondary clusters}, \ref{restricioin: secondary covering cluster} and \ref{restriction: few vertices in secondary clusters} are immediate from the description of the algorithm.
We now establish Property \ref{restriction: regualr vertex in few secondary clusters}.

Recall that our algorithm maintains a partition $(S_0,\ldots,S_r)$ of the set $V$ of regular vertices of $H$ into classes. For all $0\leq j\leq r$, we denote by $S_{\geq j}=S_j\cup\cdots\cup S_r$.
We will use the following observation in order to establish Property \ref{restriction: regualr vertex in few secondary clusters}. The proof is essentially identical to the arguments that were used in the proof of \Cref{claim: cutting G}, and is delayed to Section \ref{appx: secondary clusters} of Appendix.

\begin{observation}\label{obs: terminal growth2}
	For all $1\leq j\leq r$, $|S_{\geq j}|\leq N^{1-j\hat \eps}$ holds throughout the algorithm.
\end{observation}

Since $r= \ceil{1/\hat \eps}+1$, we conclude that set $S_r$ of vertices remains empty throughout the algorithm. Therefore, for every regular vertex $x\in V$, the number of clusters $C\in \cset_2$ to which $x$ ever belonged is bounded by:

\[ r\cdot N^{2\hat \eps}\leq \frac{2N^{2\hat \eps}}{\hat \eps}\leq \frac{64N^{\eps'/16}}{\eps'}\le W_2^{\eps'/4}.  \]

(we have used the fact that $\hat \eps=\eps'/32$, $N\leq W_1\leq W_2^{1.5}$, and $\frac{1}{(\log W_2)^{1/4}}\leq \eps'\leq 1/400$). This establishes Property \ref{restriction: regualr vertex in few secondary clusters}.

It remains to establish Properties \ref{restriction: few primary clusters} and \ref{restruction: few regular vertices in primary clusters}. Recall that, at the beginning of the algorithm, $\beta=\beta(H)=N^{1+\eps'}$. 
From Invariant \ref{inv: budgets}, as the algorithm progresses, the total budget $\beta=\sum_{C\in \cset_1}\beta(C)$ does not grow, and, for every cluster $C\in \cset_1$, $\beta(C)\geq W_2$ holds. Therefore, at the end of the algorithm:

\[|\cset_1|\leq \frac{\beta}{W_2}\leq \frac{N^{1+\eps'}}{W_2}\leq \frac{W_1^{1+\eps'}}{W_2}. \]

Since clusters may be added to set $\cset_1$, but never deleted from it, this establishes Requirement \ref{restriction: few primary clusters}.

It now remains to establish Requirement \ref{restruction: few regular vertices in primary clusters}. Consider any cluster $C\in \cset_1$, and recall that we have denoted by $N^0(C)$ the number of regular vertices in $C$ when $C$ was added to $\cset_1$. Recall that, when $C$ is added to $\cset_1$, we have set $\beta(C)=(N^0(C))^{1+\eps'}$. Denote by $\wset(C)=\set{C_1,C_2,\ldots,C_q}$ the clusters of $\cset_1$, that were created by our algorithm by applying the cluster-splitting operation to cluster $C$, and assume that they were added to $\cset_1$ in the order of their indices. For all $1\leq i\leq q$, denote $M_i=N^0(C_i)$. It is easy to verify from our algorithm that, if $M'_i$ denotes the number of regular vertices in cluster $C$ just before cluster $C_i$ was split from it, then $M_i\leq M'_i/2\leq N^0(C)/2$ holds. When cluster $C_i$ was split from $C$, we set $\beta(C_i)=M_i^{1+\eps'}$. Since, after each such cluster $C_i$ was split from $C$, the total budget $\beta$ did not increase, it is easy to verify that $\sum_{i=1}^{q}M_i^{1+\eps'}\leq (N^0(C))^{1+\eps'}$.

We construct a partitioning tree $\tset$, whose set of vertices is $\set{v(C)\mid C\in \cset_1}$ (where we consider the set $\cset_1$ of clusters at the end of the algorithm). Let $C^0$ be the cluster of that was added to $\cset_1$ first (that is, when $C^0$ was added to $\cset_1$, $C^0=H$ held, but subsequently vertices and edges may have been deleted from $C^0$). The root of the tree is vertex $v(C^0)$, and, for every pair of clusters $C,C'\in \cset_1$ with $C'\in\wset(C)$, vertex $v(C')$ becomes the child vertex of $v(C)$ in $\tset$.
For every vertex $v(C)$, the weight of the vertex is defined to be $w(C)=(N^0(C))^{1+\eps'}$.

From the above discussion, if $C'\in \wset(C)$, then $N^0(C')\leq N^0(C)/2$, so the depth of the tree $T$ is bounded by $d=\ceil{\log W_1}$. For $0\leq i\leq d$, we let $\cset^i\subseteq \cset_1$ denote the set of all clusters $C$, for which the distance from $v(C)$ to $v(C^0)$ in the tree is exactly $i$. From our discussion so far, for every cluster $C\in \cset$, $\sum_{C'\in \wset(C)}w(C')\leq w(C)$, and $w(C^0)\leq W_1^{1+\eps'}$. It follows that, for all $0\leq i\leq d$, $\sum_{C\in \cset^i}w(C)\leq W_1^{1+\eps'}$, and so $\sum_{C\in \cset_1}w(C)=\sum_{i=0}^d\sum_{C\in \cset^i}w(C)\leq W_1^{1+\eps'}\cdot 4\log W_1\leq W_1^{1+2\eps'}$.

Note that, for every cluster $C\in \cset_1$, $w(C)\geq N^0(C)$. Therefore, overall, $\sum_{C\in \cset_1}N^0(C)\leq W_1^{1+2\eps'}$.
This establishes Requirement \ref{restruction: few regular vertices in primary clusters}.

It now remains to bound the running time of the algorithm. Assume that the algorithm for the Flag-Lowering operation was executed $q$ times. For all $1\leq i\leq q$, we denote by $C_i$ the cluster that was processed during the $i$th execution of the Flag-Lower operation, by $m_i$ the number of edges that were deleted from $C_i$ during the operation, and by $k_i$ the total number of clusters that were added to set $\cset_1$ over the course of the operation. Then the running time of the $i$th Flag-Lowering operation is bounded by $O(|E_i|\cdot (W_1\cdot \mu)^{\eps'}\cdot \log (W_1\cdot \mu) + k_i(W_1\cdot \mu)^{1+\eps'}\log (W_1\cdot \mu))$. The total running time of all Flag-Lowering operations is bounded by:

\[ O\left(\sum_{i=1}^q |E_i|\cdot (W_1\cdot \mu)^{\eps'}\cdot \log (W_1\cdot \mu)\right )+O\left(\sum_{i=1}^q k_i(W_1\cdot \mu)^{1+\eps'}\log (W_1\cdot \mu)\right ) \]

Since the number of edges in every cluster $C\in \cset_1$ is bounded by $W_1\mu$, and, from Property \ref{restriction: few primary clusters} $|\cset_1|\leq \frac{W_1^{1+\eps'}}{W_2}$, we get that the first summand is bounded by: $O\left(\frac{W_1^{2+2\eps'}\cdot \mu^{1+2\eps'}}{W_2}\cdot \log (W_1\cdot \mu)\right )$. Since $\sum_{i=1}^q k_i\leq |\cset_1|\leq \frac{W_1^{1+\eps'}}{W_2}$, the second summand is also bounded by $O\left(\frac{W_1^{2+2\eps'}\cdot \mu^{\eps'}}{W_2}\cdot \log (W_1\cdot \mu)\right )$. Therefore, the total time that all Flag Lowering operations take is bounded by $O\left(\frac{W_1^{2+2\eps'}\cdot \mu^{1+2\eps'}}{W_2}\cdot \log (W_1\cdot \mu)\right )\leq O\left(\frac{W_1^{2+3\eps'}\cdot \mu^{2}}{W_2}\right )$ (we have used the fact that $\frac{1}{(\log W_2)^{1/4}}\leq \eps'\leq 1/400$).

Note that the total number of edges that are ever present in graph $H$ over the course of the algorithm is bounded by $W_1\cdot \mu$. When an edge is deleted from $H$, or inserted into $H$ (via supernode-splitting), we may need to perform the same update in a a number of clusters in $\cset_1$. Since $|\cset_1|\leq \frac{W_1^{1+\eps'}}{W_2}$, the total time that is required in order to maintain the clusters of $\cset_1$ (excluding the running time of the Flag Lowering operations) is bounded by  $O\left(\frac{W_1^{2+\eps'}\cdot \mu}{W_2}\right )$.

Overall, the total update time of the algorithm is bounded by $O\left(\frac{W_1^{2+3\eps'}\cdot \mu^{2}}{W_2}\right )$.

\section{Algorithm for the \maintaincluster Problem - Proof of \Cref{thm: main maintain cluster algorithm}}
\label{sec: balanced pseudocut}

For convenience, we will denote the graph $C$ that is given as input by $H$. Therefore,
throughout this section, we assume that we are given a valid input structure $\iset=\left(H=(V,U,E),\set{\ell(e)}_{e\in E(H)},D\right )$, where $H$ is a connected graph, that serves as input to the \maintaincluster problem. 
Graph $H$ undergoes an online sequence $\Sigma$ of valid update operations, and we are given its dynamic degree bound $\mu$. Additionally, we are given a parameter $\hat W\geq N^0(H)$, where $N^0(H)$ is the number of regular vertices of $H$ at the beginning of the algorithm.
Lastly, we are given a precision parameter  $\frac{1}{(\log W)^{1/24}}\leq \eps\leq 1/400$, where $1/\eps$ is an integer, and a distance parameter $D^*\geq D$, that is an integral power of $2$.
Our goal is to design an algorithm that supports queries $\spquery(H,v,v')$: given a pair $v,v'\in V(H)$ of regular vertices of $H$, return a path $P$ of length at most $\alpha'\cdot D^*$ connecting them in $H$, in time $O(|E(P)|)$. The algorithm may, at any time, raise a flag $F_H$, at which time it must  supply a pair  $\hat v,\hat v'$ of regular vertices of $H$ (called a witness pair), with $\dist_H(\hat v,\hat v')> D^*$.
Once flag $F_H$ is raised, the algorithm will obtain, as part of its input update sequence $\Sigma$, a valid flag-lowering sequence $\Sigma'$. Flag $F_H$ is lowered after the updates from $\Sigma'$ are processed by the algorithm. Queries $\spquery$ may only be asked when flag $F_H$ is down.
We will assume in the remainder of the proof that $\mu<\hat W^{1/10}$ holds, since otherwise we can simply use the algorithm from \Cref{thm: maintain cluster algorithm2}.

Throughout this subsection, we also assume that, for some parameter $W<\hat W$ and $c'=3$, Assumption \ref{assumption: alg for recdynnc} holds. We denote the approximation factor from \Cref{assumption: alg for recdynnc} by $\alpha=\alpha(W)$. Recall that the desired approximation factor $\alpha'$ for the \maintaincluster problem is $\alpha'=\max\set{2^{O(1/\eps^6)},(8\alpha)^{2}}$.
Throughout the algorithm, we use two parameters: $\rho=\frac{4\hat W}{W}$ and $\hat D=\frac{2^{14}\cdot D^*}{\eps^2}$. From the statement of \Cref{thm: main maintain cluster algorithm}, $\rho\leq 4\sqrt{W}$.

\iffalse
Note that, since $\frac{1}{(\log W)^{1/24}}\leq \eps\leq 1/400$, we get that:

\begin{equation}\label{eq: large W}
W>2^{1/\eps^{24}}\geq 2^{400^{24}},
\end{equation}

so in particular:

\begin{equation}\label{eq: large W 2}
W^{\eps}>2^{1/\eps^{23}}.
\end{equation}

Since, from \Cref{eq: large W}, parameter $W$ must be sufficiently large, we get that:

\[ \frac{1}{\eps^{12}}\leq (\log W)^{1/2}<\frac{\log W}{2\log\log W}, \]

and so:

\begin{equation}\label{eq: large W3}
W^{\eps^{12}}>\log^2 W.
\end{equation}
\fi

As in \cite{APSP-previous}, the main tool that our algorithm uses is  a \emph{balanced pseudocut}. This notion was defined in \cite{APSP-previous}; we use  a very similar but slightly different definition. This modified definition allows us to somewhat simplify parts of the algorithm. Additionally, we will replace expanders, that were used in the construction of \cite{APSP-previous}, with well-connected graphs. This allows us to improve the approximation factor that our algorithm achieves.

We start by defining balanced pseudocuts, and introducing some tools for them.

\subsection{Balanced Pseudocut}

The notion of a balanced pseudocut, that can be viewed as a generalization of  the notion of a balanced cut, was first introduced in \cite{APSP-previous}, and it is also a central notion that we use in our algorithm. 
Intuitively, following a rather standard definition, for a given balance parameter $\rho$, a \emph{balanced multicut} in a graph $G$ can be defined as  a subset $T\subseteq V(G)$ of vertices of $G$, such that every connected component of $G\setminus T$ contains at most $|V(G)|/\rho$ vertices. We use instead a notion of \emph{balanced pseudocuts}, that can be viewed  as a relaxation of the notion of a balanced multicut. One advantage for using this notion is that, as was shown in \cite{APSP-previous}, there is an efficient algorithm that, given a graph $G$, computes a balanced pseudocut $T$ in $G$, and also embeds an expander $X$ into $G$, where $V(X)\subseteq T$, and $V(X)$ contains a large fraction of vertices of $T$. We slightly modify the definition of balanced pseudocut from \cite{APSP-previous}. We will exploit the idea of \cite{APSP-previous} of embedding an expander defined over a large subset of vertices of the pseudocut, but we will replace the expander with a well-connected graph. 
We now formally define a pseudocut.

\begin{definition}[Pseudocut]
	Let $\iset=\left(H=(V,U,E),\set{\ell(e)}_{e\in E(H)},D\right )$ be a valid input structure, let $\hat W$ be a parameter, such that the total number of regular vertices in $H$ is at most $\hat W$, let $\hat D>D$ be a distance parameter, and let $\rho>1$ be a balance parameter. A subset $T\subseteq V$ of regular vertices of $H$ is a $(\hat D,\rho)$-\emph{pseudocut}, if, for every regular vertex $x\in V\setminus T$, the total number of regular vertices lying in $B_{H\setminus T}(x,\hat D)$ is at most $\frac{\hat W}{\rho}$. %For convenience, we denote $\rho(W,\epsx)=W^{\epsx^2/(\cpseudo\log\log W)}$.
\end{definition}

Note that, if $T$ is a $(\hat D,\rho)$-pseudocut for $H$, then  every set $T'$ of regular vertices of $H$ with $T\subseteq T'$ is also a $(\hat D,\rho)$-pseudocut for $H$. Note also that, as graph $H$ undergoes a sequence of valid update operations, set $T\cap V$ remains a valid $(\hat D,\rho)$-pseudocut. This is since valid update operations may not insert regular vertices into $H$, and since, from \Cref{obs: no dist increase}, distances in graph $H$ may only increase as it undergoes valid update operations.

As mentioned earlier, we use a subroutine that, given a valid input structure $\iset=\left(H,\set{\ell(e)}_{e\in E(H)},D\right )$, together with parameters $\rho>1$ and $\hat D>D$, computes a $(\hat D,\rho)$-pseudocut $T$ in $H$, and embeds a well-connected graph $X$ with $V(X)$ containing a large fraction of vertices of $T$, into $H$. This subroutine is implemented as follows: we start with an arbitrary pseudocut $T$ in $H$ (for example, it may contain all regular vertices of $H$), and then iterate. In every iteration, we will either embed a well-connected graph $X$ defined over a large enough subset of vertices of $T$ into $H$ as desired, or we will compute another $(\hat D,\rho)$-pseudocut $T'$ in $H$, whose cardinality is significantly lower than the cardinality of $T$. We can think of $T'$ as an ``improved'' pseudocut. We then replace $T$ with $T'$ and continue to the next iteration. In order to implement this idea, we will use the following observation that, under some conditions, allows us to improve a given pseudocut $T$.

\begin{observation}\label{obs: improving pseudocut}
		Let $\iset=\left(H=(V,U,E),\set{\ell(e)}_{e\in E(H)},D\right )$ be a valid input structure, let $\hat W$ be a parameter, such that the total number of regular vertices in $H$ is at most $\hat W$, and let $S$ be a  $(\hat D,\rho)$-pseudocut for $H$. Suppose we are given a partition $(S',S'')$ of $S$ into two subsets, such that, if we denote by $H'=H\setminus S''$, then the number of regular vertices in $B_{H'}(S',2\hat D)$ is at most $\hat W/\rho$. Then $S''$ is a valid  $(\hat D,\rho)$-pseudocut for $H$.
\end{observation}
\begin{proof}
	Let $H''=H\setminus S$. From the definition of a pseudocut, for every regular vertex $x\in V(H'')$, the number of regular vertices in $B_{H''}(x,\hat D)$ is at most $\hat W/\rho$.
	
	Assume now for contradiction that $S''$ is not a valid $(\hat D,\rho)$-pseudocut for $H$. Then there is some regular vertex $x\in V(H')$, such that $B_{H'}(x,\hat D)$ contains more than $\hat W/\rho$ regular vertices. 
	
	We consider two cases. The first case happens if $B_{H'}(x,\hat D)$ contains at least one vertex of $S'$. If this is the case, then $B_{H'}(x,\hat D)\subseteq B_{H'}(S',2\hat D)$, and so $B_{H'}(x,\hat D)$ may contain at most $\hat W/\rho$ regular vertices, a contradiction.
	
	Consider now the second case, where $B_{H'}(x,\hat D)$ contains no vertices of $S'$. Then deleting the vertices of $S'$ from $H'$ does not affect the subgraph of $H'$ induced by  $B_{H'}(x,\hat D)$. In other words,  $B_{H'}(x,\hat D)=B_{H''}(x,\hat D)$. But from the definition of a pseudocut, $B_{H''}(x,\hat D)$ contains at most $\hat W/\rho$ regular vertices, a contradiction.
\end{proof}

We now turn to the algorithm for the \maintaincluster problem, and the proof of \Cref{thm: main maintain cluster algorithm}. Recall that we denoted $\rho=\frac{4\hat W}{W}$.
Throughout the algorithm, we use a parameter $\hat D=\frac{2^{14}\cdot D^*}{\eps^2}$.
 Our algorithm will maintain a $(\hat D,\rho)$-pseudocut $T$ for graph $H$, that may change overtime. The algorithm is partitioned into three stages. The first stage continues as long as $|T|>\max\set{\frac{W}{4\mu^2},64W^{\eps}\rho\mu^2}$ holds, and it is partitioned into phases. At the beginning of every phase, we compute a pseudocut $T$, a well-connected graph $X$ defined over a large subset of vertices $T$, and an embedding of $X$ into $H$, so that the paths of the embedding are short, and the embedding causes a relatively low congestion. We then employ the algorithm for \APSP on well-connected graphs from \Cref{thm: APSP in HSS full}. Recall that the algorithm maintains a large decremental set $S'\subseteq V(X)$ of vertices, and supports \shortpath\ queries in graph $X$ between pairs of vertices $x,y\in S'$. We will also maintain an \EST in graph $H$, that is rooted at the vertices of $S'$, and has depth $\Theta(D^*)$. These data structures are sufficient in order to support $\spquery$ queries between pairs of regular vertices in $H$. Additionally, whenever some vertex leaves the \EST, we can obtain the desired pair $x,y$ of vertices of $H$ that are sufficiently far from each other. Intuitively, the number of edge deletions from $H$ that the algorithm for \APSP on well-connected graphs can accommodate is roughly comparable to $|T|^{1-O(\eps)}$. On the other hand, the algorithm that we employ at the beginning of the phase, in order to compute the pseudocut $T$, the well-connected graph $X$, and its embedding into $H$, has running time that is roughly bounded by $(\hat W\cdot \mu)^{1+O(\eps)}$. Therefore, as long as $|T|$ is sufficiently large, we can maintain the data structures efficiently. Once $|T|\leq \max\set{\frac{W}{4\mu^2},64W^{\eps}\rho\mu^2}$ holds, the first stage of the algorithm terminates. Over the course of the remainder of the algorithm, the pseudocut $T$ remains unchanged (except when vertices of $T$ are deleted from $H$ -- all such vertices are deleted from $T$ as well).
 The second stage of the algorithm is only executed if $64W^{\eps}\rho\mu^2>\frac{W}{4\mu^2}$, and it only lasts as long as $T\cap V(H)>\frac{W}{4\mu^2}$. We partition the second stage into phases as well. At the beginning of every phase, we select an arbitrary vertex $t\in T\cap V(H)$, and initialize an \EST data structure in graph $H$, with source vertex $t$, and depth parameter $\hat D$. This data structure will allow us to support \shortpath\ queries, and to identify regular vertices of $H$ whose distance from $t$ becomes too large. Once vertex $t$ is deleted from $H$, the phase ends and a new phase begins. Note that the number of phases in the second stage is bounded by $64W^{\eps}\rho\mu^2$, and every phase takes time $O(\hat W\cdot\mu\cdot  \hat D\cdot \log \hat W)$. Once $|T|\leq \frac{W}{4\mu^2}$ holds, the second stage terminates.
   We employ a different strategy for stage $3$, which is similar to the one employed by \cite{APSP-previous}. From the definition of a pseudocut, in graph $H\setminus T$, for every regular vertex $v$, $B_{H\setminus T}(v,\hat D)$ contains at most $W/4$ regular vertices. Intuitively, we compute an initial neighborhood cover $\cset$ in graph $H\setminus T$, with distance parameter $D^*$, and we ensure that the diameter of every cluster $C\in \cset$ is at most $\hat D$. This guarantees that every cluster $C\in \cset$ contains fewer than $W/4$ regular vertices of $H$. We can now employ the algorithm from \Cref{assumption: alg for recdynnc} in order to maintain, for every cluster $C\in \cset$ and distance scale $D_i\leq \hat D$, a neighborhood cover $\wset_i^C$. We construct a contracted graph $\hat H$, with the set $T$ of regular vertices, and supernodes corresponding to the clusters of $\bigcup_{C\in \cset}\bigcup_i\wset_i^C$. We show that distances between vertices of $T$ in $\hat H$ are approximately equal to those in $H$. We also design an algorithm for maintaining graph $\hat H$ via a sequence valid update operations. Since $|T|$ is sufficiently small, we can apply the algorithm from \Cref{assumption: alg for recdynnc} to compute and maintain a neighborhood cover for the contracted graph $\hat H$. This, in turn, allows us to respond to \shortpath\ queries between pairs of vertices in $T$, and to correctly establish when a pair of such vertices become too far from each other in $H$. We also maintain an \EST data structure in graph $H$, rooted at the set $T$ of vertices, that allows us to support \shortpath\ queries between pairs of regular vertices of $H$, and to correctly establish when any regular vertex of $H$ becomes too far from the vertices of $T$. Once all vertices of $T$ are deleted from $H$, we will exploit the clusters in $\set{\wset_i^C}$ for $C\in \cset$ and $i=\log D^*$, in order to continue to support \shortpath\ queries.
In the following three subsections, we provide our algorithm for each of the three stages.
\subsection{Stage 1: Large Pseudocut Regime}
\label{subsec: stage1: large pseudocut}

In this subsection we provide the first part of the algorithm for the \maintaincluster problem, that deals with the large-pseudocut setting.

Recall that we are given a valid input structure $\iset=\left(H,\set{\ell(e)}_{e\in E(H)},D\right )$, where $H$ is a connected graph, that serves as input to the \maintaincluster problem. 
Graph $H$ undergoes an online sequence $\Sigma$ of valid update operations, and we are given its dynamic degree bound $\mu$. Additionally, we are given a parameter $\hat W\geq N^0(H)$, where $N^0(H)$ is the number of regular vertices of $H$ at the beginning of the algorithm. We also use the parameter $W$ from \Cref{assumption: alg for recdynnc}, and a parameter $\rho=\frac{4\hat W}{W}$. Recall that, from the statement of \Cref{thm: main maintain cluster algorithm}, $\rho\leq \sqrt W$.

Throughout the algorithm, we will maintain a $(\hat D,\rho)$-pseudocut $T$ in graph $H$, that may change overtime. At the beginning of the algorithm, we let $T$ contain all regular vertices of $H$, so it is clearly a valid $(\hat D,\rho)$-pseudocut. As the algorithm progresses, we will compute new pseudocuts, and we will ensure that the cardinality of the pseudocut that we maintain never increases. Once we obtain a pseudocut $T$ with $|T|\leq \max\set{\frac{W}{4\mu^2},64W^{\eps}\rho\mu^2}$, the first stage of the algorithm terminates. Following is a key theorem, that allows us to compute a pseudocut $T'$ in graph $H$, and, if $|T'|>\max\set{\frac{W}{4\mu^2},64W^{\eps}\rho\mu^2}$, it also allows us to compute a well-connected graph $X$ defined over a large subset of the vertices of $T'$, together with a \HSS for $X$, and an embedding of $X$ into $H$ via a set of short paths, that cause a relatively small congestion. We will use this algorithm repeatedly. This algorithm is our main technical contribution to the improved algorithm for the \maintaincluster problem.

\begin{theorem}\label{thm: new pseoducut w well connected outer}
	There is a deterministic algorithm, whose input consists of:

\begin{itemize}
	\item a valid input structure $\iset=\left(H,\set{\ell(e)}_{e\in E(H)},D\right )$, such that the maximum vertex degree in $H$ is at most $\mu$;
	
	\item parameters $W<\hat W\leq W^{1.5}$, such that, if $N(H)$ is the number of regular vertices in $H$, then $N(H)\leq \hat W$; % and $\mu\leq W^{\eps/2}$;
	
	\item distance parameter $\hat D>D$, and precision parameter $\frac{1}{(\log W)^{1/24}}\leq \eps\leq 1/400$, such that $1/\eps$ is an integer; and
	
	\item a $(\hat D,\rho)$-pseudocut $T$ in $H$, with $|T|>\max\set{\frac{W}{4\mu^2},64W^{\eps}\rho\mu^2}$, where $\rho=\frac{4\hat W}{W}$.
\end{itemize}

The output of the algorithm is a $(\hat D,\rho)$-pseudocut $T'$ in $H$, with $|T'|\leq |T|$. Moreover, if $|T'|>\max\set{\frac{W}{4\mu^2},64W^{\eps}\rho\mu^2}$, then, additionally, the algorithm computes 
a graph $X$ with maximum  vertex degree at most   $\hat W^{32\eps^3}$ and $V(X)\subseteq T$, such that $|V(X)|=N^{1/\eps}$ for some parameter  $N\geq \max\set{\frac{W^{\eps}}{128\rho^{\eps}\mu^{2\eps}},\frac{W^{\eps^2}}{32}}$. In the latter case, the algorithm also returns an embedding $\pset$ of $X$ into $H$ via paths of length at most $D''=\frac{512 \hat D}{\eps^2}$, that cause congestion at most $\eta=2^{20}\hat W^{64\eps^3}\rho^2D''$, and a level-$(1/\eps)$ \HSS for $X$, such that $X$ is $(\eta',\td)$-well-connected with respect to the set $S(X)$ of vertices defined by the support structure, where $\eta'=N^{6+256\eps}$, and $\td=2^{\tilde c/ \eps^5}$, with $\tilde c$ being the constant used in the definition of the \HSS.
The running time of the algorithm is bounded by $O\left (\hat W^{1+O(\eps)} \cdot \hat D^2\cdot \rho^4\cdot \mu\right )$.
\end{theorem}

The proof of the theorem is somewhat technical, and we delay it to Section \ref{subsec: inner pseudocut or well connected}, after we complete the description of Stage 1 of the algorithm using it.

Throughout the algorithm, we maintain a $(\hat D,\rho)$-pseudocut $T$ for graph $H$. Over the course of the algorithm, the cardinality of the pseudocut that we maintain may only decrease. Once we obtain a pseudocut $T$ whose cardinality is at most $\max\set{\frac{W}{4\mu^2},64W^{\eps}\rho\mu^2}$, we terminate Stage 1 of the algorithm.
At the beginning of the algorithm, we compute an initial $(\hat D,\rho)$-pseudocut $T$, that contains all regular vertices of the input graph $H$. 
Let $\Lambda'=\frac{W^{1-20\eps}}{\hat D\rho^3\mu^2}$. 
	
The execution of the algorithm is partitioned into phases. At the beginning of the algorithm, the first phase starts.
Each phase continues as long as the number of edges that were deleted from graph $H$ over the course of the phase is at most $\Lambda'$. Once $\Lambda'$ edges are deleted from $H$ since the start of the phase, a new phase commences. Since the total number of edges that are ever present in graph $H$ is bounded by $N^0(H)\cdot \mu\leq \hat W\cdot \mu\leq \frac{W\rho\mu}4$, we get that the number of phases in the algorithm is bounded by:

\[\frac{W\rho\mu}{4\Lambda'}\leq W^{20\eps}\cdot \rho^4\cdot \mu^3\cdot \hat D.  \]

We now describe the execution of a single phase.

\subsubsection{Execution of a Single Phase}

At the beginning of a phase, we are given a set $T$ of regular vertices, with $|T|>\max\set{\frac{W}{4\mu^2},64W^{\eps}\rho\mu^2}$, that is a $(\hat D,\rho)$-pseudocut for graph $H$; at the beginning of the first phase, $T$ is the set of all regular vertices of $H$. At the beginning of a phase, we apply the algorithm from \Cref{thm: new pseoducut w well connected outer} to the current graph $H$ and pseudocut $T$; parameters $\hat D,W,\hat W$ and $\eps$ remain unchanged. 
Recall that the running time of the algorithm is $O\left (\hat W^{1+O(\eps)} \cdot \hat D^2\cdot \rho^4\cdot \mu\right )$.

If the algorithm returns a $(\hat D,\rho)$-pseudocut $T'$ with $|T'|\leq \max\set{\frac{W}{4\mu^2},64W^{\eps}\rho\mu^2}$, then we terminate Stage 1, and continue to Stage 2 of the algorithm.

Therefore, we assume from now on, that the algorithm returned a 
$(\hat D,\rho)$-pseudocut $T'$ in $H$, with $|T'|>\max\set{\frac{W}{4\mu^2},64W^{\eps}\rho\mu^2}$, together with a graph $X$ with maximum  vertex degree at most   $\hat W^{32\eps^3}$, and $V(X)\subseteq T$, such that $|V(X)|=N^{1/\eps}$ for some parameter $N\geq \max\set{\frac{W^{\eps}}{128\rho^{\eps}\mu^{2\eps}},\frac{W^{\eps^2}}{32}}$
 Notice that $|V(X)|=N^{1/\eps}\geq \frac{W}{2^{7/\eps}\cdot \rho\cdot \mu^2}$. The algorithm also returns an embedding $\pset$ of $X$ into $H$ via paths of length at most $D''=\frac{512 \hat D}{\eps^2}$, that cause congestion at most $\eta=2^{20}\hat W^{64\eps^3}\rho^2D''$, and a level-$(1/\eps)$ \HSS for $X$, such that $X$ is $(\eta',\td)$-well-connected with respect to the set $S(X)$ of vertices defined by the support structure, where $\eta'=N^{6+256\eps}$, and $\td=2^{\tilde c/ \eps^5}$, with $\tilde c$ being the constant used in the definition of the \HSS.
We now describe the data structures that the algorithm for a single phase maintains.

\subsubsection*{Data Structures}

For every edge $e\in E(X)$, we denote by $P(e)\in \pset$ the path that serves as the embedding path of edge $e$. For each edge $e'\in E(H)$, we initialize the set $S(e')$ of all edges $e\in E(X)$, such that $e'\in P(e)$. Since the paths in $\pset$ cause congestion at most 
$\eta$, for each edge $e'\in E(H)$, $|S(e')|\leq \eta$ holds. As updates from the input sequence $\Sigma$ of valid update operations are applied to the input graph $H$, whenever an edge $e'\in E(H)$ is deleted from $H$, we consider every edge $e\in S(e')$. Each such edge $e$ is deleted from graph $X$, and we also delete $e$ from the lists $S(e'')$ of every edge $e''\in P(e)$. Since, for every edge $e'\in E(H)$, an edge $e\in E(X)$ may only be added to set $S(e')$ at the beginning of the algorithm, and afterwards it may only be deleted from $S(e')$ once, the total time that is required in order to initialize and maintain the sets $\set{S(e')\mid e'\in E(H)}$ of edges is asymptotically bounded by the time that was required in order to compute the embedding $\pset$ of $X$ into $H$, which, in turn, is bounded by the running time of the algorithm from \Cref{thm: new pseoducut w well connected outer}.

Recall that, over the course of a single phase, graph $H$ may undergo at most $\Lambda'=\frac{W^{1-20\eps}}{\hat D\rho^3\mu^2}$ edge deletions, and, for every edge $e'\in E(H)$: 

$$|S(e')|\leq \eta=2^{20}\hat W^{64\eps^3}\rho^2D''\leq \frac{2^{29}}{\eps}\cdot \hat D\cdot \hat W^{64\eps^3}\rho^2\leq W^{2\eps}\hat D\rho^2,$$

since  $\hat W\leq W^{1.5}$, $D''=\frac{512 \hat D}{\eps^2}$, and $\hat W^{\eps}\geq W^{\eps}>2^{8/\eps}$ from inequality \ref{eq: large W 2}.

Therefore, the deletion of a single edge
from $H$ may result in the deletion of at most $W^{2\eps}\cdot \hat D\cdot \rho^2$ edges from $X$. Overall, the number of edges that may be deleted from $X$ over the course of a single phase is bounded by:

\[\Lambda=\Lambda'\cdot W^{2\eps}\cdot \hat D\cdot \rho^2\le  \frac{W^{1-20\eps}}{\hat D\rho^3\mu^2}\cdot W^{2\eps}\cdot \hat D\cdot \rho^2\leq  \frac{W^{1-18\eps}}{\rho\mu^2}. \]

(since $\Lambda'=\frac{W^{1-20\eps}}{\hat D\rho^3\mu^2}$.)

Recall that $|V(X)|\geq \frac{W}{2^{7/\eps}\cdot \rho\cdot \mu^2}\geq \frac{W^{1-\eps}}{\rho\cdot \mu^2}$. It is now easy to verify that $\Lambda\leq |V(X)|^{1-10\eps}$.

We apply the algorithm for \APSP in well-connected graphs from \Cref{thm: APSP in HSS full} to graph $X$, with parameters $N$ and $\eps$ remaining unchanged, and the \HSS for graph $X$ that we have computed. In order to be able to use the theorem, we need to verify that $\frac{N^{\eps^4}}{\log N}\geq 2^{128/\eps^6}$ holds.

Recall that $N\geq \max\set{\frac{W^{\eps}}{128\rho^{\eps}\mu^{2\eps}},\frac{W^{\eps^2}}{32}}$. Also, since $N\leq |V(X)|^{\eps}\leq \hat W^{\eps}\leq W^{1.5\eps}$, we get that $\log N\leq 1.5\eps\cdot \log W\leq \log W$. Therefore:

\[\frac{N^{\eps^4}}{\log N}\geq \frac{W^{\eps^6}}{32\log W}\geq \frac{W^{\eps^7}}{\log W}\geq W^{\eps^{8}} \geq 2^{1/\eps^{10}}\geq 2^{128/\eps^6}\] 

(we have used inequalities \ref{eq: large W3} and  \ref{eq: large W 2}).
We can now apply the algorithm from \Cref{thm: APSP in HSS full} to graph $X$, the \HSS for graph $X$ that we have computed, and parameters $N$ and $\eps$ that remain unchanged. As observed already, over the course of a single phase, graph $X$ undergoes a sequence of  $\Lambda\leq |V(X)|^{1-10\eps}$ edge deletions. Recall that the algorithm from  \Cref{thm: APSP in HSS full} maintains, over the course of the phase, a non-empty set $S'(X)$ of vertices of $X$, that is decremental: that is, after the initialization, vertices may leave $S'(X)$ but they may not join it.  The algorithm supports \shortpath\  queries between vertices of $S'(X)$: given a pair $x,y\in S'(X)$ of such vertices, return a path $P$ connecting $x$ to $y$ in the current graph $X$, whose length is at most $2^{O(1/\eps^6)}$, in time $O(|E(P)|)$. 
The total update time of the algorithm is $O( |E(X)|^{1+O(\eps)})$. 
Since the maximum vertex degree in $X$ is bounded by $\hat W^{32\eps^3}$, and $|V(X)|\leq \hat W$, we get that the total update time of the algorithm from \Cref{thm: APSP in HSS full}  is bounded by $O(\hat W^{1+O(\eps)})$.

The last data structure that we maintain is a modified \EST in graph $H$, rooted at the set $S'(X)$ of vertices. More precisely, we construct a graph $H'$, that is obtained from graph $H$ by adding a source vertex $s$ to $H$, and connecting it to every vertex of $S'(X)$ with an edge of length $1$. Once graph $H'$ is initialized, whenever graph $H$ is updated with a valid update operation $\sigma\in \Sigma$, we apply the same update operation to graph $H'$. Since a supernode-splitting operation is a special case of a vertex-splitting operation, graph $H'$ undergoes an online sequence of edge-deletion, isolated vertex-deletion, and vertex-splitting updates. We apply the algorithm from \Cref{thm: ES-tree} to maintain a modified \EST in graph $H'$, with the source vertex $s$, and a bound $2\hat D$ on the tree depth. The running time of the algorithm is $O(m^*\cdot \hat D\cdot \log m^*)$, where $m^*$ is the total number of edges that ever belonged to graph $H'$. It is easy to verify that $m^*\leq O(\hat W\cdot \mu)\leq O(\hat W^{1.1})$, and so the running time of the algorithm is bounded by $O(\hat W\cdot \mu\cdot \hat D\cdot \log W)$.
Recall that the algorithm also maintains a set $A$ of vertices of $H$, such that, for all $v\in A$, $\dist_{H'}(v,s)>2\hat D$. Whenever a new regular vertex $v$ is added to set $A$, we let $x$ be any vertex of $S'(X)$. Note that $\dist_H(x,v)>\hat D$ must hold. We raise the flag $F_{H}$, and we provide the pair $x,v$ of vertices as a witness.

We now bound the total running time of a single phase. 
Recall that the running time of the algorithm from \Cref{thm: new pseoducut w well connected outer}  is $O\left (\hat W^{1+O(\eps)} \cdot \hat D^2\cdot \rho^4\cdot \mu\right )$, the total update time of the algorithm from \Cref{thm: APSP in HSS full}  is bounded by $\hat W^{1+O(\eps)}$, and the running time of the algorithm from \Cref{thm: ES-tree} is bounded by $O(\hat W\mu\cdot \hat D\cdot \log W)\leq O(\hat W^{1+O(\eps)}\cdot \hat D\cdot \mu)$ (since  $\log W\leq W^{\eps^{12}}\leq \hat W^{O(\eps)}$ from Inequality \ref{eq: large W3}).
Therefore, the total running time of the algorithm for a single phase is at most $O\left (\hat W^{1+O(\eps)} \cdot \hat D^2\cdot \rho^4\cdot \mu\right )$. Since the number of phases in Stage 1 is bounded by $ W^{20\eps}\cdot \rho^4\cdot \hat D\cdot \mu^3$, we get that the total running time 
of Stage 1 is bounded by $O\left (\hat W^{1+O(\eps)} \cdot \hat D^3\cdot \rho^8\cdot \mu^4\right )$.

\subsubsection*{Responding to Queries}

We now provide an algorithm for supporting queries $\spquery(H,v,v')$ over the course of Stage 1. Suppose we are given a query  $\spquery(H,v,v')$, where $v,v'\in V(H)$ are regular vertices of $H$. Recall that our goal is to return a path $P$ of length at most $\alpha'\cdot D^*$ connecting $v$ to $v'$ in $H$, in time $O(|E(P)|)$. 

Recall that a $\spquery$ cannot be asked while the flag $F_H$ is up. Therefore, flag $F_H$ is currently down, and the set $A$ of vertices that the algorithm from  \Cref{thm: ES-tree} maintains contains no regular vertices. Vertices $v,v'$ must then lie in the \EST that the algorithm maintain. We perform query $\shortestpath(v)$ and $\shortestpath(v')$ in the data structure maintained by the algorithm from  \Cref{thm: ES-tree}, that must return a path $P_1$ connecting $v$ to $s$ in $H'$, and a path $P_2$ connecting $v'$ to $S$ in $H'$ of length at most $2\hat D$ each, in time $O(|E(P_1)|+|E(P_2)|)$. Let $s_1$ be the penultimate vertex on path $P_1$, and let $s_2$ be defined similarly for $P_2$. From the construction of graph $H'$, $s_1,s_2\in S'(X)$ must hold. If $s_1=s_2$, then we let $P$ be the path obtained by concatenating paths $P_1$ and $P_2$, and we return path $P$ as response to the query. The length of the path is at most $4\hat D\leq \alpha'\cdot\hat D$, and the processing time of the query is $O(|E(P)|)$. Assume now that $s_1\neq s_2$. 
We execute query $\shortpath(s_1,s_2)$ in the data structure maintained by the algorithm from  \Cref{thm: APSP in HSS full}, obtaining a path $Q$ connecting $s_1$ to $s_2$ in $X$, whose length is  at most $2^{O(1/\eps^6)}$, in time $O(|E(Q)|)$. Let $(e_1,e_2,\ldots,e_z)$ denote the sequence of the edges on path $Q$. For all $1\leq i\leq z$, let $P(e_i)\in \pset$ be the path that serves as the embedding of edge $e_i$; recall that the length of the path is bounded by $D''=\frac{512\hat D}{\eps^2}$. Let $P'$ be the path in graph $H$, obtained by concatenating the paths $P(e_1),\ldots,P(e_z)$. Then path $P'$ connects $s_1$ to $s_2$ in graph $H$, and its length is bounded by $2^{O(1/\eps^6)}\cdot \frac{512 \hat D}{\eps^2}\leq 2^{O(1/\eps^6)}\cdot \hat D$. Let $P$ be the path in graph $H$ that is obtained by concatenating paths $P_1,P'$ and $P_2$. Then path $P$ connects $v$ to $v'$, and its length is bounded by $4\hat D+2^{O(1/\eps^6)}\cdot \hat D\leq 2^{O(1/\eps^6)}\cdot\hat D\leq \frac{2^{O(1/\eps^6)}\cdot D^*}{\eps}\leq 2^{O(1/\eps^6)}\cdot D^*\leq \alpha'\cdot  D^*$. We return path $P$ as the response to the query. It is immediate to verify that the running time of the algorithm for processing the query is bounded by $O(|E(P)|)$.

In order to complete the algorithm for Stage 1, it remains to provide the proof of  \Cref{thm: new pseoducut w well connected outer}, which we do next.

\subsubsection{Proof of \Cref{thm: new pseoducut w well connected outer}}
\label{subsec: inner pseudocut or well connected}

The proof of the theorem easily follows from the following, slightly weaker, theorem.

\begin{theorem}\label{thm: new pseoducut w well connected}
	There is a deterministic algorithm, whose input consists of:
	
\begin{itemize}
	\item a valid input structure $\iset=\left(H,\set{\ell(e)}_{e\in E(H)},D\right )$, such that the maximum vertex degree in $H$ is at most $\mu$;
	\item parameters $W<\hat W\leq W^{1.5}$, such that, if $N(H)$ is the number of regular vertices in $H$, then $N(H)\leq \hat W$;
	
	\item distance parameter $\hat D>D$, and precision parameter $\frac{1}{(\log W)^{1/24}}\leq \eps\leq 1/400$, such that $1/\eps$ is an integer; and
	
	\item a $(\hat D,\rho)$-pseudocut $T$ in $H$, with $|T|>\max\set{\frac{W}{4\mu^2},64W^{\eps}\rho\mu^2}$, where $\rho=\frac{4\hat W}{W}$.
\end{itemize}

The algorithm returns either a $(\hat D,\rho)$-pseudocut $T'$ in $H$, with $|T'|<|T|/2$; or  a graph $X$ with maximum  vertex degree at most   $\hat W^{32\eps^3}$ and $V(X)\subseteq T$, such that $|V(X)|=N^{1/\eps}$ for some parameter $N\geq \frac{|T|^{\eps}}{32\rho^{\eps}}$. In the latter case, the algorithm also returns an embedding $\pset$ of $X$ into $H$ via paths of length at most $D''=\frac{512 \hat D}{\eps^2}$, that cause congestion at most $\eta=2^{20}\hat W^{64\eps^3}\rho^2D''$, and a level-$(1/\eps)$ \HSS for $X$, such that $X$ is $(\eta',\td)$-well-connected with respect to the set $S(X)$ of vertices defined by the support structure, where $\eta'=N^{6+256\eps}$, and $\td=2^{\tilde c/ \eps^5}$, with $\tilde c$ being the constant used in the definition of the \HSS.
The running time of the algorithm is bounded by $O\left (\hat W^{1+O(\eps)} \cdot \hat D^2\cdot \rho^4\cdot \mu\right )$.
\end{theorem}

We provide the proof of \Cref{thm: new pseoducut w well connected} below, after we complete the proof of \Cref{thm: new pseoducut w well connected outer} using it. Let $T$ be the input $(\hat D,\rho)$-pseudocut for graph $H$. We perform iterations, as long as the current pseudocut $T$ has cardinality greater than $\max\set{\frac{W}{4\mu^2},64W^{\eps}\rho\mu^2}$. In every iteration, we apply the algorithm from \Cref{thm: new pseoducut w well connected} to the input structure $\iset$, the current pseudocut $T$, and parameters $W,\hat W,\hat D$ and $\eps$ that remain unchanged. If the algorithm returns  a $(\hat D,\rho)$-pseudocut $T'$ in $H$, with $|T'|<|T|/2$, then we replace $T'$ with $T$, and continue to the next iteration. Assume now that the algorithm from \Cref{thm: new pseoducut w well connected} returned a graph $X$ with maximum  vertex degree at most   $\hat W^{32\eps^3}$, with $V(X)\subseteq T$, such that $|V(X)|=N^{1/\eps}$ for some parameter $N\geq \frac{|T|^{\eps}}{32\rho^{\eps}}$, together with an embedding $\pset$ of $X$ into $H$ via paths of length at most $D''=\frac{512 \hat D}{\eps^2}$, that cause congestion at most $\eta=2^{20}\cdot \hat W^{64\eps^3}\cdot \rho^2\cdot D''$, and a level-$(1/\eps)$ \HSS for $X$, such that $X$ is $(\eta',\td)$-well-connected with respect to the set $S(X)$ of vertices defined by the support structure.
In this case, we terminate the algorithm, and return the graph $X$, its correponding \HSS, and the embedding $\pset$ of $X$ into $H$.
Since $|T|\geq \frac{W}{4\mu^2}$, we are guaranteed that $N\geq \frac{|T|^{\eps}}{32\rho^{\eps}}\geq \frac{W^{\eps}}{128\rho^{\eps}\mu^{2\eps}}$, and, since $|T|\geq 64W^{\eps}\rho\mu^2$, we also get that $N\geq \frac{W^{\eps^2}}{32}$. Therefore, $N\geq \max\set{\frac{W^{\eps}}{128\rho^{\eps}\mu^{2\eps}},\frac{W^{\eps^2}}{32}}$, as required.

Since, in every iteration, $|T|$ decreases by at least factor $2$, the number of iterations is bounded by $O(\log \hat W)\leq O(\log W)\leq W^{O(\eps)}$ (the last inequality follows from Inequality \ref{eq: large W3}). The running time of a single iteration of the algorithm is bounded by $O\left (\hat W^{1+O(\eps)} \cdot \hat D^2\cdot \rho^4\cdot \mu\right )$, and so the running time of the whole algorithm is bounded by $O\left (\hat W^{1+O(\eps)} \cdot \hat D^2\cdot \rho^4\cdot\mu\right )$.

In the remainder of this subsection we focus on the proof of \Cref{thm: new pseoducut w well connected}.

\begin{proofof}{\Cref{thm: new pseoducut w well connected}}
The proof of the theorem uses the following claim.

\begin{claim}\label{claim: pseudocut inner}
	There is a deterministic algorithm, whose input is the same as that in \Cref{thm: new pseoducut w well connected}. The algorithm returns one of the following: 
	
	\begin{itemize}
		\item either a collection $E'$ of at most $\frac{|T|}{16}$ edges of $H$, such that, for every vertex $t\in T$, $B_{H\setminus E'}\left (t,\frac{32\hat D}{\eps}\right )$ contains at most $\frac{|T|}{16\rho}$ vertices of $T$; or
		
		\item a graph $X$ with maximum vertex degree at most   $\hat W^{32\eps^3}$ and $V(X)\subseteq T$, such that $|V(X)|=N^{1/\eps}$ for some parameter $N\geq \frac{|T|^{\eps}}{32\rho^{\eps}}$, together with an an embedding $\pset$ of $X$ into $H$ via paths of length at most $D''=\frac{512 \hat D}{\eps^2}$, that cause congestion at most $\eta=2^{20}\cdot \hat W^{64\eps^3}\cdot \rho^2\cdot D''$, and a level-$(1/\eps)$ \HSS for $X$, such that $X$ is $(\eta',\td)$-well-connected with respect to the set $S(X)$ of vertices defined by the support structure, where $\eta'=N^{6+256\eps}$, and $\td=2^{\tilde c/ \eps^5}$, with $\tilde c$ being the constant used in the definition of the \HSS.
	\end{itemize} 
	The running time of the algorithm is bounded by $O\left (\hat W^{1+O(\eps)} \cdot \hat D^2\cdot \rho^4\cdot \mu\right )$.
\end{claim}

\iffalse
\begin{claim}\label{claim: pseudocut inner}
	There is a deterministic algorithm, whose input is the same as that in \Cref{thm: new pseoducut w well connected}. The algorithm returns one of the following: 
	
	\begin{itemize}
		\item either a collection $E'$ of at most $\frac{|T|}{16}$ edges of $H$, such that, for every vertex $t\in T$, $B_{H\setminus E'}\left (t,\frac{32\hat D}{\eps}\right )$ contains at most $\frac{|T|}{16\rho}$ vertices of $T$; or
		
		\item graph $X$ with maximum vertex degree at most   $\hat W^{32\eps^3}$, with $V(X)\subseteq T$, such that $|V(X)|=N^{1/\eps}$ for some parameter $N\geq \frac{|T|^{\eps}}{32\rho^{\eps}}$, together with an an embedding $\pset$ of $X$ into $H$ via paths of length at most $D''=\frac{512 \hat D}{\eps}$, that cause congestion at most $\eta=2^{20}\cdot \hat W^{64\eps^3}\cdot \rho^2\cdot D''$, and a level-$(1/\eps)$ \HSS for $X$, such that $X$ is $(\eta',\td)$-well-connected with respect to the set $S(X)$ of vertices defined by the support structure, where $\eta'=N^{6+256\eps}$, and $\td=2^{c/ \eps^5}$, with $c$ being the constant used in the definition of the \HSS.
	\end{itemize} 
The running time of the algorithm is bounded by $O\left (\hat W^{1+O(\eps)} \cdot \hat D^2\cdot \rho^4\right )$.
\end{claim}
\fi

We provide the proof of \Cref{claim: pseudocut inner} below, after we complete the proof of 	\Cref{thm: new pseoducut w well connected} using it. Let $D'=\frac{32\hat D}{\eps}$.
We apply the algorithm from \Cref{claim: pseudocut inner} to the input graph $H$, with parameters $W,\hat W,\mu$, and $\eps$ remaining unchanged. If the algorithm from \Cref{claim: pseudocut inner} returns a graph $X$, with its embedding $\pset$ into $H$, and a \HSS for $X$, then we return this graph, its embedding and the \HSS as the outcome of the algorithm. Therefore, we assume from now on that the algorithm from  \Cref{claim: pseudocut inner} returned a collection $E'$ of at most $\frac{|T|}{16}$ edges of $H$, such that, for every vertex $t\in T$, $B_{H\setminus E'}(t,D')$ contains at most $\frac{|T|}{16\rho}$ vertices of $T$.

Let $S'$ be the set of regular vertices that serve as endpoints of the edges of $E'$, so $|S'|\leq \frac{|T|}{16}$. 

Let $T^*=T\cup S'$. Clearly, $T^*$ remains a $(\hat D,\rho)$-pseudocut in $H$. Note that, if we identify a subset $S\subseteq T^*\setminus S'$ of vertices, with $B_{H\setminus (T^*\setminus S)}(S,2\hat D)$ containing fewer than $\hat W/\rho$ regular vertices, then, from \Cref{obs: improving pseudocut}, $T^*\setminus S$ is a valid $(\hat D,\rho)$-pseudocut in $H$. We can then delete $S$ from $T^*$, and continue. Notice that  $B_{H\setminus (T^*\setminus S)}(S,2\hat D)\subseteq B_{H\setminus S'}(S,2\hat D)$, and so, in order to make progress, it is enough to compute a subset $S\subseteq T\setminus S'
$ of terminals, such that $B_{H\setminus S'}(S,2\hat D)$ contains fewer than $\hat W/\rho$ regular vertices. The main challenge is to execute such an algorithm efficiently. Let $H'=H\setminus S'$. Recall that we have defined a parameter $D'=\frac{32\hat D}{\eps}$.

Our algorithm will consist of a number of iterations. In every iteration, we will select some terminal $t\in T$, and we will perform a weighted BFS in graph $H'$ from $t$, up to some depth (that is at least $2\hat D$, and at most $D'/2$). If, over the course of this BFS procedure, we encounter fewer than $\hat W/\rho$ regular vertices, then we can let $S\subseteq T\setminus S'$ be the set of all terminals that the BFS discovered, except for those that appear in the last few layers of the BFS. We are then guaranteed that  $B_{H'}(S,\hat D)$ contains fewer than $\hat W/\rho$ regular vertices, so we can update the pseudocut $T^*$, by removing the vertices of $S$ from it. We will also remove some edges from graph $H'$, which no longer lie in sets $B_{H'}(t,2\hat D)$ of the remaining terminals $t\in T^*$. Alternatively, if the BFS discovers more than $\hat W/\rho$ regular vertices, then we are guaranteed, on the one hand, that $B_{H'}(t,D'/2)$ contains more than $\hat W/\rho$ regular vertices, while on the other hand, the number of vertices of $T$ lying in $B_{H'}(t,D'/2)$ is at most $\frac{|T|}{16\rho}$, from \Cref{claim: pseudocut inner}. We put the terminal $t$ aside, and we will remove all terminals of $B_{H'}(t,D')$ from the pool of terminals that need to be processed (but we do not update the pseudocut $T^*$). Intuitively, since $B_{H'}(t,D'/2)$ contains more than $\hat W/\rho$ regular vertices, we will eventually put aside at most $\rho$ terminals (more precisely, we will associate, with each terminal $t$ that we put aside, a set $V(t)\subseteq B_{H'}(t,D'/2)$ of regular vertices of cardinality at least $\hat W/\rho$, and we will ensure that all resulting sets $\set{V(t)\mid t\in A}$ are mutually disjoint, where $A$ is the set of terminals that we put aside). This ensures that the total number of terminals of $T$ that are thus eliminated from the pool of terminals to be considered (while not being removed from the pseudocut $T^*$), is at most $|T|/4$. In turn, this ensures that the final pseudocut that we obtain has cardinality at most $|T|/4+|S'|\leq |T|/2$, as required.

We now describe our algorithm in more detail. Denote $k=|T|$.
For convenience, we remove from $T$ terminals that also lie in $S'$. Note that this decreases the cardinality of $T$ by at most $k/16$.
Throughout the algorithm, we maintain a set $T^*\subseteq T\cup S'$ of vertices, such that $T^*$ is a valid $(\hat D,\rho)$-pseudocut in $H$. We also maintain a partition of the set $T$ of terminals into three subsets: set $T_0$ containing the terminals that were removed from $T^*$ (so $T^*\cap T_0=\emptyset$); set $T_1\subseteq T\cap T^*$ containing the terminals that were removed from the pool of terminals to be processed; and set $T_2\subseteq T\cap T^*$ containing all remaining terminals. We also maintain subgraph $H''\subseteq H'$, a set $A\subseteq T_1$ of \emph{special terminals}, and for every special terminal $t\in A$, two subsets $V(t)$ and $V'(t)$ of regular vertices of $H$, such that the following invariants hold.

\begin{properties}{I}
	\item for all $t\in A$, $V(t)\subseteq B_{H'}(t,D'/2)$, $V(t)\cap V(H'')=\emptyset$, and $|V(t)|\geq \hat W/\rho$; \label{inv: small ball around special terminal}
	\item for all $t\in A$, $V'(t)\subseteq B_{H'}(t,D')$; \label{inv: large ball around special terminal}
	\item for all $t,t'\in A$, if $t\neq t'$, then $V(t)\cap V(t')=\emptyset$; \label{inv: disjoint balls}
	\item $T_1\subseteq \bigcup_{t\in A}V'(t)$; and \label{inv: union of balls}
	\item for every terminal $t\in T_2$, the subgraph of $H'$ induced by $B_{H'}(t,2\hat D)$ is contained in $H''$. \label{inv: preserving grpah}
\end{properties}

At the beginning of the algorithm, we let $T^*=T\cup S'$, $T_0=T_1=A=\emptyset$, $T_2=T$, and $H''=H'$. It is easy to verify that all invariants hold. The algorithm then performs iterations, until $T_2=\emptyset$ holds. We now describe a single iteration of the algorithm. We assume that all invariants hold at the beginning of the iteration.

\paragraph{Iteration description.}
We let $t\in T_2$ be any terminal. We apply the algorithm from 
\Cref{lem: ball growing} to graph $H''$, terminal $t$, distance parameter $2\hat D$ and precision parameter $\eps$ remaining unchanged (so $k=0$ and we do not supply subsets $T_1,\ldots,T_k$ of vertices). Let $1< i \leq \frac{2}{\eps}$
be the integer that the algorithm returns. Denote $\hat B(t)=B_{H''}(t,4(i-1)\hat D)$, and $\hat B'(t)=B_{H''}(t,4i\hat D)$. We now consider two cases. 

The first case happens if the number of regular vertices in set $\hat B'(t)$ is at most $\frac{\hat W}{\rho}$. We then say that the current iteration is a type-1 iteration. Let $\hat T$ be the collection of all terminals that lie in $T_2\cap B_{H''}(t,(4i-2)\hat D)$. Then we are guaranteed that $B_{H''}(\hat T,2\hat D)$ contains at most $\frac{\hat W}{\rho}$ regular vertices. Notice also that, from Invariant \ref{inv: preserving grpah}, 
for every terminal $t'\in T_2$, the subgraph of $H'$ induced by $B_{H'}(t',2\hat D)$ is contained in $H''$. Therefore, $B_{H'}(\hat T,2\hat D)\subseteq B_{H''}(\hat T,2\hat D)$, and $B_{H'}(\hat T,2\hat D)$ contains at most $\frac{\hat W}{\rho}$ regular vertices. From 
\Cref{obs: improving pseudocut}, $T^*\setminus \hat T$ is a valid pseudocut. We update $T^*=T^*\setminus \hat T$, add the vertices of $\hat T$ to $T_0$, and delete them from $T_2$. Notice that, for every vertex $t'$ that remains in $T_2$, $B_{H''}(t',2\hat D)$ is disjoint from $B_{H''}(t,4(i-1)\hat D)$. Therefore, we can delete all edges of $E_{H''}(\hat B(t))$ and all vertices of $\hat B(t)$ from graph $H''$, without violating Invariant \ref{inv: preserving grpah}. If we denote by $E'(t)=E_{H''}(\hat B(t))$, then the running time of the algorithm from \Cref{lem: ball growing} is bounded by $O(|E'(t)|\cdot |E(H'')|^{\eps}\cdot \log \hat W)$. Since the edges of $E'(t)$ are deleted from graph $H''$, we can charge these edges for the running time of the current iteration. It is easy to verify that the remaining invariants continue to hold.

We now consider the second case, where the number of regular vertices in set $\hat B'(t)$ is greater than $\frac{\hat W}{\rho}$. We then say that the current iteration is a type-2 iteration.
In this case, we let $V(t)$ be the set of all regular vertices lying in $B_{H''}(t,D'/2)$, and we let $V'(t)$ be the set of all regular vertices lying in $B_{H''}(t,D')$. Since $D'=\frac{32\hat D}{\eps}$, we are guaranteed that $\hat B'(t)\subseteq V(t)$, and so $|V(t)|\geq \frac{\hat W}{\rho}$. Since graph $H''$ is obtained from graph $H'$ by deleting edges and vertices from it, we are guaranteed that $V(t)\subseteq B_{H'}(t,D'/2)$ and $V'(t)\subseteq B_{H'}(t,D')$. We add terminal $t$ to set $A$, and we move all terminals in set $V'(t)\cap T_2$ from $T_2$ to $T_1$. Lastly, we delete from $H''$ all vertices of $V(t)$ and edges that are incident to them. From our discussion, it is immediate to verify that Invariants \ref{inv: small ball around special terminal}, \ref{inv: large ball around special terminal} and \ref{inv: union of balls} continue to hold. Since, for every terminal $t'$ that lied in set $A$ at the beginning of the iteration, $V(t')\cap V(H'')=\emptyset$ held from Invariant \ref{inv: small ball around special terminal}, it is immediate to verify that Invariant \ref{inv: disjoint balls} continues to hold. It now remains to verify that Invariant \ref{inv: preserving grpah} also continues to hold. Consider any terminal $t'$ that lies in set $T_2$ at the end of the current itertation. From Invariant \ref{inv: preserving grpah},  the subgraph of $H'$ induced by $B_{H'}(t',2\hat D)$ was contained in $H''$ at the beginning of the iteration. Since terminal $t'$ remains in set $T_2$, at the beginning of the iteration, $B_{H''}(t,D')$ did not contain $t'$. Since we only deleted from $H''$ vertices that lie in $B_{H''}(t,D'/2)$ and their incident edges, and since $D'>8\hat D$, we did not delete any vertices of $B_{H''}(t',2\hat D)=B_{H'}(t',2\hat D)$, or edges connecting them from $H''$. Therefore,  the subgraph of $H'$ induced by $B_{H'}(t',2\hat D)$ remains contained in $H''$. The running time of a type-2 iteration is bounded by $O(|E(H)|)$.

The algorithm terminates once $T_2=\emptyset$ holds. Since, from Invariant \ref{inv: disjoint balls}, the sets $\set{V(t)\mid t\in A}$ of vertices are all mutually disjoint, and since, from Invariant \ref{inv: small ball around special terminal}, each such set contains at least $\frac{\hat W}{\rho}$ regular vertices, at the end of the algorithm $|A|\leq \rho$ must hold. Moreover, since, from Invariant  \ref{inv: large ball around special terminal}, for every terminal $t\in A$, $V'(t)\subseteq B_{H'}(t,D')$, and since, from \Cref{claim: pseudocut inner}, 
$B_{H'}\left (t,D'\right )$ contains at most $\frac{|T|}{16\rho}$ vertices of $T$, from Invariant \ref{inv: union of balls}, we conclude that $|T_1|\leq |A|\cdot \frac{k}{16\rho}\leq \frac{k}{16}$. Therefore, at the end of the algorithm, $|T_0|\geq |T|-\frac{k}{16}\geq\frac{14|T|}{16}$ holds, and so:

\[|T^*|\leq |T|+|S'|-|T_0|\leq \frac{|T|}{8}+\frac{|T|}{4}\leq \frac{|T|}{2}. \]

It now remains to analyze the running time of the algorithm. The running time of the algorithm from \Cref{claim: pseudocut inner} is bounded by $O\left (\hat W^{1+O(\eps)} \cdot \hat D^2\cdot \rho^4\cdot \mu\right )$. Since, at the end of the algorithm $|A|\leq \rho$ holds, the number of type-2 iterations is bounded by $\rho$, and each such iteration has running time $O(|E(H)|)$. If we consider a type-1 iteration in which a terminal $t$ was processed, and we let $E_t$ be the set of edges that were deleted from $H''$ over the course of the iteration, then the running time of the iteration is $O(|E_t|\cdot |E(H)|^{\eps}\cdot \log \hat W)$. Therefore, the total running time of all type-1 iterations is at most $O(|E(H)|^{1+\eps}\cdot \log \hat W)$, and the total running time of all type-2 iterations is at most $O(|E(H)|\cdot \rho)$. Overall, the running time of the algorithm (excluding the running time of the algorithm from \Cref{claim: pseudocut inner}) is bounded by $O(|E(H)|^{1+\eps}\cdot \rho \cdot \log \hat W)\leq O( \hat W^{1+O(\eps)}\rho)$, and the total running time of the algorithm is bounded by $O\left (\hat W^{1+O(\eps)} \cdot \hat D^2\cdot \rho^4\cdot \mu\right )$.

In order to complete the proof of \Cref{thm: new pseoducut w well connected}, it now remains to prove \Cref{claim: pseudocut inner}.

\subsection*{Proof of \Cref{claim: pseudocut inner}}
Denote $D'=\frac{32\hat D}{\eps}$ and $D''=\frac{16D'}{\eps}=\frac{512\hat D}{\eps^2}$.
Throughout the proof, we refer to vertices of $T$ as terminals.
The main idea of the algorithm is simple: we will maintain a set $E'$ of edges of $H$, starting from $E'=\emptyset$. We will then iteratively identify terminals $t\in T$ with $B_{H\setminus E'}(t,D'')$ containing more than $\frac{|T|}{16\rho}$ terminals, and apply the algorithm from \Cref{cor: HSS witness} to the current graph $H\setminus E'$ and the set $T'=T\cap B_{H\setminus E'}(t,D'')$ of terminals, with parameter $d=D''$. If the algorithm computes a graph $X$, its embedding into $H$, and a \HSS for $X$, then we return this graph, its embedding, and the \HSS as the outcome of the algorithm. In this case, we say that the application of the algorithm from \Cref{cor: HSS witness}  was successful. Otherwise, the algorithm from \Cref{cor: HSS witness} must return a relatively small subset $\tilde E$ of edges,  and two relatively large subsetes $T_1,T_2\subseteq T'$ of terminals, such that, if we add the edges of $\tilde E$ to set $E'$, then, $\dist_{H\setminus E'}(T_1,T_2)>D''$ holds. In this case, we say that the application of the algorithm from \Cref{cor: HSS witness}  was unsuccessful. We prove that the number of unsuccessful applications of the algorithm from \Cref{cor: HSS witness}  is relatively small; this is since, in each such application, there is a large number of terminals $t\in T$, for which $|T\cap B_{H\setminus E'}(t,D'')|$ decreases significantly. This, in turn, ensures that, at the end of the algorithm, $|E'|$ remains sufficiently small.

The main challenge in implementing this idea is efficiency: if we implement this algorithm in a straightforward way, where in every iteration, we select some terminal $t\in T$, and check whether $B_{H\setminus E'}(t,D')$ contains many terminals, we may obtain an inefficient algorithm, since it is possible that for many such terminals, there are many edges connecting the vertices of $B_{H\setminus E'}(t,D')$, but there are few terminals in $B_{H\setminus E'}(t,D')$.

In order to overcome this difficulty, we will maintain a partition $(T^A,T^I)$ of the set $T$ of terminals, where the terminals in $T^I$ are called \emph{inactive}, and the terminals in $T^A$ are called \emph{active}. We also maintain a set $E'$ of edges, and a graph $H''\subseteq H\setminus E'$. We ensure that the following invariants hold throughout the algorithm.

\begin{properties}{I'}
	\item for every inactive terminal $t\in T^I$, $B_{H\setminus E'}(t,D')$ contains at most $\frac{|T|}{16\rho}$ terminals of $T$;\label{prop: inactive settled}
	
	\item for every active terminal $t\in T^A$, if we denote by $S_t=B_{H\setminus E'}(t,D')$, then the subgraph of $H\setminus E'$ induced by $S_t$ is contained in $H''$; and
	\label{prop: active ball preserved}
	
	\item for all $i\geq 0$, after $i$ applications of the algorithm from \Cref{cor: HSS witness}, $|E'|\leq \frac{i\cdot |T|}{2^{20}\cdot\hat W^{32\eps^3}\cdot \rho^2}$ holds. \label{prop: few deleted edges}
\end{properties}

Throughout, we denote $H'=H\setminus E'$. At the beginning of the algorithm, we let $E'=\emptyset$, $H''=H$, $T^A=T$ and $T^I=\emptyset$. It is easy to verify that all invariants hold. We then perform iterations, as long as $T^A\neq \emptyset$. We now describe the execution of a single iteration.

\paragraph{Execution of an iteration.}
We select an arbitrary terminal $t\in T^A$. If $t$ is an isolated vertex in $H''$, then we move $t$ from $T^A$ to $T^I$. It is immediate to verify that all invariants continue to hold. From now on, we assume that $t$ is not an isolated vertex of $H''$. 
 We apply the algorithm from 
\Cref{lem: ball growing} to graph $H''$, terminal $t$, distance parameter $2D'$, and precision parameter $\eps$ remaining unchanged (so $k=0$ and we do not supply subsets $T_1,\ldots,T_k$ of vertices). Let $1< i \leq \frac{2}{\eps}$
be the integer that the algorithm returns. Denote $\hat B(t)=B_{H''}(t,4(i-1) D')$, and $\hat B'(t)=B_{H''}(t,4iD')$. We now consider two cases. 

The first case happens if the number of terminals  in set $\hat B'(t)$ is at most $\frac{|T|}{16\rho}$. We then say that the current iteration is a type-1 iteration. Let $\hat T$ be the collection of all terminals that lie in $T^A\cap B_{H''}(t,(4i-2) D')$. Then we are guaranteed that, for every terminal $t'\in \hat T$, 
$B_{H''}(t',D')$ contains at most $\frac{|T|}{16\rho}$ terminals of $T$. Moreover, since, from 
Invariant \ref{prop: active ball preserved}, for each such terminal $t'$, $B_{H''}(t,D')=B_{H\setminus E'}(t,D')$, we get that $B_{H\setminus E'}(t',D')$, contains at most $\frac{|T|}{16\rho}$ terminals of $T$. We add the terminals of $\hat T$ to set $T^I$ of inactive terminals. Note that, for every terminal $t'\in T^A$ that remains active, $B_{H''}(t',D')$ is disjoint from $\hat B(t)$. Therefore, we can delete from $H''$ all edges whose both endpoints lie in $\hat B(t)$, and all vertices that lie in $\hat B(t)$, without violating Invariant \ref{prop: active ball preserved}. From our discussion, Invariant \ref{prop: inactive settled} continues to hold, and it is easy to verify that Invariant \ref{prop: few deleted edges} does as well.
 If we denote by $E'(t)=E_{H''}(\hat B(t))$ (where $H''$ is the graph at the beginning of the iteration), then the running time of the algorithm from \Cref{lem: ball growing} is bounded by $O(|E'(t)|\cdot |E(H'')|^{\eps}\cdot \log \hat W)$. Since the edges of $E'(t)$ are deleted from graph $H''$, we can charge these edges for the running time of the current iteration.

We now consider the second case, where the number of terminals in set $\hat B'(t)$ is greater than $\frac{|T|}{16\rho}$. We then say that the current iteration is a type-2 iteration.  Let $\hat T$ be the collection of all terminals of $T$ that lie in $B_{H''}(t,D''/2)$. Since $D''=\frac{16D'}{\eps}$, while the integer $i$ that the algorithm from \Cref{lem: ball growing} returned is bounded by $2/\eps$, it is easy to verify that $\hat B'(t)\subseteq B_{H''}(t,D''/2)$, and so $|\hat T|\geq \frac{|T|}{16\rho}$. Notice that, for every terminal $t'\in \hat T$, $|B_{H''}(t',D'')\cap T|\geq \frac{|T|}{16\rho}$.
We denote $k=|\hat T|$, so that $k\geq \frac{|T|}{16\rho}$.

Next, we will apply the algorithm from \Cref{cor: HSS witness} to graph $H''$, set $\hat T$ of termminals, distance paramter $d=D''$, and parameter $\eps$ remaining unchanged. In order to be able to use the algorithm, we need to verify that 
$\frac{2}{(\log k)^{1/12}}< \eps<\frac{1}{400}$ holds.

Recall that, from the statement of \Cref{thm: new pseoducut w well connected},  $\frac{1}{(\log W)^{1/24}}\leq \eps\leq \frac 1{400}$ holds. Recall also that 
$|\hat T|=k\geq \frac{|T|}{16\rho}\geq W^{\eps}$ (since $|T|\geq 64W^{\eps}\rho\mu^2$).
%$|\hat T|=k\geq \frac{|T|}{16\rho}\geq \frac{W}{64\rho\mu}\geq \frac{W^{1/2}}{16}\geq W^{1/2-2\eps}$ (since $|T|\geq \frac{W}{4\mu}$, $\rho=\frac{4\hat W} W\leq \sqrt{4W}$, $\mu\leq W^{\eps}$, and $64<W^{\eps}$ from Inequality \ref{eq: large W 2}.)
Therefore, $\log k \geq \eps\log W\geq \frac{1}{\eps^{23}}$. We conclude that $(\log k)^{1/12}>\frac 2{\eps}$, as required.

We apply the algorithm from  \Cref{cor: HSS witness} to graph $H''$, the set $\hat T$ of terminals, and parameter $\eps$ remaining unchanged. We also use distance parameter $d= D''$, and congestion parameter $\eta^*=\frac{\eta}{\hat W^{32\eps^3}}$.
The running time of the algorithm is bounded by:

\[
\begin{split}
O\left (|T|^{1+O(\eps)}+|E(H)|\cdot |T|^{O(\eps^3)}\cdot(\eta^*+ D''\log \hat W)\right )& \leq O\left (\hat W^{1+O(\eps)}\cdot  \mu\cdot \frac{\eta}{\hat W^{32\eps^3}}\cdot  D''\cdot \log W\right )\\
&\leq O\left (\hat W^{1+O(\eps)} \cdot\mu\cdot  (D'')^2\cdot \rho^2 \cdot \log W\right )\\
&\leq O\left (\hat W^{1+O(\eps)} \cdot \mu\cdot \hat D^2\cdot \rho^2\right ).
\end{split}
\]

since $\eta=2^{20}\cdot\hat W^{64\eps^3}\cdot \rho^2\cdot D''$ and  $D''=\frac{512 \hat D}{\eps}$. We have also used the fact that $512/\eps<\log W$, and that $W^{\eps}\ge \log^2W$ from Inequality \ref{eq: large W3}.

We denote $N=\floor{k^{\eps}}$, so $N\geq \floor{\left(\frac{|T|}{16\rho}\right)^{\eps}}\geq \frac{|T|^{\eps}}{32\rho^\eps}$ holds. 
We now consider two cases. The first case happens if the algorithm from 
\Cref{cor: HSS witness} returned a graph $X$ with $V(X)\subseteq \hat T$, $|V(X)|=N^{1/\eps}$,  and maximum vertex degree at most   $k^{32\eps^3}\leq \hat W^{32\eps^3}$, together with an embedding $\pset$ of $X$ into $H$ via paths of length at most $D''$, and cause congestion at most $\eta^*\cdot k^{32\eps^3}\leq \frac{\eta}{\hat W^{32\eps^3}}\cdot  k^{32\eps^3}\leq \eta$, and a level-$(1/\eps)$ \HSS for $X$, such that $X$ is $(\eta',\td)$-well-connected with respect to the set $S(X)$ of vertices defined by the support structure, where $\eta'=N^{6+256\eps}$, and $\td=2^{\tilde c/ \eps^5}$, with $\tilde c$ being the constant used in the definition of the \HSS.
In this case, we say that the current iteration is \emph{successful}. We terminate the algorithm, and return the graph $X$, its embedding $\pset$, and the \HSS for $X$. As observed already, $N\geq \frac{|T|^{\eps}}{32\rho^{\eps}}$ holds, so this is a valid output of the algorithm.

We assume from now on that the algorithm from \Cref{cor: HSS witness} returned a pair $T_1,T_2\subseteq \hat T$ of disjoint subsets of terminals, and a set $\tilde E$ of edges of $H''$, such that $|T_1|=|T_2|$, and:

\[|T_1|\geq \frac{k^{1-4\eps^3}}{4}\geq \frac{|T|^{1-4\eps^3}}{64\rho}.\] 

(since $k\geq\frac{|T|}{16\rho}$). We say that the current iteration is unsuccessful.
Recall that the algorithm also guarantees that: 

\[\begin{split}
|\tilde E|&\leq \frac{D''\cdot |T_1|}{\eta^*}\\
&\leq \frac{|T|\cdot D''\cdot \hat W^{32\eps^3}}{\eta}\\
&\leq \frac{|T|}{2^{20}\cdot\hat W^{32\eps^3}\cdot \rho^2},
\end{split}\] 

(since $\eta=2^{20}\cdot\hat W^{64\eps^3}\cdot \rho^2\cdot D'')$.

Recall also that, for every pair  $t\in T_1,t'\in T_2$ of terminals, $\dist_{H''\setminus \tilde E}(t,t')>D''$ holds.
We add the edges of $\tilde E$ to set $E'$, delete them from $H''$, and continue to the next iteration.
It is immediate to verify that Invariant \ref{prop: inactive settled} continues to hold. Since the changes to graphs $H\setminus E'$ and $H''$ were identical, Invariant \ref{prop: active ball preserved} also continues to hold. Lastly, since $|\tilde E|\leq \frac{|T|}{2^{20}\cdot\hat W^{32\eps^3}\cdot \rho^2}$, Invariant \ref{prop: few deleted edges} continues to hold.

This completes the description of a single iteration, and of the algorithm. As mentioned already, the algorithm terminates once $|T^A|=\emptyset$ holds. From Invariant \ref{prop: inactive settled}, at the end of the algorithm, for every terminal $t\in T$, $B_{H\setminus E'}(t,D')$ contains at most $\frac{|T|}{16\rho}$ terminals of $T$.

In order to analyze the algorithm, we need the following observation.

\begin{observation}\label{obs: type-2 iterations}
	The number type-2 iterations of the algorithm is bounded by $2^{13}\rho^2\cdot |T|^{8\eps^3}$.
\end{observation}

\begin{proof}
	Recall that at the beginning of each type-2 iteration $i$, we are given a set $\hat T\subseteq T$ of terminals, such that $|\hat T|\geq \frac{|T|}{16\rho}$, and,  for every terminal $t'\in \hat T$, $\hat T\subseteq B_{H''}(t',D'')\cap T$.
	
At the end of iteration $i$, we obtain two subsets $T_1,T_2\subseteq \hat T$ of terminals, such that, once graph $H''$ is updated (by deleting the edges of $\tilde E$ form it), $\dist_{H''}(T_1,T_2)> D''$ holds. Recall also that 
$|T_1|,|T_2|\geq \frac{|T|^{1-4\eps^3}}{64\rho}$. Therefore, during iteration $i$, there are at least $\frac{|T|^{1-4\eps^3}}{64\rho}$ terminals $t'\in T_1$, such that, for each terminal $t'\in T_1$, $|B_{H''}(t',D'')\cap T|$ decreased by at least $\frac{|T|^{1-4\eps^3}}{64\rho}$.

Let $t\in T$ be a terminal, and let $i$ be any type-2 iteration. We say that terminal $t$ is hit in iteration $i$, if, over the course of iteration $i$, $|B_{H''}(t',D'')\cap T|$ decreased by at least $\frac{|T|^{1-4\eps^3}}{64\rho}$. Clearly, for a fixed terminal $t$, the number of iterations in which $t$ may be hit is bounded by $64\rho\cdot |T|^{4\eps^3}$. Let $\Pi$ be the collection of pairs $(i,t)$, where $i$ is a type-2 iteration, and terminal $t$ was hit during iteration $i$. Then $|\Pi|\leq 64|T|^{1+4\eps^3}\cdot \rho$. From the above discussion, in each unsuccessful type-2 iteration, the number of terminals that are hit is at least $\frac{|T|^{1-4\eps^3}}{64\rho}$. Therefore, the total number of unsuccessful type-2 iterations is bounded by:

\[\frac{|\Pi|}{|T|^{1-4\eps^3}/(64\rho)}\leq \frac{2^{12}\cdot |T|^{1+4\eps^3}\cdot \rho^2}{|T|^{1-4\eps^3}}\leq 2^{12}\rho^2\cdot |T|^{8\eps^3}. \]

Since there is at most one successful type-2 iteration, we get that the total number of type-2 iterations is bounded by $2^{13}\rho^2\cdot |T|^{8\eps^3}$.	
\end{proof}

From Invariant \ref{prop: few deleted edges}, at the end of the algorithm: 

$$|E'|\le \frac{|T|}{2^{20}\cdot\hat W^{32\eps^3}\cdot \rho^2}\cdot 2^{13}\rho^2\cdot |T|^{8\eps^3}\leq \frac{|T|}{16},$$

as required.

It now remains to bound the running time of the algorithm. Recall that the running time of a single type-2 iteration is bounded by $O\left (\hat W^{1+O(\eps)} \cdot \hat D^2\cdot \rho^2\cdot \mu\right )$.
Since the number of type-2 iterations is at most $O(\rho^2\cdot |T|^{8\eps^3})\leq O(\rho^2\cdot \hat W^{O(\eps)})$, we get that the total time spent on type-2 iterations is at most 
$O\left (\hat W^{1+O(\eps)} \cdot \hat D^2\cdot \rho^4\cdot\mu\right )$.

Recall that in each type-1 iteration, we delete some subset $E'(t)$ of edges from graph $H''$, and the running time of the iteration is bounded by $O(|E'(t)|\cdot |E(H'')|^{\eps}\cdot \log \hat W)\leq O(|E'(t)|\cdot \hat W^{O(\eps)}\cdot \log \hat W)$. Therefore, the total running time spent on type-1 iterations is bounded by $O\left (|E(H'')|\cdot  \hat W^{O(\eps)}\right )\leq O\left (\hat W^{1+O(\eps)}\cdot \mu\right )$. We conclude that the total running time of the algorithm is bounded by: $$O\left (\hat W^{1+O(\eps)} \cdot \hat D^2\cdot \rho^4\cdot\mu\right ).$$
\end{proofof}

\subsection{Stage 2: Medium-Size Pseudocut}
\label{subsec: stage 2 mid size pseudo}

The second stage of the algorithm is only executed if $64W^{\eps}\rho\mu^2>\frac{W}{4\mu^2}$, and it only lasts as long as $|T\cap V(H)|>\frac{W}{4\mu^2}$.  Recall that, at the beginning of Stage 2, $|T|\leq 64W^{\eps}\rho\mu^2$ holds.
Throughout the algorithm, we denote by $T'=T\cap V(H)$.
We partition the second stage into phases. At the beginning of every phase, we select an arbitrary vertex $t\in T\cap V(H)$, and initialize the modified \EST data structure from \Cref{thm: ES-tree} in graph $H$, with source vertex $t$, and depth parameter $D^*$. Since supernode-splitting operations are a special case of vertex-splitting, we can think of graph $H$ as undergoing an online sequence of edge-deletion, isolated vertex-deletion and vertex-splitting updates. Notice that vertex-splitting may not be applied to vertex $t$, since it is not a supernode. Recall that the total update time of the algorithm from \Cref{thm: ES-tree} is 
 $O(m^*\cdot D^*\cdot \log m^*)$, where $m^*$ is the total number of edges that ever belonged to graph $H$. Since $m^*\leq \hat W\cdot \mu$ and $\mu\leq \hat W^{0.1}\leq W^{0.2}$, the total update time is bounded by $O(\hat W\cdot \mu\cdot D^*\cdot \log W)$.
 
 Recall that the algorithm from  \Cref{thm: ES-tree} maintains a set $S^*$ of vertices of $H$, such that, for every vertex $x\in S^*$, $\dist_H(t,x)> D^*$ holds. Whenever a regular vertex $v$ is added to set $S^*$, we raise the flag $F_H$, and provide $(t,v)$ as a witness pair. Notice that it is possible that, during the flag-lowering update sequence, vertex $t$ is deleted from $H$. In any case, the phase terminates once $t$ is deleted from $H$. When $|T\cap V(H)|$ falls below $\frac{W}{4\mu^2}$, the second stage terminates.
 
 It is easy to verify that the number of phases in the second stage is bounded by $64W^{\eps}\rho\mu^2$, and, since the total update time of every phase is bounded by $O(\hat W\cdot \mu\cdot D^*\cdot \log W)$, we get that the total update time of the algorithm for Stage 2 is bounded by $O\left(\hat W^{1+O(\eps)}\cdot  D^*\cdot \rho\cdot \mu^3\right )$.

\subsubsection*{Responding to Queries}

We now provide an algorithm for supporting queries $\spquery(H,v,v')$ over the course of Stage 2. Suppose we are given a query  $\spquery(H,v,v')$, where $v,v'\in V(H)$ are regular vertices of $H$. Recall that our goal is to return a path $P$ of length at most $\alpha'\cdot D^*$ connecting $v$ to $v'$ in $H$, in time $O(|E(P)|)$. 

Recall that a $\spquery$ cannot be asked while the flag $F_H$ is up. Therefore, flag $F_H$ is currently down, and the set $S^*$ of vertices that the algorithm from  \Cref{thm: ES-tree} maintains contains no regular vertices. Vertices $v,v'$ must then lie in the \EST that the algorithm maintain. We perform query $\shortestpath(v)$ and $\shortestpath(v')$ in the data structure maintained by the algorithm from  \Cref{thm: ES-tree}, that must return a path $P_1$ connecting $v$ to $t$ in $H$, and a path $P_2$ connecting $v'$ to $t$ in $H$ of length at most $D^*$ each, in time $O(|E(P_1)|+|E(P_2)|)$. By concatenating the two paths, we obtain a path $P$ in graph $H$, connecting $v$ to $v'$, of length at most $2 D^*\leq \alpha'\cdot D^*$, in time $O(|E(P)|)$.

\subsection{Stage 3: Small Pseudocut Regime}
\label{subsec: stage3: small pseudocut}

Recall that, at the beginning of Stage 3, we are given a $(\hat D,\rho)$-pseudocut $T$, with $|T|\leq \frac{W}{4\mu^2}$. This pseudocut will remain fixed over the course of Stage 3, except that, whenever vertices of $T$ are deleted from $H$, we delete them from $T$ as well. Throughout the remainder of the algorithm, we refer to the vertices of $T$ as terminals, and we denote $T'=T\cap V(H)$. Clearly, set $T'$ remains a $(\hat D,\rho)$-pseudocut for $H$ in the remainder of the algorithm.

%Assume first that $64W^{\eps}\rho\mu^2\leq \frac{W}{4\mu^2}$, and so $|T|\leq\frac{W}{4\mu^2}$
The algorithm for Stage 3 is almost identical to that from \cite{APSP-previous}, with only slight technical differences, due to the slightly different definition of a pseudocut, and the specifics of the statement of \Cref{assumption: alg for recdynnc}.
We start by providing a high-level overview of the algorithm.
We define a modified graph $\tH$, that will be more convenient for us to use. Intuitively, graph $\tH$ is obtained from graph $H$ by ``splitting'' every terminal $t\in T$ into a number of ``fake terminals'', that we refer to as \emph{copies of $t$}. We will ensure that, for every pair $t,t'\in T$ of terminals, if there is a short path $P$ connecting $t$ to $t'$ in $H$, such that no inner vertices of $P$ are terminals, then there is a path $P'$ connecting a copy of $t$ to a copy of $t'$ in $\tH$, of a similar length. However, graph $\tH$ will have one additional useful property: for every regular vertex $v$ of $\tH$, $B_{\tH}(v,\hat D)$ only contains a relatively small number of regular vertices (from the definition of a pseudocut, this property holds for graph $H\setminus T$, but it does not necessarily hold for graph $H$). This property will allow us to construct a collection $\cset$ of vertex-induced subgraphs of $\tH$, each of which contains at most $W$ regular vertices, such that, for every fake terminal $\tilde t$, there is some graph $C\in \cset$, with $B_{\tH}(\tilde t,  \hat D)$ contained in $C$. While the graphs in $\cset$ are not necessarily disjoint, we will ensure that every regular vertex of $H$ belongs to few such graphs. 
For all $0\le i\leq \log D^*$, let $D_i=2^i$.
Since each graph $C\in \cset$ contains at most $W$ regular vertices,
for all $0\leq i\leq \log D^*$, we can use the algorithm from \Cref{assumption: alg for recdynnc} in order to solve the \recdynnc problem on graph $C$, with distance parameter $D_i$. We denote by $  \rset_i^C$ the resulting collection of clusters that the algorithm maintains.

Next, we construct another dynamic graph $\hat H$, that we call a \emph{contracted} graph. The regular vertices of $ \hat H$ correspond to the terminals of $T'$, while supernodes correspond to the clusters of $\bigcup_{C\in \cset}\bigcup_{i\geq 0}  \rset_i^C$. We will show that distances between pairs of terminals in $\hat H$ are approximately equal to the distances between the same terminals in graph $H$.
Once graph $\hat H$ is initialized, as the clusters in the neighborhood covers $\set{  \rset_i^C}_{C\in \cset,i\geq 0}$ evolve, graph $\hat H$ will undergo corresponding updates, that will be implemented via valid update operations.
 Since $|T|<W$, we can again use the algorithm from \Cref{assumption: alg for recdynnc}, in order to maintain a neighborhood cover $\hat  \rset$ for the contracted graph $\hat H$, with distance parameter roughly comparable to $D^*$. This data structure allows us to correctly establish when the distance between a pair of terminals becomes sufficiently large, so we can raise flag $F_H$ whenever this happens. The data structure also allows us to respond to \spquery queries between pairs of terminals. 
 
 In addition to the above data structures, we will maintain an \EST in graph $H$, that is rooted at the vertices of $T'$. This data structure will allow us to correctly establish when a regular vertex is too far from the terminals, and to support \spquery queries between pairs of regular vertices of $H$.
 
 This completes the high-level description of the algorithm for Stage 3.
 We now describe all data structures that we maintain. We start with the modified graph $\tilde H$, and then describe its decomposition into a collection $\cset$ of clusters. We then define the contracted graph $\hat H$ and analyze its properties. Lastly, we provide all additional data structures that we maintain, analyze the total update time of the algorithm, and provide an algorithm for responding to \spquery queries. We let $\tset$ denote the time horizon associated with Stage 3, and whenever we refer to ``initial'' graph $H$, we refer to graph $H$ at the beginning of Stage 3.

\subsubsection{Modified Graph $\tH$ and Its Properties}
At the beginnning of Stage 3, we construct a modified graph $\tH$ from the initial graph $H$, as follows. We process every terminal $t\in T$ one by one. When terminal $t$ is processed, for ever edge $e\in \delta_H(t)$, we subdivide $e$ with a new vertex $\tilde t_e$, that we call a \emph{copy} of $t$; the lengths of the two resulting edges are both set to be $\ell_H(e)$. Once very edge incident to $t$ in $H$ is subdivided, we obtain a collection $F(t)=\set{\tilde t_e\mid e\in \delta_H(t)}$ of copies of $t$; since the dynamic degree bound for $H$ is $\mu$, $|F(t)|\leq \mu$ must hold. We denote $F=\bigcup_{t\in T}F(t)$, and we refer to the vertices of $F$ as \emph{fake terminals}.  We then delete all terminals of $T$ from the resulting graph, obtaining the initial modified graph $\tilde H$. 
The fake terminals and the regular vertices of $V(H)\setminus T$ become regular vertices in graph $\tH$, while the supernodes of $H$ remain supernode vertices for graph $\tH$. Clearly, every edge of $\tH$ connects a regular vertex to a supernode.

As graph $H$ undergoes valid update operations, we perform similar valid update operations in graph $\tilde H$, that we describe below. Throughout, we will ensure that, for every terminal $t\in T$, for every edge $e=(t,u)$ that is incident to $t$ in the current graph $H$, there is a corresponding edge $e'$ in the current graph $\tH$ connecting a copy of $t$ to $u$, whose length is $\ell_H(e)$. This correspondence is a bijection: every edge incident to a terminal in $H$ corresponds to a unique edge incident to a copy of the same terminal in $F$ and vice versa. We also maintain bi-directional pointers between such pairs of edges. 
Additinally, if $e$ is an edge of $H$ whose endpoints are non-terminal vertices, then $e$ must also be present in $\tilde H$. Similarly, every edge of $\tilde H$ whose endpoints are not in $F$ must be present in $H$.
At the beginning of the algorithm, if $e=(t,u)$ is an edge of $H$ that is incident to a terminal $t\in T$, then its corresponding edge in $\tH$ is the edge $(\tilde t_e,u)$, that was obtained by subdividing $e$. We now show an algorithm that, given a valid update operation $\sigma\in \Sigma$ for graph $H$, performs a corresponding update in graph $\tH$.

Let $\sigma\in \Sigma$ be a valid update operation for graph $H$.
Assume first that $\sigma$ is the deletion of an edge $e=(u,v)$, where $u$ is a supernode. If $v\not\in T$, then we delete edge $e$ from $\tH$ as well. Otherwise, we locate the unique edge $e'$ that corresponds to $e$ in $\tH$; this edge must connect a copy of $v$ to $u$. We delete edge $e'$ from $\tH$.
Assume now that $\sigma$ is the deletion of an isolated vertex $v$ from $H$. If $v\not \in T$, then we delete $v$ from $\tH$. Otherwise, we delete every copy of terminal $v$ from $\tH$.

Lastly, assume that $\sigma$ is a supernode-splitting operation, that is applied to supernode $u\in V(H)$, with a set $E'\subseteq \delta_H(u)$ of edges. We will apply the corresponding supernode-splitting operation to supernode $u$ in $\tH$, but we may need to slightly modify the set $E'$ of edges. In order to do so, we construct a set $E''$ of edges in graph $\tH$, that correspond to the edges of $E'$. We consider the edges $e\in E'$ one by one. Let $e=(u,v)$ be any such edge, where $u$ is a supernode. If $v\not\in T$, then we include edge $e$ in $E''$. Otherwise, we include in $E''$ the unique edge $\tilde e$ of $\tH$ corresponding to $e$, that must connect a copy of $v$ to $u$. Once we process every edge in $E'$, we obtain the final set $E''$ of edges, and we apply a supernode-splitting operation to vertex $u$ in graph $\tH$, with the set $E''$ of edges. Recall that, as the result, a new supernode $u'$ is inserted into $\tH$. 
Similarly, the supernode-splitting operation in graph $H$ inserts the same supernode $u'$ into $H$. Consider now any edge $e=(u,v)\in E'$.
If $v\not\in T$, then edge $(u',v)$ is inserted into both $H$ and $\tH$. Assume now that $v\in T$. Recall that edge $\tilde e$ of $\tH$, that corresponds to edge $e$, was added to set $E''$; this edge connects some copy $t'$ of terminal $v$ to $u$. Edge $e'=(v,u')$ was inserted into graph $H$, and edge $\tilde e'=(t',u')$ was inserted into graph $\tH$ via the supenode-splitting operations. We make $\tilde e'$ be the edge of $\tilde H$ corresponding to edge $e'$ of $H$, and we add pointers between the two edges in both directions.

This completes the definition of the dynamic graph $\tilde H$. Note that $\tilde H$ is a bipartite graph, where one side of the bipartition contains all supernodes of $H$, and the other side contains regular vertices of $V(H)\setminus T$, and the fake terminals. In other words, we will view the fake terminals as regular vertices of $\tilde H$.
Therefore, we can view $\tilde \iset=\left(\tH,\set{\ell(e)}_{e\in E(\tH)},D\right )$ as a valid input structure, with graph $\tH$ undergoing an online sequence $\tilde \Sigma$ of valid update operations. It is easy to verify that the dynamic degree bound of $(\tH,\tilde \Sigma)$ remains at most $\mu$. The number of regular vertices in $\tH$ is bounded by the number of regular vertices in $H$ plus $|F|\leq \mu\cdot |T|\leq \frac{W}{4\mu}$. Therefore, the total number of regular vertices in $\tH$ is at most $N^0(H)+ \frac{W}{4\mu}$.
Since $(H,\Sigma)$ has dynamic degree bound $\mu$, the number of edges that are ever present in $H$ is at most $N^0(H)\cdot \mu\leq \hat W\cdot \mu$, and so the number of edges that are ever present in $\tH$ is also at most $\hat W\cdot \mu$. %The total number of vertices that are ever present in $H$ is therefore bounded by $2\hat W$.
%Since every edge of $\tH$ corresponds to some edge of $H$, the total number of edges that are ever present in $H$ is at most $W'\mu$. The total number of vertices that are ever present in $\tH$ is then bounded by $2W'\mu+|F|\leq 2W'\mu+|T|\mu$. 

From the discussion so far, there is a deterministic algorithm, that, given the initial graph $H$, initializes the initial graph $\tH$, in time $O(|E(H)|)$. Given the online sequence $\Sigma$ of valid update operations for $H$, the algorithm also produces the corresponding online update sequence $\tilde \Sigma$ for $\tH$, in time that is asymptotically bounded by the length of the description of $\Sigma$.
%The following two observations explore some useful properties of the modified graph $\tH$.

\iffalse
Consider any time $\tau \in \tset$ during the time horizon. We can equivalently obtain a graph $\tilde H\attime$ from graph $H\attime$, as follows. We process every terminal $t\in T$ one by one. Consider any such terminal $t$, and let $\set{e_1,\ldots,e_r}$ be the set of all edges that were incident to terminal $t$ in the original graph $H\attime[0]$. For all $1\leq i\leq r$, denote $e_i=(t,u_i)$. From the definition of valid update operations, if $u$ is a supernode that is a neighbor of $t$ in graph $H\attime$, then the ancestor-supernode of $u$ at time $0$ must be one of the vertices $u_1,\ldots,u_r$. This is since edges may only be inserted into $H$ via supernode-splitting operations. Let $U\attime(t)$ denote the set of all supernodes that are the neighbors of terminal $t$ in $H\attime$. Then we can define a partition $(U_1,\ldots,U_r)$ of $U\attime(t)$, where for all $1\leq i\leq r$, set $U_i$ contains all supernodes $u$ with $\anc\attime[0](u)=u_i$. We delete terminal $t$ from graph $H\attime$, and insert $r$ copies $\tilde t_{e_1},\ldots,\tilde t_{e_r}$ of $t$ instead. For all $1\leq i\leq r$, for every supernode $u\in U_i$, we add an edge $(\tilde t_{e_i},u)$ of length $\ell_H(e_i)$ to the graph. Once we process every terminal $t\in T$, we obtain graph $\tH\attime$, which is identical to the dynamic graph that is obtained from $\tilde H\attime[0]$ by applying the update sequence $\tilde \Sigma$ to it, up to time $\tau$. 
\fi
The following two observations easily follow the definition of graph $\tH$, and the bijection between the edges of $E(H)$ and $E(\tilde H)$ that we defined.

\begin{observation}\label{obs: short paths in modified graph}
Let $\tau\in \tset$ be any time during the time horizon, let $t,t'\in T$ be a pair of terminals, and let $P$ be a path in graph $H\attime$ connecting $t$ to $t'$, whose length is denoted by $D'$, such that no inner vertex of $T$ is a terminal. Then there is a path $P'$ in graph $\tH\attime$ connecting a copy of $t$ to a copy of $t'$, such that the length of $P'$ is at most $D'$.
\end{observation}
\begin{proof}
	Denote $P=(t,z_1,z_2,\ldots,z_a,t')$. Observe that $z_1$ and $z_2$ must be supernodes, that we denote by $u$ and $u'$, respectively. For all $1\leq i< a$, we let $e_i=(z_i,z_{i+1})$. From the definition of graph $\tH$, each such edge $e_i$ is present in graph $\tH\attime$. Moreover, there is an edge $e_0$ in $\tilde H\attime$, connecting some copy $\tilde t$ of $t$ to supernode $u=z_1$, and there is an edge $e_r$, connecting supernode $u'=z_a$ to some copy $\tilde t'$ of $t'$. By concatenating edges $e_0,e_1,\ldots,e_r$, we obtain a path $P'$ in graph $\tH$, that connects a copy $\tilde t$ of terminal $t$ to a copy $\tilde t'$ of terminal $t'$. The length of the path is the same as the length of path $P$.
\end{proof}

The next observation is immediate from our discussion.

\begin{observation}\label{obs: paths from modified back}
	Let $\tau\in \tset$ be any time during the time horizon, let $t,t'\in T$ be a pair of terminals, and let $P'=(\tilde t,z_1,z_2,\ldots,z_a,\tilde t')$ be a path in graph $\tilde H\attime$ connecting a copy $\tilde t$ of terminal $t$, to a copy $\tilde t'$ of  terminal $t'$, whose length is denoted by $D'$. Consider a sequence $S=(t,z'_1,z'_2,\ldots,z'_a, t')$  of vertices of $H\attime$, that is obtained from  the sequence $\tilde S=(\tilde t,z_1,z_2,\ldots,z_a,\tilde t')$ of vertices of $ \tilde H\attime$, by replacing every fake terminal with the corresponding original terminal from $T$. Then sequence $S$ defines a path in graph $H\attime$, that connects $t$ to $t'$, and has length $D'$.
\end{observation}

Lastly, the following observation bounds the number of regular vertices in balls of radius $\hat D/2$ in graph $\tilde H$.

\begin{observation}\label{obs: few regular vertices in balls}
Let $v$ be a regular vertex in $\tH\attime[0]$. If $v\not\in F$, then the number of regular vertices in $B_{\tH\attime[0]}(v,\hat D)$ is bounded by $W/2$, and if $v\in F$, then the number of regular vertices in $B_{\tH\attime[0]}(v,\hat D/2)$ is bounded by $W/2$.
\end{observation}
\begin{proof}
	Let $v$ be a regular vertex of $\tH\attime[0]$. Assume first that $v\not\in F$. Denote $S=B_{\tH\attime[0]}(v,\hat D)$, and let $G$ be the subgraph of $\tH\attime[0]$ induced by $S$. Let $F'=F\cap S$. Since $|T|\leq \frac{W}{4\mu^2}$ holds, we get that $|F'|\leq |F|\leq |T|\cdot \mu\leq \frac{W}{4\mu}$.
	
	From the definition of graph $\tH\attime[0]$, every vertex of $F$ has degree $1$ in $\tH\attime [0]$. 
	Let $G'=G\setminus F$. Then the distance from $v$ to every vertex of $G'$ remains at most $\hat D$ in $G'$. 
	
	Let $H'=H\attime[0]\setminus T$. From the definition of graph $\tH$, $G'\subseteq H'$, and so $V(G')\subseteq B_{H'}(v, \hat D)$.
	
	From the definition of a pseudocut, $B_{H'}(v,\hat D)$ contains at most $\frac{\hat W}{\rho}\leq \frac{W}{4}$ regular vertices. We conclude that the number of regular vertices in $S\setminus F'=V(G')$ is bouded by $\frac{W}{4}$. Since $|F'|\leq \frac{W}{4}$, we conclude that the number of regular vertices in  $S$ is bounded by $\frac W 2$.
	
	Assume now that $v\in F$, and let $u$ be the unique neighbor of $v$ in $\tH$. If all neighbors of $u$ in $\tH$ lie in $F$, then $B_{\tH}(v,\hat D/2)\subseteq F$ (as all vertices of $F$ have degree $1$ in $\tH$), and so the number of regular vertices in $B_{\tH}(v,\hat D/2)$ is bounded by $|F|\leq \frac{W}{4}$. Otherwise, at least one regular vertex $v'\not\in F$ is a neighbor of $u$. Since the length of every edge in $\tH$ is bounded by $D\leq D^*<\hat D/8$, we get that $B_{\tH}(v,\hat D/2)\subseteq B_{\tH}(v',\hat D)$, and so the number of regular vertices in $B_{\tH}(v,\hat D)$ is bounded by $W/2$.
\end{proof}

%==============================================
%==============================================
%==============================================
%==============================================
%==============================================
%==============================================
\subsubsection{Constructing the Collection $\cset$ of Subgraphs of $\tilde H$}

We provide an algorithm that computes a collection $\cset$ of vertex-induced subgraphs (that we call clusters) of the initial graph $\tH\attime[0]$, that have some useful properties. First, we ensure that the number of regular vertices in each cluster of $\cset$ is bounded by  $W$. We also ensure that every regular vertex of $\tH\attime[0]$ lies in at most $W^{\eps/8}$ such clusters. Lastly, we ensure that, for every regular vertex $v\in V(\tH\attime[0])$, some cluster $C\in \cset$ contains all vertices of $B_{\tH\attime[0]}(v,D^*)$. The main result of this subsection is summarized in the following claim.

\begin{claim}\label{claim: cutting tilde H}
	There is a deterministic algorithm that computes a collection $\cset$ of vertex-induced subgraphs (clusters) of graph $\tH\attime[0]$ with the following properties:
	
	\begin{itemize}
		\item for every cluster $C\in \cset$, the number of regular vertices (including fake terminals) in $C$ is at most $W$;
		
	%	\item the total number of regular vertices in all clusters $C\in \cset$ is at most $(W'\cdot \mu)^{1+\eps}$; 
		
		\item every regular vertex $v\in V(\tH\attime[0])$ belongs to at most $W^{\eps/8}$ graphs in $\cset$; and
		
		\item for every regular vertex $v$ of $\tH\attime[0]$, there is at least one cluster $C\in \cset$, with $B_{\tH\attime[0]}(v,D^*)\subseteq V(C)$.
	\end{itemize}

Additionally, the algorithm computes, for every regular vertex $v$ of $\tH\attime[0]$ a cluster $C(v)\in \cset$ with $B_{\tH\attime[0]}(v,D^*)\subseteq V(C(v))$.
The running time of the algorithm is $O\left ((\hat W\cdot \mu)^{1+\eps}\right )$.
\end{claim}

\begin{proof}
	We apply the algorithm from \Cref{claim: cutting G} to graph $G=\tH\attime[0]$, where the set of terminals is the set of regular vertices of $\tH\attime[0]$ (which includes the set $F$ of fake terminals). The precision parameter $\eps$ remains unchanged, and the distance parameters are $D=D^*$ and $D'=\hat D/2$. Since $\hat D=\frac{2^{14}\cdot D^*}{\eps^2}$, we are guaranteed that $D'\geq \frac{2^{13}\cdot D}{\eps^2}$. Let $\cset$ be the collection of clusters that the algorithm from  \Cref{claim: cutting G} computes.
	
	The algorithm from \Cref{claim: cutting G} ensures that, for every cluster $C\in \cset$, there is some regular vertex $v$ of $\tH\attime[0]$, with $V(C)\subseteq B_{H\attime[0]}(v,\hat D/2)$. From \Cref{obs: few regular vertices in balls}, the number of regular vertices in each cluster $C\in \cset$ is bounded by $W/2$. Recall that the number of regular vertices in graph $\tH$ is bounded by $|F|+N^0(H)\leq \hat W\cdot \mu\leq W^{1.6}$. The algorithm from \Cref{claim: cutting G} guarantees that every regular vertex of $\tH$ belongs to at most $\frac{64 (W^{1.6})^{\eps/16}}{\eps}\leq W^{\eps/8}$ clusters of $\cset$. It also computes, for every regular vertex $v$ of $\tH\attime[0]$, a cluster $C(v)\in \cset$, with $B_{\tH\attime[0]}(v,D^*)\subseteq V(C(v))$. Since the number of regular vertices in $\tH\attime[0]$ is bounded by $\hat W\cdot \mu$, and $|E(\tH\attime[0])|\leq \hat W\cdot \mu$, the running time of the algorithm is bounded by 
	$O\left ((\hat W\cdot \mu)^{1+\eps}\right )$.
\end{proof}

Let $\cset$ be the collection of clusters that the algorithm from \Cref{claim: cutting tilde H} computed. As graph $\tH$ undergoes the sequence $\tilde \Sigma$ of valid update operations, we update each of the clusters $C\in \cset$ accordingly, as described in Section \ref{subsubsec: updating clusters}. Therefore, online sequence $\tilde \Sigma$ of valid update operations to graph $\tilde H$ naturally defines, for every cluster $C\in \cset$, an online sequence $\Sigma_C$ of valid update operations to cluster $C$. Furthermore, since every regular vertex of $H$ belongs to at most $W^{\eps/8}$ graphs in $\cset$, it is easy to see that there is a deterministic algorithm that, given an update $\sigma$ ot graph $\tilde H$, produces the corresponding updates to graphs of $\cset$, whose total running time is bounded by the length of the description of sequence $\tilde \Sigma$ times $O(W^{\eps/8})$. 

Consider now some cluster $C\in \cset$. We have now defined a valid input structure  $\iset_C=\left(C,\set{\ell(e)}_{e\in E(C)},D \right )$ associated with cluster $C$, and an online sequence $\Sigma_C$ of valid update operations that cluster $C$ undergoes. It is easy to verify that, since the dynamic degree bound of $(H,\Sigma)$ is bounded by $\mu$, the dynamic degree bound of $(C,\Sigma_C)$ is bounded by $\mu$ as well. Recall that the number of regular vertices in $C$ is bounded by $W$. Notice that, from \Cref{obs: maintain ball covering property},
for every regular vertex $v$ of $\tH\attime[0]$, if $C(v)\in \cset$ is the cluster computed by \Cref{claim: cutting tilde H}, then 
 at all times $\tau\in \tset$, $B_{\tH}(v,D^*)\subseteq V(C(v))$ holds.

Consider again some cluster $C\in \cset$.
For all $0\leq i\leq \log D^*$, we can define a valid input structure 
$\iset^C_i=\left(C_i,\set{\ell(e)}_{e\in E(C_i)},D_i \right )$ associated with graph $C$ and distance parameter $D_i=2^i$ in a natural way: initially, graph $C_i$ is obtained from $C$ by deleting all edges whose length is greater than $D_i$. As graph $C$ undergoes valid update operations, we perform similar update operations in graph $C_i$, but we ignore all edges whose length is greater than $D_i$; such edges do not need to be deleted from $C_i$, and we never insert such edges via supernode-splitting operations.  We can now view $\iset_i^C$, together with the online sequence of valid update operations that graph $C_i$ undergoes, as an input to the \recdynnc problem. Since the number of regular vertices in $C_i$ is bounded by $W$, we can apply the algorithm from \Cref{assumption: alg for recdynnc} to this instance of \recdynnc. We denote by $\rset_i^C$ the collection of clusters (the neighborhood cover of $C_i$ with distance parameter $D_i$) that this data structure maintains, and we denote the data structure itself by $\DS_i^C$. For every regular vertex $v\in V(C)$, we denote by $\coveringcluster_i^C(v)$ the cluster $\coveringcluster(v)$ that the data structure maintains. We also refer to queries $\spquery$ that the data structure supports as $\spquery_i^C$.
We denote all data structures $\set{\DS_i^C}_{0\leq i\leq \log D^*}$ by $\DS^C$, and $\rset^C=\bigcup_{i=0}^{\log D^*}\rset_i^C$.
Lastly, we denote by $\DS$ all data structures in $\set{DS^C\mid C\in \cset}$, and we denote by $\rset=\bigcup_{C\in \cset}\rset^C$.

Let $N^0(C)$ denote the number of regular vertices in the initial graph $C$.
From \Cref{assumption: alg for recdynnc}, for all $0\leq i\leq \log D^*$ the total update time 
of data structure $\DS_i^C$ is bounded by $O\left (N^0(C)\cdot W^{\delta}\cdot D_i^3\cdot \mu^c\right )$. The total update time of data structure $\DS^C$ is then bounded by  $O\left(N^0(C)\cdot W^{\delta}\cdot (D^*)^3\cdot \mu^c\right )$. Since every regular vertex of $H$ may lie in at most $W^{\eps/8}$ graphs in $\cset$, we get that $\sum_{C\in \cset}N^0(C)\leq \hat W\cdot W^{\eps/8}$, and so the total update time of data structure $\DS$ is bounded by:

\[ \sum_{C\in \cset}O\left(N^0(C)\cdot W^{\delta}\cdot (D^*)^3\cdot \mu^c\right )\leq O\left(\hat W\cdot W^{\delta+\eps/8}\cdot (D^*)^3\cdot \mu^c\right ). \]

For convenience, we denote $\Delta(W)$ by $\Delta$.
Recall that the algorithm from \Cref{assumption: alg for recdynnc} ensures that, for all $C\in \cset$ and $0\leq i\leq \log D^*$, every regular vertex of $C$ may lie in at most $\Delta$ clusters of $\rset^C_i$ during the time horizon $\tset$. Therefore, every regular vertex of $C$ may belong to at most $2\Delta\log D^*$ clusters of $\rset^C$ over the course of the time interval $\tset$. Since every regular vertex of $\tH$ lies in at most $W^{\eps/8}$ clusters of $\cset$, we conclude that, for every regular vertex $v\in V(\tH)$, the number of clusters of $\rset$ to which it ever belongs during time horizon $\tset$ is bounded by $2W^{\eps/8}\cdot\Delta\log D^*$. 

\subsubsection{Contracted graph $\hat H$}
\label{subsubsec contracted}

In this subsection we define a dynamic graph $\hat H$, that we call a \emph{contracted graph}, and analyze some of its properties. The definition of the graph and the analysis are very similar to those in \cite{APSP-previous}, with some small technical differences.

Consider any time $\tau\in \tset$ during the time horizon. Graph $\hat H\attime$ is defined as follows. The set of vertices of $\hat H\attime$ is the union of two subsets: set of regular vertices and set of supernodes. The set of regular vertices of $\hat H\attime$ is the set $T'=T\cap V(H\attime)$ of terminals. The set of supernodes is $\set{u(R)\mid R\in \rset\emph{ and } V(R)\cap F\neq\emptyset}$. 
For every cluster $R\in \rset$, if $R\in \rset^C_i$ for some cluster $C\in \cset$ an integer $0\leq i\leq \log D^*$, then we say that the \emph{scale} of cluster $R$ is $i$, and we denote $\scale(R)=i$.

We now define the edges of $\hat H\attime$. Consider some cluster $R\in \rset$, and assume that $\scale(R)=i$. For every terminal $t\in T$, such that some copy of $t$ lies in $R$, we add an edge $(t,u(R))$ of length $2^i$ to graph $\hat H\attime$.
For every edge $e=(t,u(R))$ of graph $\hat H$, we maintain a list $L(e)$ of all fake terminals $\tilde t\in V(R)$ that are copies of terminal $t$, and a counter $n_e=|L(e)|$.

\paragraph{Maintaining graph $\hat H$.}
At the begining of Stage 3, after we compute the modified graph $\tilde H$, the initial set $\cset$ of clusters of $\tilde H$, and the initial collection $  \rset$ of clusters, we initialize the graph $\hat \hset\attime[0]$, the lists $L(e)$ and the counters $n_e$ for all edges $e$ of $\hat \hset\attime[0]$ in a straightforward way. The time required to initialize the graph is asymptotically bounded by the time that is required to compute the initial collections $\cset$, $  \rset$ of clusters.

Next, we show that graph $\hat H$ can be maintained correctly via valid update operations. Graph $\hat H$ may only need to be updated in one of the following cases: (i) some terminal $t\in T$ is deleted from $H$ via isolated vertex deletion update; or (ii) some fake terminal $\tilde t$ is deleted from some cluster $R\in \rset$; or (iii) a new cluster is inserted into set $\rset$ via a cluster-splitting update. We now consider each of these events in turn.

Assume first that some terminal $t\in T$ is deleted from $H$ via isolated vertex deletion update. Then terminal $t$ is currently an isolated vertex in $H$, and so all fake terminals that are copies of $t$ are also isolted vertices in $\tilde H$. All such fake terminals are deleted from graph $\tilde H$, and from clusters of $\cset$, which, in turn, leads to the deletion of all copies of $t$ from the clusters of $\rset$. We delete all edges incident to $t$ from graph $\hat H$, and then delete terminal $t$ via isolated vertex deletion update operation.

Assume now that some fake terminal $\tilde t$ is deleted from some cluster $R\in \rset$, and assume that $\tilde t$ is the copy of some terminal $t\in T$. Notice that edge $e=(t,u(R))$ currently lies in graph $\hat H$. If $n_e>1$, then another copy of $t$ still lies in cluster $R$. In this case, no further updates to graph  $\hat H$ are needed, but we decrease the counter $n_e$ by $1$, and we delete terminal $t$ from list $L(e)$. Otherwise, we delete the edge $(t,u(R))$ from graph $\hat H$. If vertex $u(R)$ becomes isolated, then we delete this vertex from $\hat H$ as well, via isolated vertex deletion update.

Lastly, assume that a new cluster $R'$ was inserted into $\rset$. From the definition of the set $\rset$ of clusters and the \recdynnc problem, this may only happen if, for some cluster $C\in \cset$, distance scale $0\leq i\leq D^*$, a cluster-splitting update was applied to cluster $R\in \rset^C_i$, so $R'\subseteq R$ must hold. If $R'$ contains no fake terminals, then no further changes to graph $\hat H$ are required. Assume  now that $R'$ contains at least one fake terminal, and let $F'=F\cap V(R')$. Then $F'\subseteq R$ must hold. Let $T'\subseteq T$ be the set of all terminals whose copies lie in $F'$. Then for every terminal $t\in T'$, edge $(t,u(R))$ is currently present in graph $\hat H$. We let $E'=\set{(t,u(R))\mid t\in T'}$ be a subset of edges incident to vertex $u(R)$. We then apply supernode-splitting update to supernode $u(R)$, with the set $E'$ of edges. As the result, we insert a new supernode $u(R')$ that corresponds to the new cluster $R'$ into $\hat H$, and we insert an edge connecting $u(R')$ to every terminal in $T'$. We also initialize the lists $L(e)$ and the counters $n_e$ of all such newly inserted edges $e$.

It is easy to verify that the algorithm maintains the graph $\hat H$, the lists $L(e)$, and the counters $n_e$ for edges $e\in E(\hat H)$ correctly. The running time of the algorithm is asymptotically bounded by the total update time of the data structure that maintains the collection $\rset$ of clusters.

We have now defined a valid input structure $\hat\iset=(\hat H,\set{\ell(e)})_{e\in E(\hat H)},D^*)$, that undergoes a sequence $\hat \Sigma$ of valid update operations. We now bound the dynamic degree bound of $(\hat \iset,\hat \Sigma)$. Recall that for every terminal $t\in T$, there are at most $\mu$ fake terminals that are copies of $t$ in $\tilde H$. As we have shown already, for every regular vertex $v\in V(\tH)$, the number of clusters of $\rset$ to which $v$ ever belongs during time horizon $\tset$ is bounded by $2W^{\eps/8}\cdot\Delta\cdot \log D^*$. Therefore, for every terminal $t\in T$, the total number of clusters of $\rset$ to which copies of $t$ ever belong during the time horizon $\tset$ is bounded by $2W^{\eps/8}\cdot\Delta\cdot \mu\cdot \log D^*$. If we denote by $\hat \mu$ the dynamic degree bound of $(\hat \iset,\hat \Sigma)$, then $\hat \mu\leq 2W^{\eps/8}\cdot\Delta\cdot \mu\cdot \log D^*$. Since the number of regular vertices in $\hat H$ is bounded by $W$, we can again apply the algorithm from  \Cref{assumption: alg for recdynnc}  to instance $(\hat \iset,\hat \Sigma)$ of the \recdynnc problem. But before we do so, we discuss some useful properties of graph $\hat H$, namely that it approximately preserves distances between terminals.

\paragraph{Distance Preservation.}

We start by showing that for every pair $t,t'\in T$ of terminals, if $\dist_H(t,t')\leq D^*$, then $\dist_{\hat H}(t,t')\leq 4D^*$.

\begin{claim}\label{claim: short distance preservation}
	Let $\tau\in \tset$ be any time during the time horizon, and let $t,t'\in T$ be a pair of terminals with $\dist_{H\attime}(t,t')= D'$, where $D'\leq D^*$. Then $\dist_{\hat H\attime}(t,t')\leq 4D'$.
\end{claim}
\begin{proof}
	We fix some time $\tau\in \tset$. Whenever we refer to dynamic graphs or data structures in this proof, we refer to the at time $\tau$, unless stated otherwise. Let $t,t'\in T$ be a pair of terminals with $\dist_{H\attime}(t,t')= D'$, for some $D'\leq D^*$, and let $P$ be a path of length $D'$ connecting $t$ to $t'$ in $H$. We can assume w.l.o.g. that path $P$ contains no terminals as inner vertices, since otherwise we can partition path $P$ into subpaths, such that the endpoints of each subpath are terminals, and no inner vertices of the subpath are terminals. By applying the claim to each subpath separately and concatenating the resulting paths, we obtain the desired path connecting $t$ to $t'$ in $\hat H$, of length at most $4D'$. Therefore, we assume from now on that $P$ contains no terminals as inner vertices.
	
	Let $e=(t,u)$ and $e'=(u',t')$ be the first and the last edges of $P$, respectively. From the construction of graph $\tilde H$, there is a copy $\tilde t$ of $t$, and an edge $\tilde e=(\tilde t,u)$ in graph $\tilde H$, whose length is $\ell_H(e)$. Similarly, there is a copy $\tilde t'$ of $t'$, and an edge $\tilde e'=(\tilde t',u')$ in $\tilde H$, whose length is $\ell_H(e')$. All other edges of $P$ are present in graph $\tilde H$. Therefore, there is a path $\tilde P$ in graph $\tilde H$, whose length is $D'$, that connects $\tilde t$ to $\tilde t'$. Since, from \Cref{obs: no dist increase},  valid update operations cannot decrease distances between regular vertices, $\dist_{\tH\attime[0]}(t,t')\leq D'\leq D^*$ held.
	From \Cref{claim: cutting tilde H}, there is a cluster $C\in \cset$ with $B_{\tH\attime[0]}(\tilde t,D^*)\subseteq C$. From \Cref{obs: maintain ball covering property}, at time $\tau$, $B_{\tH\attime}(\tilde t,D^*)\subseteq C\attime$ continues to hold. Therefore, $P'\subseteq C$ holds at time $\tau$, and $\dist_{C\attime}(\tilde t,\tilde t')\leq D'$.
	
	Let $0\leq i\leq \log D^*$ be the smallest integer for which  $D_i\geq D'$ holds, so  $D'\leq D_i\leq 2D'$. Since $D^*$ is an integral power of $2$, $D_i\leq D^*$ holds. Consider the cluster $R=\coveringcluster^C_i(\tilde t)$. This cluster must contain $B_{\tH}(\tilde t,D_i)$, and so it must contain both $\tilde t$ and $\tilde t'$. Therefore, graph $\hat H$ contains edges $(t,u(R))$ and $(u(R),t')$ of length $D_i$ each. By concatenating the two edges, we obtain a path in graph $\hat H$, connecting $t$ to $t'$, whose length is $2D_i\leq 4D'$.
\end{proof}

Notice that, from the above claim, whenever we identify a pair $t,t'$ of terminals with $\dist_{\hat H}(t,t')>4D^*$, we can raise flag $F_H$ with the pair $t,t'$ of witness vertices, since we are then guaranteed that $\dist_H(t,t')>D^*$ holds.

Next, we provide an algorithm that, given a path connecting a pair $t,t'$ of terminals in graph $\hat H$, returns a path of comparable length connecting $t$ to $t'$ in graph $H$. This algorithm will allow us to support $\spquery$ queries between pairs of vertices in $H$.
Recall that we denoted by $\alpha=\alpha(W)$ the approximation factor achieved by the algorithm from \Cref{assumption: alg for recdynnc}.

\begin{claim}\label{claim: transform path from contracted}
	There is a deterministic algorithm, that, given a pair $t,t'\in T$ of terminals, and a path $\hat P$ connecting $t$ to $t'$ in the current graph $\hat H$ of length $D'$, returns a path $P$ of length at most $\alpha\cdot D'$, connecting $t$ to $t'$ in $H$, in time $O(|E(P)|)$.
\end{claim}
\begin{proof}
	We assume that we are given a pair $t,t'\in T$ of terminals, and a path $\hat P$ connecting $t$ to $t'$ in the current graph $\hat H$ of length $D'$. We denote the sequence of vertices on path $\hat P$ by $(t=t_1,u(R_1),t_2,u(R_2),\ldots,t_{q},u(R_q),t_{q+1}=t')$.
	For all $1\leq j\leq q$, we denote $i_j=\scale(R_j)$. Then $D'=2\sum_{j=1}^q2^{i_j}$. We now provide an algorithm that, for all $1\leq j\leq q$, computes a path $Q_j$ in graph $H$, connecting $t_j$ to $t_{j+1}$, such that the length of $Q_j$ is at most $\alpha\cdot D_{i_j}$.
	
	Consider an integer $1\leq j\leq q$. Since edges $e=(t_j,u(R_j)),e'=(t_{j+1},u(R_j))$ are present in graph $\hat H$, there must be a copy $\tilde t_j$ of terminal $t_j$, and a copy $\tilde t_{j+1}$ of terminal $t_{j+1}$, that lie in cluster $R_j$. We can find vertices $\tilde t_j,\tilde t_{j+1}$ in time $O(1)$ using lists $L(e)$ and $L(e')$. We let $C\in \cset$ be the cluster for which $R_j\in \rset^C_{i_j}$. We execute query $\spquery^C_{i_j}$ in data structure $\DS^C_{i_j}$, to obtain a path $Q'_j$ in graph $C$, that connects $\tilde t_j$ to $\tilde t_{j+1}$, and has length at most $\alpha\cdot D_{i_j}$. The time required for processing the query is $O(|E(Q'_j)|)$. Since $C\subseteq \tilde H$, path $Q'_j$ also lies in graph $\tilde H$. We let $\tilde S$ be the sequence of vertices on path $Q'_j$, and we let $S$ be the sequence of vertices obtained from $\tilde S$ by replacing every fake terminal with the corresponding original terminal from $T$. From \Cref{obs: paths from modified back}, sequence $S$ of vertices defines a path in graph $H$ that connects $t_j$ to $t_{j+1}$, and has length at most $\alpha\cdot D_{i_j}$. We denote the resulting path by $Q_j$. Clearly, the time required to compute the path $Q_j$ is bounded by $O(|E(Q_j)|)$.
	
	By concatenating the paths $Q_1,\ldots,Q_r$, we obtain a path $P$ in graph $H$ that connects $t$ to $t'$. The length of the path is at most $\sum_{j=1}^q\alpha\cdot D_{i_j}\leq \alpha\cdot D'$. The running time of the algorithm is $O(|E(P)|)$.
\end{proof}
\subsubsection{Additional Data Structures}

We now describe the data structures that the algorithm for Stage 3 maintains, in addition to the modified graph $\tilde H$, the collections $\cset$ and $\rset$ of clusters, and the contracted graph $\hat H$.

\paragraph{Maintaining short distances between the terminals.}
Recall that we have defined a valid input structure
 $\hat\iset=(\hat H,\set{\ell(e)})_{e\in E(\hat H)},D^*)$, that undergoes a sequence $\hat \Sigma$ of valid update operations with dynamic degree bound $\hat \mu\leq 2W^{\eps/8}\cdot\Delta\cdot \mu\cdot \log D^*$, and that the number of regular vertices in the initial graph $\hat H$ is at most $W/(4\mu^2)$.
 We view $(\hat\iset, \hat \Sigma)$ as an instance of the \recdynnc problem, but we replace the distance parameter $D^*$ with $4D^*$. 
 
 We apply the algorithm from \Cref{assumption: alg for recdynnc} to this instance of \recdynnc, and we denote by $\hat \rset$ the collection of clusters of graph $\hat H$ that the algorithm maintain.
 For every cluster $\hat R\in \hat \rset$, we also maintain a list $\lambda_{\hat R}$ of all terminals $t\in T$, for which $\coveringcluster(t)=\hat R$. We also maintain two counters: a counter $\hat n_{\hat R}=|\lambda_{\hat R}|$, and a counter $\hat n'_{\hat R}=|T\cap V(\hat R)|$. We initialize the lists and the counters at the beginning of the algorithm, when the initial collection $\hat\rset$ of clusters is computed. Whenever, for some terminal $t\in T$, the algorithm from \Cref{assumption: alg for recdynnc} changes $\coveringcluster(t)$ from cluster $\hat R$ to cluster $\hat R'$, we delete $t$ from $\lambda_{\hat R}$, and we insert it into $\lambda_{\hat R'}$, and update the counters $\hat n_{\hat R}$ and $\hat n_{\hat R'}$ accordingly. Whenever a terminal $t$ is deleted from some cluster $\hat R$, we decrease the counter $\hat n'_{\hat R}$. Whenever a new cluster $\hat R''$ is inserted into $\hat \rset$, we set $\lambda_{\hat R''}=\emptyset$, $\hat n_{\hat R''}=0$, and we initialize the counter $\hat n'_{\hat R''}$ to $|T\cap V(\hat R'')|$. We denote by $\DS'$ the data structure that the algorithm from \Cref{assumption: alg for recdynnc} maintains, augmented with the algorithm for maintaining the lists $\lambda_{\hat R}$ and counters $\hat n_{\hat R}$, $\hat n'_{\hat R}$  for clusters $\hat R\in \hat\rset$. It is easy to see that the lists and the counters can be maintained without increasing the asymptotic total update time of the algorithm. Therefore, the total update time needed to maintain data structure $\DS'$ is bounded by:

 \[O\left (W^{1+\delta}\cdot (D^*)^{3}\cdot \hat \mu^c\right )
 \leq O\left (W^{1+\delta+c\eps/8}\cdot \Delta^c\cdot \mu^c \cdot (D^*)^{3}\cdot (\log D^*)^c\right ),
 \]

since $\hat \mu=2W^{\eps/8}\cdot\Delta\cdot \mu\cdot \log D^*$.

Recall that the algorithm from \Cref{assumption: alg for recdynnc}
ensures that every terminal $t\in T$ may lie in at most $\Delta=\Delta(W)$ clusters of $\hat \rset$ over the course of Stage 3 of the algorithm. Our goal is to ensure that, as long as $T\cap V(H)\ne \emptyset$, there is some cluster $\hat R\in \hat \rset$, that contains all terminals of $T\cap V(H)$, and that such a cluster can be computed in time $O(1)$. We denote by $T'=T\cap V(H)$ the set of terminals that currently lie in graph $H$.
Consider a cluster $\hat R\in \hat \rset$. We say that cluster $\hat R$ is \emph{special} if $\frac{\hat n'_{\hat R}}{\hat n_{\hat R}}\leq 2\Delta$. Note that, if $\hat R$ is a special cluster, then $\hat n_{\hat R}\neq 0$ and $\lambda_{\hat R}\neq \emptyset$ must hold. We use the following observation.

\begin{observation}\label{obs: special cluster exists}
	For all $\tau\in \tset$, if $T'\neq \emptyset$ holds at time $\tau$, then at least one cluster $\hat R\in \hat \rset$ is special.
\end{observation}
\begin{proof}
	Consider any time $\tau\in \tset$, such that $T'\neq \emptyset$ holds at time $\tau$, and assume for contradiction that no cluster of $\hat \rset$ is special. Observe that $\sum_{\hat R\in \hat \rset}\hat n_{\hat R}=|T'|$, since for every terminal $t\in T'$, there is a single cluster $\hat R=\coveringcluster (t)$. Since no cluster is special, we get that, for every cluster $\hat R\in \hat \rset$, $\hat n'_{\hat R}>2\Delta\cdot \hat n_{\hat R}$, and so $\sum_{\hat R\in \hat \rset}\hat n'_{\hat R}>2\Delta\cdot \sum_{\hat R\in \hat \rset}\hat n_{\hat R}\geq 2\Delta\cdot |T'|$. However, every terminal of $T'$ may lie in at most $\Delta$ sets of $\hat \rset$, so $ \sum_{\hat R\in \hat \rset}\hat n'_{\hat R}\leq \Delta\cdot |T'|$ must hold, a contradiction.
\end{proof}

Notice that we can maintain, for every cluster $\hat R\in \hat \rset$, a bit $b_{\hat R}$ that indicates whether cluster $\hat R$ is currently special, and we can also maintain a list $\hat \rset^S\subseteq \hat \rset$ of special clusters, without increasing the asymptotic running time of data structure $\DS'$. We will also maintain the cardinality of the set $T'=T\cap V(H)$ of terminals, that we denote by $\hat N$. From \Cref{obs: special cluster exists}, the list $\hat \rset^S$ of special clusters is always non-empty. Consider now a special cluster $\hat R\in \hat \rset^S$. We say that cluster $\hat R$ is \emph{heavy} if $\hat n'_{\hat R}\geq \frac{|T'|}{2}$, and we say that it is \emph{light} otherwise. We need the following observation.

\begin{observation}\label{obs: few heavy clusters}
	The total number of clusters $\hat R$ that ever belonged to $\hat \rset$, such that $\hat R$ was ever a heavy special cluster, is bounded by $4\Delta\log W$.
\end{observation}
\begin{proof}
	We partition the second stage of the algorithm into phases. Each phase lasts until the number of terminals in $T'$ decreases by at least factor $2$ since the beginning of the phase. Therefore, the number of phases is bounded by $\log |T|\leq \log W$.
	
	Consider now some phase, and denote the number of terminals in $T'$ at the beginning of the phase by $N'$. 
	Let $\Pi$ be the collection of pairs $(\hat R,t)$, where $t\in T$, and $\hat R$ is a cluster that belonged to $\hat \rset$ at any time during the current phase, such that $t\in \hat R$ held at any time during the current phase.
	Since every terminal may belong to at most $\Delta$ clusters of $\hat \rset$ over the course of Stage 3, we get that $|\Pi|\leq \Delta\cdot N'$. On the other hand, if any cluster $\hat R$ was heavy at any time during the current stage, then it must have contained at least $N'/4$ terminals at that time. Therefore, if $z$ is the number of clusters of $\hat \rset$ that were ever heavy during the current phase, then $|\Pi|\ge z\cdot N'/4$ must hold. We conclude that $z\leq 4\Delta$, and so the total number of clusters 
	$\hat R$ that ever belonged to $\hat \rset$, such that $\hat R$ was ever a heavy special cluster, is bounded by $4\Delta\log W$.
\end{proof}

If, at any time $\tau$ during Stage 3, a cluster $\hat R$ becomes a heavy special cluster, then, from time $\tau$, we maintain a set $S(\hat R)$ that contains all terminals that do not lie in cluster $\hat R$. Maintaining set $S(\hat R)$ over the course of the remainder of Stage 3 takes time at most $O(|T|)$: after initializing set $S(\hat R)$, we simply need to add to this set any terminal that is deleted from cluster $\hat R$, and delete from it any terminal that was deleted from graph $\hat H$. Since the total number clusters of $\rset$ that are heavy and special at any time during Stage 3 is bounded by $O(\Delta\log W)$, maintaining all such lists $S(\hat R)$ for all such clusters $\hat R$ takes at most $O(W\Delta\log W)$ time. For every heavy and special cluster $\hat R$, whenever the list $S(\hat R)$ is non-empty, we raise the flag $F_H$. As a witness pair, we supply a pair of terminals $(t,t')$, where $t\in \lambda_{\hat R}$ and $t'\in S(\hat R)$. Since $t\in \lambda_{\hat R}$, $\hat R=\coveringcluster(t)$ holds, and so cluster $\hat R$ must contain $B_{\hat H}(t,4D^*)$. Since $t'\in S(\hat R)$, it must be the case that $\dist_{\hat H}(t,t')>4D^*$, and so from \Cref{claim: short distance preservation}, $\dist_H(t,t')>D^*$. 

Whenever a new cluster $\hat R$ is added to the list $\hat \rset^S$ of special clusters, and cluster $\hat R$ is light, we construct two sets of terminals: set $T_1$ containing all terminals in list $\lambda_{\hat R}$; and set $T_2$, containing $|T_1|$ terminals that do not lie in $\hat R$. Both lists can be constructed in time $O(\hat n'_{\hat R})\leq O(\Delta\cdot \hat n_{\hat R})$, by going over all terminals that lie in $\hat R$. Observe that, since cluster $\hat R$ is not heavy, the number of terminals of $T'$ that do not lie in $\hat R$ is at least $|T_1|$.
Moreover, since, for every terminal $t\in T_1$, $B_{\hat H}(t,4D^*)\subseteq V(\hat R)$, for every pair $t\in T_1$, $t'\in T_2$ of terminals, $\dist_{\hat H}(t,t')> 4D^*$ holds, and so $\dist_H(t,t')> D^*$. As long as $T_1$ and $T_2$ are non-empty, we raise the flag $F_H$, and provide a pair $(t,t')$ of terminals with $t\in T_1$ and $t'\in T_2$ as a witness pair. Once either of the sets $T_1$ or $T_2$ become empty, we re-evaluate whether cluster $\hat R$ remains special and light, and if so, we repeat the algorithm. Notice that we have spent time $O(\Delta\cdot \hat n_{\hat R})$ in order to compute the sets $T_1$ and $T_2$ of the terminals, but then at least $\hat n_{\hat R}$ terminals were subsequently deleted from graph $H$. Therefore, the total time that is required in order to process light special clusters is bounded by $O(\Delta\cdot |T|)\leq O(W\cdot \Delta)$. Overall, the time that we spent on processing special clusters is asymptotically bounded by the total update time of data structure $\DS'$. Our algorithm now guarantees that, at all times, at least one cluster lies in set $\hat \rset^S$, and, if flag $F_H$ is down, then every cluster in $\hat \rset^S$ contains all terminals that currently lie in graph $H$.

%We raise flag $F_H$, and we provide the pair $t,t'$ of terminals as the witness pair.

\paragraph{Maintaining short distances from regular vertices to the terminals.}

As long as $T\cap V(H)\neq \emptyset$, we will  maintain a modified \EST in graph $H$, rooted at the vertices of $T$, with depth $D^*$, in order to ensure that the distances between regular vertices and terminals are small. Specifically, let $ H'$ be the graph that is obtained from $H$ by adding a source vertex $s$, and connecting it to every vertex in $T'$ with an edge of length $1$. Whenever graph $ H$ undergoes updates, we also update graph $ H'$ accordingly. Whenever a terminal $t\in T$ is deleted from graph $ H$ via an isolated vertex deletion operation, we delete the edge $(s,t)$ and we delete termian $t$ from $ H'$. It is easy to verify that graph $ H'$ only undergoes edge-deletion, isolated vertex-deletion and supernode-splitting (which is a special case of vertex-splitting) updates. We use the algorithm from \Cref{thm: ES-tree} to maintain a modified \EST in graph $H'$, with distance bound $D^*+1$. Recall that the total update time of the algorithm is $O(m^*\cdot D^*\cdot \log m^*)$, where $m^*$ is the total number of edges that ever belonged to graph $H'$. Since $m^*\leq N^0(H)\cdot \mu+|T|\leq 2\hat W$, the total update time of the algorithm is bounded by $O(\hat W\cdot \mu \cdot D^*\cdot \log(\hat W\mu))\leq O(\hat W\cdot \mu\cdot D^*\cdot \log W)$. Recall that the algorithm from \Cref{thm: ES-tree}  maintains a collection $S$ of vertices of $H'$, such that, for every vertex $v\in S$, $\dist_{H'}(v,s)>D^*+1$ holds. Whenever a regular vertex $v$ of $H$ joins the set $S$, we raise a flag $F_H$, and provide a witness pair $(v,t)$, where $t$ is any terminal vertex. We are then guaranteed that $\dist_{H}(v,t)>D^*$ holds.

The \EST data structure is maintained as long as not all vertices of $T$ are deleted from $H$. Let $\tau^*$ be the first time when all vertices of $T$ are deleted from $H$. From time $\tau^*$ onwards, we will maintain another data structure, that will allow us to support \spquery queries, which we describe next.

\paragraph{Maintaining short distances between regular vertices.}

Once all terminals are deleted from $H$, we will employ additional data structures, in order to ensure that, at all times $\tau\geq \tau^*$, there is a cluster $R\in \rset$, that contains all regular vertices. %Let $\tau^*$ denote the first time when all vertices of $T$ are deleted from $H$.
Starting from time $\tau^*$ onwards, we discard all clusters from $\rset$, except for those whose scale is $\log D^*$. In other words, we now set $\rset=\bigcup_{C\in \cset}\rset^C_{\log D^*}$. Since the algorithm from  \Cref{claim: cutting tilde H} guarantees that every regular vertex $v\in V(\tilde H)$ belongs to at most $W^{\eps/8}$ clusters in $\cset$, in the remainder of the algorithm, for every regular vertex $v$ of $H$, the total number of clusters in $\rset$ that $v$ can ever belong to is bounded by $\Delta'=W^{\eps/8}\cdot \Delta$.

We initialize a counter $N$, that counts the number of regular vertices currently in graph $H$.
For every regular vertex $v\in V(H)$, we maintain a cluster $R^*(v)\in \rset$, that is defined as follows. Recall that the algorithm from \Cref{claim: cutting tilde H} computes a cluster $C(v)\in \cset$ with $B_{\tH\attime[0]}(v,D^*)\subseteq V(C(v))$, and from \Cref{obs: maintain ball covering property}, $B_{\tH}(v,D^*)\subseteq V(C(v))$ continues to hold throughout Stage 3. We let $R^*(v)=\coveringcluster^{C(v)}_{\log D^*}(v)$. Note that $B_{\tH}(v,D^*)\subseteq V(R^*(v))$ must hold at all times, though cluster $R^*(v)$ may change over the course of the algorithm.
 Moreover, once all terminals are deleted from $H$, starting from time $\tau^*$ onwards, $\tilde H=H$ holds, and so $B_H(v,D^*)=B_{\tH}(v,D^*)\subseteq V(R^*(v))$ must hold. Therefore, for all times $\tau\geq \tau^*$, if $R=R^*(v)$ holds for a regular vertex $v$, then $B_H(v,D^*)\subseteq R$ holds.

For every cluster $R\in \rset$, we will maintain a list $\lambda(R)$ of all regular vertices $v\in V(H)$ with $R^*(v)=R$, together with two counters: counter $ n_{ R}=|\lambda_{ R}|$, and a counter $ n'_{R}$, that is equal to the number of regular vertices in cluster $R$. We initialize all the lists and the counters at time $\tau^*$. Notice that we can keep track of the clusters $R^*(v)$ for regular vertices $v\in V(H)$ using the data structure $\DS$, without increasing its asymptotic running time. Whenever, for any regular vertex $v\in V(H)$, cluster $R^*(v)$ changes from  $R$ to $R'$, we delete $t$ from $\lambda_{ R}$, insert it into $\lambda_{ R'}$, and update the counters $ n_{ R}$ and $ n_{R'}$ accordingly. Whenever a regular vertex $v$ is deleted from some cluster $ R\in \rset$, we decrease the counter $ n'_{ R}$. Whenever a new cluster $ R''$ is inserted into $ \rset$, we set $\lambda_{ R''}=\emptyset$, $ n_{ R}=0$, and we initialize the counter $ n'_{ R}$ to the number of regular vertices that lie in $R''$. It is easy to verify that these additional modifications of data structure $\DS$ do not increase its asymptotic total update time, and that the time required to initialize the lists $\lambda_R$ and conters $n_R,n'_R$ for all clusters $R\in \rset$ is also asymptitcally bounded by the total update time of data structrure $\DS$.

%Recall that the algorithm from \Cref{assumption: alg for recdynnc}
%ensures that every terminal $t\in T$ may lie in at most $\Delta=\Delta(W)$ clusters of $\hat \rset$ over the course of Stage 3 of the algorithm. Our goal is to ensure that, as long as $T\cap V(H)\ne \emptyset$, there is some cluster $\hat R\in \hat \rset$, that contains all terminals of $T\cap V(H)$, and that such a cluster can be computed in time $O(1)$. We denote by $T'=T\cap V(H)$ the set of terminals that currently lie in graph $H$.

Consider a cluster $R\in \rset$. We say that cluster $R$ is special if $\frac{  n'_{  R}}{  n_{  R}}\leq 2\Delta'$. Since, for every regular vertex $v$ of $H$, the total number of clusters in $\rset$ that $v$ can ever belong to is bounded by $\Delta'$, we obtain the following observation, whose proof is essentially identical to the proof of \Cref{obs: special cluster exists} and is omitted here.
 
\begin{observation}\label{obs: special cluster exists 2}
	For all  $\tau\geq \tau^*$, at time $\tau$, at least one cluster $R\in   \rset$ is special.
\end{observation}

\iffalse
\begin{proof}
	Consider any time $\tau\in \tset$, such that $T'\neq \emptyset$ holds at time $\tau$, and assume for contradiction that no cluster of $\hat \rset$ is special. Observe that $\sum_{\hat R\in \hat \rset}\hat n_{\hat R}=|T'|$, since for every terminal $t\in T'$, there is a single cluster $\hat R=\coveringcluster (t)$. Since no cluster is special, we get that, for every cluster $\hat R\in \hat \rset$, $\hat n'_{\hat R}>2\Delta\cdot \hat n_{\hat R}$, and so $\sum_{\hat R\in \hat \rset}\hat n'_{\hat R}>2\Delta\cdot \sum_{\hat R\in \hat \rset}\hat n_{\hat R}\geq 2\Delta\cdot |T'|$. However, every terminal of $T'$ may lie in at most $\Delta$ sets of $\hat \rset$, so $ \sum_{\hat R\in \hat \rset}\hat n_{\hat R}\leq \Delta\cdot |T'|$ must hold, a contradiction.
\end{proof}
\fi

As before, we can maintain, for every cluster $R\in \rset$, a bit $b_{R}$ that indicates whether cluster $R$ is currently special, and we can also maintain a list $   \rset^S\subseteq    \rset$ of special clusters, without increasing the asymptotic running time of data structure $\DS$. Recall that we denoted by $N$ the total number of regular vertices currently in $H$. From \Cref{obs: special cluster exists 2}, the list $   \rset^S$ of special clusters is always non-empty. Consider now a special cluster $R\in \rset^S$. We say that cluster $R$ is heavy if $   n'_{R}\geq \frac{N}{2}$, and we say that it is light otherwise. 
Using the same arguments as in the proof of \Cref{obs: few heavy clusters}, we conclude that, the total number of clusters $R$ that ever belonged in $\rset$, such that $R$ was ever a heavy special cluster is bounded by $4\Delta'\log W$.

\iffalse
\begin{observation}\label{obs: few heavy clusters}
	The total number of clusters $   R$ that ever belonged to $   \rset$, such that $   R$ was ever a heavy special cluster, is bounded by $2\Delta\log W$.
\end{observation}
\begin{proof}
	We partition the second stage of the algorithm into phases. Each phase lasts until the number of terminals in $T'$ decreases by at least factor $2$ since the beginning of the phase. Therefore, the number of phases is bounded by $\log |T|\leq \log W$.
	
	Consider now some phase, and denote the number of terminals in $T'$ at the beginning of the phase by $N'$. 
	Let $\Pi$ be the collection of pairs $(\hat R,t)$, where $t\in T$, and $\hat R$ is a cluster that belonged to $\hat \rset$ at any time during the current phase, such that $t\in \hat R$ held at any time during the current phase.
	Since every terminal may belong to at most $\Delta$ clusters of $\hat \rset$ over the course of Stage 3, we get that $|\Pi|\leq \Delta\cdot N'$. On the other hand, if any cluster $\hat R$ was heavy at any time during the current stage, then it must have contained at least $N'/2$ terminals at that time. Therefore, if $z$ is the number of clusters of $\hat \rset$ that were ever heavy during the current phase, then $|\Pi|\ge z\cdot N'/2$ must hold. We conclude htat $z\leq 2\Delta$, and so the total number of clusters 
	$\hat R$ that ever belonged to $\hat \rset$, such that $\hat R$ was ever a heavy special cluster, is bounded by $2\Delta\log W$.
\end{proof}
\fi

If, at any time $\tau>\tau^*$, a cluster $   R$ becomes a heavy special cluster, then, from time $\tau$, we maintain a set $S(   R)$ that contains all regular vertices that do not lie in cluster $   R$. Maintaining set $S(   R)$ over the course of the remainder of Stage 3 takes time at most $ O(\hat W)$: after initializing set $S(   R)$, we simply need to add to this set any regular vertex that is deleted from cluster $   R$, and delete from it any regular vertex that was deleted from graph $   H$. Since the total number clusters of $\rset$ that are heavy and special at any time during Stage 3 is bounded by $O(\Delta'\log W)\leq O\left (W^{\eps/8}\cdot \Delta\right )$, maintaining all such lists $S(   R)$ for all such clusters $   R$ takes at most $O(W^{1+O(\eps)}\Delta)$ time. For every heavy and special cluster $R$, whenever the list $S(   R)$ is non-empty, we raise the flag $F_H$. As a witness pair, we supply a pair of regular vertices $(v,v')$, where $v\in \lambda_{R}$ and $v'\in S(R)$. Recall that, if $v\in \lambda_R$, then $B_H(v,D^*)\subseteq R$ must hold. Since $v'\not\in R$, we conclude that $\dist_H(v,v')>D^*$ currently holds, so $(v,v')$ is a valid witness pair.

Whenever a new cluster $R$ is added to the list $\rset^S$ of special clusters, and cluster $R$ is light, we construct two sets of regular vertices: set $V_1$ containing all regular vertices in list $\lambda_{R}$; and set $V_2$, containing $|V_1|$ regular vertices that do not lie in $ R$. Both lists can be constructed in time $O(n'_{R})\leq O(\Delta'\cdot  n_{R})\leq O(W^{\eps/8}\cdot \Delta\cdot n_R)$, by going over all regular vertices that lie in $R$. Observe that, since cluster $R$ is not heavy, the number of regular vertices of $H$ that do not lie in $R$ is at least $|V_1|$. 
Moreover, since, for every vertex $v\in V_1$, $B_{H}(v,D^*)\subseteq V(R)$, for every pair $v\in V_1$, $v'\in V_2$ of regular vertices, $\dist_{H}(v,v')> D^*$ holds. As long as $V_1$ and $V_2$ are non-empty, we raise the flag $F_H$, and provide a pair $(v,v')$ of regular vertices with $v\in V_1$ and $v'\in V_2$ as a witness pair. Once either of $V_1$ or $V_2$ become empty, we re-evaluate whether cluster $R$ remains special and light, and if so, we repeat the algorithm. Notice that we have spent time $O(W^{\eps/8}\cdot \Delta\cdot n_{R})$ in order to construct the sets $V_1$ and $V_2$ of the regular vertices, but then at least $n_{R}$ regular vertices were subsequently deleted from graph $H$ via flag-lowering sequences. Therefore, the total time that is required in order to process light special clusters is bounded by $O(N\cdot W^{\eps/8}\cdot \Delta)\leq O(\hat W^{1+O(\eps)}\cdot \Delta)$. Overall, the time that we spent on processing special clusters is asymptotically bounded by the total update time of data structure $\DS$ plus $O(\hat W^{1+O(\eps)}\cdot \Delta)$. Our algorithm now guarantees that, at all times $\tau\geq \tau^*$, at least one cluster lies in set $ \rset^S$, and, if flag $F_H$ is down, then every cluster in $\rset^S$ contains all regular vertices that currently lie in graph $H$.

% It is easy to extend data structure $\DS$ that maintains the clusters of $\rset$ in order to also maintain the counters $N,\set{   n_R}_{R\in \rset}$, and the special clustesr $R^*(v)$ for regular vertices $v\in V(H)$, without increasing its total update time.

This completes the description of the data structures that the algorithm for Stage 3 maintains. We now bound the total update time of the algorithm.

Recall that the running time of the algorithm from \Cref{claim: cutting tilde H} is bounded by:

\[O\left ((\hat W\cdot \mu)^{1+\eps}\right );\]

total update time of data structure $\DS$ is bounded by:

\[O\left(\hat W\cdot W^{\delta+\eps/8}\cdot (D^*)^{3}\cdot \mu^c\right ).\]

The total update time of data structure $\DS'$ is bounded by:

\[O\left (W^{1+\delta+c\eps/8}\cdot \Delta^c\cdot \mu^c \cdot (D^*)^3\cdot (\log D^*)^c\right ),
\]

and the total update time of the data structure from \Cref{thm: ES-tree} that maintains the modified \EST is bounded by:

\[O(\hat W\cdot \mu \cdot D^*\cdot \log (\hat W\mu))\leq O(\hat W\cdot \mu \cdot D^*\cdot \log W).\]

Additionally, maintaining and processing the special clusters of $\rset$ takes $O(\hat W^{1+O(\eps)}\cdot \Delta)$ time, in addition to the total update time of data structure $\DS$.

The remaining time, that is needed in order to maintain the modified graph $\tilde H$, the clusters of $\cset$, and the contracted graph $\hat H$ are asymptotically bounded by the above.

Overall, the total update time of the algorithm from Stage 3 is bounded by:

\[O\left (\hat W\cdot W^{\delta+c\eps/8}\cdot \Delta^c\cdot \mu^c \cdot (D^*)^{3}\cdot (\log D^*)^c+(\hat W\mu)^{1+O(\eps)}\cdot \Delta\right ),
\]

Since the total update time of the algorithm from Stage 1 is bounded by:

\[O\left (\hat W^{1+O(\eps)} \cdot \hat D^3\cdot \rho^8\mu^4\right ),\]

the total update time of the algorithm from Stage 2 is bounded by:

\[ O\left(\hat W^{1+O(\eps)}\cdot D^*\cdot \rho\cdot \mu^2\right ),\]

and since $\hat D=O(D^*/\eps^2)$ and $\rho=\frac{4\hat W}{W}$, we get that the total running time of the algorithm for the \maintaincluster problem is bounded by:

\[O\left (\hat W\cdot W^{\delta+c\eps/8}\cdot\Delta^c\cdot \mu^c \cdot (D^*)^{3}\cdot (\log D^*)^c\right )+O\left (\hat W^{1+O(\eps)}\cdot (D^*)^3\cdot \mu^4\cdot \Delta\cdot \left(\frac{\hat W}{W}\right )^8\right ).
\]

\subsubsection{Responding to Queries}

We now provide an algorithm for supporting queries $\spquery(H,v,v')$ over the course of Stage 3. Suppose we are given a query  $\spquery(H,v,v')$, where $v,v'\in V(H)$ are regular vertices of $H$. Recall that our goal is to return a path $P$ of length at most $\alpha'\cdot D^*$ connecting $v$ to $v'$ in $H$, in time $O(|E(P)|)$. 

We employ different algorithms, depending whether $T\cap V(H)=\emptyset$ holds. Assume first that at least one terminal $t\in T$ lies in graph $H$.  In this case, we query the \EST data structure to obtain a path $P_1$ connecting $v$ to $s$ in graph $H'$, whose lenghth is at most $D^*+1$, in time $O(|E(P_1)|)$. By deleting the last vertex from path $P_1$, we obtain a path in graph $H$, connecting $v$ to some terminal $t\in T$, such that the length of the path is at most $D^*$. Similarly, we compute a path $P_2$, connecting $v$ to some terminal $t'\in T$, of length at most $D^*$, in time $O(|E(P_2)|)$. If $t=t'$, then we obtain a path $P$ connecting $v$ to $v'$ in $H$, by concatenating $P_1$ to $P_2$. The length of the path is at most $2D^*\leq \alpha'\cdot D^*$, and the running time of the algorithm is $O(|E(P)|)$.

Assume now that $t\neq t'$. Recall that we maintain a list $\hat \rset^S$ of special clusters of $\hat \rset$, that is always non-empty. Let $\hat R\in \hat \rset^S$ be any such cluster. Since the queries cannot be asked when flag $F_H$ is up, $T'\subseteq V(\hat R)$ must hold. We run query $\spquery(\hat R,t,t')$ in data structure $\DS'$, to obtain a path $Q$ in graph $\hat H$ that connects $t$ to $t'$, and has length at most $4\alpha\cdot D^*$, in time $O(|E(Q)|)$. We then apply the algorithm from \Cref{claim: transform path from contracted} to path $Q$, in order to obtain a path $P'$ in graph $H$, connecting $t$ to $t'$, whose length is at most $4\alpha^2\cdot D^*$, in time $O(|E(P')|)$. By concatenating paths $P_1,P'$ and $P_2$, we obtain a path $P$ in graph $H$, connecting $v$ to $v'$. The length of the path is bounded by $2D^*+4\alpha^2\cdot D^*\leq 8\alpha^2D^*\leq \alpha'$. The running time of the algorithm is $O(|E(P)|)$.

Assume now that all terminals of $T$ have already been deleted from $H$. Recall that we maintain a set $\rset^S\subseteq \rset$ of special clusters, which is non-empty. We let $R\in \rset^S$ be any such cluster. Since flag $F_H$ is currently down, every regular vertex of $H$ lies in cluster $R$. Recall also that $\scale(R)=\log D^*$. Assume that $R\in \rset^C_{\log D^*}$ for some cluster $C\in \cset$. We apply query $\spquery(R,v,v')$ to data structure $\DS^C_{\log D^*}$, to obtain a path $P$ connecting $v$ to $v'$ in graph $H$, of length at most $\alpha\cdot D^*$, in time $O(|E(P)|)$. We return this path as the outcome of the query.

\newpage
\appendix

\section{Proof of \Cref{cor: fully APSP}}
\label{appx: proof main APSP}
 We assume that we are given an instance $G$ of fully dynamic \APSP problem, with lengths $\ell(e)>0$ on edges $e\in E(G)$ and $|V(G)|=n$. We assume without loss of generality that the shortest edge length is $1$ and longest edge length is $\Lambda$. At the cost of losing factor $2$ in the approximation ratio, we can assume that all edge lengths are integral. For all $1\leq i\leq \ceil{\log \Lambda}$, we denote $D_i=2^i$. %Let $m^*$ be the total number of edges and vertices that are ever present in $G$. We assume w.l.o.g. that our algorithm is given an estimate $m^*\leq m\leq 2m^*$. Otherwise, we can start with estimate $m$ that is the maximum of $2|E(G\attime[0])|$,
% and the smallest value for which  $\frac{1}{(\log m)^{1/200}}<\eps<1/400$ holds. We can then
% wait until the number of edges and vertices that ever appeared in graph $G$ reaches $m$, and then start the algorithm from scratch, while doubling the estimate $m$.  
  
  Notice that \Cref{thm: main final dynamic NC algorithm} implies that \Cref{assumption: alg for recdynnc2} holds, for   $\alpha(\hat W)=(\log\log \hat W)^{2^{O(1/\eps^2)}}$.
 
For all $1\leq i\leq \ceil{\log \Lambda}$, we use  \Cref{thm: main fully-dynamic APSP single scale} to obtain Algorithm $\alg_i$ for the $D_i$-restricted \APSP problem on graph $G$, with parameters $m$ and $\eps$ remaining unchanged. We denote by $\DS_i$ the data structure that algorithm $\alg_i$ maintains. Recall that the amortized update time of Algorithm $\alg_i$ is bounded by $n^{O(\eps)}$ per operation. Therefore, the amortized update time  of all algorithms $\alg_i$, for $1\leq i\leq \ceil{\log \Lambda}$ is  at most
 $O\left (n^{O(\eps)}\cdot \log \Lambda\right )$ per operation.

Recall that each such algorithm $\alg_i$ achieves approximation factor $(\alpha(n^3))^{O(1/\eps)}\leq (\log\log n)^{2^{O(1/\eps^2)}}=\alpha'$.

Next, we provide an algorithm for responding to query \distquery between  a pair $x,y$ of vertices of $G$. We perform a binary search on integers $1\leq i\leq \ceil{\log \Lambda}$, to find an integer $i'$, such that the response of data structure $\DS_{i'}$ to query $\shortpath(x,y)$ is ``YES'', and the response of data structure $\DS_{i'-1}$ to the same query is ``NO''. Since query time for a single such query is  $O\left(2^{O(1/\eps)}\cdot \log n\right )$, the time required to compute the index $i'$ is bounded by $O\left(2^{O(1/\eps)}\cdot \log n\log\log \Lambda\right )$. We then return $D_{i'}=2^{i'}$ as the estimate on $\dist_G(x,y)$. Since the response of data structure $\DS_{i'-1}$ to the query is ``NO'', we are guaranteed that $\dist_G(x,y)\geq D_{i'}/2$. On the other hand, since data structure $\DS_{i'}$ responded ``YES'', we are guaranteed that there exists a path in graph $G$, connecting $x$ to $y$, of length at most $D_{i'}\cdot \alpha'$. %Since, for each $1\leq i\leq \ceil{\log \Lambda}$, the time required to respond to query $\shortpath$ in data structure $\DS_i$ is bounded by $O\left(2^{O(1/\eps)}\cdot \log m\right )$, the total time required to respond to \distquery is bounded by  $O\left(2^{O(1/\eps)}\cdot \log m\log \Lambda\right )$.

Finally, we provide an algorithm for responding to query \pathquery between a pair $x,y$ of vertices of $G$. We start by executing the algorithm for responding to $\distquery$ between $x$ and $y$, and obtain an integer $i'$ as described above, such that $\dist_G(x,y)\geq D_{i'}/2$, and algorithm $\alg_{i'}$, in response to query $\shortpath(x,y)$ responds ``YES''. We then ask the same algorithm to compute a path $P$ connecting $x$ to $y$ in the current graph $G$, of length at most $D_{i'}\cdot \alpha'$, in time $O(|E(P)|)$, and return the resulting path $P$. Clearly, the running time of the algorithm is $O\left (|E(P)|+2^{O(1/\eps)}\cdot \log n\cdot \log\log\Lambda \right )$.

\section{Proofs Omitted from Section \ref{sec: prelims}}

\subsection{Proof of \Cref{lem: ball growing}}
\label{sec: ball-growing}

For all $i\geq 1$, we denote by $L_i$ the set of all vertices of $G$ that lie at distance $2(i-1)D+1$ to $2iD$ from $v$ in $G$. In other words:

\[L_i=B_G(v,2iD)\setminus B_G(v,2(i-1)D).   \]

We refer to the vertices of $L_i$ as \emph{layer $i$ of the BFS}. We denote by $m_i$ the total number of edges $e\in E(G)$, such that both endpoints of $e$ lie in $ L_1\cup\cdots\cup L_i$. For all $1\leq j\leq k$, we also denote by $N^j_i$ the total number of vertices of $T_j$ that lie in $L_1\cup\cdots\cup L_i$. The following definition is central to our algorithm.

\begin{definition}[Eligible Layer]
	Let $i>1$ be an integer. We say that layer $L_i$ of the BFS is \emph{eligible}, if the following conditions hold:
	
	\begin{properties}{C}
		\item $m_i\leq m_{i-1}\cdot |E(G)|^{\eps}$;\label{cond: few edges} 
		
		\item for all $1\leq j\leq k$, $N^j_i\leq N^j_{i-1}\cdot |T_j|^{\eps}$. \label{cond: few type-j vertices}
	\end{properties}
\end{definition}

We use the following simple observation, that follows from standard arguments.

\begin{observation}\label{obs: eligible layer}
	There is an integer $1<i\leq \frac{2k+2}{\eps}$, such that layer $L_i$ is eligible.
\end{observation}
\begin{proof}
	Let $z=\ceil{1/\eps}$. We start by showing that there are at most $z$ layers $L_i$ with $i>1$, for which condition \ref{cond: few edges}  is violated. Indeed, assume otherwise. Let $1<i_1<i_2<\cdots<i_{z}$ be indices of layers for which condition \ref{cond: few edges} is violated. Since we have assumed that vertex $v$ is not isolated in $G$, $m_1\geq 1$ must hold. From Condition \ref{cond: few edges}, for all $1<a\leq z$, $m_{i_a}> m_{i_{a}-1}\cdot |E(G)|^{\eps}\geq m_{i_{a-1}}\cdot |E(G)|^{\eps}$. Therefore, $m_z>m_1\cdot |E(G)|^{z\eps}\geq |E(G)|$, which is impossible
	
	Next, we show that, for all $1\leq j\leq k$, the number of layers $L_i$ with $i>1$, for which the condition $N^j_i\leq N^j_{i-1}\cdot |T_j|^{\eps}$ is violated, is bounded by $z+1$. Indeed, assume otherwise, and let $1<i'_1<i'_2<\cdots<i'_{z+1}$ be indices of layers for which the condition is violated.
	Then $N^j_{i'_1}\geq 1$ must hold, and,  for all $1<a\leq z+1$, $N^j_{i_a}> N^j_{i_{a}-1}\cdot |T_j|^{\eps}\geq N^j_{i_{a-1}}\cdot |T_j|^{\eps}$. Therefore, $N_{z+1}> |T_j|^{z\eps}\geq |T_j|$, which is impossible.
	
	We conclude that the number of layers $L_i$ with $i>1$ that are not eligible is bounded by $z+k\cdot (z+1)\leq \frac{2k+1}{\eps}$, since $z=\ceil{\frac 1{\eps}}$, and $\eps<1/4$.
\end{proof}

We run the weighted BFS algorithm from vertex $v$ in graph $G$, until we reach the first layer $1<i\leq \frac{2k+2}{\eps}$ that is an eligible layer. We then  return integer $i$. if we denote by $S=B_G(v,2(i-1)D)$, and $S'=B_G(v,2iD)$, then the running time of the algorithm is bounded by $O(|E_G(S')|\cdot \log n)\leq O(|E_G(S)|\cdot |E(G)|^{\eps}\cdot \log n)$, as required. From the definition of an eligible layer, it is easy to verify that the output $i$ has all required properties.

\subsection{Modified Even-Shiloach Trees -- Proof of \Cref{thm: ES-tree}}
\label{sec: appx-ES-tree}

The proof uses standard techniques, and is essentially identical to the proof of Theorem 3.2 in \cite{APSP-previous}. The main difference is that we allow vertex-splitting operations to be applied to any vertex of $G$ except for the source $s$, while in  \cite{APSP-previous}, the input graph $G$ is bipartite, and vertex-splitting could only be applied to vertices on one side of the bipartition. 

\paragraph{Data Structures.}
We maintain the graph $G$ as an adjacency list: for every vertex $v$, we maintain a linked list of its neighbors. 
Throughout the algorithm, we will maintain the following data structures:

\begin{itemize}
	\item A shortest-path tree $T$ rooted at vertex $s$, that contains all vertices $x\in V(G)$ with $\dist_G(s,x)\leq D$. For every such vertex $x$, its correct distance $\lambda (x)=\dist_G(s,x)$ from $s$ is stored together with $x$.
	
	\item A set $S^*$ containing all vertices of $G$ that do not lie in $T$.
	
	\item For every vertex $x\in V(T)$, let $\pred(x)$ be the set of all vertices $y\in V(T)$ with $(x,y)\in E(G)$. We maintain a heap $\heap(x)$ in which all elements of $\pred(x)$ are stored, where the key associated with each element $y\in \pred(x)$ is $\lambda (y)+\ell(y,x)$.

	\item For every edge $e=(y,x)$ with $y,x\in V(T)$, we store, together with vertex $y$, a pointer to its corresponding element in the heap $\heap(x)$ and vice versa.
\end{itemize}

%We will implement edge-length increase update as follows. Suppose $e$ is an edge, whose lenght is increased from $\ell(e)$ to a new value $2\ell(e)\leq \ell'\leq D^* $. We will insert a new edge $e'$ into the graph $H$, that is a copy of the edge $e$, but whose length is $\ell'(e)$. Then we will delete the edge $e$ from $H$, using an edge-deletion update operation that is described below. Note that, for every edge $e$ that ever belongs to the input dynamic graph $H$, we may create up to $\log D^*$ copies that are inserted into the data structure.
Throughout,  we denote by $m^*$ the total number of edges that are ever present in $G$.

\paragraph{Initialization.}
We run Dijkstra's algorithm on graph $G$ time $O(|E(G)|\log m^*)\leq O(m^*\cdot \log m^*)$, up to distance threshold $D$, to construct the initial tree $T$. For every vertex $x\in V(T)$, we initialize $\lambda(x)=\dist_G(s,x)$, and for all $x\not\in V(T)$, we set $\lambda(x)=\infty$. We then process all vertices of $V(T)$ in the order of their distance from $s$, and insert each such vertex $x$ into all heaps $\heap(y)$ of all vertices $y\in V(G)$ that are neighbors of $x$ in $G$. Lastly, we initialize the set $S^*$ of vertices to contain every vertex $x\in V(G)\setminus V(T)$. Clearly, initialization takes time at most $O(m^*\cdot \log m^*)$.

We now assume that we are given a valid data structure, and describe an algorithm for handling updates.

\paragraph{Edge Deletion.}
The procedure for processing edge deletions is completely standard. The description provided here is due to Chechik \cite{chechik}. It is somewhat different but equivalent to the standard description.
Suppose an edge $e=(x,y)$ is deleted from the graph $G$.  We start by updating the heaps $\heap(x)$, $\heap(y)$ with the deletion of this edge.  If $e\not\in E(T)$, then no further updates are necessary.

We assume from now on that $e=(x,y)$ is an edge of the tree $T$, and we assume w.l.o.g. that $y$ is the parent of $x$ in $T$. Let $S$ be the set of all vertices of $T$ that lie in the subtree $T_x$ of $T$ that is rooted at $x$. We delete the edge $(x,y)$ from $T$, thereby disconnecting all vertices of $S$ from $T$.  The remainder of the algorithm consists of two phases. In the first phase, we identify the set $R\subseteq S$ of all vertices, whose distance from $s$ has increased, and connect all remaining vertices of $S$ to $T$. In the second phase, we attempt to reconnect vertices of $R$ to $T$.

In order to implement the first phase, we maintain a heap $Q$ of all vertices that need to be examined, where the key associated with each vertex $a$ in $Q$ is $\lambda(a)$.
We initialize $Q$ to maintain a single vertex -- the vertex $x$, and we also initialize $R=\emptyset$. Heap $Q$ will have the property that, if some vertex $a$ belongs to $Q$, and vertex $a'$ was the parent of $a$ in $T$, then $a'$ was added to $R$ (or, if $a=x$, then $a'\in T$). Moreover, if $a$ is a vertex of $Q$ with smallest key $\lambda(a)$, and  $a'$ is the element lying at the top of $\heap(a)$, then either $\lambda(a')\geq \lambda(a)$, or $a'\in T$ must hold. Both these invariants hold at the beginning.

The algorithm iterates, as long as $Q\neq \emptyset$. Let $a$ be a vertex of $Q$ with smallest key $\lambda(a)$.
 Let $b$ be the element lying at the top of $\heap(a)$. 
 If $b\in V(T)$, then
 we check whether connecting vertex $a$ to the tree $T$ via vertex $b$ will allow us to keep $\lambda(a)$ unchanged, or, equivalently, whether $\lambda(a)=\lambda(b)+\ell(a,b)$. If this is the case, then we connect $a$ to the tree $T$ via $b$, delete $a$ from $Q$, and proceed to the next iteration. Otherwise, we are guaranteed that $\lambda(a)$ must increase. We then add $a$ to $R$, add all children of $a$ in the original tree $T$ to the heap $Q$, and delete $a$ from $\heap(b')$ for all neighbors $b'$ of $a$. 
Alternatively, if $b\not\in V(T)$, then, from our invariant $\lambda(b)\geq \lambda(a)$ currently holds, which again means that $\lambda(a)$ must increase. In this case, we also add $a$ to $R$, add all 
 children of $a$ in the original tree $T$ to the heap $Q$, and delete $a$ from $\heap(b')$ for all neighbors $b'$ of $a$. 

Note that the algorithm examines vertices of $S$ in the non-decreasing order of their label $\lambda(a)$ (except that, when some vertex $a$ is reconnected to the tree $T$, then its descendants will never be examined). Once a vertex $a$ is examined, it is either connected to the tree, or it is added to $R$. In the latter case, all children of $a$ are added to $Q$. This ensures that, throughout the algorithm, if $a$ is a vertex of $Q$ with smallest key $\lambda(a)$, and  $a'$ is the element lying at the top of $\heap(a)$, then either $\lambda(a')\geq \lambda(a)$, or $a'\in T$ must hold. Indeed, if $\lambda(a')< \lambda(a)$ but $a'\not\in T$, then we should have examined $a'$ before, and, if it was added to $R$, then it would have been deleted from $\heap(a)$.

The first phase terminates once $Q=\emptyset$. It is not hard to see, using standard analysis, that its running time is bounded by $O(1+\sum_{v\in R}\deg_G(v)\cdot \log m^*)$.

In the second phase, we run Dijkstra's algorithm on the vertices of $R$, up to distance $D$, trying to reconnect them to the tree $T$. We also update the heaps of all vertices of $R$, and of their neighbors, accordingly. This step can also be implemented in time $O(\sum_{v\in R}\deg_G(v)\log m^*)$. Every vertex of $R$ that was not reconnected to tree $T$ is added to set $S^*$.

To summarize, in processing edge-deletion operations, whenever, for any vertex $v$, its distance from $s$ increases, we may need to pay $O(\deg_G(v)\log m^*)$ in running time. It is then easy to see that the total update time due to processing edge deletions is bounded by $O(m^*\cdot D\cdot \log m^*)$.

\paragraph{Deletion of Isolated Vertices.}
Deletion of isolated vertices is straightforward, and takes time $O(1)$  per vertex. Note that an isolated vertex may not belong to $T$ (unless that vertex is $s$ and $T$ only contains the vertex $s$), so apart from deleting the vertex from $G$ and from $S^*$ (if it belongs to $S^*$), no further updates are necessary.

\paragraph{Vertex Splitting.}
Recall that in a vertex-splitting update, we are given a vertex $v\in V(G)\setminus\set{s}$, and a non-empty set $E'\subseteq \delta_G(e)$ of edges. %Set $E_1$ of edges is listed explicitly, while set $E_2$ of edges is not given explicitly; instead we are given a set $\lambda_H(u)\setminus E_2$ of edges. %We denote by $S(u)$ the set of all neighbors of $u$ in $H$, and by $E^*$ the set of all edges incident to $u$ in $H$. 
We need to add a new vertex $v'$ to the graph, and, for every edge $e=(v,u)\in E'$, insert an edge $e'=(v',u)$ of length $\ell(e)$ into $G$. %Additionally, we need to delete all edges in $E^*\setminus E_2$ from $H$. 

If $v\not \in V(T)$, then we simply set $\lambda(v')=\lambda(v)=\infty$, add $v'$ to set $S^*$, and terminate the update algorithm.

Assume now that $v\in V(T)$. Let $\hat v$ be the parent of the vertex $v$ in the tree $T$. We now proceed as follows:

\begin{enumerate}
	\item Add a new vertex $v'$ as the child of $\hat v$ to the tree $T$ (it is convenient to think of it as a copy of $v$); set $\ell(v',\hat v)=\ell(v,\hat v)$, $\lambda(v')=\lambda(v)$, add $\hat v$ to $\heap(v')$, and add $v'$ to $\heap(\hat v)$.
	
	\item For every edge $e=(v,u)\in E'$ with $u\neq \hat v$, add an edge $e'=(v',u)$ of length $\ell(e)$ to the graph $G$, add $u$ to $\heap(v')$, and add $v'$ to $\heap(u)$. Notice that the insertion of these edges does not decrease the distance of any vertex from $s$, since  $\lambda(v)=\lambda(v')$, and every vertex that serves as an endpoint of a newly inserted edge is also a neighbor of $v$.
	
%	\item For every edge $e\in \lambda(u)\setminus E_2$, delete $e$ from the graph $H$, and from the \EST data structure, using the edge-deletion update operation.
	
	\item If edge $(v,\hat v)\not\in E'$, delete the edge $(v',\hat v)$ from the graph $G$, and update the \EST data structure accordingly.
\end{enumerate}

%If $u=s$, then the update procedure is almost identical, except that we initially insert $u'$ as a child of $u$, and set the length of the edge $(u,u')$ to be $1/2$. This ensures that, as edges corresponding to the edge set $E'$ are inserted into the graph, the distances from vertices of $H$ to $s$ do not decrease, and, since the lengths of all edges in $H$ are at least $1$, $T$ remains a valid shortest-path tree. 
%In the last step, we delete the edge $(u,u')$ from our data structure using the edge-deletion update operation.

The processing time of this update procedure, excluding the calls to the edge-deletion updates in the \EST data structure, is $O(|E'|\cdot \log m^*)$. 
The insertion of edge $(v',\hat v)$ into $T$, if this edge does not lie in $E'$, increases the total number of edges inserted into $G$. However, since $|E'|\geq 1$ must hold, this new inserted edge can be charged to some edge of $E'$, increasing the total number of edges that are ever present in $G$ by at most factor $2$. 
From the above discussion, if we denote the length of the input update sequence $\Sigma$ by $k$, the total update time of the \EST data structure is bounded by $O((m^*\cdot D+k)\cdot\log m^*)$.

Lastly, observe that each update operation in $\Sigma$ either inserts at least one edge into $H$, or deletes at least one vertex or an edge from $H$. Since the initial graph $G$ is connected,  $k\leq O(m^*)$ holds, and the total running time of the algorithm is at most $O(m^*\cdot D\cdot \log m^*)$.

\paragraph{Responding to Queries $\shortestpath$.} Recall that in  $\shortestpath$, we are given a vertex $x\in V(G)$, and our goal is to either correctly establish, in time $O(1)$, that $\dist_G(s,x)>D$, or to return a shortest $s$-$x$ path $P$, in time $O(|E(P)|)$. Given a query vertex $x$, we check whether $\lambda(x)=\infty$. If so, we report that $\dist_G(s,x)>D$. Otherwise, we retrace the unique path $P$ connecting $x$ to $s$ in the tree $T$, and return it, in time $O(|E(P)|)$.

%\section{Proofs Omitted from \Cref{sec: valid inputs}}
\section{Completing the Proof of \Cref{thm: main final dynamic NC algorithm}}
\label{subsec: reduce dependence on D}

In this subsection we prove \Cref{thm: main final dynamic NC algorithm} using the algorithm from \Cref{thm: main final dynamic NC algorithm inner}. The proof is practically identical to a proof of a similar statement that appeared in \cite{APSP-previous}.
We can assume that graph $H$  has no isolated vertices, as all such vertices can be ignored (e.g. each such vertex can be placed in a separate cluster). We will also assume w.l.o.g. that $\mu\geq 3$ holds, since otherwise we can simply increase $\mu$ to $\mu=3$; this does not affect the asymptotic bounds in the theorem statement.  Using the arguments from Section \ref{subsubsec: bounding D}, at the cost of losing a factor $2$ in the approximation ratio, we can assume that $D\leq 3\hat W/\mu\leq \hat W$.
As in \cite{APSP-previous}, the main idea of the proof is to apply the algorithm from \Cref{thm: main final dynamic NC algorithm inner} recursively, for smaller and smaller distance bounds. Specifically, we prove the following lemma by induction.

\begin{lemma}\label{lem: inductive dynamic NC algorithm}
	There is a universal constant $\hat c$, and a deterministic algorithm for the \recdynNC problem, that,  given a valid input structure $\iset=\left(H=(V,U,E),\set{\ell(e)}_{e\in E(H)},D \right )$ undergoing a sequence of valid update operations with dynamic degree bound $\mu$, and parameters $\hat W$, $1/(\log \hat W)^{100}\leq \eps<1/400$,  and $1\leq i\le \ceil{1/\eps}$ such that, if we denote by $N^0(H)$ the number of regular vertices in $H$ at the beginning of the algorithm, then $N^0(H)\cdot \mu\leq  \hat W$ and $D \leq 6\hat W^{\eps i}$, achieves approximation factor $\alpha_i=(\log\log \hat W)^{\hat c i\cdot 2^{\hat c /\eps^2}}$, and has total update time at most: $\left (\hat c^i\cdot \hat W^{1+\hat c\eps}\cdot \mu^{\hat c/\eps} \right )$.
	Moreover, the algorithm ensures that for every regular vertex $v\in V$, the total number of clusters $C\in \cset$, to which vertex $v$ ever belongs over the course of the algorithm, is bounded by $\hat W^{4i\eps^4}$.
\end{lemma}

\Cref{thm: main final dynamic NC algorithm} follows immediately from \Cref{lem: inductive dynamic NC algorithm}, by using $i=\ceil{1/\eps}$.
In the remainder of this subsection, we focus on proving \Cref{lem: inductive dynamic NC algorithm}.

The proof is by induction on $i$. The base case is when $i=1$, so $D\leq 6\hat W^{\eps}$.
We use the algorithm from \Cref{thm: main final dynamic NC algorithm inner}, whose approximation factor is $(\log\log \hat W)^{2^{O(1/\eps^2)}}\leq (\log\log \hat W)^{\hat c\cdot 2^{\hat c /\eps^2}}=\alpha_1$ (if constant $\hat c$ is large enough), and total update time is  bounded by:

\[O\left ((N^0(H))^{1+O(\eps)}\cdot \mu^{O(1/\eps)}\cdot D^3\right )\leq \left (\hat c \cdot \hat W^{1+\hat c\eps}\cdot \mu^{\hat c/\eps} \right ),\]

as required.

For the induction step, we consider some integer $1< i\leq \ceil{1/\eps}$, and assume that the lemma holds for all integers below $i$.
Consider the input graph $H$ and the distance threshold $D\leq  6\hat W^{\eps i}$.
We use a parameter $D'=\floor{\hat W^{\eps (i-1)}/4}$.
Clearly, $D'\geq \hat W^{\eps(i-1)}/8$. 
We say that an edge $e$ of $H$ is \emph{long} if $\ell(e)>D'$, and we say that it is \emph{short} otherwise.

Let $H'$ be a dynamic graph that is  obtained from $H$ by deleting all long edges from it. We can generate a valid update sequence for graph $H'$ from the valid update input sequence $\Sigma$ for graph $H$ in a natural way, by ignoring all updates concerning long edges. 
Recall that the input to the supernode-splitting update is a supernode $u \in U$ and a nonempty set of edges $E' \subseteq \delta_H(u)$. By deleting from $E'$ all long edges, we get a valid input for supernode-splitting operation in $H'$ (if $E'$ only contains long edges, then we ignore this update).
We can now define a valid input structure  $\iset'=\left(H',\set{\ell(e)}_{e\in E(H')}, 4D' \right )$, that undergoes a sequence $\Sigma'$ of valid update operations with dynamic degree bound $\mu$. If $N^0(H')$ denotes the number of regular vertices in the initial graph $H'$, then $N^0(H')=N^0(H)$, and $N^0(H')\cdot \mu\leq \hat W$ holds. Since $4D'\leq \hat W^{\eps (i-1)}$, we can apply the algorithm from the induction hypothesis to input structure $\iset'$, and sequence $\Sigma'$ of valid update operations, distance bound $4D'$. Recall that the approximation factor of the algorithm is $\alpha_{i-1}$, and its total update time is bounded by 
$\left (\hat c^{i-1}\cdot \hat W^{1+\hat c\eps}\cdot \mu^{\hat c/\eps} \right )$.
Additionally, the algorithm ensures that for every regular vertex $v\in V$, the total number of clusters $C\in \cset$, to which vertex $v$ ever belongs over the course of the algorithm, is bounded by $\Delta=\hat W^{4(i-1)\eps^4}$.
We denote by $\dset(H')$ the corresponding data structure, and by
$\cset'$ the collection of clusters that the algorithm maintains.
We denote the algorithm that we have described so far, that maintains the data structure $\dset(H')$, by $\alg_1$.

We use the neighborhood cover $\cset'$ in order to define another dynamic graph $\hat H$, as follows.
We start with letting $\hat H=H$, and then round all edge lengths up to the nearest integral multiple of $D'$, denoting the resulting new length of each edge $e$ by $\hat \ell(e)$. 
%except that we may increase the edge lengths: for an edge $e\in E(\hat H)$, let $z$ be the smallest integer, such that $z\cdot D' \geq \ell(e)$. We set the length of $e$ to be $\hat \ell(e)=z\cdot D'$.  
Notice that, if $e$ is a short edge, then $\ell(e)\leq \hat\ell(e)= D'$, and if $e$ is a  long edge, then $\ell(e)\leq \hat \ell(e)\leq 2 \ell(e)$. Additionally, for every cluster $C\in \cset'$, we add a supernode $u(C)$ to graph $\hat H$, and connect $u(C)$ with an edge to every {\bf regular} vertex $v\in V(H')$ that lies in $C$; the length of this edge is $4D'$. This ensures that the length of every edge in $\hat H$ is an integral multiple of $D'$. For each time point $\tau$, we denote by $\hat H\attime$ the graph $\hat H$ obtained immediately after the $\tau$th update in sequence $\Sigma$ to graph $H$ is processed.

We now proceed as follows. First, we show an algorithm to construct an initial graph $\hat H\attime[0]$, and an algorithm to produce an online sequence of valid update operations $\hat \Sigma$ for this graph, so that, at every time $\tau$, the resulting graph that we obtain is precisely $\hat H\attime$. We will also show that, for every pair of regular vertices $v,v'\in V(H)$, the distance between them in $\hat H$ is close to that in $H$. We will use the algorithm from \Cref{thm: main final dynamic NC algorithm inner} on graph $\hat H$, to maintain a neighborhood cover $\hat \cset$ of regular vertices in graph $\hat H$. Lastly, we show that we can use the resulting dynamic neighborhood cover $\hat C$ for graph $\hat H$ in order to maintain the desired neighborhood cover for graph $H$.

\paragraph{Maintaining Graph $\hat H$.}
Recall that, from the definition of the \recdynnc problem, initially, the  set $\cset'$ of clusters consists of a single cluster $H'$. As the time progresses, the clusters in $\cset'$ may only be updated via allowed change operations: $\delvertex$, $\addsupernode$, and $\csplit$.
We define the initial graph $\hat H\attime[0]$ as follows. We set $\hat H\attime[0]=H$, except that we set the edge lengths $\set{\hat \ell(e)}_{e\in E(\hat H\attime[0])}$ as described above. Additionally, we add a single supernode $u(H')$, that connects to every regular vertex of $H'$ with an edge of length $4D'$. We use the following claim in order to produce input update sequence $\hat \Sigma$ for $\hat H$.

\begin{claim}\label{claim: maintain hat H}
	There is a deterministic algorithm that, given, at each time $\tau\geq 1$, the $\tau$-th update $\sigma_{\tau}$ in the input update sequence $\Sigma$ for $\iset$, produces a sequence $\hat \Sigma_{\tau}$ of valid update operations for graph $\hat H$, such that, for all $\tau\geq 0$, the graph obtained from $\hat H\attime[0]$ by applying the update sequence $(\hat \Sigma_1\circ \cdots\circ \hat \Sigma_{\tau})$ to it is precisely $\hat H\attime[\tau]$. 
	The dynamic degree bound for the resulting dynamic graph $\hat H$ is at most $\mu + \hat W^{4(i-1)\eps^4}$.
	The total update time of the algorithm is bounded by $O\left (\hat c^{i-1}\cdot \hat W^{1+\hat c\eps}\cdot \mu^{\hat c/\eps} \right )$.
\end{claim}

\begin{proof}
	We run Algorithm $\alg_1$, that solves the \recdynnc problem on graph input structure  $\iset'=\left(H',\set{\ell(e)}_{e\in E(H')}, 4D' \right )$, that undergoes a sequence of valid update operations with dynamic degree bound $\mu$. Recall that the algorithm 
	maintains the neighborhood cover $\cset'$, achieves approximation factor $\alpha_{i-1}$, and has total update time at most $\left (\hat c^{i-1}\cdot \hat W^{1+\hat c\eps}\cdot \mu^{\hat c/\eps} \right )$. 
	We are also guaranteed that for every regular vertex $v$ of $H'$, the total number of clusters in $\cset'$ that ever contain $v$ is bounded by $\hat W^{4(i-1)\eps^4}$.
	
	Recall that initially, $\cset'=\set{H'}$. Consider now some time $\tau>0$, when an update $\sigma_\tau\in \Sigma$ for input structure $\iset$ arrives. We initialize $\hat \Sigma_{\tau}=\emptyset$, and we start by updating the data structure $\dset(H')$ with the update operation $\sigma_{\tau}$,  which may result in some changes to the clusters of $\cset'$. We consider the resulting changes to  the clusters one by one. 
	If the change is $\addsupernode$, in which a new supernode is added to some cluster $C\in \cset'$, then we ignore this update.
	If the change is $\delvertex$, where, for some cluster $C\in \cset'$, and a vertex $x\in V(C')$, vertex $x$ is deleted from $C$, and $x$ is a regular vertex, then we add to the sequence $\hat \Sigma_t$ an edge-deletion operation for the edge  $(x,u(C))$; if $x$ is a supernode then we ignore this change. 
	If the change is $\csplit$, where, for some cluster $C\in \cset'$, we create a new cluster $C'\subseteq C$, then we add a supernode-splitting operation to $\hat \Sigma_{\tau}$, defined as follows. The supernode-splitting is applied to  supernode $u(C)$, and the corresponding set $E'$ of edges contains all edges $(u(C),v)$, where $v$ is a regular vertex of $H$ lying in $C'$. The new supernode that is added to graph $\hat H$ is $u(C')$, where $C'$ is the new cluster. Note that the time that is needed in order to compute the set $E'$ of edges is $O(|V(C')|)$; since the algorithm for the \recdynnc problem on $\iset'$ needs to explicitly store the new cluster $C'$,  this time is subsumed by the time that Algorithm $\alg_1$ takes in order to execute the cluster-splitting update.

	Lastly, we consider the update operation $\sigma_{\tau}$ itself, and add additional updates to $\hat \Sigma_{\tau}$ as follows. If $\sigma_{\tau}$ is the deletion of an isolated vertex $x$ from graph $H$, then $x$ must also currently be an isolated vertex of $\hat H$ (as it was just deleted from all clusters of $\cset'$ containing it). We then add isolated vertex deletion operation for vertex $x$ to $\hat \Sigma_{\tau}$. If $\sigma_{\tau}$ is the deletion of an edge $e$, then add to $\hat \Sigma_{\tau}$ the deletion of $e$.
	Lastly, if $\sigma_{\tau}$ is a supernode-splitting operation, for a supernode $u$ of graph $H$, whose corresponding set of edges is $E' \subseteq \delta_H(u)$, then we perform the same supernode-splitting operation in graph $\hat H$, with edge set $E'$.
	This completes the description of the algorithm for producing the sequence $\hat \Sigma_{\tau}$.
	
	%\mynote{So each update operation $\delta_{\tau}$ will trigger a series of updates on $\hat{H}$. Do we need to specify the ordering that we do these updates by? actually no, because updates on $\hat{H}$ caused by changes in $\mathcal{C}'$ only concern newly added supernodes, not supernodes already present in $H$.} 

	It is immediate to verify that  for all $\tau>0$, the graph obtained by applying update sequence $(\hat \Sigma_1\circ \cdots\circ \hat \Sigma_{\tau})$ to graph $\hat H\attime[0]$ is precisely $\hat H\attime[\tau]$.
	Next, we bound the dynamic vertex degree for the resulting dynamic graph $\hat H$. Recall that, from the statement of \Cref{lem: inductive dynamic NC algorithm}, for every regular vertex $v$ of graph $H'$, the total number of clusters in $\cset'$ that ever contain $v$ is bounded by $\hat W^{4(i-1)\eps^4}$. Therefore, for every regular vertex $v$ of $\hat H$, there may be at most $\hat W^{4(i-1)\eps^4}$ clusters $C\in \cset'$, such that vertex $v$ is connected with an edge to supernode $u(C)$. Since the dynamic degree bound of every regular vertex in $H$ is at most $\mu$, we get that the dynamic degree bound for $\hat H$ is $\mu + \hat W^{4(i-1)\eps^4}$.
	
	The total update time of this algorithm is subsumed by the total update time of Algorithm $\alg_1$, and is bounded by  $\left (\hat c^{i-1}\cdot \hat W^{1+\hat c\eps}\cdot \mu^{\hat c/\eps} \right )$.
\end{proof}

\paragraph{Distance Preservation.}

The following two lemmas were proved in \cite{APSP-previous}. Since our construction of graph $\hat H\attime[\tau]$ from graph $H\attime$ for all $\tau\geq 0$ is identical to that of \cite{APSP-previous}, the proofs of both lemmas are identical to those in \cite{APSP-previous}, and are omitted here. The only difference is that the value of the parameter $\alpha_{i-1}$ that we use is different from that in \cite{APSP-previous}, but it plays no role in the proofs.
%We now show that distances between regular vertices in graph $\hat H$ are not much larger than the corresponding distances in graph $H$.

\begin{lemma}[Lemma C.2 in \cite{APSP-previous}]\label{lem: distance preservation}
	Throughout the algorithm, for every pair $v,v'\in V(H)$ of regular vertices, if $\dist_H(v,v')\leq D$, then  $\dist_{\hat H}(v,v')\leq 20\cdot \dist_H(v,v')+8D'$.
\end{lemma}

\begin{lemma}[Claim C.3 in \cite{APSP-previous}]\label{claim: path transforming}
	There is a deterministic algorithm, that we call \newline \algtransformpath, that, given a path $P$ in graph $\hat H$, connecting a pair $v,v'$ of regular vertices, computes a path $P'$ in graph $H$, connecting the same pair of vertices, such that $\ell_{H}(P')\leq \alpha_{i-1}\cdot \hat \ell_{\hat H}(v,v')$. The running time of the algorithm is $O(|E(P')|)$.
\end{lemma}

\iffalse
\begin{proof}
	We process every supernode $u(C)$ on path $P$, with $C\in \cset'$ one by one. Consider any such supernode, and let $v_C,v'_C$ be the regular vertices of $\hat H$ appearing immediately before and immediately after $u(C)$ on path $P$.  We perform query $\spquery(C,v_C,v'_C)$ to the data structure $\dset(H')$, that maintains a solution to \recdynnc problem on graph $H'$, and obtain a path $P_C$, of length at most $\alpha_{i-1}\cdot D'$, connecting $v$ to $v'$ in $H$, in time $O(|E(P_C)|)$. Notice that, since the lengths of the edges $(v_C,u(C)),(v'_C,u(C))$ are $4D'$ each, after this step is completed for every supernode $u(C)$ on path $P$ with $C\in \cset''$, we obtain a path $P'$ in graph $H$, connecting $v$ to $v'$, whose length is at most $\alpha_{i-1}\cdot \hat \ell_{\hat H}(v,v')$. %Recall that every edge $e\in E(H)$ was subdivided by two vertices, $u_e,v_e$, in order to obtain graph $H''$. In our second step we suppress all such vertices $u_e,v_e$ lying on path $P'$, to obtain a path $P''$ in graph $H$, connecting $v$ to $v'$. This step does not increase the length of the path, so the final lenght of the path that we obtain is at most $\alpha''\cdot \hat \ell_{\hat H}(v,v')$.	
\end{proof}
\fi

\paragraph{Remainder of the Algorithm.}
Consider now the dynamic graph $\hat H$. 
Recall that the length of every edge in graph $\hat H$ is an integral multiple of  $D'$. %, or it is equal to $D'=\floor{(3m)^{\eps (i-1)}}\geq \floor{m^{\eps(i-1)}}$. Denote $\rho=\floor{m^{\eps(i-1)}/10}$.

Let $\hat H'$ be a graph that is identical to $\hat H$, except that for every edge $e\in E(\hat H)$, we set its new length $\hat \ell'(e)=\hat \ell(e)/D'$ (recall that $\hat \ell(e)$ is the length of $e$ in graph $\hat H$). %It is immediate to verify that for every edge $e$:

% \[ \hat \ell(e)/(2\rho)\leq \hat \ell'(e)\leq \hat \ell(e)/\rho.\]

We set $\hat D=50D/D'$. Since $D\leq 6\hat W^{\eps i}$, while $D'=\floor{\hat W^{\eps (i-1)}/4}$, we get that $\hat D\leq \Theta(\hat W^{\eps})$. 
Notice that, for every edge $e$ of $\hat H'$,  $\hat{\ell}'(e) \le \hat{D}$. Therefore, we have now defined a valid input structure $\hat \iset'=\left(\hat H',\set{\hat \ell'(e)}_{e\in E(\hat H')}, \hat D \right )$.
Using the algorithm from \Cref{claim: maintain hat H}, we obtain an online sequence $\hat \Sigma=(\hat \Sigma_1\circ \hat \Sigma_2\circ\cdots)$ of valid update operations for graph $\hat H'$, with dynamic degree bound  $\mu' \leq \mu + \hat W^{4(i-1)\eps^4}$ where $\mu$ is the dynamic degree bound of graph $H$.
Let $N^0(\hat H')=N^0(H)$ be the number of regular vertices that belonged to graph $\hat H'$ initially. We set $\hat W'=\hat W^{1+4(i-1)\eps^4}$. 
Since $N^0(H)\cdot \mu\leq \hat W$, we get that $N^0(\hat H')\cdot \mu'\leq \hat W'$. Clearly $\eps\geq 1/(\log \hat W')^{100}$ continue to hold.

We apply the algorithm from \Cref{thm: main final dynamic NC algorithm inner} in order to maintain a solution to the \recdynnc problem on graph $\hat H'$, with distance bound $\hat D$, and parameter $\hat W'$ replacing $\hat W$, with parameter $\eps$ remaining unchanged. Recall that the approximation factor that the algorithm achieves is:

$$\hat \alpha=(\log\log \hat W')^{2^{O(1/\eps^2)}}\leq (\log\log \hat W)^{2^{O(1/\eps^2)}}.$$

The total update time of the algorithm is bounded by:

\[O\left ((N^0(H))^{1+O(\eps)}\cdot (\mu')^{O(1/\eps)}\cdot \hat D^3\right )\leq \hat W^{1+\hat c\eps}\cdot \mu^{\hat c/\eps},\]

since $\hat D\leq \Theta(\hat W^{\eps})$, and $\mu'\leq \mu\cdot \hat W^{4(i-1)\eps^4}\leq \mu\cdot \hat W^{4\eps^3}$.

 Moreover, the algorithm ensures that for every regular vertex $v\in V$, the total number of clusters in the weak neighborhood cover $\cset$ that the algorithm maintains, to which vertex $v$ ever belongs over the course of the algorithm, is bounded by:

\[(\hat W')^{4\eps^4}\leq \left(\hat W^{1+4(i-1)\eps^4}\right )^{4\eps^4}\leq \hat W^{4i\eps^4}.\]

We denote the corresponding data structure by $\dset(\hat H')$. The corresponding neighborhood cover is denoted by $\hat \cset$, and the algorithm that we have just described, for maintaining the data structure $\dset(\hat H')$, by $\alg_2$. For every regular vertex $v$ of $\hat H'$, we denote the cluster $\coveringcluster(x)$ that the algorithm maintains by $\coveringcluster'(x)$.

The neighborhood cover $\cset$ that we maintain for graph $H$ is defined as follows.
For every cluster $\hat C\in \hat \cset$, we define the corresponding cluster $C\in \cset$. Cluster $C$ is a subgraph of $H$ induced by the set $S(\hat C)=V(\hat C)\cap V(H)$ of vertices. 
Once set $\cset$ of clusters is initialized from set $\hat \cset$ at the beginning of the algorithm, all updates to the clusters of $\cset$ can be implemented via allowed update operations. Indeed, if $\hat C$ is a cluster in $\hat \cset$, and $C$ is the corresponding cluster in $\cset$, then cluster $C$ may only need to be updated if cluster $\hat \cset$ undergoes $\delvertex$ or $\addsupernode$ update, and the affected vertex lies in graph $H$. In such a case, we perform an identical update to cluster $C$. if cluser $\hat C$ undergoes a cluster-splitting operation, which results in the addition of a new cluster $\hat C'\subseteq \hat C$ to set $\hat \cset$, then we let $C'$ be the subgraph of $H$ induced by the set $V(\hat C')\cap V(H)$ of vertices; since $\hat C'\subseteq \hat C$, it is easy to verify that $C'\subseteq C$, so we can create a new cluster $C'$ via the cluster-splitting update applied to cluster $C$, and add it to $\cset$.

Since every regular vertex $v$ of $\hat H'$ may ever belong to at most $\hat W^{4i\eps^4}$ clusters of $\hat \cset$ over the course of the entire algorithm, it is immediate to verify that every regular vertex $v$ of $H$ may ever belong to at most $\hat W^{4i\eps^4}$ clusters of $ \cset$ over the course of the algorithm.

For every regular vertex $v$, and for all time points $\tau$, we set $\coveringcluster(v)$ as follows. Let $\hat C$ be the cluster $\coveringcluster'(v)$ that data structure $\dset(\hat H')$ maintains at time $\tau$, and let $C$ be the cluster of $\cset$ corresponding to $\hat C$. We then set $\coveringcluster(v)=C$.
We will use the following observation to show that, if $C=\coveringcluster(v)$, then $B_H(v,D)\subseteq V(C)$.

\begin{observation}\label{obs: covering transfer}
	Let $v$ be a regular vertex of $H$, $\hat C$ a cluster of $\hat \cset$, and $C$ the cluster of $\cset$ corresponding to $\hat C$. Assume that, at some time $\tau$, $B_{\hat H'}(v,\hat D)\subseteq V(\hat C)$. Then, at time $\tau$, $B_{H}(v,D)\subseteq V(C)$.
\end{observation}
\begin{proof}
	Consider any regular vertex $v\in V(H)$, a cluster $\hat C$ of $\hat \cset$, and the corresponding cluser $C$ of $\cset$.
	We fix some time $\tau$, and whenever we refer to graphs, data structures, or other dynamic objects in this proof, we refer to them at time $\tau$.

	Denote $X=B_H(v,D)$. Recall that, from \Cref{lem: distance preservation}, for every regular vertex $v'\in X$, $\dist_{\hat H}(v,v')\leq 20\cdot \dist_H(v,v')+8D'$, and so $\dist_{\hat H'}(v,v')\leq \frac{20\dist_H(v,v')+8D'}{D'}\leq \hat D$.
	Similarly, if $u\in X$ is a supernode, then there is a regular vertex $v'\in X$, that is a neighbor of $u$, with $\dist_H(v,v')\leq \dist_H(v,u)-\ell_H(v',u)$. 
	We then get that $\dist_{\hat H}(v,v')\leq 20\cdot \dist_H(v,v')+8D'$, and $\dist_{\hat H}(v,u)\leq 20\cdot \dist_H(v,v')+9D'+\ell_H(v',u)\leq 20D+9D'$. Therefore, $\dist_{\hat H'}(v,u)\leq (20D+9D')/D'\leq \hat D$.
	We conclude that
	$X=B_H(v,D)\subseteq B_{\hat H'}(v,\hat D)$. 

Since $B_{\hat H'}(v,\hat D)\subseteq V(\hat C)$, we get that $X\subseteq V(\hat C)$ also holds. Since $V(C)=V(\hat C)\cap V(H)$, and $X\subseteq V(H)$, we get that $X\subseteq V(C)$, as required.
\end{proof}

Consider now some regular vertex $v\in V(H)$, and assume that, at  time $\tau$, $\coveringcluster(v)=C$ held. Let $\hat C\in \hat\cset$ be the corresponding cluster of $\cset$, so $\coveringcluster'(v)=\hat C$. Then $B_{\hat H'}(v,\hat D)\subseteq V(\hat C)$ must hold, and, from \Cref{obs: covering transfer}, $B_H(v,D)\subseteq V(C)$ holds.

It is also easy to see that our assignment of covering clusters obeys the Consistent Covering property. Indeed, consider any regular vertex $v$ of $H$, and two time  points $\tau'<\tau$. Let $C=\coveringcluster(v)$ at time $\tau$, and let $C'=\anc\attime[\tau'](C)$. It is enough to show that, at time $\tau'$, $B_H(v,D)\subseteq V(C')$ held.

Let $\hat C,\hat C'$ be the clusters of $\hat \cset$ corresponding to $C$ and $C'$, respectively. Then, at time $\tau$, $\hat C=\coveringcluster'(v)$, and $\anc\attime[\tau'](\hat C)=\hat C'$. From the Consistent Covering property of data structure $\dset(\hat H')$, at time $\tau'$, $B_{\hat H'}(v,\hat D)\subseteq V(\hat C')$ held. From \Cref{obs: covering transfer}, at time $\tau'$,  $B_H(v,D)\subseteq V(C')$ held. This establishes the Consistent Covering property of our algorithm.

\paragraph{Responding to queries}
We now show an algorithm for responding to queries $\spquery(C,v,v')$, where $C$ is a cluster in $\cset$, and $v,v'\in V(C)$
are regular vertices lying in $C$. Denote $S'=V(C)$, and let $\hat C\in \hat \cset$ be the corresponding cluster (with $S'\subseteq V(\hat C)$). We run query $\spquery(\hat C,v,v')$ in data structure $\dset(\hat H')$, obtaining a path $P$ in graph $\hat H'$, connecting $v$ to $v'$, of length at most $\hat D \cdot \hat \alpha\leq \frac{50D}{D'}\cdot (\log\log \hat W)^{2^{O(1/\eps^2)}}$ (we have used the fact that $\hat D=50D/D'$ and $\hat \alpha\leq (\log\log \hat W)^{2^{O(1/\eps^2)}}$). Note that the length of path $P$ in graph $\hat H$ is at most $50D  \cdot (\log\log \hat W)^{2^{O(1/\eps^2)}}\leq D \cdot (\log\log \hat W)^{2^{\tilde c/\eps^2}}$. The time required to process query $\spquery(\hat C,v,v')$ in data structure $\dset(\hat H')$ is $O(|E(P)|)$. Lasly, we apply Algorithm  \algtransformpath from \Cref{claim: path transforming} to path $P$ in graph $\hat H$, to obtain a path $P'$ in graph $H$, connecting $v$ to $v'$, whose length is bounded by:

\[
\begin{split}
\alpha_{i-1}\cdot \hat \ell_{\hat H}(v,v')&\leq \alpha_{i-1}\cdot D\cdot (\log\log \hat W)^{2^{\tilde c/\eps^2}}\\
&\leq D\cdot (\log\log \hat W)^{\hat c (i-1)\cdot 2^{\hat c /\eps^2}}\cdot  (\log\log \hat W)^{2^{\tilde c/\eps^2}}\\
&\leq D\cdot (\log\log \hat W)^{\hat c i\cdot 2^{\hat c /\eps^2}}\\
&=D\cdot \alpha_i. 
\end{split}
\]

The running time of the algorithm is $O(|E(P')|)$.

%Since $\alpha'=(\log m)^{(i-1)c/\eps^3}$, and since we can choose $c$ to be a large enough constant, we get that the length of the path is bounded by $D\cdot  (\log (3m))^{ic/\eps^3}=\alpha D$, as required.

From the above discussion, and the fact that our algorithm supports $\spquery$ queries, it is immediate to verify that, throughout the algorithm, $\cset$ is a weak $(D,\alpha_i\cdot D)$-neighborhood cover of the regular vertices of $H$.

\paragraph{Total Update Time.}
We now bound the total update time of the algorithm. The update time is dominated by the update time of the algorithm from \Cref{claim: maintain hat H}, and the algorithm $\alg_2$. The former has update time $O\left (\hat c^{i-1}\cdot \hat W^{1+\hat c\eps}\cdot \mu^{\hat c/\eps} \right )$, while the latter has update time $O\left (\hat W^{1+\hat c\eps}\cdot \mu^{\hat c/\eps}\right )$. Since we can assume that $\hat c$ is a sufficiently large constant, the total update time of the algorithm is bounded by $\left(\hat c^{i}\cdot \hat W^{1+\hat c\eps}\cdot \mu^{\hat c/\eps} \right )$.

%---------------------------------------
%---------------------------------------
%---------------------------------------
%---------------------------------------

%---------------------------------------
%---------------------------------------
%---------------------------------------
%---------------------------------------

%---------------------------------------
%---------------------------------------
%---------------------------------------
%---------------------------------------

%---------------------------------------
%---------------------------------------
%---------------------------------------
%---------------------------------------

%---------------------------------------
%---------------------------------------
%---------------------------------------
%---------------------------------------

%---------------------------------------
%---------------------------------------
%---------------------------------------
%---------------------------------------

%---------------------------------------
%---------------------------------------
%---------------------------------------
%---------------------------------------

\section{Proof of \Cref{lemma: inductive}}
\label{subsec: proof of inductive lemma 1}

Throughout, we use a parameter $\eps'=\eps^4$. Since $1/(\log N)^{1/50}\leq \eps<1/400$, it is easy to verify that $1/(\log N^{\eps})^{1/24}\leq \eps'<1/400$, and so in particular, $1/(\log(N^0(H)))^{1/24}\leq \eps'<1/400$ holds.
Note that, if $\mu\geq (N^0(H))^{1/10}$, then \Cref{recdynnc simple}, with parameter $\eps'$ replacing $\eps$, provides an algorithm with the desired guarantees. We will assume from now on that $\mu<(N^0(H))^{1/10}$.

The proof is by induction on $i$.
The base case is $i=3$. 
We assume that we are given as input  a valid input structure $\iset=\left(H,\set{\ell(e)}_{e\in E},D \right )$ undergoing a sequence of valid update operations with dynamic degree bound $\mu$, such that, if $N'$ denotes the number of regular vertices in $H$ at the beginning of the algorithm, then $N'\leq  N^{3\eps}$.
Using the arguments from \Cref{subsubsec: bounding D}, at the cost of losing a factor $2$ in the approximation ratio, we can assume that $D\leq 3N'$. 
Let $V'$ denote the set of all regular vertices that lie in graph $H$ at te beginning of the algorithm. We start with $\cset=\set{H}$, and then, over the course of $|V'|$ iterations, we create clusters $C_v=H$ for all $v\in V'$, by splitting them off from $H$. In the remainder of the algorithm, we will maintain the collection $\cset=\set{C_v\mid v\in V'}$ of clusters. As $H$ undergoes valid update operations, we perform similar updates in the clusters of $\cset$, as described in Section \ref{subsubsec: updating clusters}.  Once a regular vertex $v$ is deleted from $V(H)$, we delete all edges and vertices from $C_v$. 
Throughout the algorithm, we also set $\coveringcluster(v)=C_v$ for all $v\in V'$. For each vertex $v\in V'$, we initialize the algorithm from \Cref{thm: ES-tree} for maintaining a modified \EST in graph $C_v$, with source vertex $v$ and distance bound $D$, as the graph undergoes valid update operations; since supernode-splitting is a special case of vertex-splitting, every update that graph $C_v$  undergoes is either edge-deletion, or isolated vertex-deletion, or vertex-splitting.
Whenever some vertex $x\in V(C_v)$ is added to the set $S^*$ of vertices (in which case $\dist_{C_v}(x,v)>D$ must hold), we delete vertex $x$ with its all incident edges from $C_v$. We denote the algorithm for maintaining the \EST in graph $C_v$ by $\aset(C_v)$. Recall that the total update time of the algorithm is bounded by  $O(m^*\cdot D\cdot \log m^*)$, where $m^*$ is the total number of edges that ever belonged to graph $C_v$. Since $m^*\leq N'\mu$, and $D\leq 3N'$, we get that the total update time of Algorithm $\aset(C_v)$ is bounded by $O((N'\mu)^3)$. Lastly, since $|V'|\leq N'$, the total update time of the whole algorithm is bounded by $O((N')^4\mu^3)\leq O(N'\cdot N^{12\eps}\mu^3)\leq N^0(H)\cdot N^{\tilde c\eps}\cdot \mu^4$, as required. It is easy to verify that lists $\clusterlist(x)$ for vertices $x\in V(H)$ and $\clusterlist(e)$ for edges $e\in E(H)$ can be maintained without increasing the asymptotic running time of the algorithm. Every regular vertex $v\in V'$ belongs to at most $N'\leq N^{3\eps}\leq \Delta_3$ clusters in $\cset$ over the course of the algorithm. Lastly, when a query $\spquery(C_z,x,y)$ arrives, where $x,y\in V(C_z)$, we perform queries $\shortestpath(x)$ and $\shortestpath(y)$ in data structure $\aset(C_z)$, to obtain a path $P_1$ of length at most $D$ connecting $x$ to $z$ in time $O(|E(P_1)|)$, and a path $P_2$ of length at most $D$ connecting $y$ to $z$ in time $O(|P_2|)$. By concatenating both paths, we obtain a path connecting $x$ to $y$ in cluster $C_z$, whose length is at most $2D\leq \alpha_3\cdot D$. The running time of the algorithm that responds to the query is $O(|E(P)|)$, and the algorithm achieves approximation factor $4\leq \alpha_3$.

We also need to consider separately the special case where $i=4$.
We assume that we are given as input  a valid input structure $\iset=\left(H,\set{\ell(e)}_{e\in E(H)},D \right )$ undergoing a sequence of valid update operations with dynamic degree bound $\mu$, such that, if $N'$ denotes the number of regular vertices in $H$ at the beginning of the algorithm, then $N'\leq N^{4\eps}$. We also assume that  $\mu\leq (N')^{1/10}$.
Using the arguments from \Cref{subsubsec: bounding D}, at the cost of losing a factor $2$ in the approximation, we can assume that $D\leq 3N'$. 
Denote $W=N^{3\eps}$ and $\hat W'=N^{4\eps}$. Clearly, 
$W<\hat W'\leq W^{1.5}$ holds. %We also use  a precision parameter $\eps'=\eps^4$. Since $1/(\log \hat W)^{100}\leq \eps<1/400$, we get that $\frac{1}{(\log W)^{1/24}}\leq \eps'\leq 1/400$.

Recall that the total update time of the algorithm for the case where $i=3$ was bounded by $O((N^0(H'))^4\cdot \mu^3)$, where $N^0(H')$ is the number of regular vertices in the input graph $H'$ at the beginning of the algorithm. By setting $\delta=3$, $c'=0$, and letting $c$ be a large enough constant, this running time can be bounded by $N^0(H')\cdot W^{\delta}\cdot \mu^c$ (since $N^0(H')\leq N^{3\eps}=W$ in this case). Therefore,  \Cref{assumption: alg for recdynnc} holds for the parameter $W$ that we just defined, with parameters $c'=0$, $\delta=3$, $\alpha(N'\mu)=\alpha_3$, and $\Delta(N')=\Delta_3$. By applying \Cref{recdynnc from assumption} with precision parameter $\eps'$, we obtain an algorithm  $\aset_4$, for instance $\iset$, that achieves approximation factor:

\[\begin{split}
&\max\set{2^{O(1/(\eps')^6)}\cdot \log\log  N',\frac{\alpha_{3}^2\cdot\log\log  N'}{(\eps')^{16}}}\\
&\quad\quad\quad\quad\quad\quad
\leq \max\set{2^{O(1/\eps^{24})}\cdot \log\log N',\frac{\left((\log\log (N^2))^{\tilde c\cdot 2^{6}/\eps^{24}}\right )^2\cdot\log\log N'}{\eps^{64}}} \\
&\quad\quad\quad\quad\quad\quad\leq \left((\log\log (N^2))^{\tilde c\cdot 2^{6}/\eps^{24}}\right )^4\\
&\quad\quad\quad\quad\quad\quad\leq (\log\log (N^2))^{\tilde c\cdot 2^{8}/\eps^{24}}=\alpha_4.
\end{split}\] 

The algorithm ensures that, for every regular vertex $v$ of $H$, the number of clusters $C\in \cset$, such that $v$ ever belongs to $C$ over the course of the algorithm is bounded by $( N')^{4\eps'}\leq N^{4\eps^4}=\Delta_4$.

We now bound the total update time of the algorithm. 
In order to bound the total update time of the algorithm, observe that $\frac{\hat W'}{W}=N^{\eps}$, and that $W^{\delta}=N^{9\eps}$. Therefore, from \Cref{recdynnc from assumption}, the total update time of the algorithm is bounded by:

\[
\begin{split}
	&O\left (N'\cdot N^{9\eps}\cdot N^{3\eps\cdot c\eps'}\cdot \Delta_3^c\cdot \mu^{c} \cdot D^{3}\cdot (\log N)^{3c}\right )+O\left(N'\cdot N^{O(\eps\cdot \eps')}\cdot D^3\cdot \mu^{4}\cdot \Delta_3\cdot N^{8\eps}\right )\\
	&\quad\quad\quad\quad\quad\quad\leq  N'\cdot N^{O(\eps)}\cdot D^3\cdot \mu^{c}\\
	&\quad\quad\quad\quad\quad\quad\leq N^0(H)\cdot N^{\tilde c\eps+4\tilde c\eps^2}\cdot \mu^{4\tilde c } \cdot D^{3},
\end{split}
\]

if $\tilde c$ is sufficiently large
(we have used the fact that $\Delta_3=N^{3\eps}$, and, since $\eps\geq 1/(\log N)^{4}$, $N^{\eps}>\log N$ holds.). %Note that we can assume that $N^0(H)\cdot \mu=N\cdot \mu>\hat W^{3\eps}$, since otherwise we can use the algorithm for $i=3$. Since $\hat N'=\hat W^{4\eps}$, we get that $\hat N'\leq \hat W^{\eps}\cdot N\cdot \mu$. Therefore, overall, the total update time of the algorithm is bounded by:

%\[\begin{split}
%&N^0(H)\cdot \hat W^{O(\eps)}\cdot D^3\cdot \mu^{O(1)}\cdot (\log \hat W)^{O(1)}\\
%&\quad\quad\quad\quad\quad\quad\leq 
%N^0(H)\cdot \hat W^{\tilde c\eps }\cdot D^3\cdot \mu^{\tilde c}.
%\end{split}\]
%
%assuming $\tilde c$ is a sufficiently large constant.

For the step of the induction, we consider an integer $4<i\leq q$. We assume that the claim holds for integer $i-1$, and prove it for $i$. 
We assume that we are given as input  a valid input structure $\iset=\left(H,\set{\ell(e)}_{e\in E(H)},D \right )$ undergoing a sequence of valid update operations with dynamic degree bound $\mu$, such that, if $N'$ denotes the number of regular vertices in $H$ at the beginning of the algorithm, then $N^{(i-1)\eps}<N'\leq N^{i\eps}$ (if $N'\leq N^{(i-1)\eps}$, then we can use the algorithm from the induction hypothesis). We also assume that $\mu\leq (N')^{1/10}$.
Using the arguments from \Cref{subsubsec: bounding D}, at the cost of losing a factor $2$ in the approximation, we can assume that $D\leq 3N'$. 
Denote $W=N^{(i-1)\eps}$ and $\hat W'=N'$. Since $i>4$, 
$W<\hat W'\leq W^{1.5}$ holds. We also use  a precision parameter $\eps'=\eps^4$, as before.

From the induction hypothesis, there is an algorithm that, given a valid input structure $\iset'$ undergoing a sequence of valid update operations with the dynamic degree bound $\mu'$, such that, if $N''$ is the number of regular vertices in the initial graph, then $N''\leq W$ holds, achieves approximation factor $\alpha_{i-1}$, and has total update time 
at most:

\[\begin{split}
N''\cdot N^{\tilde c\eps+\tilde c(i-1)\eps^2}\cdot (\mu')^{\tilde c (i-1)} \cdot D^{3}.
\end{split}
\]

The algorithm also ensures that every regular vertex belongs to at most $\Delta_{i-1}$ clusters over the course of the algorithm.

We choose a parameter $\delta$, so that $N^{\tilde c\eps+\tilde c(i-1)\eps^2}=W^{\delta}$. Since $W=N^{(i-1)\eps}$, it is easy to see that $\delta=\frac{\tilde c+\tilde c(i-1)\eps}{i-1}$.

%We set $\delta=\frac{\tilde c}{i-1}+\tilde c\eps$, so that:

%\[N^{\tilde c\eps+\tilde c(i-1)\eps^2}\leq N^{(i-1)\eps\delta}\leq (N'')^{\delta}.\]

%Notice that, since  $W=N^{(i-1)\eps}$,
%we get that $W^{\delta}=N^{\tilde c\eps+\tilde c(i-1)\eps^2}$.

 By setting $c=\tilde c\cdot (i-1)$, we get that the total update time of the algorithm is bounded by $N''\cdot W^{\delta}\cdot D^3\cdot (\mu')^c$. Therefore,  \Cref{assumption: alg for recdynnc} holds for the parameter $W$ that we just defined, $c'=3$, $\alpha(N'')=\alpha_{i-1}$, and $\Delta(N'')=\Delta_{i-1}$,
and the parameters $\delta$ and $c$ as defined above. By using \Cref{recdynnc from assumption} with precision parameter $\eps'$, we obtain an algorithm  $\aset_i$, for instance $\iset$, that achieves approximation factor:

\[\begin{split}
&\max\set{2^{O(1/(\eps')^6)}\cdot \log\log N,\frac{\alpha_{i-1}^2\cdot\log\log N}{(\eps')^{16}}}\\
&\quad\quad\quad\quad\quad\quad
\leq \max\set{2^{O(1/\eps^{24})}\cdot \log\log N,\frac{\left((\log\log  (N^2))^{\tilde c\cdot 2^{2(i-1)}/\eps^{24}}\right )^2\cdot\log\log N}{\eps^{64}}} \\
&\quad\quad\quad\quad\quad\quad\leq \left((\log\log (N^2))^{\tilde c\cdot 2^{2(i-1)}/\eps^{24}}\right )^4\\
&\quad\quad\quad\quad\quad\quad\leq (\log\log (N^2))^{\tilde c\cdot 2^{2i}/\eps^{24}}=\alpha_i.
\end{split}\] 

The algorithm ensures that, for every regular vertex $v$ of $H$, the number of clusters $C\in \cset$, such that $v$ ever belongs to $C$ over the course of the algorithm is bounded by $(N')^{4\eps'}\leq N^{4\eps^4}=\Delta_i$.

It now remains to bound the total update time of the algorithm. Recall that $W^{\delta}=N^{\tilde c\eps+\tilde c(i-1)\eps^2}$, $\frac{\hat W}{W}=N^{\eps}$,  $W=N^{(i-1)\eps}$, $\eps'=\eps^4$, $c'=3$, and $c=\tilde c\cdot (i-1)$. Recall also that $i-1\leq 1/\eps$. From \Cref{recdynnc from assumption}, the total update time of the algorithm is bounded by:

\[\begin{split}
&O\left (N'\cdot W^{\delta+c\eps'}\cdot \Delta_{i-1}^c\cdot \mu^{c} \cdot D^{3}\cdot (\log D)^c\right )+O\left (N'\cdot W^{O(\eps')}\cdot D^3\cdot \mu^{4}\cdot \left(\frac{\hat W}{W}\right )^8\cdot \Delta_{i-1}\right )\\
&\leq O\left (N'\cdot N^{\tilde c\eps+\tilde c(i-1)\eps^2}\cdot N^{c(i-1)\eps^5}\cdot N^{4c\eps^4}\cdot \mu^{c} \cdot D^{3}\cdot (\log N)^c\right )+O\left (N'\cdot N^{O((i-1)\eps^5)}\cdot D^3\cdot \mu^{4}\cdot N^{9\eps}\right )\\
%&O\left ((N'\mu)^{1+\tilde c(i-1)\eps^4}\cdot N^{\tilde c\eps+\tilde c(i-1)\eps^2}\cdot (\Delta_{i-1})^{\tilde c\cdot (i-1)}\cdot \mu^{\tilde c\cdot (i-1)+1} \cdot D^{3}\cdot  (\log N)^{3\tilde c\cdot (i-1)}\right )\\
%&\quad\quad\quad\quad\quad+O\left((N'\mu)^{1+O(\eps^4)}\cdot D^3\cdot \mu^{10}\cdot N^{7\eps}\right )\\
&\leq O\left (N'\cdot N^{\tilde c\eps+\tilde c(i-1)\eps^2}\cdot N^{5\tilde c \eps^3}\cdot \mu^{\tilde c(i-1)} \cdot D^{3}\cdot (\log N)^{\tilde c(i-1)}\right )\\
%&\quad\quad\quad\leq O\left ((N')^{1+\tilde c(i-1)\eps^4}\cdot N^{\tilde c\eps+\tilde c(i-1)\eps^2}\cdot N^{4\tilde c\cdot (i-1)\eps^4}\cdot \mu^{\tilde c\cdot i} \cdot D^{3}\cdot  (\log N)^{3\tilde c\cdot (i-1)}\right )\\
&\leq N'\cdot N^{\tilde c\eps+\tilde c i\eps^2} \mu^{\tilde c i} \cdot D^{3}.
\end{split}
\]

(We have used the facts that 
$\Delta_{i-1}=N^{4\eps^4}$, and, since $\eps \ge 1/(\log N)^{1/24}$, $\log N<N^{\eps^4}$ holds. In particular, $(\log N)^{\tilde c(i-1)} \leq N^{\tilde c(i-1)\eps^4})\leq N^{\tilde c\eps^3}$.)
%\end{proof}

%\input{appx-reduction}
%\input{rec-comp.tex}
%\input{proofs-basic-alg}

\section{Proof of \Cref{obs: terminal growth2}}
\label{appx: secondary clusters}

Throughout the proof, when we refer to \emph{copies} of a regular vertex $x\in V$, we only refer to copies of $x$ that lie in the clusters of $\cset_2\cup \set{H'}$; we ignore copies of $x$ that belong to clusters of $\cset_1$.
	Since, for all $1\leq j\leq r$, vertices may join set $S_{\geq j}$ over the course of the algorithm, but they may never leave it,
	it is enough to prove that, at the end of the algorithm, $|S_{\geq j}|\leq N^{1-j\hat \eps}$ holds for all $1\leq j\leq r$. The proof is by induction on $j$.
	
	The base is $j=1$. Consider any regular vertex $x\in V$. Note that a new copy of vertex $x$ may only be created over the course of Phase 1 of a Flag-Lowering Operations, during one of the iterations when some terminal $t$ was processed, via the
	algorithm from 
	\Cref{lem: ball growing}. Let $1<i_t\leq\frac{2r+6}{\eps'}$ be the integer that the algorithm from \Cref{lem: ball growing} returned. Recall that we have denoted $B_t=B_C(t,4(i_t-1)D)$ and $B'_t=B_C(t,4i_tD)$, where $C$ is the cluster to which the Flag Lowering operation is applied. The algorithm from   \Cref{lem: ball growing} ensured that $|S_0\cap B'_t|\leq |S_{0}\cap B_t|\cdot N^{\hat \eps}$. The vertices of $B_t$ are then deleted from graph $H'$, and the only vertices for which new copies are created are the vertices of $B'_t\setminus B_t$. We assign, to every regular vertex $y\in B_t$, a charge of $N^{\hat \eps}$. Notice that the total charge to all regular vertices of $B_t$ is at least as large as $|S_0\cap  (B'_t\setminus B_t)|$, so the charge is at least as large as the number of new copies of vertices in $S_0$ that were created. Since the vertices of $B_t$ are deleted from graph $H'$ during the current iteration, they will never be charged again. Overall, we get that the total charge to all vertices of $V$, over the course of the algorithm, is bounded by $N^{1+\hat \eps}$. Note that a vertex $x$ may only be added to set $S_{\geq 1}$ when $n_x\geq N^{2\hat \eps}+1$, so at least $N^{2\hat \eps}$ copies of vertex $x$ have been created. Since the total number of copies of all vertices in $S_0$ that are ever created is bounded by $N^{1+\hat \eps}$, we get that $|S_{\geq 1}|\leq \frac{N^{1+\hat \eps}}{N^{2\hat \eps}}\leq N^{1-\hat \eps}$ holds.

	Consider now some integer $j>1$, and assume that the claim holds for $j-1$. Consider some vertex $x\in V$, that was added to set $S_{\geq j}$ at some time $\tau$ during the algorithm's execution. Let $\tau'<\tau$ be the time when vertex $x$ was added to set $S_{\geq (j-1)}$. Then at time $\tau'$, $n_x=(j-1)\cdot N^{2\hat \eps}$ held, and at time $\tau$, $n_x=j\cdot N^{2\hat \eps}$ held. Therefore, between time $\tau'$ and $\tau$, vertex $x$ belonged to class $S_{j-1}$, and during that time, $N^{2\hat \eps}$ new copies of this vertex were created.
	
	As before, new copies of regular vertices may only be created  
	over the course of Phase 1 of a Flag-Lowering operations, during one of the iterations when some terminal $t$ was processed, via the
	algorithm from 
	\Cref{lem: ball growing}. Let $1<i_t\leq\frac{2r+6}{\eps'}$ be the integer that the algorithm from \Cref{lem: ball growing} returned. Recall that we have denoted $B_t=B_C(t,4(i_t-1)D)$ and $B'_t=B_C(t,4i_tD)$, where $C$ is the cluster to which the Flag Lowering operation is applied. The algorithm from   \Cref{lem: ball growing} ensured that $|S_{j-1}\cap B'_t|\leq |S_{j-1}\cap B_t|\cdot N^{\hat \eps}$. The vertices of $B_t$ are then deleted from graph $H'$, and the only vertices for which new copies are created are the vertices of $B'_t\setminus B_t$. We assign, to every regular vertex $y\in B_t\cap S_{j-1}$, a charge of $N^{\hat \eps}$. Notice that the total charge to all regular vertices of $B_t\cap S_{j-1}$ is at least as large as $|S_{j-1}\cap (B'_t\setminus B_t)|$, so the charge is at least as large as the number of new copies of vertices in $S_{j-1}$ that were created. Since the vertices of $B_t\cap S_{j-1}$ are deleted from graph $H'$ during the current iteration, they will never be charged again for the vertices of $S_{j-1}$. Therefore, a vertex that ever belonged to set $S_{j-1}$ may only be charged at most once for creating new copies of vertices of $S_{j-1}$, and the amount of the charge is $N^{\hat \eps}$. Since, from the induction hypothesis, at the end of the algorithm, $|S_{\geq (j-1)}|\leq N^{1-(j-1)\hat \eps}$ holds, the total number of copies of vertices of $S_{j-1}$ that were ever created during the algorithm is bounded by $|S_{\geq (j-1)}|\cdot N^{\hat \eps}\leq J^{1-j\hat \eps+2\hat \eps}$. As discussed already, in order for a vertex of $S_{j-1}$ to join set $S_{j}$, we need to create at least $N^{2\hat \eps}$ new copies of that vertex. We conclude that the total number of vertices that ever belonged to set $S_j$ over the course of the algorithm is bounded by $N^{1-j\hat \eps}$. If vertex $v$ belongs to $S_{\geq j}$ at the end of the algorithm, then it must have belonged ot $S_j$ at some time during the algorithm. Therefore, at the end of the algorithm, $|S_{\geq j}|\leq N^{1-j\hat \eps}$ holds.

\newpage
\bibliographystyle{alpha}
\bibliography{APSP-fully-dynamic}

%\bibliographystyle{alpha}
%\bibliographystyle{plain}

%\bibliography{APSP-p3}

\end{document}